\documentclass[12pt]{article}

\usepackage{scicite}
\usepackage{graphicx}
\usepackage{amssymb,amsmath}

\usepackage{times}
\usepackage{mathptmx}
\usepackage{txfonts}

\usepackage{rotating,booktabs} 

\newenvironment{sciabstract}{%
\begin{quote} \bf}
{\end{quote}}

\newcounter{lastnote}

\def\onesentencesummary{%
Novel, nonmajoritarian voting methods based on conditional commitments can achieve fairness without major welfare costs.
}
\title{%
Efficient democratic decisions via nondeterministic proportional consensus\\
\normalsize~\\{\bf One sentence summary:} \onesentencesummary
}

\author
{Jobst Heitzig$^{1\ast}$ and Forest W Simmons$^{2}$\\
\\
\normalsize{$^{1}$FutureLab on Game Theory and Networks of Interacting Agents,}\\ 
\normalsize{Potsdam Institute for Climate Impact Research,}\\
\normalsize{PO Box 60 12 03, D-14412 Potsdam, Germany}\\
\normalsize{$^{2}$Liberal Arts \& Mathematics Division, Portland Community College,}\\
\normalsize{Cascade Campus, 705 N. Killingsworth Street, TH 220, Portland, OR 97217, USA}\\
\normalsize{$^\ast$To whom correspondence should be addressed; E-mail: heitzig@pik-potsdam.de.}
}

\date{}

\usepackage{setspace}

\usepackage{hyperref}

\newcommand{\naturals}{\mathbb{N}}
\newcommand{\rationals}{\mathbb{Q}}
\newcommand{\reals}{\mathbb{R}}

\begin{document} 




\maketitle 


\begin{sciabstract}
Abstract:
Are there voting methods which 
(i) give everyone, including minorities, 
an equal share of effective power even if voters act strategically,
(ii) promote consensus rather than polarization and inequality,
and 
(iii) do not favour the status quo or rely too much on chance?

We show the answer is yes by describing two nondeterministic voting methods,
one based on automatic bargaining over lotteries,
the other on conditional commitments to approve compromise options.
Our theoretical analysis and agent-based simulation experiments suggest that with these,
majorities cannot consistently suppress minorities as with deterministic methods,
proponents of the status quo cannot block decisions as in consensus-based approaches,
the resulting aggregate welfare is comparable to existing methods,
and average randomness is lower than for other nondeterministic methods.
\end{sciabstract}

\noindent
Majority rule, considered a cornerstone of democracy, 
allows the oppression of minorities --- Tocqueville's `tyranny of the majority' \cite{lewis2013direct} ---
which may lead to separatism or violent conflict \cite{Collier2004,Cederman2010}. 
One way to address this is the fundamental fairness principle of proportionality (e.g., \cite{Cohen1997,Cederman2010}).
But if proportionality is only used to elect a representative body that then uses majority voting after all,
the problem remains \cite{Zakaria1997}.
Why? 
Proportional representation does not imply proportional power:
even a 49 percent faction may not be able to influence any decision.
For example, given the strong polarization in the US Senate \cite{mccarty2016polarized}, 
the Democratic Party currently appears to have zero effective power in it, 
according to the Banzhaf and Shapley--Shubik power indices \cite{dubey1979mathematical}.
But can power be distributed proportionally at all?

Smaller groups often try to overcome the problem by seeking consensus,
but that is difficult in strategic contexts \cite{davis1992some}. 
Status quo supporters may block consensus indefinitely,
or, if the fallback is majority voting, a majority can simply wait for that to be invoked.
Hence common consensus procedures are either not neutral about the options
or effectively majoritarian like most common voting methods if voters act strategically.

Judging from social choice theory, the formal science of group decision making,
such nonproportional effective power distribution seems unavoidable \cite{may1952set},
whether in a political or everyday context.
But this is only so if the employed decision methods are required to be essentially {\em deterministic,}
only using chance to resolve ties.
In fact, {\em nondeterministic} methods cannot only distribute power proportionally, which is obvious,
but at the same time support consensus and thus lead to efficient outcomes \cite{Heitzig2010a}.

It may seem outlandish to use a decision method that employs chance on a regular basis, producing uncertain outcomes.
But real-world problems typically involve quite some unavoidable stochastic risk and other forms of uncertainty anyway, 
e.g., due to lacking information, complexity, or dependence on others \cite{Carnap1947}. 
Also, routine use of nondeterministic procedures in contexts such as 
learning \cite{cross1973stochastic}, 
optimization \cite{kingma2014adam}, 
strategic interactions \cite{harsanyi1973games}, 
or the allocation of indivisible resources as in school choice \cite{troyan2012comparing}
shows that using chance can be quite beneficial, efficient, and acceptable.
Those examples also demonstrate that carefully using chance must not be confused with outright randomness. 

~

\noindent
In this article, we adopt the working hypothesis that at least in everyday situations
in which people often say ``let's have a vote'',
many groups might try a nondeterministic voting method if that has clear advantages.
For such situations, we study two such methods, one of which is novel, 
that achieve fairness by distributing power proportionally
and increase efficiency by supporting not just full but also partial consensus and compromise.

\begin{figure}\centering
\includegraphics[width=0.3\textwidth,trim=0 0 0 0,clip]{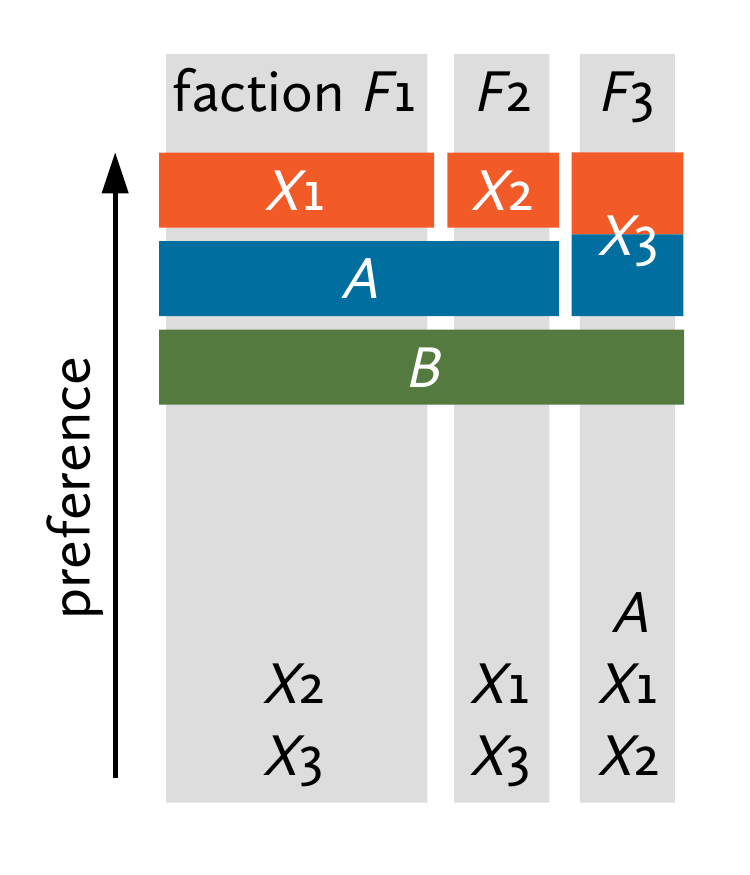}
\caption{\label{fig:example}\small\onehalfspacing
    Archetypical group decision problem 
    with potential for suppression of minorities, partial, or full consensus. 
    Each of three factions of different size (column width)
    has a unique favourite (topmost).
    There might also be a potential `partial consensus' option $A$
    and/or a potential `full consensus' $B$.
    With strategic voters, common deterministic methods pick $X_1$ for sure. 
    Our methods {\em Nash Lottery} and {\em MaxParC} pick $B$ for sure if present (green);
    they pick one of $X_{1,2,3}$ with probabilities (colored area) proportional to faction size if neither $A$ nor $B$ is present (orange);
    and they pick $A$ or $X_3$ with proportional probabilities if $A$ but not $B$ is present (blue).
}
\end{figure}

As an illustrative test case (Fig.\,\ref{fig:example}), 
consider a group of three factions $F_1$, $F_2$, $F_3$ with sizes $S_{1,2,3}$ (in percent), 
each of which has a favourite option, $X_{1,2,3}$, 
not liked by the other two factions, respectively.
Assume there is a fourth option $A$ not liked by $F_3$ 
but liked by $F_{1,2}$ almost as much as $X_{1,2}$.
We call $A$ a potential `partial consensus' for $F_{1,2}$ together.
While efficiency requires that $A$ gets a good chance of winning, 
proportionality requires that also $X_3$ gets some chance of winning.
Accordingly, our methods will assign winning probabilities of $S_1 + S_2$\,\% to $A$ and $S_3$\,\% to $X_3$,
even if voters vote strategically.
If we add a fifth option $B$ which $F_{1,2}$ like slightly less than $A$, and which $F_3$ likes almost as much as $X_3$,
both our methods will pick this potential `full consensus' $B$ for sure. 
In contrast, if $S_1 > S_2 + S_3$ and voters act strategically,  
virtually all existing voting methods will either pick $X_1$ with certainty, 
or will assign probabilities of $S_{1,2,3}$\,\% to $X_{1,2,3}$,
in both cases ignoring $A$ (and $B$) and thus producing much less overall welfare.
No deterministic voting method can let $F_{1,2}$ together make sure that $A$ gets a chance
without allowing them to render $F_3$'s votes completely irrelevant.
This can only be achieved by employing a judicious amount of chance.

But how exactly? 
How to design a nondeterministic voting method that is both efficient and proportional,
even when voters act strategically, 
and also fulfills other basic consistency requirements 
like those typically studied in social choice theory
--- such as anonymity, neutrality, monotonicity, and clone-proofness ---
that make it plausible and hard to manipulate?

Our first method, the {\em Nash Lottery (NL)}, 
is basically what is known as `Nash Max Product' or `Maximum Nash Welfare'
in the literature on fair division of resources.
As suggested in \cite{Aziz2019}, 
we translate it to our voting context by interpreting {\em winning probability} as
a ``resource'' to be divided fairly, and study the strategic implications of this.
NL can be interpreted as a form of {\em automatic bargaining} 
by means of the Nash bargaining solution. 
Similar to score-based methods such as Range Voting (RV) \cite{laslier2010handbook},
it asks each voter, $i$, to give a {\em rating,} $r_{ix}\ge 0$, for each option $x$.
It then assigns winning probabilities, $p_x$, 
that maximize a certain function, $f(r,p)$.
RV maximizes $f(r,p) = \sum_i\sum_x r_{ix} p_x$,
resulting in a very efficient majoritarian method
that is deterministic (usually $p_x=1$ for some $x$)
but neither distributes power proportionally nor supports consensus when voters are strategic.
NL instead maximizes
\begin{equation}\label{eqn:NL}
    f(r,p) = \sum_i\log\left(\sum_x r_{ix} p_x\right),
\end{equation}
resulting in a nondeterministic method that supports both full and partial consensus.
In the {\em Supplementary Text,} 
we prove that in situations similar to Fig.\,\ref{fig:example}, a full consensus will be the sure winner,
and that using the logarithm rather than any other function of $\sum_x r_{ix} p_x$
is the unique way to achieve a proportional power distribution.

NL is conceptually simple and has some other desirable properties shown in Fig.\,\ref{fig:properties}
such as being immune to certain manipulations, e.g., cloning options or adding bad options.
But it has three important drawbacks.
Its tallying procedure is intransparent, requiring numerical optimization.
It lacks certain intuitive `monotonicity' properties:
when a new option is added or a voter increases some existing option's rating, 
some other option's winning probability may increase rather than decrease.
And NL often employs much more randomness than necessary.

\begin{figure}
\includegraphics[width=\textwidth,trim=0 0 0 0,clip]{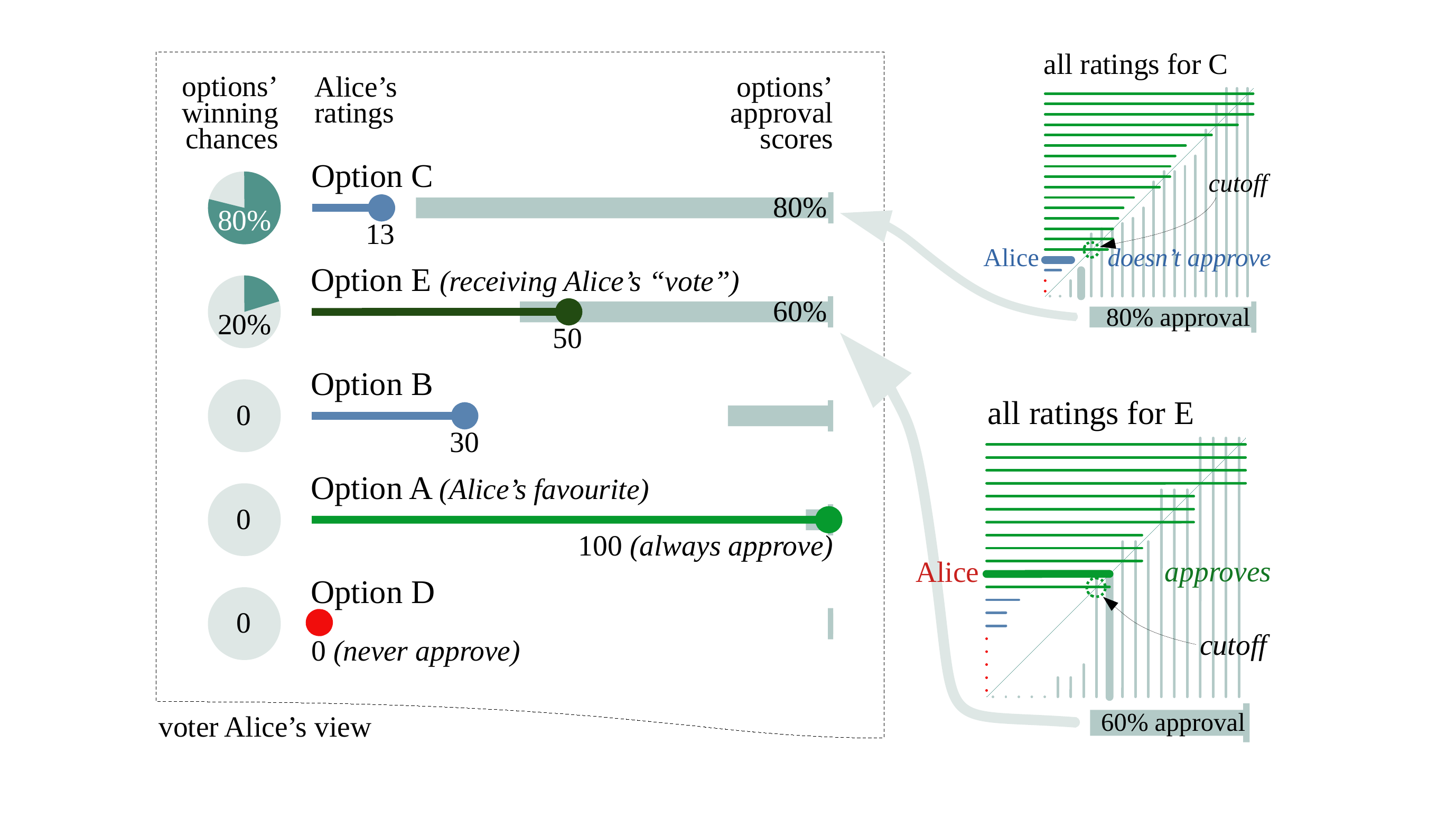}
\caption{\label{fig:method}\small\onehalfspacing
    Voting method MaxParC from the view of some voter Alice (left).
    Rating 13 for option C is interpreted as saying that Alice approves of C if less than 13 percent 
    of voters do {\em not} approve of C.
    Resulting approval scores can be found graphically in a way similar to ref.\,\cite{Granovetter1978} (right).
    }
\end{figure}
All three drawbacks are overcome by our second method, 
the novel {\em Maximal Partial Consensus (MaxParC),}
which is conceptually more complex, but strongly monotonic, much easier to tally, and produces less entropy.
Based on the idea of {\em conditional commitments,}
it lets each voter safely transfer ``their'' share of the winning probability to potential consensus options 
if enough other voters do so as well. 
That is done in a way inspired by Granovetter's famous `threshold model' \cite{Granovetter1978,Wiedermann2020}.
Again, voters assign numerical ratings, $0 \leq r_{ix} \leq 100$.
This is interpreted as a `willingness to approve', stating that 
``$i$ will approve of $x$ if strictly less than $r_{ix}$ percent of all voters do {\em not} approve of $x$.'' 
To solve this recursive definition of `approval' for any given option $x$, 
MaxParC sorts the ballots ascendingly w.r.t.\ their rating of option $x$, 
then finds the first ballot $i$ in this ordering
such that strictly less than $r_{ix}$ percent of the ballots precede it (i.e., have $r_{jx} < r_{ix}$).
This ballot $i$ and all later ballots $j$ (those with $r_{jx}\ge{}$ the cutoff $r_{ix}$) are said to approve of $x$.
After thus determining which ballots approve which options, 
MaxParC then proceeds like the `Conditional Utilitarian Rule' from \cite{Duddy2015,Aziz2019}:
one ballot is drawn at random, and from the options approved by this ballot,
that with the largest overall approval wins.
If one compromise option is rated positive by everyone, it will win for sure.
Fig.\,\ref{fig:method} illustrates the MaxParC procedure.

\begin{figure}
\includegraphics[width=\textwidth,trim=60 480 80 30,clip]{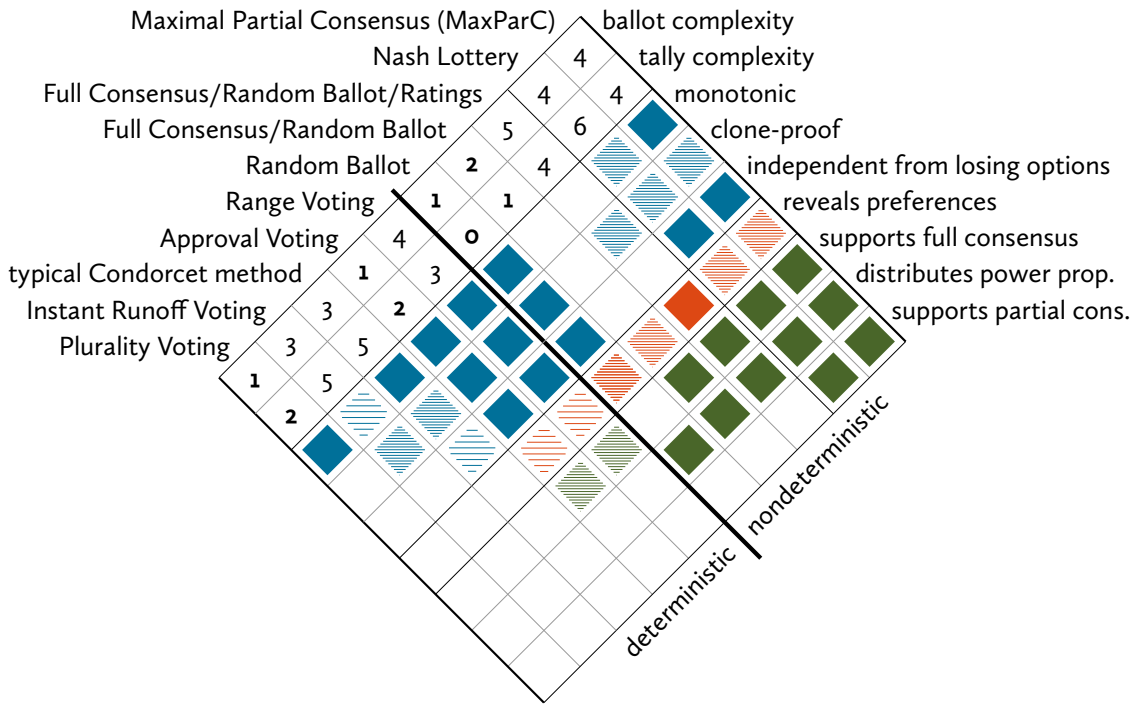}
\caption{\label{fig:properties}\small\onehalfspacing
    Properties of common group decision methods, Nash Lottery, and MaxParC.
    Solid and dashed diamonds indicate full and partial fulfillment,
    numbers are qualitative complexity assessments by the authors,
    color only distinguishes different groups of criteria
    (see {\em Supplementary Text} details and proofs).}
\end{figure}

Fig.\,\ref{fig:properties} summarizes our theoretical analysis of the formal properties 
of NL and MaxParC as compared to typical voting methods from the literature,
validating that the latter perform well in terms of these qualitative criteria.  

To assess the potential costs of achieving fairness and supporting consensus
in more quantitative terms of welfare, voter satisfaction, and entropy,
we finally performed a large agent-based simulation experiment.
In over 2.5 million hypothetical group decision problems, 
we compared NL and MaxParC's performance to that of 
five deterministic majoritarian and three nondeterministic proportional methods: 
Plurality Voting (PV), Approval Voting (AV), RV, Instant-Runoff Voting (IRV), Simpson--Kramer (a simple Condorcet method, SC);
`Random Ballot' (RB), and two methods from \cite{Heitzig2010a} (FC and RFC).
To generate the decision problems,
we used random combinations of 
the number and compromise potential of options
and the number, individual preference distributions, and risk-attitudes of voters.
For each combination of decision problem and voting method, 
we simulated several opinion polls, a main voting round, 
and an interactive phase where ballots could be modified continuously for strategic reasons.
In this, we assumed various mixtures of behavioural types of voters:
lazy voting, sincere voting, individual heuristics, trial and error, and coordinated strategic voting.
For each decision problem, we computed several metrics of social welfare, randomness, and voter satisfaction for all voting methods, 
and which voters would prefer which voting methods
(see Materials and Methods for details).

As can be expected, typically a majority of the simulated voters preferred the results of the majoritarian methods 
over those of the proportional ones. 
On average, voters preferred MaxParC over the other proportional methods;
among the majoritarian methods, there was no predominant preference. 
Individual voters' satisfaction, normalized to zero for their least-preferred option and one for their favourite, 
averaged around 67\,\% for PV, AV, RV, and IRV; 61\,\% for SC, NL, MaxParC; and still 57\,\% for RB, FC, RFC. 

MaxParC produced about 60\,\% of the entropy of RB, NL about 80\,\%. 
In MaxParC, the largest winning probability was about 65\,\% on average, in NL only about 53\,\%. 

\begin{figure}
\includegraphics[width=\textwidth]{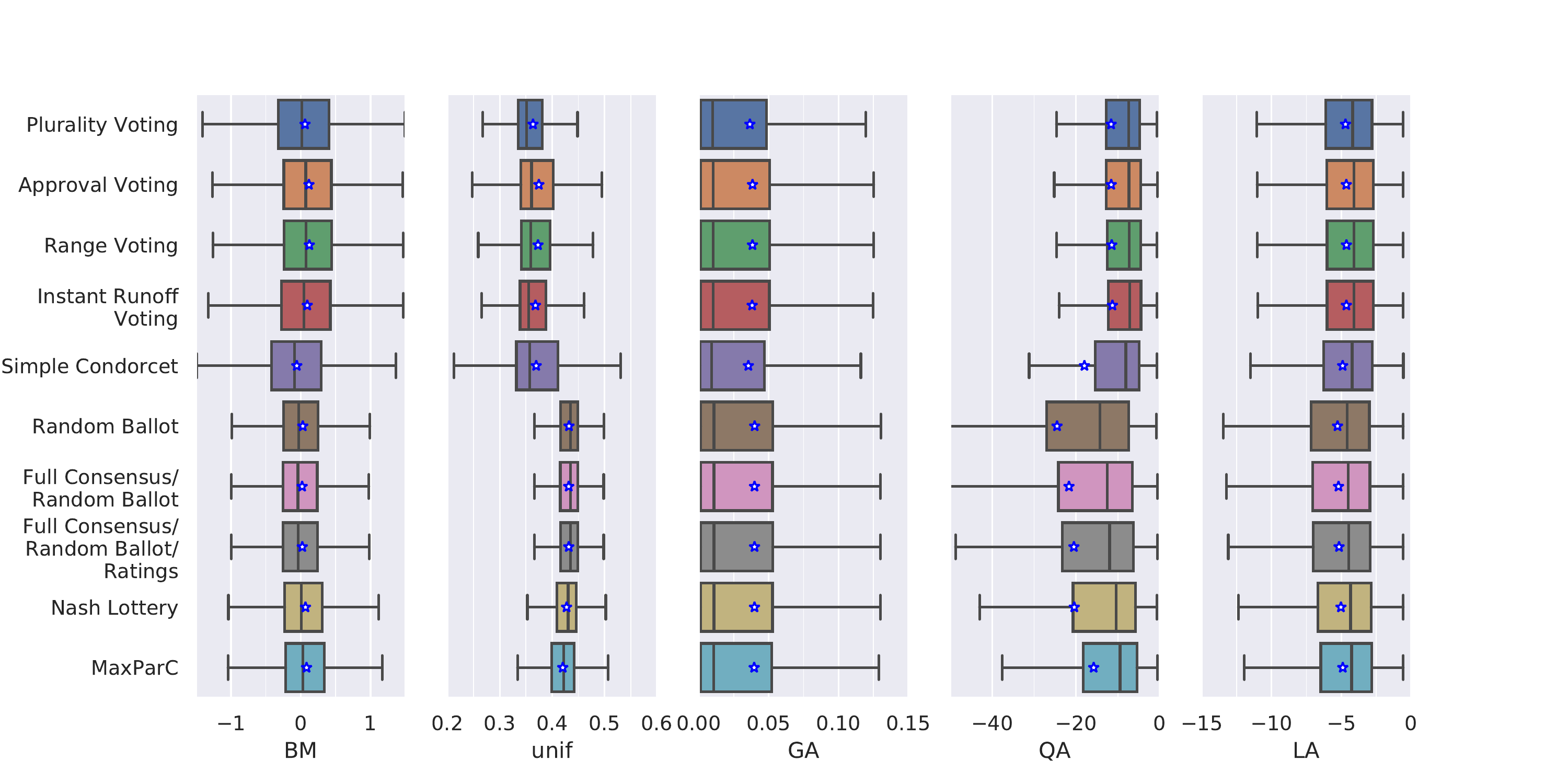}
\caption{\label{fig:welfare}\small\onehalfspacing
    Distribution of final Gini-Sen welfare across 2.5 mio.\ agent-based simulations by voting method (rows), 
    for five different models of how voter preferences might be distributed (columns). 
    See {\em Supplementary Text} for definitions and more results.}
\end{figure}

The deterministic methods produced somewhat higher welfare on average, 
but for some preference models and welfare metrics, 
the nondeterministic methods matched or outperformed them (Fig.\ \ref{fig:welfare}). 
In more than 75\,\% of cases, 
the utility difference between the average and the worst-off voter under RV
was at least seven times the difference in average voter utility between RV and MaxParC. 
This can be interpreted as saying that the welfare costs of fairness and consensus
are small compared to the inequality costs of majoritarianism.

On most results, preference distributions had a larger effect than behavioural type or the amount of interaction. 
Surprisingly, strategic voters gained no clear advantage over lazy voters, and also risk-attitudes played a minor role.

~

\noindent
In 2007, one of us (Heitzig) asked the election methods electronic mailing list \cite{Lanphier1996}
what method would elect the compromise rather than the majority option 
in a situation similar to Fig.\,\ref{fig:example}, even when voters acted strategically. 
Soon it became obvious that no deterministic method would do, 
but several lottery methods were quickly found that elected the compromise with certainty. 
So why do election methods experts show little enthusiasm for nondeterministic methods? 
Perhaps because their primary interest is in periodic high-stakes public elections every several years. 
The proportional fairness of lottery methods is due to their average proportionality over many individual decisions. 
Few would suggest deciding which of two newlyweds shall be the household’s dictator by flipping a coin. 
Using coin flips for their many everyday decisions would be better ---
because stakes are lower and advantages level out over time --- 
but would still not lead to a single consensus. 
Using the two methods presented here would likely make them 
agree on some compromise in most situations and toss a coin only rarely. 
Both the splitting-up into many decisions and the incentives for agreement lower the resulting overall entropy. 

While this seems to imply that such consensus-supporting proportional methods are best used for everyday decisions only,
they might even be applied to larger decisions such as allocating some budget or electing a parliament.
This is because the asset distributed by these methods need not be `winning probability' in a single-outcome decision as in this article.
For example, suppose NL or MaxParC instead of one of the common simple proportional methods 
was used for allocating parliamentary seats to party lists,
based on voters' ratings of all parties.
Would not this method take better advantage of opportunities for consensus without sacrificing proportionality? 
Since the seat distribution would on average have a lower entropy than usual, 
would it not avoid unnecessary balkanization or fragmentation of parliament without sacrificing representation of minorities? 
We hope that this discussion serves to stimulate the reader's imagination to some of the possibilities of application, 
as well as avenues for further exploration. 

\bibliography{library}

\bibliographystyle{Science}

{\bf Acknowledgements:} We'd like to thank 
Marius Amrhein,
Markus Brill,
Pascal F\"uhrlich,
Anne-Marie George,
Ulrike Kornek,
Fabrizio Kuruc,
Adrian Lison,
E.\ Keith Smith,
Lea Tamberg,
and the members of the election-methods list \cite{Lanphier1996}
for fruitful discussions and comments.
Funding: this work received no external funding.
Author contributions: 
J.H.\ and F.W.S.\ conceptualized the study, developed theory and methodology, performed the formal analysis, and wrote the manuscript.
J.H.\ developed the software and performed the simulations. 
The authors declare no competing interests.\\
{\bf Data and materials availability:}
All data is available in the manuscript or the supplementary materials.
An online voting tool based on the MaxParC method presented here is under open-source development at 
\href{https://github.com/mensch72/maxparc-ionic}{https://github.com/mensch72/maxparc-ionic}.\\
{\bf Supplementary Materials:}
Materials and Methods, Supplementary Text, References (22--44), Table S1--S3, Fig S1--10.

\clearpage

\renewcommand{\contentsname}{Supplementary Materials}
\tableofcontents

\section{Materials and Methods}

\subsection{Summary}

\subsubsection{Context}
A finite group of {\em voters} must collectively pick exactly one {\em winning option} 
out of a given finite number of {\em options} of any kind 
(e.g., certain kinds of objects, places, time-points, actions, strategies, people, etc.).
The menu of options is already {\em given} at the beginning of the situation we consider,
and all options are mutually exclusive and {\em feasible,} i.e.,
each one could be implemented without violating any relevant constraint 
(e.g., budgets, applicable laws, basic rights, time constraints etc.).
Although in the real world, the option menu might sometimes change during a group decision procedure,
we consider the composition of the option menu as a separate process here 
which has been completed before the situation we consider.
We assume the voters will apply some {\em formalized} method to pick one option
that may be described as some form of {\em protocol} or {\em game form}
which requires the voters to provide some kind of information in a step we term {\em voting} 
and then determines a winning option from this information in a step we term {\em tallying,}
using some kind of {\em algorithm} that may or may not involve some form of randomization.
We call the information a voter provides this voter's {\em ballot} 
and the method that turns sets of ballots into winning options a {\em voting method.}

In addition, we assume each voters possesses some form of {\em preferences} regarding the options 
and regarding possible probability distributions of options.
Again, even though in the real world, preferences might sometimes change during a group decision procedure,
in particular in certain forms of deliberation or consensus finding,
we consider here also the formation of preferences as a separate process 
which has been completed before the voting.
This assumption is in line with the established approach taken in social choice theory.
We do {\em not,} however, assume that there is a simple deterministic relationship between voters' preferences and their ballots,
but rather assume that voters may use different kinds of heuristics or strategies to decide how to fill in their ballots.

\subsubsection{Problem statement}
In this context, we aim at finding a voting method that fulfills certain consistency, fairness, and efficiency {\em criteria} 
as stated in the main text and detailed further in Section \ref{suppl} of this document.

\subsubsection{Transdisciplinary methodological approach}
To this end, we study the qualitative and quantitative properties of different voting methods,
some well-known from the social choice literature, one adapted from the theory of fair budget allocation, 
and one designed newly.
To study the {\em qualitative properties} of voting methods, we apply a mixture of methods from social choice theory and game theory.
For the {\em quantitative properties} we apply large-scale (Monte-Carlo) numerical simulations of an agent-based model
whose assumptions are partially based on the spatial theory of voting from political science 
and on insights from the study of risk attitudes and bounded rationality in behavioural economics.
We analyze simulation results by means of metrics adapted from welfare economics and information theory.
 
\paragraph*{Note on terminology.} 
Because we assume preferences have been determined before voting, 
we do not in this study distinguish linguistically between the terms {\em `consensus', `consent',} and {\em `compromise',}
but rather use a very pragmatic working definition of {\em potential consensus} here.
In this study, consensus does not mean that all voters consider the exact same option from the option menu their favourite option.
Informally, we rather say there is (full or partial) {\em potential consensus} whenever there is some option which (all or some group of) voters would prefer 
to having one randomly drawn voter make the choice.

\subsection{Ballot types and voting methods}

In this section we introduce a formal mathematical framework for comparing quite different voting methods
and then use it to define our versions of a number of common group decision methods,
at which point the motivations for the various abstract notions should become clear.

We assume 
an infinite universe of potential voters $\cal I$
and an infinite universe of potential options $\cal X$,
leading to 
a universe of potential finite electorates ${\cal E} = \{E\subset{\cal I}: 1\leq |E|<\infty\}$
and a universe of potential finite choice sets ${\cal C} = \{C\subset{\cal X}: 1\leq |C|<\infty\}$.
For each choice set $C\in\cal C$, 
let $L(C) = \{\ell\in[0,1]^C: \sum_{x\in C}\ell(x) = 1\}$
be the set of all {\em lotteries} on $C$,
and let $\ell_x\in L(C)$ with $\ell_x(x)=1$ be the {\em sure-thing} lottery that picks $x\in C$ for sure.
In most of what follows we deal with fixed sets $E$ and $C$ 
and denote their sizes by $N$ and $k$,
but for some proofs we have to consider all of ${\cal E}$ and ${\cal C}$.

\subsubsection{Ballot types}

Since we will deal with voting methods using quite different types of ballots,
some only letting the voter mark a single option, others many, still others requiring a strict ranking
or asking for quantitative ratings or the like,
we need a formal framework general enough to cover all relevant cases 
and make them comparable in those respects important for the assessment of the method's properties.
In particular, we need to make clear for each ballot type what it means when we say
that a ballot states a preference for one option over another
or a ballot results from another ballot by advancing an option to a certain degree.
The following abstract definition will prove useful in these tasks:

A {\em ballot type} is a tuple $(B,P,Q)$ with the following properties:
\begin{itemize}
    \item $B$ is a function such that for all choice sets $C\in\cal C$, 
        $B(C)$ is a nonempty set 
            representing the different ways $b$ in which a voter might fill in a ballot for the choice set $C$,
        and $B(C)\cap B(C')=\emptyset$ if $C\neq C'$.
    \item $P$ is a function such that for all $C\in\cal C$ and all filled-in ballots $b\in B(C)$, 
        $P_b$ is a strict partial ordering relation on $C$ (i.e., irreflexive, asymmetric, and transitive, but not necessarily complete)
            representing that part of ballot $b$ that will be interpreted as {\em stated preferences},
            with $x\,P_b\,y$ meaning that $x$ is put in a strictly ``better place'' (ranking, rating, threshold, etc.) on $b$ than $y$ is.
    \item $Q$ is a function such that for all $C\in\cal C$ and all options $x\in C$,
        $Q^C_x$ is a strict partial ordering relation on $B(C)$
            with $b\,Q^C_x\,b'$ meaning that the two filled-in ballots $b,b'\in B(C)$ 
                only differ in the fact that $b$ puts $x$ in a strictly ``better place'' than $b'$
                while each other option is treated the same on $b$ and $b'$,
                so that $b$ can be seen as resulting from $b'$ by ``advancing'' $x$ in some way and changing nothing else.
\end{itemize}

Note that $P_b = P_{b'}$ does not imply $b = b'$ in general 
since some ballot types (e.g., ratings ballots) also contain other information than just a binary preference relation.
 
If $C\in\cal C$, $b\in B(C)$, and $x\,P_b\,y$ for all $y\in C\setminus\{x\}$, 
we call $x$ the {\em stated favourite} on $b$ and write $F(b) = x$.
Since $P_b$ may be incomplete but is asymmetric, 
a ballot contains either no stated favourite (in which case we write $F(b)=\emptyset$) or exactly one.
If $F(b)=x$ and $\neg y\,P_b\,z$ for any $y,z\neq x$, we say that $b$ is a {\em bullet vote} for $x$.

\subsubsection{Ballot profiles and voting methods.}

A {\em ballot profile} of type $(B,P,Q)$ for electorate $E\in\cal E$ and choice set $C\in\cal C$
is a function $\beta:E\to B(C)$ specifying a filled-in ballot $\beta_i$ for each voter $i\in E$.

A (potentially probabilistic) {\em voting method} is a tuple $(B,P,Q,M)$ such that
$(B,P,Q)$ is a ballot type and $M$ is a function 
such that for all $E\in\cal E$ and $C\in\cal C$, and all ballot profiles $\beta$ of type $(B,P,Q)$ for $E$ and $C$,
it specifies a {\em winning lottery} $M(\beta) \in L(C)$.
Of course, $M(\beta)$ may be a sure-thing lottery with $M(\beta)_x=1$ for some $x\in C$.

\paragraph{Plurality Voting (PV).}

We formalize the ballot type of {\em Plurality Ballot} as follows:
\begin{itemize}
    \item $B(C) = C$, 
        i.e., each voter has to {\em vote for} exactly one option $x\in C$ by putting $b = x$.
    \item $x\,P_b\,y$ iff $b = x \neq y$, i.e., 
        voting for $x$ is interpreted as stating a strict preference for $x$ over all other options.
    \item $b\,Q^C_x\,b'$ iff $b = x \neq b'$,
        i.e., ``advancing'' $x$ means converting a vote for a different option into a vote for $x$. 
\end{itemize}

Using this ballot type, the voting method of {\em Plurality Voting}
now puts $M(\beta)_x = 1_A(x) / |A|$,
where $1_A$ is the indicator function of the set $A = \{ x: p(x)\ge p(y)$ for all $y\in C\}$ of {\em plurality winners,}
and $p(x) = |\{ i\in E : \beta_i = x \}|$ is $x$'s {\em plurality score}.

Note that for simplicity, in our version one cannot abstain under plurality voting, and hence every ballot is a bullet vote.
Generally $|A|=1$ except for ties, which means that this is a ``deterministic'' method in our terminology.

\paragraph{Approval Voting (AV).}

We formalize the ballot type of {\em Approval Ballot} as follows:
\begin{itemize}
    \item $B(C) = \{0,1\}^C$, 
        i.e., each voter can either {\em approve} (by putting $b(x) = 1$) or {\em disapprove} (by putting $b(x) = 0$) 
        of each option $x\in C$ individually.
    \item $x\,P_b\,y$ iff $b(x) > b(y)$, i.e., iff $b(x) = 1$ and $b(y) = 0$,
        meaning that approving $x$ is interpreted as stating a strict preference for $x$ over all non-approved options.
    \item $b\,Q^C_x\,b'$ iff $b(x) > b'(x)$ and $b(y) = b'(y)$ for all $y\in C\setminus\{x\}$,
        i.e., ``advancing'' $x$ means converting a non-approval of $x$ into an approval of $x$. 
\end{itemize}
Using this ballot type, the voting method of {\em Approval Voting}
now puts $M(\beta)_x = 1_A(x) / |A|$,
where $1_A$ is the indicator function of the set $A = \{ x: a(x)\ge a(y)$ for all $y\in C\}$ of {\em approval winners}
and $a(x) = \sum_{i\in E}\beta_i(x)$ is $x$'s {\em approval score}.

There are two equivalent ways to ``abstain'' under approval voting: 
putting $b(x)\equiv 0$ for all $x\in C$ 
or putting $b(x)\equiv 1$ for all $x\in C$.
Bullet voting for $x$ means putting $b(x)=1$ and $b(y)=0$ for all other $y$.

\paragraph{Range Voting (RV).}

In our version of range voting, the ballot type of {\em Range Ballot} has:
\begin{itemize}
    \item $B(C) = [0,100]^C$, 
        i.e., one can assign any real-valued {\em rating} $0\leq b(x)\leq 100$ 
        to each option $x\in C$ individually.\footnote{%
            Voter $i$'s choice of $b(x)$ is what was denoted $r_{ix}$ in the main text.}
    \item $x\,P_b\,y$ iff $b(x) > b(y)$ as before,
        meaning that a higher rating states a strict preference.
    \item $b\,Q^C_x\,b'$ iff $b(x) > b'(x)$ and $b(y) = b'(y)$ for all $y\in C\setminus\{x\}$,
        i.e., ``advancing'' $x$ means raising its rating. 
\end{itemize}
Using this ballot type, the voting method of {\em Range Voting} 
puts $M(\beta)_x = 1_R(x) / |R|$,
where $1_R$ is the indicator function of the set $R = \{ x: r(x)\ge r(y)$ for all $y\in C\}$ of {\em range winners}
and $r(x) = \sum_{i\in E}\beta_i(x)$ is $x$'s {\em range score}.

There are infinitely many equivalent ways to ``abstain'' under range voting: 
choose some $\alpha\in[0,100]$ and put $b(x)\equiv \alpha$ for all $x\in C$.
There are also infinitely many ways to ``bullet vote'' for $x$ under range voting, 
and they are not (!) equivalent: 
choose $0\leq\alpha<\gamma\leq 100$ and put $b(x)=\gamma$ and $b(y)=\alpha$ for all other $y\in C$.

~

\noindent
In addition to the above three ``scoring methods'', 
we consider the following two more complicated ``ranking methods''.

\paragraph{Instant-Runoff Voting (IRV).}

In our version of instant-runoff voting, 
we use the following ballot type of {\em Truncated Ranking Ballot}:
\begin{itemize}
    \item $B(C) = \left\{ b\in (\naturals\cup\{\infty\})^C: b[b^{-1}[\naturals]] = \{1,\ldots,|b^{-1}[\naturals]|\} \right\}$.
        In other words, one has to assign consecutive and distinct integer {\em ranks} $1,2,\ldots$ 
        to any empty or nonempty subset of the options, 
        leaving the other options unranked (here formally encoded by ``rank'' $\infty$).
    \item $x\,P_b\,y$ iff $b(x) < b(y)$ (!) since a smaller rank number indicates a ``better place''.  
    \item $b\,Q^C_x\,b'$ iff $P_b|_{C\setminus\{x\}} = P_{b'}|_{C\setminus\{x\}}$,\footnote{%
        If $R\subseteq X\times X$ is a binary relation on some set $X$ and $S\subseteq X$ is a subset of $X$,
        then $R|_S = R\cap(S\times S)$ is the restriction of $R$ to $S$.}
        $x\,P_b\,y$ whenever $x\,P_{b'}\,y$,
        $y\,P_{b'}\,x$ whenever $y\,P_b\,x$,
        but $P_b \neq P_{b'}$.
        In other words, advancing $x$ means changing the ranks so that the resulting ordering $P$ doesn't change
        except that some options ranked better than $x$ before are now either ranked lower than $x$ or not ranked at all, 
        and/or $x$ was not ranked before and is now ranked better than at least one option.
\end{itemize}
Using this ballot type, our simple version of the voting method of {\em Instant-Runoff Voting} 
(aka Single Transferable Vote, or Alternative Vote)
runs like this:
Initialize $D = C$ and repeat the following as long as $|D|>1$:
For each option $x\in D$, calculate the score 
$s(x) = \Big|\Big\{ i\in E : \beta_i(x) < \beta_i(y)$ for all $y\in D\setminus\{x\} \Big\}\Big|$.
From the set of worst-scored options,
$W = \{x\in D: s(x)\leq s(y)$ for all $y\in D\}$,
draw a random member and remove it from $D$.
The remaining member of $D$ wins.

Abstention means not ranking any option (i.e., putting $b_x = \infty$ for all $x$).
A bullet vote ranks exactly one option (i.e., puts $b_x=1$ for some $x$ and $b_y = \infty$ for all other $y$).

\paragraph{Simple Condorcet (SC).} 

In our version of the Simple Condorcet method, 
we use the following ballot type of {\em Weak Ranking Ballot:}
\begin{itemize}
    \item $B(C) = (\naturals\cup\{\infty\})^C$.
        In other words, one can assign arbitrary integer {\em ranks} $1,2,\ldots$ 
        to any empty or nonempty subset of the options, 
        leaving the other options unranked (again encoded by $\infty$).
    \item $x\,P_b\,y$ iff $b(x) < b(y)$ as in IRV.  
    \item $b\,Q^C_x\,b'$ iff $P_b|_{C\setminus\{x\}} = P_{b'}|_{C\setminus\{x\}}$,
        $x\,P_b\,y$ whenever $x\,P_{b'}\,y$,
        $y\,P_{b'}\,x$ whenever $y\,P_b\,x$,
        but $P_b \neq P_{b'}$.
        In other words, advancing $x$ means changing the ranks so that the resulting ordering $P$ doesn't change
        except that some options ranked better than $x$ before are now either ranked equal to or lower than $x$ or not ranked at all, 
        and/or some options ranked equal to $x$ before are now ranked lower than $x$ or not ranked at all, 
\end{itemize}
Note that we allow both that some options are ranked equal and some rank numbers are skipped
since the only information our version of the {\em Simple Condorcet} method (aka the Minimax Condorcet or Simpson--Kramer method with pairwise opposition as score) is the ordering $P_b$.
We put $M(\beta)_x = 1_O(x) / |O|$,
where $1_O$ is the indicator function of the set $O = \{ x: o(x)\le o(y)$ for all $y\in C\}$ of {\em weak condorcet winners},
$o(x) = \max_{y\in C} o(x,y)$ is $x$'s {\em worst opposition value},
and $o(x,y) = |\{ i\in E: y\,P_{\beta_i}\,x \}|$ for all $x,y\in C$.

There are infinitely many equivalent ways to ``abstain'' under the Simple Condorcet method: 
choose some $\alpha\in\naturals\cup\{\infty\}$ and put $b(x)\equiv \alpha$ for all $x\in C$.
The most straightforward is to not rank any option at all.
There are also infinitely many equivalent ways to ``bullet vote'' for $x$ under the Simple Condorcet method: 
choose $\infty\geq\alpha>\gamma\in\naturals$ and put $b(x)=\gamma$ and $b(y)=\alpha$ for all other $y\in C$.

~

\noindent
After these five ``deterministic'' methods, we now turn to five ``non-deterministic'' methods,
beginning with three from the literature.

\paragraph{Random Ballot (RB).}

The {\em Random Ballot} (aka Random Dictator) method uses Plurality Ballots 
but puts $M(\beta)_x = p(x) / N$.
The interpretation is that one ballot is drawn uniformly at random to decide.

\paragraph{Full Consensus / Random Ballot (FC).}

An {\em FC Ballot} is basically a combination of two Plurality Ballots:
\begin{itemize}
    \item $B(C) = C\times C$, 
        i.e., each voter specifies one ``proposed consensus'' $b^1\in C$ and one ``fall-back'' option $b^2\in C$.
    \item $x\,P_b\,y$ iff $b^2 = x \neq y$, i.e., 
        only the fall-back part of the ballot is interpreted as stating a strict preference for $b^2$ over all other options,
        while the consensus part is interpreted as inherently strategic.
    \item $b\,Q^C_x\,b'$ iff $b\neq b'$ and ($b^1 = x \neq b'^1$ or $b^1 = b'^1$) and ($b^2 = x \neq b'^2$ or $b^2 = b'^2$),
        i.e., ``advancing'' $x$ means advancing it in at least one of the two ballot parts. 
\end{itemize}
The method of {\em Full Consensus / Random Ballot (FC)} is now defined as in \cite{Heitzig2010a} (there called ``Voting method 1''):
$M(\beta)_x = 1$ if $\beta^1_i = x$ for all $i\in E$; 
otherwise $M(\beta)_x = p_2(x) / N$ for all $x\in C$ (``fall-back lottery''),
where $p_2(x) = |\{ i\in E : \beta^2_i = x \}|$ is $x$'s {\em fall-back score}.
The interpretation is that if all voters propose the same consensus, that option wins, otherwise the fall-back lottery applies.

A bullet vote is a bullet vote on both ballot parts.
There is no way to abstain.

\paragraph{Full Consensus / Random Ballot / Ratings (RFC).}

Similarly, an {\em RFC Ballot} is a combination of two Plurality Ballots and a Range Ballot:
\begin{itemize}
    \item $B(C) = C\times C\times [0,100]^C$, 
        i.e., each voter specifies one proposed consensus $b^1\in C$, one fall-back option $b^2\in C$,
        and a vector of ratings $b^3(x)\in[0,100]$ for all $x\in C$.
    \item $x\,P_b\,y$ iff $b^3(x) > b^3(y)$, i.e., 
        only the ratings part of the ballot is interpreted as stating preferences, 
        while the other two parts are interpreted as inherently strategic.
    \item $b\,Q^C_x\,b'$ iff $b\neq b'$ and ($b^1 = x \neq b'^1$ or $b^1 = b'^1$) and ($b^2 = x \neq b'^2$ or $b^2 = b'^2$)
        and $b^3(x) \ge b^3(y)$,
        i.e., ``advancing'' $x$ means advancing it in at least one of the three ballot parts. 
\end{itemize}
The method of {\em Full Consensus / Random Ballot / Ratings (RFC)} is also defined as in \cite{Heitzig2010a} (there called ``Voting method 2'').
For all $j\in E$, let $r_j = \sum_{y\in C} p_2(y) \beta^3_j(y) / N$ be the rating of the fall-back lottery by voter $j$. 
Then put $M(\beta)_x = \frac{|A_x|}{N} + (1 - \frac{|A|}{N}) \frac{p_2(x)}{N}$, where  
$A_x$ is the set of all $i\in E$ for which $\beta^1_i = x$ and $\beta^3_j(x) \geq r_j$ for all $j\in E$ 
(i.e., whose proposed consensus is $x$ and is preferred to the fall-back lottery by all voters according to their ratings),
and $A = \bigcup_{y\in C} A_y$.
The interpretation is that a voter $i$ is drawn uniformly at random,
and if $i$'s proposed consensus $\beta^1_i$ is unanimously preferred to the fall-back lottery $p_2(x)/N$, it wins, 
otherwise the fall-back lottery is applied.

A bullet vote is a bullet vote on all three ballot parts.
There is no way to abstain.

\paragraph{Nash Lottery (NL).}

The {\em Nash Lottery} method uses Range Ballots.
Given $\beta$, $i\in E$, and $\ell\in L(C)$,
let $r_i(\ell) = \sum_{x\in C} \ell(x) \beta_i(x)$
and $S(\ell) = -\sum_{i\in E}\log r_i(\ell)$.
If there is a unique $\ell\in L(C)$ with $S(\ell) > S(\ell')$ for all $\ell'\in L(C)\setminus\{\ell\}$,
we put $M(\beta) = \ell$.
In the rare cases where $\arg\max_{\ell\in L(C)} S(\ell)$ is not a singleton,
we use that $\ell$ which our numerical optimizer 
(the {\tt minimize} function from the {\tt scipy.optimize} Python package with method `SLSQP') returns.

For formal theoretical analyses, one can use the following tie-breaker instead.
Put $r^k_i = \sum_x \ell(x) \sqrt[k]{\beta_i(x)}$ for all $k=1,2,3,\dots$,
and $S^k(\ell) = \sum_i\log r^k_i(\ell) \in[-\infty,\infty)$.
Note that all $S^k$ are continuous, continuously differentiable,
and weakly concave functions of $\ell$.
Hence $S^1$ has a global maximum that is attained on a non-empty compact convex
set $T^1\subseteq L(C)$,
and for all $k\ge 2$, $S^k$ restricted to $T^{k-1}$ has a global maximum that
is attained on a non-empty compact convex set $T^k\subseteq T^{k-1}$.
Then also $T = \bigcap_{k=1}^\infty T^k$ is non-empty compact convex,
hence Lebesgue-measurable, and hence has a well-defined unique centre of mass
$\ell = \int_T\ell d\ell / \int_T d\ell$ with $\ell\in T$ because of the convexity.
We now put $M(\beta) = \ell$.
The rationale for using concave functions of ratings to break ties is
that in this way lotteries with lower entropy are preferred. 
The rationale for using the $k$-th square roots for this task is that in this
way the tie-breaking is complete except in the case of clones (see below).

A bullet vote is to rate one option at $>0$ and all others at $0$,
abstention means rating all options at the same value $>0$.

In the context of ``dichotomous preferences'', 
a similar method based on Approval Ballots was studied in \cite{Aziz2019} 
under the name ``Nash Max Product''.
The same idea is also common in the literature on fair division \cite{Moulin2004}.

\paragraph{Maximal Partial Consensus (MaxParC, MPC).}

For our simulations, we use this version of {\em MaxParC ballots:}
\begin{itemize}
    \item $B(C) = [0,100]^C$,
        i.e., one can assign any real-valued {\em willingness to approve} $0\leq b(x)\leq 100$
        to each option $x\in C$ individually.
    \item $x\,P_b\,y$ iff $b(x) = 100 > b(y)$ or $b(x) > 0 = b(y)$,
        i.e., we only interpret the special values 100 and 0 as ``stated preferences''
        and treat all intermediate values as inherently strategic since their interpretation
        relates to other voters' willingnesses.
    \item $b\,Q^C_x\,b'$ iff $b(x) > b'(x)$ and $b(y) = b'(y)$ for all $y\in C\setminus\{x\}$,
        i.e., ``advancing'' $x$ means raising the willingness to approve of it.
\end{itemize}
The {\em Maximal Partial Consensus (MaxParC, MPC)} method now works as follows.
A voter approves of an option if enough other voters do so as well;
a non-abstaining voter $i$ is drawn uniformly at random; 
then from the highest-scoring options approved by $i$, one is drawn uniformly at random.
Once it is decided who approves of what, 
the procedure corresponds to what is described in \cite{Duddy2015}, page 4 (last paragraph of section 3),
and analysed in \cite{Aziz2019} under the name `Conditional Utilitarian Rule'. 

To define this formally, we will introduce the following mathematical objects.
The set $A(x)$ will be the set of voters $i$ that turn out to approve of $x$ 
since their willingness to approve of $x$, $\beta_i(x)$, is properly larger than $100\times(1 - |A(x)|/N)$.
The quantity $a'(x)$ will be $x$'s approval score plus a fractional part used for tiebreaking.
The set $A_i$ will be the set of options approved by $i$, and $A'_i$ the set of highest-scoring options in $A_i$.
Finally, $A(\emptyset)$ will be those voters who don't approve of any option and thus ``effectively abstain''.

Formally, their definition is this:
For all $x\in C$, let $A(x)$ be the largest subset $A\subseteq E$ such that $|A|/N + \beta_i(x)/100 > 1$ for all $i\in A$.
Let $a'(x) = |A(x)| + \sum_{i\in E}\beta_i(x) / 100 N$. 
For all $i\in E$, put $A_i = \{ x\in C: i\in A(x) \}$ and $A'_i = \arg\max_{y\in A_i} a'(y)$.
Finally, put $A(\emptyset) = E - \bigcup_{x\in E} A(x)$.
Then $M(\beta)_x = \sum\{ 1/|A'_i| : i\in E$ with $x\in A'_i\} / (N - |A(\emptyset)|)$.

A bullet vote is to rate one option at $100$ and all others at $0$, 
while abstention means rating all options at $0$.

Note: since $A(x)$ can be found in $N\log N$ time, 
the total tallying complexity is $O(k N \log N)$.

\subsection{Agent-based simulations}

\subsubsection{Modeling individual preferences}

We simulate voter preferences 
by using one of several different models to generate a profile of individual utility functions over options, $u_i:C\to\reals$,
and then derive individual utility functions over lotteries depending on each voters risk attitude type.
Our {\em utility models} are the following.

\paragraph{Uniform model (Unif)}

In this simplest non-spatial utility model,
each value $u_i(x)$ is drawn uniformly at random from the unit interval $[0,1]$.
The resulting preference orderings form what is usually called the {\em impartial culture} model \cite{Laslier2010}.

\paragraph{Block model (BM)}

In this non-spatial utility model, there are $r \ge 1$ {\em voter blocks}
whose expected relative sizes $s_1,\dots,s_r$ 
are drawn independently from a log-normal distribution such that $\ln s_j \sim N(0,h)$,
where $h\ge 0$ is a {\em block size heterogeneity} parameter.
In particular, if $h = 0$, all blocks are of similar size,
while larger values of $h$ will lead to ever smaller minorities.

For each voter $i$ independently, 
the probability to belong to block $j$ is then $s_j / \sum_{j'=1}^r s_{j'}$.
Let $J(i)$ be $i$'s block.
Then the utility $u_i(x)$ that voter $i$ would get from option $x$ is now a sum of a block-dependent component and an individual component,
\begin{equation}
    u_i(x) = U_{J(i)}(x) + \iota \varepsilon_i(x),
\end{equation}
where all $U_{J(i)}(x)$ and $\varepsilon_i(x)$ are independent standard normal variables
and $\iota > 0$ is an {\em individuality} parameter.

\paragraph{Spatial preference models}

In the spatial theory of voting \cite{Carroll2013} (also called ``spatial cultures'' in \cite{Laslier2010}), 
voters $i$ and options $x$ are represented by {\em ideal points} (or bliss points) $\eta_i$ and {\em positions} $\xi_x$
in a low-dimensional {\em policy space} $\reals^d$, $d\geq 1$,
and the utility $u_i(x)$ that voter $i$ would get from option $x$ 
depends in a monotonically decreasing fashion on the distance between $\eta_i$ and $\xi_x$.
We distinguish the following spatial voting models:

\subparagraph{Linear homogeneous (LH) model.}
Utilities are decreasing linearly with distance, 
\begin{equation}
    u_i(x) = - ||\eta_i - \xi_x||_1
\end{equation}
where the $\xi_x$ are distributed independently and uniformly on the cube $[-1,1]^d$,
and the $\eta_i$ are distributed independently and uniformly on the cube $[-\omega,\omega]^d$,
where $\omega > 0$ is a {\em voter heterogeneity} parameter.

\subparagraph{Quadratic homogeneous (QH) model.}
Utilities decrease quadratically with distance, 
\begin{equation}
    u_i(x) = - ||\eta_i - \xi_x||_2^2,
\end{equation}
the $\xi_x$ are distributed independently according to the multivariate standard normal distribution,
and the $\eta_i$ according to the symmetric multivariate normal distribution with zero mean and standard deviation $\omega$.

\subparagraph{Gaussian homogeneous (GH) model.}
As in the quadratic homogeneous model, but with Gaussian utilities 
\begin{equation}
    u_i(x) = e^{- ||\eta_i - \xi_x||_2^2 / 2 \sigma^2}
\end{equation}
for some $\sigma > 0$.

~

\noindent
In addition to the above, rather classical spatial models, we also use the following three variants,
which introduce some idea borrowed from what \cite{Laslier2010} calls ``distributive cultures'':

\subparagraph{Gaussian allotment (GA) model.}
As in the Gaussian homogeneous model, but with each option having a different standard deviation $\sigma_x > 0$,
so that
\begin{equation}
    u_i(x) = e^{- ||\eta_i - \xi_x||_2^2 / 2 \sigma_x^2} / (\sqrt{2 \pi} \sigma_x)^d.
\end{equation}
The interpretation is that each option $x$ allots a unit amount of total utility to all potential ideal points of voters
using a symmetric multivariate normal distribution 
whose standard deviation $\sigma_x$ represents the {\em broadness} of option $x$'s ``platform''.
Because of the normalization factor $\sigma_x^{-d}$, if two options have very close positions but different broadness,
voters close to their position will prefer the ``narrower'' option
and voters farther away will prefer the ``broader'' option.

\subparagraph{Quadratic allotment (QA) model.}
As in the Gaussian allotment model, but with log-transformed utilities, resulting in a quadratic functional form:
\begin{equation}
    u_i(x) = - ||\eta_i - \xi_x||_2^2 / 2 \sigma_x^2 - d \ln (\sqrt{2 \pi} \sigma_x).
\end{equation}
The interpretation is that here option $x$ allots a unit amount of total wealth instead of a unit amount of total utility,
and voters' utility is logarithmic in wealth.

\subparagraph{Linear allotment (LA) model.}
As in the quadratic allotment model, but 
with $\xi_x$ and $\eta_i$ distributed on cubes as in the linear homogeneous model,
and with a linearly decreasing utility:
\begin{equation}
    u_i(x) = - ||\eta_i - \xi_x||_1 / \sigma_x - d \ln (2 \sigma_x).
\end{equation}  
The interpretation is that each option $x$ allots a unit amount of total wealth to all potential ideal points of voters
using a symmetric multivariate exponential distribution with density
$e^{- ||\eta_i - \xi_x||_1 / \sigma_x} / (2 \sigma)^d$,
and that utility is logarithmic in wealth.

\subparagraph{Distribution of options' broadnesses.}
In the three allotment models, we draw the options' broadnesses $\sigma_x$ independently 
from a log-normal distribution such that $\ln\sigma_x \sim N(\ln\sigma_0,\varrho)$,
where $\sigma_0 > 0$ is the {\em median broadness} and $\varrho\ge 0$ is a {\em broadness heterogeneity} parameter.
The three homogeneous models are then equivalent to the case $\varrho = 0$ of the allotment models.

\paragraph{Utility of uncertain prospects}

Regarding their preferences over uncertain prospects, 
represented as proper lotteries $\ell\in L(C)$ over options,
we assume each voter is of one of three {\em risk attitude types}
that determine how they derive utility functions over lotteries from their utility functions over options.

\subparagraph{Expected utility theory.}
We assume voters of {\em expected utility theory (EUT)} type 
evaluate the utility of a lottery of options $\ell\in L(C)$
by taking the expected value of the individual options' utilities,
$u_i(\ell) = \sum_{x\in C} \ell(x) u_i(x)$.
To see a major qualitative difference between the above linear, quadratic, and Gaussian models,
consider the one-dimensional case of three options 
placed symmetrically at $\xi_A = -1$, $\xi_B = 1$, $\xi_C = 0$ with $\sigma_x\equiv 1$, 
and compare the utility a non-central voter at $\eta_i \ge 2$ will assign to 
the ``compromise'' option $C$ and to the ``polar'' lottery $\ell = (A+B)/2$ (tossing a coin to decide between $A$ and $B$).
In the LH model, $u_i(C) = u_i(\ell)$, i.e., the voter is indifferent between the compromise option and the polar lottery.
In the QH model, $u_i(C) > u_i(\ell)$, i.e., the voter prefers the compromise.
In the GH model, $u_i(C) < u_i(\ell)$, i.e., the voter prefers the polar lottery.
More generally, this means that the quadratic/Gaussian models tend to have 
a larger/smaller number of potential consensus options than the linear models, respectively.

\subparagraph{Cumulative prospect theory.}
Motivated by recent empirical evidence \cite{Bruhin2010}, 
we assume that only about 20 percent of voters are of EUT type,
while the remaining 80 percent evaluate $\ell$ instead as follows.

40 percent are of {\em low-expectations cumulative prospect theory (LCP)} type.
Such a voter $i$ treats all options she prefers to her least-desired one as ``gains'', 
hence sorts the options by descending utility,
$u_i(x_1) \ge u_i(x_2) \ldots \ge u_i(x_k)$,
calculates the cumulative probabilities
$c_j = \sum_{j'=1}^j \ell(x_{j'})$,
so that $c_0 = 0$ and $c_k = 1$,
and then evaluates $\ell$ as
\begin{equation}
    u_i(\ell) = \sum_{j=1}^k w_j u_i(x_j),
\end{equation}
where $w_j = W(c_j) - W(c_{j-1})$,
$W(c) = \delta c^\gamma / (\delta c^\gamma + (1-c)^\gamma)$ is the {\em probability weighting function}
with $W(0) = 0$ and $W(1) = 1$,
and we choose $\delta = 0.926$ and $\gamma = 0.377$ following the pooled group estimates for gains from \cite{Bruhin2010}.

The remaining 40 percent are of {\em high-expectations cumulative prospect theory (HCP)} type.
Such a voter $i$ treats all options except her favourite one as ``losses'', 
hence sorts the options by {\em ascending} utility,
$u_i(x_1) \le u_i(x_2) \ldots \le u_i(x_k)$,
calculates the cumulative probabilities
$c_j = \sum_{j'=1}^j \ell(x_{j'})$,
so that $c_0 = 0$ and $c_k = 1$,
and then evaluates $\ell$ as
\begin{equation}
    u_i(\ell) = \sum_{j=1}^k w_j u_i(x_j),
\end{equation}
where $w, W$ are as above,
but now with $\delta = 0.991$ and $\gamma = 0.397$ following the pooled group estimates for losses from \cite{Bruhin2010}.

\subparagraph{Generic utilities.}
Note that due to the involvement of independent continuously distributed utility components, 
in all our utility models the resulting utility functions $u_i$ will be generic with probability one in the following sense.
Different rational-valued lotteries $\ell \neq \ell'\in L(C)\cap\rationals^C$
will have different utilities $u_i(\ell) \neq u_i(\ell')$,
so that each voter will have strict preferences over all pairs of rational-valued lotteries.
In particular, each voter will have a unique {\em favourite} option $f_i = \arg\max_{x\in C}u_i(x) \in C$.

\subsubsection{Voting behaviour}

\paragraph{Polling and final voting rounds.}
In our simulation model, the actual voting round is preceded by a number $R > 0$ of polling rounds.
In each polling round, voters are asked to name their favourite and all options they ``approve'' of,
and the total {\em favourite polling scores} $f^p(x)$ and {\em approval polling scores} $a^p(x)$ are published
so that voters can base their voting behaviour in later polling rounds and the final voting round on this information.
The actual voting round is then assumed to consist of an initial ballot
that can then be changed for some time in an interactive phase as a response to the current ballots' tallying results,
so that our setup allows for the simulation of the emergence of strategic equilibria.

We assume that each voter $i$ is of either of five {\em behavioural types} $\tau(i)$:
sincere, lazy, heuristic, trial-and-error, or factional.
Heuristic and factional voters together form the set of ``strategic'' voters, 
which we assume to make up about half the electorate.\footnote{%
    One of the few countries in which the election outcome can be used for a rough assessment 
    of the percentage of voters who take into account strategic reasoning is Germany
    because of the strategic incentive to ``split vote'' by voting for different parties with your first and second votes.
    Several studies show that in recent parliamentary elections 
    about half of the voters who had an incentive to ``split vote'' 
    because their favoured party had no chance of winning a direct mandate
    actually did split their vote (\cite{Behnke2004}, p.17).
    Using an elaborate methodology, 
    \cite{Sommer2015} classified voters in Germany's 2013 parliamentary elections into several behavioural types
    and found that about 15.9 per cent had an incentive to split and did split,
    while 11.9 per cent had an incentive to split and didn't split (\cite{Sommer2015} Table 1),
    i.e., 57 per cent of those with an incentive to vote strategically did so.  
    }
Trial-and-error and factional voters together form the set of ``interactive'' voters
who will potentially change their ballots during the interactive phase.
For lack of empirical data on interactive voting systems,
we assume that these also make up about half of the electorate, 
while the rest will stick to their ballots during the interactive phase.
Lazy voters, filling in their ballots in the simplest possible and non-strategic way,
are assumed to make up about one sixth of the electorate.\footnote{%
    One of the few systems in which the election outcome can be used for a rough assessment
    of the percentage of voters who supply less information on their ballot than would be advisable 
    is the Single Transferable Vote system used in Ireland's parliamentary elections
    because voters may keep their submited ranking so short that during the iterative tallying process
    their vote gets ``exhausted'' and thus essentially wasted.
    Election outcomes suggest that between 10 and 25 per cent of voters are ``lazy'' in this sense
    (the number of exhausted votes can be calculated easily from public data,
    e.g., on 
    \url{https://en.wikipedia.org/wiki/Dublin_Central_\%28D\%C3\%A1il_\%C3\%89ireann_constituency\%29},
    by comparing the elected candidates total vote turnout with the number of cast ballots) 
    }
We assume that about another sixth is sincere and fills in their ballots non-strategically to best represent their actual preferences.
Together with the above assumptions, 
this implies that also about one sixth is heuristic, using a simple form of strategic reasoning, 
about one sixth is of trial-and-error type, 
starting sincerely but testing simple modifications during the interactive phase,
and about one third is of factional type, 
starting heuristically and following best-response strategies proposed by their faction leaders during the interactive phase.
Although some studies also suggest that some voters do not sufficiently understand elections in order 
to vote sincerely or at least properly lazily, but will rather vote more or less erratically, 
we do not include an erratic type here since it would only increase the noise in the data. 

In addition to this ``middle''scenario, we test two further scenarios, 
one ``strategic'' 
and one ``lazy'',\footnote{%
    In many US elections, voters may use a so-called ``straight ticket'' 
    which might be interpreted as indicating a certain level of lazyness.
    As there are often up to or even more than half of all voters using straight ticket voting, 
    we assume that in the ``lazy'' scenario half of the voters are lazy.
    }
with behavioural types distributed according to the probabilities listed in Table~\ref{tbl:bt}.

We now specify our behavioural assumptions for the five types.

\paragraph{Sincere voters ($\tau(i) = S$).}

A sincere voter fills in her ballot $b$ in a certain way that represents her ``true'' preferences, 
in particular so that the stated preferences $P_b$ are compatible with her true preferences
in the sense that $x\,P_b\,y$ implies $u_i(x) > u_i(y)$.
Since many ballot types allow for more than one way of sincere voting,
we make the following explicit assumptions for the different voting methods.
Several of these make use of the {\em benchmark lottery} $\ell\in L(C)$ whose
winning probabilities are proportional to the latest favourite polling scores, $\ell(x) = f^p(x)/N$,
and on its expected utility $u_i(\ell) = \sum_{x\in C} \ell(x) u_i(x)$.
The MaxParC sincere strategy also makes use of an estimate $\alpha\in[0,1]$ of the proportion of lazy voters in the electorate.
\begin{itemize}
  \item In the first polling round, she names her true favourite and approves of all $x$ with above-average utility, 
        using equal weights for all options.
  \item In later polling rounds, she names her true favourite and approves of all $x$ with above-average utility, 
        using weights based on the latest favourite polling scores, as in Approval Voting.
  \item In the actual voting round, her initial and final ballot is determined like this:
    \begin{itemize}
      \item In Plurality and Random Ballot, she marks her true favourite: $b = f_i$.
      \item In Approval Voting, she marks all $x$ with at-least-average utility, 
            where the average is weighted with the latest favourite polling scores
            so that ``approval'' is with respect to the benchmark of the currently most relevant seeming options:
            $b(x) = 1$ iff $u_i(x) \geq u_i(\ell)$.
      \item In Range Voting and the Nash Lottery, she assigns ratings from 0 to 100 proportional to utility:
            $b(x) = 100 \frac{u_i(x) - \min_{y\in C} u_i(y)}{\max_{y\in C} u_i(y) - \min_{y\in C} u_i(y)}$.
      \item In IRV and Simple Condorcet, she ranks all $x$ with at-least-average utility as in Approval Voting, 
            in correct order of preference:
            $b(x) = |\{ y\in C : u_i(y)\ge u_i(x) \}|$ iff $u_i(x) \geq u_i(\ell)$, else $b(x) = \infty$.
      \item In FC, she marks her true favourite as ``favourite'' and marks that option as ``consensus''
            which has the highest approval polling score among those options she herself approves of:
            $b = (\arg\max_{x\in C}u_i(x), \arg\max_{x\in C,\,u_i(x) \geq u_i(\ell)} a^p(x))$.\footnote{%
                In the rare cases where several $a^p(x)$ are equal, we use $f^p(x)$ as a first-order tie-breaker
                and $u_i(x)$ as a second-order tie-breaker.}
      \item In RFC, she combines a sincere FC ballot with a sincere Range Voting ballot.
      \item In MaxParC, she assigns a willingness of $0$ to all non-approved options, 
            and willingness values from $100 \alpha$ to 100 scaling linearly with utility for all other options:
            $b(x) = 0$ if $u_i(x) < u_i(\ell)$, else 
            $b(x) = 100\left(\alpha + (1-\alpha) \frac{u_i(x) - u_i(\ell)}{\max_{y\in C} u_i(y) - u_i(\ell)}\right)$.\footnote{%
                This is the simplest sincere voting heuristic for MaxParC 
                that (i) guarantees that my share of winning probability goes to an option which I prefer to the benchmark lottery,
                that (ii) leads to full consensus if applied by all and if a potential full consensus exists,
                and that (iii) otherwise leads to partial consensus with a high probability 
                if the utility-by-distance curves are rather concave (as in the LH, QH, LA and QA models) 
                than convex (as in the GH and GA models).
                See Section~\ref{sec:mpsincere} for a more detailed discussion of this 
                and for alternative heuristic formulas for sincere voting under MaxParC.} 
    \end{itemize}
\end{itemize}

\paragraph{Lazy voters ($\tau(i) = L$).}

A lazy voter 
marks or ranks (only) her true favourite in Plurality, Random Ballot, Approval Voting, IRV, and Simple Condorcet,
marks the same option as consensus in FC and RFC, 
and gives it a rating/willingness of 100 and all others a rating/willingness of zero in Range Voting, the Nash Lottery, RFC, and MaxParC
(``bullet voting'').

\paragraph{Heuristic voters ($\tau(i) = H$).}

Heuristic voters try to adjust their voting behaviour to that of the other voters in order
to increase the chances of preferred options and avoid ``wasting their vote''.
But since their information is restricted to polling scores, 
they can only act boundedly rational. 
In addition, we assume they do not employ full optimization given that data
but rather use more or less simple or moderately complex ``heuristic'' strategies \cite{Gigerenzer2011a}
mainly based on the idea of ``exaggerating'' their stated preferences 
regarding the two options between which a nip-and-tuck race seems most likely \cite{VanderStraeten2010,Bower-Bir2013},
and possibly taking into account next-most likely nip-and-tuck races as well.
For the more complex voting methods, 
we do however allow for heuristics that require basic computational tasks 
such as forming sums, products and ratios and following simple decision trees.

\begin{itemize}
  \item In polling rounds, she acts as in Plurality and Approval Voting,
        while in the actual voting round, her initial and final ballot is determined as follows.
  \item In Plurality, she marks her preferred option among the two best-placed in the latest favourite polling scores:
        $b = y$ if $u_i(y) > u_i(z)$, else $b = z$, 
        where $y = \arg\max_{x\in C} f^p(x)$ and $z = \arg\max_{x\in C\setminus\{y\}} f^p(x)$.\footnote{%
            The rationale is that your vote is most relevant in a nip-and-tuck race,
            and the most likely nip-and-tuck race is between the favourite poll's leader and runner-up, 
            so that you should vote for your preferred one among those two.}
  \item In Approval Voting, she marks all $x$ she prefers to the option $y$ leading the latest approval polling scores,
        and marks $y$ iff she prefers $y$ to the runner-up $z$ in the latest approval polling scores:
        $b(x) = 1$ iff ($u_i(x) > u_i(y)$ or $x=y$ and $u_i(y) > u_i(z)$), 
        where $y = \arg\max_{x\in C} a^p(x)$ and $z = \arg\max_{x\in C\setminus\{y\}} a^p(x)$.\footnote{%
            This is called the ``leader rule'' in \cite{Laslier2009}, see also \cite{Myerson1993,Dellis2010}.
            Since the most likely nip-and-tuck race is between the approval poll's leader and runner-up, 
            you should approve only your preferred among those two. 
            Since the next likely nip-and-tuck race is between the leader and some other option, 
            you should also approve of all options you prefer to the leader.}
  \item In Range Voting, she applies the same strategy as in Approval Voting to find her ``approved'' options, 
        then assigns a rating of $100$ to approved options and a rating of $0$ to the other options.\footnote{Following the same rationale.}
  \item In IRV, she denotes the options in descending order of their latest approval polling scores
        as $x_0,x_1,\dots,x_{k-1}$ and then constructs her ranking as follows:
        In rank 1 she puts either $x_0$ or $x_1$ depending on which she prefers, and labels the other option as $y$.
        Then, for each rank $r=2,3,\ldots,k-1$, 
        she puts either $y$ or $x_r$ in rank $r$, depending on which she prefers, and labels the other option as the new $y$.\footnote{%
            The rationale here is that the most likely nip-and-tuck race is between $x_0$ and $x_1$,
            in which case her vote must go to the better of those from beginning on.
            Her 2nd ranked option only becomes relevant when the 1st ranked gets eliminated during the tally,
            in which case the most likely race is between the other ($y$) and $x_2$, so she should rank the better of those 
            two 2nd. Her $r$-th ranked option only becomes relevant when all higher ranked get eliminated,
            in which case the most likely race is between the one option among $x_0,\ldots,x_{r-1}$ not yet ranked,
            which is the current $y$, and $x_r$, so she should rank the better of those two next.
            See also \cite{Moulin1979}, Fig.~3, and \cite{Bag2009}.
        }  
  \item In Simple Condorcet, she finds $y,z$ as in Approval Voting, 
        assigns a tied rank of one to her preferred option among $y,z$ and all options she prefers to both,
        assigns sincere ranks to those other options she prefers to at least one of $y,z$,
        and doesn't rank the less preferred option among $y,z$ and all she considers even less desirable. 
  \item In Random Ballot, FC and RFC, she acts like a sincere voter.
  \item In the Nash Lottery, she first uses the latest favourite polling scores to compute the utility $\upsilon$
        of the benchmark lottery with probabilities $f^p(x) / N$.
        She then computes her rating for any $x$ based on $x$'s apparent chances as estimated by $f^p(x) / N$ 
        and on the difference between $u_i(x)$ and $\upsilon$ as follows.
        If she is of EU type, she has $\upsilon = \sum_{x\in C} u_1(x) f^p(x) / N$ and uses 
        \begin{equation}
            b(x) = 1 + \frac{f^p(x) ( u_i(x) - \upsilon )}{\max_{y\in C} f^p(y) ( \upsilon - u_i(y) )},
        \end{equation}
        where the denominator is chosen so that the smallest resulting rating is exactly zero.\footnote{%
            The rationale is that if some options appear to have higher chances than others,
            exaggerating one's preferences regarding these options will increase one's influence 
            (see Section \ref{heurnashlott} for a formal derivation of this heuristic strategy).
            In the special case where all $f^p(x)$ are equal, the strategy reduces to voting sincerely.}
        If she is LCP or HCP type, she similarly uses
        \begin{equation}
            b(x_j) = 1 + \frac{w_j ( u_i(x_j) - \upsilon )}{\max_{j'=1}^k w_{j'} ( \upsilon - u_i(x_{j'}) )},
        \end{equation}
        with $x_j,w_j$ as described in the LCP and HCP models above.
  \item In MaxParC, she applies the same strategy as in Approval Voting to find her ``approved'' options, 
        then assigns to an approved option $x$ a willingness that is at least as large as her sincere MaxParC willingness for $x$ 
        and large enough to make sure she is counted as approving of $x$ should $x$'s approval score be as predicted by $a^p(x)$.
        More precisely, she puts 
        \begin{equation}
            b(x) = \max\{b^s(x), 101 - 100 a^p(x)/N\},
        \end{equation}
        where $b^s(x) = \max\{0, 100 \frac{u_i(x) - u_i(\ell)}{\max_{y\in C} u_i(y) - u_i(\ell)}\}$.
        To a non-approved option $x$, she assigns a willingness at most her sincere willingness 
        and small enough so that she is counted as not approving of $x$:
        \begin{equation}
            b(x) = \min\{b^s(x), 99 - 100 a^p(x)/N\}.        
        \end{equation}
\end{itemize}

\paragraph{Interactive voting.}

We assume that in the final voting round 
voters have to submit a ballot but can then still change their ballots continuously over some time interval 
during which they can observe the resulting tally statistics in real-time,
thus introducing the possibility to interactively test voting strategies and react on others' strategies.
We assume there are two additional types of voters who will change their votes during this interval:
``trial-and-error'' voters and ``factional'' voters,
while all other voters submit an initial ballot as described above and don't change it afterwards.
This interactive phase is simulated long enough so that in typical situations a strategic equilibrium can emerge. 
Before the interactive phase starts, these voters behave like heuristic voters.
In the interactive phase, they vote as follows:

\paragraph{Trial-and-error voters ($\tau(i) = T$).}

The interactive phase consists of a large number of consecutive time points, 
at each of which some percentage of the trial-and-error voters will update their ballots.
When a trial-and-error voter $i$ updates her ballot, she picks a random modification
out of a set of elementary modifications that depend on the ballot type (see below),
submits the modified ballot, observes the resulting change in utility $u_i$ due to all simulateneous modifications,
and either sticks to or undoes the modification.
She undoes the modification if either $u_i$ has decreased 
or if $u_i$ has stayed constant and the modification was towards a strictly less ``sincere'' ballot (see below).

We assume these elementary modifications:
\begin{itemize}
  \item On a Plurality, FC, or RFC Ballot, either the favourite or the consensus option may be replaced by any other option.
  \item On an Approval Ballot, one can add or remove approval for a single option.
  \item On a Range, MaxParC, or RFC Ballot, one can replace the rating or willingness value for a single option by any value in $[0,100]$.
  \item A Truncated or Weak Ranking Ballot $b$ may be replaced by another such ballot $b'$ if for some option $x$,
        $P_b|_{C\setminus\{x\}} = P_{b'}|_{C\setminus\{x\}}$, i.e.,
        a single option $x$ may be moved to an arbitrary new position in the ranking,
        making place for it by shifting the other ranks if necessary.
\end{itemize}
In Random Ballot, trial-and-error voters will always vote sincerely since that is a dominant strategy.

A modified ballot $b'$ is strictly less sincere than $b$ if: 
\begin{itemize}
  \item In Plurality and FC: $u_i(b') < u_i(b)$ (resp. $u_i(b'_1) < u_i(b_1)$ for FC).
  \item In Approval Voting, IRV, Simple Condorcet, and MaxParC: 
        $e(b') > e(b)$ with $e(b) = |\{ (x,y)\in C^2: u_i(x) > u_i(y)$ but $y\,P_b\,x \}|$ 
        (number of wrongly stated binary preferences).
  \item In Range Voting and the Nash Lottery: $||b' - b^s|| > ||b - b^s||$, where $b^s$ is the sincere ballot described above.
  \item In RFC: $u_i(b'_1) \le u_i(b_1)$ and $||b'_3 - b^s|| \ge ||b_3 - b^s||$, but not both equal. 
\end{itemize}

Trial-and-error voters behave as sincere voters during polling and 
also start the interactive phase with a sincere ballot.

\paragraph{Factional voters ($\tau(i) = F$).}

Since strategic voting can be much more effective when coordinating with other voters having similar preferences,
we assume that voters of this type change their ballots as follows during the interactive phase,
starting it with a heuristic ballot as described above,
and after voting as heuristic voters in the polling rounds, too.

For each $x\in C$, we consider the ``faction'' $F_x$ 
of all voters $i$ with $\tau_i = F$ favouring $x$, 
$F_x = \{i \in E: \tau_i = F, f_i = x \}$.
Each faction $F_x$ is assumed to possess enough information and computing capabilities to calculate a 
{\em best unanimous response} to all other voters' current ballots,
which is a voting behaviour where all $i\in F_x$ submit the same filled-in ballot
and no other unanimous voting behaviour of all $i\in F_x$ 
would generate a strictly higher total utility $U = \sum_{i\in F_x} u_i$
given that all other voters $j\in E\setminus F_x$ submit the same ballots as before.
The assumption that factions cannot coordinate their members to vote differently 
even if that might be better than all voting the same way
can be interpreted as a form of bounded rationality. 

During the interactive phase, each faction, whether small or large, 
has the same constant probability rate for updating their ballots,
leading to a Poisson process of updates by randomly picked factions.
When a faction $F_x$ updates their ballots, 
they replace their current ballots by a best unanimous response to all other voters' current ballots as follows:

\begin{itemize}
  \item In Plurality, they find the plurality scores $p(y)$ resulting from all other voters' ballots,
        find the set $A$ of options less than $|F_x|$ many votes behind the leader,
        $A = \{ y\in C: p(y) + |F_x| > \max_{z\in C}p(z) \}$,
        and vote for that $y\in A$ which maximizes $U$:
        $b = \arg\max_{y\in A} U(y)$ with $U(y) = \sum_{i\in F_x} u_i(y)$.
  \item In Approval Voting, they find $y$ in the same way as in Plurality, only using approval scores instead of plurality scores,
        and then bullet vote for it: $b(y) = 1$, $b(z) = 0$ for all $z\neq y$.
  \item In Range Voting, they find $y$ in the same way as in Plurality, only using Range Voting scores divided by 100 
        instead of plurality scores,
        and then bullet vote for it: $b(y) = 100$, $b(z) = 0$ for all $z\neq y$.
  \item In IRV, they find the best response truncated ranking ballot by constructing a set $A$ of ``candidate'' truncated rankings $(x_1,x_2,\dots,x_\ell)$ 
        that cover all possible results they can effect by submitting identical ballots, 
        and then select the member of $A$ that gives the best result.
        $A$ is constructed iteratively by adding ever longer truncated rankings as follows.
        Given all other voters' ballots, 
        they start by finding the set $Y$ of options $y\in C$ for which $y$ survives the elimination process during the tally strictly longer
        when they rank $y$ 1st than when they submit a blank ballot.
        They put $A = \{ (y) : y\in Y \}$. 
        Then, for each ranking $(x_1,x_2,\dots,x_\ell)\in A$ with $\ell < k-1$, 
        they find the set $Y$ of options $y$ for which $y$ survives the elimination process during the tally strictly longer
        when they submit the longer ranking $(x_1,x_2,\dots,x_\ell,y)$ than when they submit the shorter ranking $(x_1,x_2,\dots,x_\ell)$.
        They add all those ballots $(x_1,x_2,\dots,x_\ell,y)$ to $A$ and iterate until no further ballots are added.
        One can show that for each possible truncated ranking ballot, there is a member of $A$ that has the same effect
        when used as the unanimous ballot of all faction members,
        and $|A| \le 2^{k}$.\footnote{See Section \ref{factionalirv} for a proof sketch.}
  \item In Simple Condorcet, they find the binary opposition values $o(y,z)$ 
        resulting from all other voters' ballots,
        and put $o(y) = \max_{z\in C} o(y,z)$, $o_0 = \min_{y\in C} o(y)$,
        and $A = \{ y\in C: o(y) < o_0 + |F_x| \}$.
        For each $y\in A$, they put $A_y = \{ z\in A: o(z) < o(y) \}$, 
        and check whether there is a function $g:A_y\to C$ such that $\{(z,g(z)):z\in A_y\}$ is acyclic
        and for all $z\in A_y$, $o(z,g(z)) + |F_x| > o(y)$.
        If this is the case, $y$ can be made the winner by ranking the options in any way that ranks $y$ first and
        ranks each $z\in A_y$ below its $g(z)$. 
        Among these $y\in A$, they find the one with the largest $U(y)$ and submit any ranking
        which ranks $y$ first, ranks each $z\in A_y$ below its $g(z)$, and doesn't rank any further options.
        If no such $y$ exists, they submit a bullet vote for $x$.
  \item In Random Ballot, they mark their true favourite since that is a dominant strategy: $b = \arg\max_{x\in C}u_i(x)$.
  \item In FC, they mark their true favourite as ``favourite'' and find an optimal option for marking as ``consensus''
        by computing the resulting $U$ for all of the $k$ many possible choices. 
  \item In RFC, they submit sincere ratings and find an optimal combination of options for marking as ``favourite'' and ``consensus''
        by computing the resulting $U$ for all of the $k^2$ many possible combinations.
  \item In the Nash Lottery, they try to find a (globally) best response by starting with a common ballot derived by
        averaging the faction members' sincere Nash Lottery ballots (see above)
        and then following a simple steepest ascent optimization algorithm until reaching a (local) optimum of $U(x)$.
        Although this local optimum might not be a globally best response, 
        we assume they use the resulting ballot anyway, which can be considered an additional form of bounded rationality.
  \item In MaxParC, they compare the results of all the $2^{k}$ many ballots $b_A$ of the form $b_A(x) = 100$ if $x\in A$ 
        and $b_A(x) = 0$ else, for some subset of options $A\subseteq C$. 
        They identify that $A$ which maximizes $U$ given all others' ballots.
        Note that the corresponding $b_A$ is a unanimous best response since only the resulting approvals matter.
        For this $A$, they calculate the approval scores $a_y$ that would result from using ballot $b_A$ in MaxParC 
        given all others' ballots.
        Then they define 
        \begin{equation}
            w(y) = \sum_{i\in F_x} \max\{0, 100 \frac{u_i(y) - u_i(\ell)}{\max_{z\in C} u_i(z) - u_i(\ell)}\} / |F_x|,
        \end{equation}
        which is the average sincere ballot of the faction members,
        and use the ballot with 
        \begin{equation} 
            b(y) = \max\{100 (1 - a_y/N), w(y)\}
        \end{equation}
        for $y\in A$
        and 
        \begin{equation} 
            b(y) = \min\{99 (1 - a_y/N), w(y)\}
        \end{equation}
        for $y\notin A$,
        which leads to the same approvals as $b_A$ and is thus also a best unanimous response.
        In other words, they use that best unanimous response which is closest to the average sincere ballot.
  \item In polling rounds, they act as in Plurality and Approval Voting.
\end{itemize}

Note that if a steady state emerges, it approximates a pure-strategy Nash equilibrium
between the trial-and-error voters as individual players and the factions as aggregate players,
which will in general however not be a strong or coalition-proof equilibrium since although we regard factions,
we do not regard inter-factional coalitional strategies.
Also, the process may also lead to cyclic or more complex attractors rather than a steady state.

\subsubsection{Experiment design}

We generated $M = 1,293,906$ many independent group decision problems,
drawing their parameters independently from the following probability distributions
(where parameter names in code are set in {\tt this font}):
\begin{description} 
\item[Number of voters $N$.] We drew odd numbers between 9 and 999 
           such that $\log_{10} N$ was approximately uniformly distributed in the interval $[1, 3]$. 
\item[Number of options $k$.] Uniformly in $\{3, \dots, 9\}$.
\item[Preference models.] Uniformly in $\{$Unif, BM, GA, QA, LA$\}$.
\item[BM parameters.] For the block model: 
  number of voter blocks {\tt Bmr} $= r\sim\text{Unif}\{2, 5, 9\}$,
  block size heterogeneity {\tt Bmh} $= h\sim\text{Unif}\{0,1\}$,
  individuality {\tt Bmiota} $= \iota\sim\text{Unif}\{0.1, 0.5\}$. 
\item[Spatial model parameters.] For GA, QA, and LA:
  policy space dimension {\tt dim} $= d\sim\text{Unif}\{1, 2, 3\}$,
  voter heterogeneity {\tt omega} $= \omega\sim\text{Unif}\{1, 2, 3, 5\}$,
  option broadness heterogeneity {\tt rho} $= \rho\sim\text{Unif}\{0, 1/3, 2/3, 1\}$,
  where $\rho = 0$ corresponds to the homogenous cases GH, QH, LH.
\item[Risk attitude scenarios.] Uniformly in $\{$all-EUT, all-LCP, all-HCP, mixed$\}$,
  where in `mixed' 20\% of the voters are EUT, 40\% LCP, and 40\% HCP.
\item[Number of polling rounds $R$.] Uniformly in $\{ 1, 2, 3, 5, 7, 10 \}$.
\item[Behavioural type scenarios.] Uniformly in $\{$lazy, middle, strat, all-L, all-S, all-T, all-H, all-F$\}$
\end{description} 
The following parameters were not varied:
\begin{description} 
\item[Length of interactive phase.] 100 time points.
\item[Trial-and-error frequency.] At each time point, 50\% of the trial-and-error voters updated their ballots.
\item[Factional update probability.] At each time point, each faction had a 10\% probability to update their ballots. 
\end{description} 

For each group decision problem, 
we constructed a second problem in which a randomly chosen option was replaced 
by a compromise option $y$ that was constructed from set $C_0$ of the remaining $k-1$ options,
to analyse the effect that a specifically designed compromise option would have.
Depending on the preference model, voters' preferences about $y$ were constructed as follows:
In Unif and BM, the compromise got the average utility of the other options,
$u_i(y) = \sum_{x\in C_0} u_i(x) / (k-1)$;
in GH, QH, and LH, the compromise's position $\xi_y$ was chosen to be a weighted average
of the other options' positions $\xi_x$, 
with weights $w_x$ proportional to first-preference support 
and inversely proportional to options' platforms' broadness $\sigma_x$:
\begin{align}
    \xi_y &= \sum_{x\in C_0} w_x \xi_x / \sum_{x\in C_0} w_x, \\
    w_x &= |\{i\in E: x=\arg\max_{z\in C}u_i(z)\}| / \sigma_x.
\end{align}
For each of these $2M$ decision problems, we simulated $R$ rounds of polling.
Finally, for each of the ten voting methods independently, 
we simulated an initial voting round and an interactive voting phase\footnote{%
    For the basically deterministic IRV, 
    we did not calculate tie probabilities since this would have been too costly due to the iterative nature of the method;
    instead, we resolved ties in IRV randomly, so that the resulting lottery always appeared to be a sure-thing lottery instead of the true tying lottery.
    For the NL method, the optimization problem $\max S(\ell)$ was solved using Sequential Least Squares Programming (SLSQP);
    to avoid a convergence failure due to singular Jacobian matrices because of zero ratings, 
    we added $10^{-5}$ to all ratings (the maximal rating always being 100).
    In the interactive phase of NL, a faction's best response ratings optimization problem was solved using Constrained Optimization By Linear Approximation (COBYLA)
    since that converged better than SLSQP.}
based on the same polling results,
and determined all options' resulting winning probabilities $\ell$ 
both after the initial voting round and after the interactive phase. 

\subsubsection{Social welfare metrics}

To measure the welfare effects of the tested voting methods, 
we use a set of metrics which are based on three different social welfare measures
(utilitarian, Gini-Sen, and egalitarian welfare),
taken either on an absulute or a relative scale, 
and either taken before or after the interactive phase of the simulations,
giving a total of twelfe different metrics per problem and method.

All these measures are aggregating the voters' individual utility $u_i(\ell)$ they get from the resulting lottery $\ell$, 
as modelled by the various utility models discussed above.
In applications where there is only a single decision taken, 
these measures must hence be interpreted as measuring the `ex ante' efficiency of the method, 
as opposed to the `ex post' efficiency that would be based on the utilities of the actual options chosen by the resulting lottery.
In applications where we imagine a sequence of decisions,
our efficiency metrics can be interpreted as measuring the long-run efficiency of the method over the whole sequence of decisions. 

\paragraph{Utilitarian welfare}

The simplest and most popular measure is the one proposed by average utilitarianism,
$W_\text{util.}(\ell) = \sum_{i\in E} u_i(\ell) / |E|$.

Since in all our utility models, 
lottery utility $u_i(\ell)$ is a linear combination of option utilities $u_i(x)$,
the lottery that maximizes $W_\text{util.}(\ell)$ is a sure-thing lottery.

\paragraph{Gini-Sen welfare}

As $W_\text{util.}(\ell)$ is insensitive to redistribution of utility across voters,
and hence to inequality between voters' utilities, we also use two inequality-averse metrics,
the first of which is the Gini-Sen welfare function
$W_\text{Gini}(\ell) = \sum_{i\in E} \sum_{j\in E\setminus\{i\}} \min\{u_i(\ell), u_j(\ell)\} / |E|(|E|-1)$.

As a motivating story one can imagine voters meet in a large sequence of bilateral meetings 
and each time evaluate the welfare status of society welfare in terms of the smaller of their two utilities,
and overall social welfare is then measured by the average of all these individual pairwise evaluations.
Another motivation for the same metric is that it can be seen as an ``inequality-adjusted'' version of $W_\text{util.}$,
since we have $W_\text{Gini}(\ell) = W_\text{util.}(\ell)(1 - I_\text{Gini}(\ell))$,
where $I_\text{Gini}(\ell)$ is the well-known Gini coefficient of inequality in utilities $u_i(\ell)$ \cite{Sen1974}.

Note that the lottery $\ell^*$ which maximizes $W_\text{Gini}(\ell)$ 
can be expected to be a proper lottery rather than a sure-thing lottery.
This is because randomization tends to reduce inequality more than it reduces average utility.

\paragraph{Egalitarian welfare}

As the most extremely inequality-averse welfare metric, we also consider the egalitarian one,
$W_\text{egal.}(\ell) = \min_{i\in E} u_i(\ell)$.
As in the case of Gini-Sen welfare, maximization of ex-ante egalitarian welfare usually requires randomization. 

\paragraph{Absolute and relative welfare metrics}

All three welfare metrics measure welfare on the same scale as individual utility, 
hence are hard to compare directly across different utility models 
since these use quite different scales. 
Also, for some models their distribution is quite skewed, having a long lower tail.
In addition to the above {\em absolute} welfare metrics, we therefore also compare
the {\em relative} metrics
\begin{align}
    relW_\text{util./Gini/egal.}(\ell) &= \frac{W_\text{util./Gini/egal.}(\ell) - \min_{x\in C}W_\text{util./Gini/egal.}(x)}%
        {\max_{x\in C}W_\text{util./Gini/egal.}(x) - \min_{x\in C}W_\text{util./Gini/egal.}(x)}
        \in [0, \infty]
\end{align} 
which rescale the welfare so that the sure-thing (!) lotteries giving the lowest and highest welfare get scores 0 and 1, respectively.
This design allows us to interpret values larger than 1 as welfare gains from randomization.

Still, as it turned out, the relative versions of Gini-Sen and egalitarian welfare often take very large values for nondeterministic methods
and thus now have a very skewed distribution with a long upper tail.
For this reason, we also study an {\em alternative relative} version of all three metrics,
defined as  
\begin{align}
    altrelW_\text{util./Gini/egal.}(\ell) 
        &= 2\frac{W_\text{util./Gini/egal.}(\ell) - \min_{x\in C}W_\text{util./Gini/egal.}(x)}%
        {W_\text{util./Gini/egal.}(\ell) + \max_{x\in C}W_\text{util./Gini/egal.}(x) - 2\min_{x\in C}W_\text{util./Gini/egal.}(x)} \\
        &= \frac{2relW_\text{util./Gini/egal.}(\ell)}{1 + relW_\text{util./Gini/egal.}(\ell)}
        \in [0, 2].
\end{align} 
These are now restricted to the interval $[0, 2]$, 
again taking a value of 0 and 1 for the sure-thing (!) lotteries giving the lowest and highest welfare.

\paragraph{``Cost of fairness''}

As an alternative to the above relative welfare metrics, 
one can also compare welfare differences between methods with 
utility differences within the electorate to assess the influence of method choice on welfare.
In analogy to the notion of a ``price of anarchy'' \cite{Koutsoupias1999}, 
we therefore define a ``relative cost of fairness'',
\begin{align}
    CF &= \frac{W_\text{util}^{RV} - W_\text{util}^{MPC}}{W_\text{util}^{RV} - W_\text{egal}^{RV}},
\end{align}
where the numerator is the absolute difference in average voter utility 
between the best deterministic method RV's result and the best proportional method MaxParC's result
(which could be termed the ``absolute cost of fairness''),
and the denominator is the difference between the average and minimum voter utility under RV
(which could be termed the ``absolute egalitarian inequality'').

\subsubsection{Randomization metrics}

To measure the degree of {\em randomization} a voting method actually applies,
we computed two established entropy measures, 
Shannon entropy and R\'enyi entropy of degree two,
and the maximal probability $\max_{x\in C}\ell_x$,
again applied to the results before and after the interactive phase.
This gives six randomization metrics in total per problem and method.

\subsubsection{Voter satisfaction metrics}

As another type of performance indicators, we computed each voter's ``satisfaction level''
\begin{align}
    \frac{u_i(\ell) - \min_{x\in C} u_i(x)}{\max_{x\in C} u_i(x) - \min_{x\in C} u_i(x)} \in [0, 1],
\end{align}
which would be zero if $i$'s least preferred option won for sure, and unity if $i$'s favourite won for sure.
Based on these, we report average satisfaction levels in the whole electorate 
and, to assess possible advantages of strategic behaviour, 
by behavioural type.

\subsubsection{Consequentialist preferences over methods}

Finally, to get an idea of which methods voters would chose if that choice was itself performed by majority voting,
we counted for each decision problem
how many voters would prefer the lottery resulting from some method $A$ to that resulting from some method $B$.

\section{Supplementary Text}
\label{suppl}

\subsection{Properties of voting methods}
\label{sec:properties}

\subsubsection{Basic consistency properties}

\paragraph{Anonymity.}

A voting method is {\em anonymous} iff it treats all voters alike, i.e.,
iff its result is invariant under permutations of voters.
All considered methods have this property.

\paragraph{Neutrality.}

A voting method is {\em neutral} iff it treats all options alike, i.e.,
iff the resulting winning probabilities of any two options $x,y$ are swapped when $x,y$ are swapped on all ballots.
All considered methods have this property.

\paragraph{Pareto-efficiency w.r.t.\ stated preferences.}

An option $y$ is {\em Pareto-dominated} w.r.t.\ stated preferences 
iff there is another option $x$ with $x\,P_{\beta_i}\,y$ for all $i\in E$.
A voting method is {\em Pareto-efficient} w.r.t.\ stated preferences 
iff all Pareto-dominated options get zero winning probability.
All considered methods except FC and RFC fulfill this.
Since in FC, only the fall-back option is interpreted as stating a preference, 
a Pareto-dominated $y$ might still be everyone's proposed consensus and win.
Similarly, since in RFC only the ratings are interpreted as preferences, 
$y$ still might be named by someone as fall-back option and thus have positive winning probability.

It is more difficult to check whether also an option which is Pareto-dominated w.r.t.\ {\em true} preferences
will have zero winning probability, since this depends on whether and how voters behave strategically.
Our numerical simulations at least suggests that all of the considered methods, including FC and RFC, 
fulfill this criterion under normal circumstances. 

\subsubsection{Monotonicity properties}

Although there are a number of variants of the `monotonicity' criterion, 
we here focus on two variants of Woodall's `mono-raise' monotonicity \cite{Woodall1997},
which differ only really for non-deterministic methods,
and one properly weaker property related to Woodall's `mono-add-plump' monotonicity.

\paragraph{Strong mono-raise monotonicity.}

A voting method is {\em strongly mono-raise monotonic}
iff the winning probability of an option $y$ cannot increase if a different option $x$ is advanced on one ballot:
$M(\beta)_y \le M(\beta')_y$ whenever $x\neq y$, $\beta_i\,Q^C_x\,\beta'_i$ for some $i\in E$,
and $\beta_j = \beta'_j$ for all $j\in E\setminus\{i\}$.

\paragraph{Weak mono-raise monotonicity.}

A voting method is {\em weakly mono-raise monotonic}
iff the winning probability of an option $x$ cannot decrease if $x$ is advanced on one ballot:
$M(\beta)_x \ge M(\beta')_x$ whenever $\beta_i\,Q^C_x\,\beta'_i$ for some $i\in E$ and
$\beta_j = \beta'_j$ for all $j\in E\setminus\{i\}$.

\paragraph{Weak mono-raise-abstention monotonicity.}

We call a voting method {\em weakly mono-raise-abstention monotonic}
iff the winning probability of an option $x$ cannot decrease if $x$ is advanced on an abstention ballot:
$M(\beta)_x \ge M(\beta')_x$ whenever $\beta_i\,Q^C_x\,\beta'_i$ for some $i\in E$,
$\beta_j = \beta'_j$ for all $j\in E\setminus\{i\}$,
and $\beta'_i$ is an abstention ballot.

Obviously, strong mono-raise monotonicity implies weak mono-raise monotonicity, 
which in turn implies weak mono-raise-abstention monotonicity.

~

\noindent
For PV, AV, RV, SC, and RB it is straightforward to prove all three forms of mono-raise monotonicity
(exercise left to the reader).
IRV is known to violate both strong and weak mono-raise monotonicity \cite{Woodall1997} 
but is easily seen to fulfill weak mono-raise-abstention monotonicity.

FC fulfills weak but not strong mono-raise monotonicity since if $z$ is everyone's proposed consensus, 
advancing $x$ on some consensus ballot destroys the consensus 
so that someone's fall-back option $y\neq x$ can get gets positive winning probability.

RFC violates both weak and strong mono-raise monotonicity. 
Consider the case of three options $x,y,z$ and two voters who both name $z$ as consensus and rate $(x,y,z)$ at $(0,3,1)$.
If one names $x$ and the other $y$ as fall-back, $x$ and $y$ both get winning probability $1/2$, 
but if both name $x$ as fall-back, $z$ wins for sure.

~

\noindent
NL violates strong mono-raise monotonicity.
Again consider three options $x,y,z$ and two voters $1,2$ who rate them as
$\beta_1(x,y,z) = (1/2, 0, 1/6)$ and $\beta_2(x,y,z) = (0, 1, 3/4)$.
Then $M(\beta)_y = 0$.
But if we increase $\beta_1(x)$ to $1$, we get $M(\beta)_y = 1/2 > 0$.
Numerical simulations suggest that NL fulfills weak mono-raise monotonicity, 
which we conjecture but were not able to prove yet unfortunately.

Regarding weak mono-raise-abstention monotonicity,
it was shown in \cite{Aziz2019} that NL (there called ``Nash Max Product'')
fulfills a roughly equivalent condition of ``Strict Participation'' 
when all ballots are ``dichotomous'' in the sense that  
all ratings are either zero or 100 (or some other common, fixed, positive number). 
Using the Envelope Theorem, we can give an alternative proof of weak mono-raise-abstention monotonicity
for arbitrary ratings.
{\em Proof.}
Let $E$, $C$ be fixed, consider some $i\in E$ and $x\in C$,
assume $\beta:E\to B(C)$ is fixed except for its entry $\beta_i(x)$,
and assume $\beta_i(y) = \epsilon$ for all $y\in C\setminus\{x\}$.
We will study the change of the NL probabilities $p^\ast(\alpha) = M(\beta)$ as a function of the parameter $\alpha = \beta_i(x)$
and show that $\alpha' > \alpha$ implies $p^\ast(\alpha') \ge p^\ast(\alpha)$, which will suffice to prove the claim.
$p^\ast(\alpha)$ is the solution of the maximization of the continuously differentiable function
$f(p,\alpha) = \sum_{j\in E}\log h_j(p,\alpha)$ with $h_j(p,\alpha) = \sum_{y\in C}\beta_j(y) p_y$ and $\beta_i(x) = \alpha$
under the constraint $g(p,\alpha) = \sum_{y\in C} p_y = 1$.
Let $V(\alpha) = \max_{p,g(p,\alpha)=1} f(p,\alpha)$ be the corresponding maximum.
Since the constraint is independent of the parameter $\alpha$,
the envelope theorem implies that
\begin{align}
    V'(\alpha) &= \frac{\partial f(p,\alpha)}{\partial\alpha}\Bigg|_{p=p^\ast(\alpha)}
    = \frac{p^\ast(\alpha)_x}{h_i(p^\ast(\alpha),\alpha)}.
\end{align}
Now assume an infinitesimal increase in $\alpha$ from $\alpha=\alpha_0$ to $\alpha = \alpha_1 = \alpha_0 + d\alpha$ with $d\alpha>0$.
Then the above implies
\begin{align}
    V(\alpha_1) &= V(\alpha_0) + V'(\alpha_0)d\alpha 
    = f(p^\ast(\alpha_0),\alpha_0) + \frac{p^\ast(\alpha_0)_x}{h_i(p^\ast(\alpha_0),\alpha_0)}d\alpha,
\end{align}
but also
\begin{align}
    V(\alpha_1) &= f(p^\ast(\alpha_1),\alpha_1) \\
    &= f(p^\ast(\alpha_1),\alpha_0) + \frac{\partial f(p,\alpha)}{\partial\alpha}\Bigg|_{p=p^\ast(\alpha_1), \alpha=\alpha_0}d\alpha \\
    &= f(p^\ast(\alpha_1),\alpha_0) + \frac{p^\ast(\alpha_1)_x}{h_i(p^\ast(\alpha_1),\alpha_0)}d\alpha.
\end{align}
Since optimization means that $f(p^\ast(\alpha_0),\alpha_0) \ge f(p^\ast(\alpha_1),\alpha_0)$, this implies
\begin{align}
    0 &\le \frac{f(p^\ast(\alpha_0),\alpha_0) - f(p^\ast(\alpha_1),\alpha_0)}{d\alpha} \\
    &= \frac{p^\ast(\alpha_1)_x}{h_i(p^\ast(\alpha_1),\alpha_0)} - \frac{p^\ast(\alpha_0)_x}{h_i(p^\ast(\alpha_0),\alpha_0)} \\
    &= \frac{p^\ast(\alpha_1)_x}{(\alpha_0-\epsilon) p^\ast(\alpha_1)_x + \epsilon} 
     - \frac{p^\ast(\alpha_0)_x}{(\alpha_0-\epsilon) p^\ast(\alpha_0)_x + \epsilon},
\end{align}
Since $\epsilon > 0$, this implies $p^\ast(\alpha_1)_x \ge p^\ast(\alpha_0)_x$.
Since this holds for all values $\alpha_0$ of $\alpha$, we have shown that $p^\ast(\alpha)_x$ 
is a weakly increasing function of $\alpha$ as claimed. 
{\em Q.E.D.}

~

\noindent
MaxParC fulfills all three forms. 
{\em Proof.} It suffices to show that if 
(i) $x$ is advanced by one voter $i$ from $\beta_i(x) = r$ to $\beta_i(x) = r' > r$,
(ii) some voter $j$ is drawn at random, 
and (iii) some option $y\neq x$ is in the set $A'_j$ after the change,
then $y$ must have been in $A'_j$ before the change and $A'_j$ can only have grown due to the change.
It is easy to see that $A(x)$ can only have grown and that no other $A(z)$ has changed,
hence $a'(x)$ has properly grown but no other $a'(z)$ has changed.
So if $y$ is in $A'_j = \arg\max_{z\in A_j} a'(z)$, 
the value of $\max_{z\in A_j} a'(z)$ has not changed,
hence $y$ must have been in this set before, 
and the only change in $A'_j$ can be that now $x$ is also in $A'_j$.
This means $A'_j$ can only have grown and thus $y$'s winning probability decreased. 
{\em Q.E.D.}

\subsubsection{Further consistency properties}

\paragraph{Independence from Pareto-dominated alternatives.}

The idea of this criterion is that the ``removal'' of a Pareto-dominated option $y$ from all ballots
should have no effect on the winning probabilities.
It was first introduced by Steve Eppley on the election-methods emailing list.\footnote{%
\url{http://lists.electorama.com/pipermail/election-methods-electorama.com//2003-March/107700.html}}
Since for some methods it is not obvious what is meant by ``removal'', 
we do not study this in a formal way here but rather discuss it verbally.

For methods using a ballot type that lets voters rate or rank all options independently 
(AV, RV, SC, NL and MaxParC), let us assume ``removal'' means 
leaving the other options' ratings unchanged. 
For IRV, let us assume ``removal'' implies decreasing the ranks of the later-ranked options by one.
Then those six methods all fulfill this criterion, 
and so do Plurality and Random Ballot whenever no voter has named $y$ as favourite (which rational voters wouldn't).

As this criterion implies Pareto-efficiency, FC and RFC do not fulfill it.

We note that in particular many other Condorcet-type methods, 
which elect a winner of all pairwise comparisons for sure if such an option exists,
including the `Ranked Pairs' method by Nicolaus Tideman \cite{Tideman1987}
and the `Beatpath' methods by Markus Schulze \cite{Schulze2011}, 
fail this criterion.

\paragraph{Independence from losing options.}

This criterion
demands that the removal of any option $y$ receiving zero winning probability
must have no effect on the winning probabilities.
This is a variation of the famous `Independence of Irrelevant Alternatives' criterion,
and which is stronger than Independence from Pareto-dominated alternatives if Pareto-efficiency is given.

It is easy to see that again AV, RV, RB, and NL fulfill this and FC and RFC do not.
For PV, some voters may have voted for $y$ and now vote for the current runner-up and make it win.

IRV and SC also do not, as can be seen from the example of three options $x,y,z$ 
and three factions $F_{1,2,3}$ of sizes $4,3,2$ and rankings
$F_1:x>y>z$, $F_2:y>z>x$, $F_3:z>x>y$. Both methods elect $x$ for sure but elect $z$ if $y$ is removed.

MaxParC also fulfills this criterion. {\em Proof.}
Removal of $y$ does not change who approves of which other options. 
If $y$ has zero winning probability, every voter who approves of $y$ also approves of some higher-scoring option.
Hence for no voter the set of highest-scoring approved options changes.
Thus all other options' winning probabilities are unaffected.
{\em Q.E.D.}

An even stronger variant of `Independence of Irrelevant Alternatives' that can also be interpreted as a form of `monotonicity' 
goes as follows: 
removing any option $y$ from $E$ must not decrease any remaining option $x$'s winning probability.
NL probably violates this while MaxParC clearly fulfills it.

\paragraph{Independence from cloned options.}

Another type of criterion deals with the addition of an option $y$, called a `clone', that is very ``similar'' to some existing option $x$.
Since ``similarity'' can be defined in different ways depending on the ballot type,
we restrict our interest here to the special case where $y$ is a unique `exact clone' of $x$,
meaning all voters are truly indifferent between $x$ and $y$ but not between these two and any further option $z$.
We demand that in that case and under plausible assumptions on voter's voting behaviour, 
the addition of $y$ shall not change the winning probability of any other option $z\notin\{x,y\}$.

Let us assume that after the addition of $y$, 
voters will assign $y$ the exact same approval, rating, or ranking (if tied rankings are allowed, otherwise an adjacent ranking) as $x$,
and will name any option $z\notin\{x,y\}$ as favourite or proposed consensus iff they named the same option before the addition,
only possibly switching from naming $x$ to now naming $y$.
It is then easy to see that AV, RV, IRV, SC, RB, RFC, NL and MaxParC all fulfill this form of `exact clone independence', 
while PV and FC do not. 

Note that there are other, stronger, forms of clone-independence, including the one discussed in \cite{Tideman1987}, 
that some variants of IRV, many Condorcet-type methods, NL, and MaxParC might not fulfill.

\paragraph{Revelation of preferences.}

Some voting methods have the property that, sometimes depending on the level of strategic behaviour,
voters' filled-in ballots reveal all or part of their preferences.
Under RB, for example, whenever a voter has an option she strictly prefers to all other options (a unique favourite),
it is a weakly dominant strategy to specify that option. 
This form of ``strategy-proofness'' can be interpreted as implying that RB ``reveals unique favourites'' 
(but nothing else about voters' preferences).

As another example, it was shown in \cite{Heitzig2010a} that under RFC, 
whenever a voter has preferences conforming to expected utility theory (see below) with some utility function $u$,
she has no strategic incentive to specify different ratings than a properly rescaled version of $u$,
and hence RFC can be said to ``reveal von-Neumann--Morgenstern utility functions''
(but no preferences that do not conform to expected utility theory).
Still, RFC is not strategy-proof in the sense that there always exist weakly dominant strategies,
since in its other two ballot components, 
a rational voter may want to name a proposed consensus option that depends on others' preferences, 
and may have incentives to name a different option as ``fall-back'' than her favourite.
This shows that full preference revelation is related to but neither implied by nor stronger than strategy-proofness.

FC reveals favourites but its consensus ballot component is strategic.
NL and MaxParC also reveal favourites in the sense that a voter has no incentive to 
not rank her favourite first or to rate it below 100,
but usually has an incentive to rate all other options strictly below 100.
Neither of them however reveal much more of a voter's preferences.

AV and RV don't reveal favourites since typically a rational voter has an incentive to approve of (or rate at 100) some additional options.
Still, in the case where voters have no information about others' preferences,
AV and RV can be said to reveal something about a voter's preferences, 
because in that case a rational expected utility theory voter 
would approve of (or rate at 100) all options she prefers to drawing an option uniformly at random (and would rate all other options at 0), 
so that one can infer that she strictly prefers each approved to each disapproved option.

Similarly, under NL, expected utility theory voters who use the zero-information heuristic derived in the end of \ref{heurnashlott} 
also reveal their full preferences, but this heuristic might not be a weakly dominant strategy under zero information, 
so rational voters may not use it.
Under MaxParC, the linear heuristic derived in \ref{sec:mpsincere} reveals the above-average part of a voter's utility function
but is also typically not weakly dominant under zero information.

IRV and SC also do not reveal favourites. 
For IRV, consider three options and six voters and 
assume voter 1 has preferences $1:A>B>C$ and the others vote $A>B$, $A$, $B>C$, $C>A$, $C$.
Then if 1 votes sincerely, $B$ is removed and a coin toss between $A$ and $C$ results.
But if 1 votes $B>A>C$, $A$ still gets probability 1/2 but now $B$ gets 1/3 and $C$ gets 1/6, which 1 strictly prefers.
For SC, consider three options and three voters and  
assume voter 1 has preferences $1:A>B>C$ and the others vote $B=C>A$, $C>A>B$.
Then 1 would want to vote $A=B>C$ or $B>A>C$ to ensure a coin toss between $B$ and $C$ 
rather than voting sincerely $A>B>C$ and getting $C$ for sure.

\subsubsection{Proportional allocation of effective power}

This criterion requires that in every situation $(C,E)$ 
and for every option $x\in C$ and group of voters $G\subseteq E$,
there must be a way of voting $\beta_G\in B(C)^G$ for $G$ 
so that for all ways of voting $\beta_{-G}\in B(C)^{E\setminus G}$ of the other voters,
the winning probability of $x$ is at least as large as $G$'s relative size:
$M(\beta_G,\beta_{-G})_x \ge |G|/|E|$.
A related criterion was discussed for the special case of ``dichotomous preferences''
under the name ``Core Fair Share'' in \cite{Aziz2019}.

\begin{figure}
\centering
\includegraphics[width=0.5\textwidth]{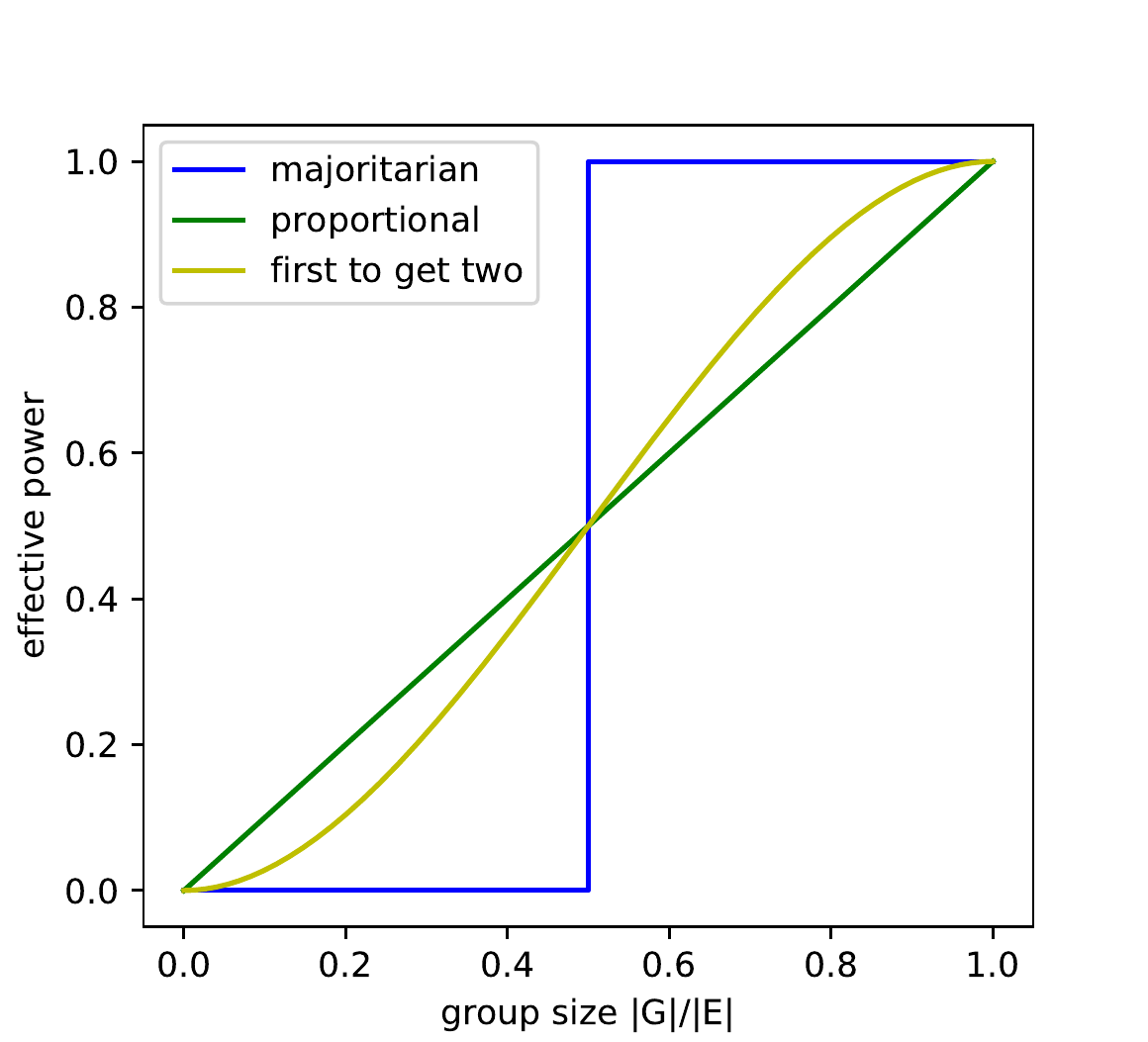}
\caption{\label{fig:power}
Effective power of a voter group $G$ as a function of group size,
for majoritarian deterministic methods PV, AV, RV, IRV, SC (blue),
proportional nondeterministic methods RB, FC, RFC, NL, MaxParC (green),
and an example of a nonproportional nondeterministic method (yellow). 
}
\end{figure}

Since all considered methods are neutral and anonymous, one can summarize the power distribution by 
drawing the maximal winning probability a group of size $|G|/|E| = s$ 
can guarantee any option $x$ of their choice under the various methods,
as is done in Fig.\,\ref{fig:power}. 
For all considered deterministic methods, this ``effective decision power'' is 
basically a step function with the value zero for $s<1/2$ and one for $s>1/2$ (blue line).
Only for $s=1/2$ the value depends on the method's detailed treatment of ties, 
which we do not discuss here.
By contrast, for all considered non-deterministic methods, 
it can easily be seen that any group $G$ can guarantee $x$ a probability at least $|G|/|E|$ by simply bullet-voting for $x$, 
so effective decision power is simply equal to $s$ (green line), 
which we call {\em proportional allocation of effective power.}

Note that of course there are also non-deterministic neutral anonymous methods with different allocations of effective power.
E.g., one could draw a sequence of plurality ballots at random until one option was named twice (``first to get two''), 
giving a smooth but S-shaped nonlinear power curve $s^2(3-2s)$.

For NL, it is the specific use of the logarithm that gives a linear power curve.
Indeed, consider a method that puts $M(\beta) = \arg\max_\ell S(\ell)$
for $S(\ell) = \sum_{i\in E} f(r_i(\ell))$,
some weakly increasing and continuously differentiable function $f$,
and $r_i(\ell) = \sum_{x\in C}\ell_x\beta_i(x)$.
If the power curve is linear, 
then whenever a group of voters $G\notin\{\emptyset,E\}$ bullet-votes for $x$ 
and the other voters $E\setminus G$ bullet-vote for $y$,
we must have $p := M(\beta)_x = |G|/|E| =: s$ and $q := M(\beta)_y = 1 - |G|/|E| = 1-s$,
hence the first-order condition
\begin{align}
    0 &= (\partial_p - \partial_q) S|_{p=s,q=1-s} = sf'(100s) + (1-s)f'(100(1-s))
\end{align}
implies $f'(100s)\propto 1/s$ for all rational numbers $s\in(0,1)$ 
and thus $f(r)\propto\log r$ for all real numbers $r\in(0,100)$.

\subsubsection{Consensus supporting properties}

In this section, we show that our two focus methods NL and MPC support both full and partial consensus
even with strategic voters. 
To do so, we show that the respective potential consensus options result 
both from sincere voting (see \ref{sec:mpsincere} for a discussion of sincere voting in MaxParC) 
and in several forms of strategic equilibrium in archetypial decision situations.

\paragraph{Nash Lottery supports full consensus}

\subparagraph{Assumptions.}

We assume two equal-sized factions $F_1,F_2$ of $m$ many voters each, and three options $A,B,D$.
Voters in $F_1$ have von-Neumann--Morgenstern utility function $u_1(A,B,D)=(1,0,u)$ and submit ratings $r_1(A,B,D)=(1,0,r)$,
those in $F_2$ have $u_2(A,B,D)=(0,1,v)$ and submit $r_2(A,B,D)=(0,1,s)$
with $r,s\in(0,1)$ and $1/2<u,v<1$, so that both factions prefer $D$ to a coin toss between $A$ and $B$.

\subparagraph{Resulting lottery and expected utilities.}

If $A,B,D$ get probabilities $p,q,1-p-q$,
the resulting expected utilities are
\begin{align}
    U_1 &= p + (1-p-q)u, \\
    U_2 &= q + (1-p-q)v,
\end{align}
and the Nash sum is 
$$ f = m\log(p+(1-p-q)r) + m\log(q+(1-p-q)s). $$
Because $f$ is concave in both $p$ and $q$, 
the unique pair $p,q$ maximizing $f$ can be found as follows.
Given $q\in[0,1]$, $f$ is maximized by that $p\in[0,1-q]$ which is closest to the point $p_0(q)$ of zero slope,
\begin{align}
    0 &= \partial_p f = \frac{1-r}{p+(1-p-q)r} + \frac{-s}{q+(1-p-q)s}, \\
    p_0(q) &= \frac{(1-q)(1-2r)s + (1-r)q}{2(1-r)s}.
\end{align}
Similarly, given $p$, $f$ is maximized by that $q\in[0,1-p]$ closest to
\begin{align}
    q_0(p) &= \frac{(1-p)(1-2s)r + (1-s)p}{2(1-s)r}.
\end{align}
If we introduce the notation $[x]_0^y = \max(0,\min(x,y))$,
the maximum of $f$ is thus attained where
\begin{align}
    p &= \left[\frac{(1-q)(1-2r)s + (1-r)q}{2(1-r)s}\right]_0^{1-q}, \\
    q &= \left[\frac{(1-p)(1-2s)r + (1-s)p}{2(1-s)r}\right]_0^{1-p}.    
\end{align}
Depending on $r,s$, the solution $(p,q)$ found by the Nash Lottery method
and resulting utilities $(U_1,U_2)$ are the following:
\begin{align}
    (p,q,U_1,U_2) = \left\{ \begin{array}{ll}
        (\frac 1 2, \frac 1 2, \frac 1 2, \frac 1 2) & r + s < 1, \\
        \frac{(1-2r, 0, 1-2r+u, v)}{2-2r} & s \le \frac 1 2, r \ge 1 - s, \\
        \frac{(0, 1-2s, u, 1-2s+v)}{2-2s} & r \le \frac 1 2, s \ge 1 - r, \\
        (0, 0, u, v) & s, r \ge \frac 1 2.
    \end{array} \right.
\end{align}

\subparagraph{Outcome with sincere voters.}

Sincere voters put $r=u>1/2$ and $s=v<1/2$ and thus get $p=q=0$, 
i.e., the consensus option $D$ wins for sure.

\subparagraph{Strategic equilibria between factions.}

To analyse strategic incentives for the two factions, 
we treat $F_1,F_2$ as the players of a two-player game in which they simultaneously choose $r,s\in(0,1)$,
and study the Nash equilibria (NE) of that game.
Given some $s$, $F_1$'s best responses are the following:
If $s<\min(\frac 1 2, 1-u)$, each $r<1-s$ is a best response.
If $1-u\le s<\frac 1 2$, each $r\ge 1-s$ is a best response.
If $s>\frac 1 2$, only $r=1-s$ is a best response.
If $s=\frac 1 2\ge u$, each $r\le \frac 1 2$ is a best response.
Finally, if $s=\frac 1 2<u$, only $r=\frac 1 2$ is a best response.  

Since we assume $u,v>\frac 1 2$, this results in the following sets of NE.
Any combination $r,s<\frac 1 2$ is a NE giving only $U_1=U_2=\frac 1 2$.
Any combination $(r,1-r)$ with $r\in(1-v,\frac 1 2)$ is a NE giving $(U_1,U_2)=\frac{(1+u-2r,v)}{2-2r}$.
Any combination $(1-s,s)$ with $s\in(1-u,\frac 1 2)$ is a NE giving $(U_1,U_2)=\frac{(u,1+v-2s)}{2-2s}$.
Finally, $s=r=\frac 1 2$ is the ``focal'' NE giving $(U_1,U_2)=(u,v)$ and the largest utility sum $U_1+U_2$ of all NE.

\subparagraph{Summary.}
The above analysis shows that in this scenario, Nash Lottery supports full consensus with both sincere voters (who would put $r=u$ and $s=v$) 
and strategic voters (who would put $s=r=\frac 1 2$).
We conjecture that similar calculations will show that the same holds with more and unequally sized factions.

\paragraph{Nash Lottery supports partial consensus}

Assume that to the above we now add a third faction $F_3$ of size $n-2m$ and a fourth option $C$,
and utilities $u_3(A,B,D,C)=(0,0,0,1)$, $u_1(C)=u_2(C)=0$.

\subparagraph{Strategic equilibria between factions.}

$F_3$ has a dominant strategy to bullet-vote for $D$, 
i.e., put $r_3(A,B,D,C)=(0,0,0,1)$,
and $F_1$, $F_2$ have no reason not to put $r_1(C)=r_2(C)=0$.
If we parameterize the probabilities of $(A,B,D,C)$ as $(p(1-w),q(1-w),(1-p-q)(1-w),w)$,
the Nash sum becomes
$$ f = m\log(p+(1-p-q)r) + m\log(q+(1-p-q)s) + 2m\log(1-w) + (n-2m)\log(w), $$
which is maximized by $w=1-2m/n$ and the same values of $(p,q)$ as above.
Since $F_1$, $F_2$'s utilities are proportional to the case above,
\begin{align}
    U_1 &= (1-w)(p + (1-p-q)u), \\
    U_2 &= (1-w)(q + (1-p-q)v),
\end{align}
the strategic analysis is the same as before, 
so putting $r=s=\frac 1 2$ is again the utility-maximizing and focal equilibrium.

\subparagraph{Outcome with sincere voters.}

Sincere voters in $F_1$ and $F_2$ still put $r=u>1/2$, $s=v<1/2$, $\beta_i(C)=0$,
and those in $F_3$ bullet-vote for $C$.
They still get $p=q=0$ and in addition $w=1-2m/n$, 
i.e., the partial consensus option $D$ now gets probability 
$1-w = 2m/n = (|F_1| + |F_2|) / |E|$,
as required.

~

\noindent
This shows that in this scenario, Nash Lottery also supports partial consensus.
We conjecture that the same holds with more and unequally sized factions,
and with several partial consensusses between different sets of factions.

\paragraph{MaxParC supports full consensus}
\label{sec:mpfull}

Note that under MaxParC, one has never an incentive to approve of a worst-liked option or to disapprove of one's favourite.
Since the following is not restricted to voters with expected utility preferences,
we don't use von-Neumann--Morgenstern utility functions $u_i$ here 
but rather state a voter's preferences over lotteries of options $\ell,\ell'$
by means of the binary relations
$\ell\,P_i\,\ell'$ (strict preference for $\ell$ over $\ell'$), 
$\ell\,R_i\,\ell'$ (weak preference), and
$\ell\,E_i\,\ell'$ (indifference),
only assuming that $R_i$ is a quasi-ordering (not necessarily complete)
and that $P_i$ and $E_i$ are its antisymmetric and symmetric parts.

\subparagraph{Assumptions.}

Assume $m \ge 2$ factions $F_j$ with sizes $N_j\ge 1$, $N=\sum_{j=1}^m N_j$,
with distinct true favourites $x_j$,
and assume voters are indifferent about pairs of other factions' favourites:
$x_{j'} E_i x_{j''}$ for all $i\in F_j$ if $j' \neq j \neq j''$.
Let $\ell_b$ be the benchmark lottery of drawing a random voter's favourite:
$\ell_b(x_j) = N_j/N$.
Assume there is just one more option $y$, and this is a potential full consensus option:
$y\,P_i\,\ell_b$ for all $i$.

Assume the MaxParC ballot profile $\beta$ has
$\beta_i(x_j) = 100$, $\beta_i(x_{j'}) = 0$, and $0 < \beta_i(y) \le 100/N$ for all $j\neq j'$ and $i\in F_j$.
Then each $i\in F_j$ approves of $x_j$ and $y$, hence all vote for $y$ and $y$ is the sure winner.

\subparagraph{Outcome with sincere voters.}

As discussed in \ref{sec:mpsincere}, there is no unique way to vote ``sincerely'' in MaxParC,
hence we rather discuss the results of voters applying one of the heuristics discussed there.

A voter applying the conservative satisficing heuristic
rates their favourites $x_j$ at 100 and the compromise $y$ at 
$100(1-u_i(\ell_b)/u_i(y)) > 0$.
Also with the informed satisficing heuristic, the linear heuristic, and the hyperbolic heuristic,
voters rate $y$ at $>0$.
So if all voters apply one of these heuristics, $y$ wins for sure.

\subparagraph{Nash equilibrium.}

Since no $i\in F_j$ can make any $i'\notin F_j$ vote for $x_j$,
the only way $i$ could only improve the result would be by making the vote of some $i'\in F_j$ go to $x_j$ instead of to $y$.
But this is only possible by lowering $\beta_i(y)$ to zero (either certainly or with some positive probability),
which will make everyone disapprove of $y$ and vote for their favourites, resulting in $\ell_b$.
Since no mixture of $\ell_b$ with the sure-thing lottery $\ell_y$ is an improvement for $i$,
$\beta$ is a {\em Nash equilibrium in pure strategies.}

Likewise, any group of voters $G\subseteq F_j$ from the same faction could only improve the result for each of them
by making the vote of some $i'\in F_j$ go to $x_j$.
As above, this is only possible if at least one $i\in G$ lowers $\beta_i(y)$ to zero,
again resulting in $\ell_b$, which is no improvement.
Hence $\beta$ remains a Nash equilibrium when some group of voters from the same faction is considered to act as one player;
in particular, if each faction is considered one player (this could be called a ``factional Nash equilibrium'').

Still, as with other voting methods, it is easy to see that there are many other Nash equilibria
(e.g., the less efficient one where everyone bullet-votes, resulting in $\ell_b$),
so the criterion of being a Nash equilibrium is not sufficiently discriminatory
and stronger game-theoretic solution concepts are called for.

\subparagraph{Strong Nash equilibrium.}

Assume a proper subgroup $G\subset E$ intersecting at least two factions,
let's say it intersects the factions
$F_1,\ldots,F_r$.
Assume the voters in $G$ change their ballots in some way that improves the result for them all.
Assume some $i\in G$ stops approving of $y$.
Then the votes of all $i'\notin F_1+\cdots+F_r$ will go to their favourites,
hence, for $j=1\ldots r$, at least $N_j+1$ votes must go to $x_j$
for this to be an improvement for all in $G$,
which is impossible since there are only $N_1+\cdots+N_r$ votes left to distribute.
Hence no $i\in G$ stops approving of $y$, and no $x_j$ gets approval by all voters,
so $y$ is still the sure winner, and there is no improvement for $G$ after all.

Finally, assume the whole electorate could improve the result for all.
Then there would be $\ell'\neq \ell_b$ with $\ell'\,P_i\,y$ for all $i$.
Hence there would be $\ell''\neq \ell_b$ with $\ell''\,P_i\,y$ for all $i$ and $\ell''(y)=0$.
But $\ell''\neq \ell_b$ implies there is $j$ with $\ell''(x_j) < \ell_b(x_j)$,
so all $i\in F_j$ would have $y\,P_i\,\ell_b\,P_i\,\ell''$, a contradiction.

This shows that $\beta$ is a {\em strong Nash equilibrium,} i.e.,
no group whether small or large, unanimous or cross-faction, has an incentive to deviate from $\beta$.

\subparagraph{Other methods.}
Under majoritarian methods, in particular PV, AV, RV, IRV, and SC,
there is usually no Nash equilibrium between the factions that would give $y$ positive winning probability,
simply because whenever $N_j > N/2$ for some $j$, faction $F_j$ will enforce that $x_j$ wins.
Also, RB does not support full consensus since for all $F_j$ it is strictly dominant to vote for $x_j$.
FC and RFC however do support full consensus, as shown in \cite{Heitzig2010a}.

With sincere voters, only AV and RV also support full consensus, 
while PV, IRV and SC would still elect $x_j$ whenever $N_j > N/2$.

\paragraph{MaxParC favours full over partial consensus}
\label{sec:mpfullpartial}

\subparagraph{Assumptions.}

As a generalization of the above,
assume now that there is an additional option $z$,
considered a potential partial consensus by the union $H=F_1+\cdots+F_h\subset E$ of some of the factions,
and considered equally bad by all others,
so that $\ell_{b/z}\,P_i\,\ell_b$ for all $i\in H$
and $z\,E_i\,x_j$ for all $i\notin H+F_j$,
where $\ell_{b/z}$ is the result of all $i\in H$ voting for $z$ and all others voting for their favourites:
$\ell_{b/z}(z) = |H|/N$ and $\ell_{b/z}(x_j) = |F_j|/N$ for all $j>h$.

Consider an extension of the above ballot profile $\beta$ with
$0 < \beta_i(z) \le 100 (1+N-|H|)/N$ and $\beta_{i'}(z)=0$ for all $i\in H$, $i'\notin H$.
Note that then all $i\in F_j$ approve of $x_j$, all $i\in H$ approve of $z$, and all approve of $y$,
hence again $y$ is the sure winner.

\subparagraph{Outcome with sincere voters.}

A voter applying one of the heuristic in \ref{sec:mpsincere}
will rate $y$ at $>0$ but will rate $z$ at 0 if she is not a member of $H$.
Hence if all voters apply some of these heuristics,
$y$ will be strictly more approved than $z$ and still win for sure.

\subparagraph{Strong Nash equilibrium.}

If all $i\in H$ consider full consensus still better than their potential partial consensus, i.e.,
$y\,P_i\,\ell_{b/z}$ for all $i\in H$,
then we will show that $\beta$ is again a strong Nash equilibrium,
at least when all $i\in H$ have von Neumann--Morgenstern expected utility functions $u_i(\ell)$ over lotteries.
Assume some group $G$ can improve the result to some lottery $\ell'$ by modifying their ballots.
If no $i\in G$ stopped approving of $y$, $y$ would remain the sure winner,
hence some $i\in G$ stops approving of $y$ and the votes of all $i'\in H - G$ go to $z$,
while those of $i\in F_j - G$ for $j>h$ go to $x_j$.
As above, for all $j$ with $F_j\cap G\neq\emptyset$,
at least $N_j+1$ votes must then go to $x_j$ for this to be an improvement for all $i\in F_j\cap G$,
hence less than $|H|$ votes are left that could go to either $z$ or some of $x_1,\ldots,x_h$.
Those $i\in F_j\cap G$, $j\le h$, have
$N u_i(\ell') = v_z u_i(z) + v_j u_i(x_j)$,
where $v_z,v_j$ are the votes going to $z$ or $x_j$, respectively,
and $v_z + \sum_{j=1}^h v_j < |H|$.
Since $N u_i(y) > N u_i(\ell_{b/z}) = |H| u_i(z)$
and $N u_i(y) > N u_i(\ell_b) = N_j u_i(x_j)$,
we have
$u_i(y) < u_i(\ell') = [ v_z u_i(z) + v_j u_i(x_j)] / N
< [ v_z / |H| + v_j / N_j ] u_i(y)$,
i.e., $v_j > N_j (1 - v_z / |H|)$ for all $j\le h$,
thus $|H| - v_z > \sum_{j=1}^h v_j > |H| (1 - v_z / |H|) = |H| - v_z$, a contradiction.
Note that the same kind of argument can be made if there are several partial potential consensus options $z$,$z'$,\ldots,
if each pair of corresponding supporting groups $H$,$H'$ is either disjoint or one contains the other
(so that they form a hierarchy).

In other words, no group has an incentive to deviate from electing a good enough full consensus,
even if a whole hierarchy of narrower and broader partial consensus options is available.

\paragraph{MaxParC supports a single partial consensus}
\label{sec:mpsinglepartial}

\subparagraph{Assumptions.}

Assume the same situation as in \ref{sec:mpfullpartial}, but without the full consensus option $y$,
so that only the partial consensus option $z$ remains besides the favourites $x_j$.
Then the same ballot profile $\beta$, just with $y$ removed,
leads to the partial consensus result $\ell(z) = |H|/N$ and $\ell(x_j) = |F_j|/N$ for $j>h$.

\subparagraph{Outcome with sincere voters.}

If the voters from $H$ apply the hyperbolic heuristic, 
they rate $z$ at $100(1-\ell(x_j)/u_i(z))$ which is by assumption larger than
$100(1 - |H|/N)$, so they all end up voting for $z$.

With the linear heuristic, however, they may rate $z$ too low for getting their votes
since $100 (u_i(z) - \ell(x_j)) / (1 - \ell(x_j))$ might be smaller than $100(1 - |H|/N)$.

\subparagraph{Strong Nash equilibrium.}

We can show this $\beta$ is again a strong Nash equilibrium when voters have von Neumann-Morgenstern utilities.
Assume some $i\in G\cap H$ stops approving of $z$.
Then, for each $j$ with $G\cap F_j\neq\emptyset$, at least $N_j+1$ votes must go to $x_j$
for this to be an improvement for all in $G$, but for each $j$ with $G\cap F_j=\emptyset$, all $N_j$ votes go to $x_j$,
a contradiction as above.
So no $i\in G\cap H$ stops approving of $z$.
If $G\cap H\neq\emptyset$, some $i\in G - H$ must vote for $z$ for those voters to profit from the deviation.
But then not enough votes are left in $G-H$ to make all $i\in G-H$ profit as well.
So $G\cap H=\emptyset$, but since $E-H$ has no potential for even partial consensus,
they cannot improve over the benchmark lottery either. This completes the proof.

\subparagraph{Other methods.}
Again, under majoritarian methods, in particular PV, AV, RV, IRV, and SC,
there is no Nash equilibrium between the factions that would give $z$ positive winning probability
if one of the factions is in a majority.
Also, RB does not support partial consensus since for all $F_j$ it is strictly dominant to vote for $x_j$.
FC and RFC also fail to support partial consensus: If $z$ wins because the fallback was not invoked,
the voters in $E\setminus H$ can cause the fallback to be invoked and have strict incentive to do so;
if the fallback is invoked, no voter in any $F_j\subset H$ will vote for $z$ 
since they have then a strict incentive to vote for $x_j$ instead.  

\paragraph{MaxParC supports disjoint partial consensuses}

If several disjoint groups of factions exist each of which has a potential partial consensus,
the situation can get a little trickier, and the canonical ballot profile might not be a strong Nash equilibrium
but only a coalition-proof equilibrium.
We treat a simple special case first to demonstrate this.

\subparagraph{Example.}

Assume $N=6$, four factions of sizes $N_1=N_4=1$ and $N_2=N_3=2$ with favourites $x_1\ldots x_4$,
and two potential partial consensus options $z,z'$, with utilities as in the following table:
\begin{center}
\begin{tabular}{rrrrrr}\hline
    & faction & $F_1$ & $F_2$ & $F_3$ & $F_4$ \\
    & size    & 1     & 2     & 2     & 1 \\\hline
    utility
    & 100   & $x_1$ & $x_2$ & $x_3$ & $x_4$ \\
    & 75    & $z$   & $z$   & $z'$  & $z'$ \\
    & 0     & rest  & rest  & rest  & rest \\\hline
\end{tabular}
\end{center}
A canonical ballot profile $\beta$ that realizes both partial consensuses is given by this table:
\begin{center}
\begin{tabular}{rrrrrr}\hline
    & faction & $F_1$ & $F_2$ & $F_3$ & $F_4$ \\
    willingness $\beta_i(x)$
    & 100   & $x_1$ & $x_2$ & $x_3$ & $x_4$ \\
    & 51    & $z$   & $z$   & $z'$  & $z'$ \\
    & 0     & rest  & rest  & rest  & rest \\\hline
\end{tabular}
\end{center}
Although this is a Nash equilibrium between the invididual voters and a Nash equilibrium between the four factions,
it is not a strong Nash equilibrium since the two middle factions can profit from approving their mutual favourites,
i.e., deviating as follows:
\begin{center}
\begin{tabular}{rrrrrr}\hline
    & faction & $F_2$ & $F_3$ \\
    willingness $\beta'_i(x)$
    & 100   & $x_2$ & $x_3$ \\
    & 51    & $z$   & $z'$  \\
    & 35    & $x_3$ & $x_2$ \\
    & 0     & rest  & rest  \\\hline
\end{tabular}
\end{center}
This will result in $F_2,F_3$ approving both $x_2,x_3$ so that
these options get a higher approval (4) than $z,z'$ (having 3),
the votes of $F_2,F_3$ now go to $x_2,x_3$ in equal shares (due to the tiebreaker),
and those of $F_1,F_4$ still go to $z,z'$.

Still, the above deviation by $F_2,F_3$ is not {\em coalition-proof}
since each of these two factions has an incentive to betray the other by not performing the agreed deviation after all,
i.e., by deviating from the planned deviation. E.g., if $F_2$ defects in this way, we have the profile
\begin{center}
\begin{tabular}{rrrrrr}\hline
    & faction & $F_1$ & $F_2$ & $F_3$ & $F_4$ \\
    willingness $\beta''_i(x)$
    & 100   & $x_1$ & $x_2$ & $x_3$ & $x_4$ \\
    & 51    & $z$   & $z$   & $z'$  & $z'$ \\
    & 35    &       &       & $x_2$ \\
    & 0     & rest  & rest  & rest  & rest \\\hline
\end{tabular}
\end{center}
which now makes $F_2$'s and $F_3$'s votes both go to $x_2$, profiting $F_2$ even more
and leaving $F_3$ with strictly less than under $\beta$.
Because of this risk of being betrayed by $F_2$,
$F_3$ has few incentives to agree with $F_2$ to perform the original deviation $\beta'$.

\subparagraph{Conjecture.} 
More generally, we conjecture that under quite general conditions, 
there will be at least a certain type of coalition-proof equilibrium (similar to \cite{Bernheim1987a})
which results in the election of a broad consensus.

More specifically, consider the following type of situation:
There are $M\ge 2$ disjoint blocks $B_1,\ldots,B_M$ of voters,
each block $B_k$ having size $N_k = |B_k|$
and consisting of $m_k\ge 2$ disjoint factions $F_{k1},\ldots,F_{km_k}$,
and we assume their sizes $N_{kj} = |F_{kj}|$ are all at least $2M$.
Each faction $F_{kj}$ has a distinct favourite option $x_{kj}$,
each block $B_k$ a potential partial consensus option $y_k$.
No other options exist.
Let $\ell_b(x_{kj}) = N_{kj} / N$ define the benchmark lottery
and $\ell_c(y_k) = N_k / N$ define the partial consensus lottery.
Each voter $i\in F_{kj}$ has a von Neumann--Morgenstern utility function with
$u_i(x_{kj}) = 1 > u_i(y_k) > N_{kj} / N_k$ and $u_i(z) = 0$ for all other options.
The expected utility for $i\in F_{kj}$ resulting from some ballot profile $\beta'$ under MaxParC is thus
$u_i(\beta') = M(\beta')(x_{kj}) + u_i(y_k) M(\beta')(y_k)$,
where $M(\beta')$ is the resulting lottery.

Now consider the following ``canonical'' ballot profile $\beta$:
For $i\in F_{kj}$,
$\beta_i(x_{kj}) = 100$, $100(N - N_k)/N < \beta_i{y_k} \le 100(1 + N - N_k)/N$, and $\beta_i(z) = 0$ for all other options.
Note that under MaxParC with this ballot profile,
each $i\in F_{kj}$ approves of $x_{kj}$ and $y_k$, hence ends up voting for $y_k$,
so that the resulting lottery is $\ell_c$ as desired.

Also assume that any group $G$ of voters can secretly plan to deviate from $\beta$,
leading to a modified profile $\beta'$ with $\beta'_i=\beta_i$ for all $i\notin G$,
but that no member of $G$ can be sure that the others will actually perform the deviation;
rather, any subgroup $H\subset G$ can secretly plan a further deviation $\beta''$ from $\beta'$,
with $\beta''_i=\beta'_i$ for all $i\notin H$.

Then we conjecture that with the above strategy profile $\beta$, 
if there is a group $G$ with a deviation $\beta'$ from $\beta$ that strictly profits all members
(i.e., $u_i(\beta') > u_i(\beta)$ for all $i\in G$),
there is a subgroup $H\subset G$ with a further deviation $\beta''$ from $\beta'$ that strictly profits all its members
(i.e., $u_i(\beta'') > u_i(\beta')$ for all $i\in H$)
and is strictly worse than $\beta$ for at least one member of $G$
(i.e., there is $i'\in G$ with $u_{i'}(\beta'') < u_{i'}(\beta)$).

\paragraph{Summary}

We have shown in this section that NL and MaxParC both support full and partial consensus in a number of archetypical decision situations,
whereas all other eight studied methods and all majoritarian methods do not.

\subsection{Results of simulation experiments}

We simulated $2M=$ 2,587,812 decision problems in total
and stored the resulting welfare, randomization, and satisfaction metrics of all ten methods
and the preferences over all method pairs.

To analyse voter behaviour during the interactive phase, we also stored
(i) the share of factional updates that led to a change in the faction's ballots (metric {\tt moverate}),
(ii) the share of trial-and-error updates that did not lead to a change in the voter's ballot (metric {\tt keeprate}),
and (ii) the share of problems in which the ballots after the interactive phase differed from before that phase (metric {\tt interactivechanged}).

Table \ref{tbl:compliance} gives an overview of all metrics' mean values.

In addition to univariate and bivariate statistics for all metrics, 
we also fitted an OLS generalized linear regression model for each metric $Y$,
separately for each preference model $U$,
using the following parameters as explanatory variables:
dummy variables for the voting method (using RV as reference method);
log-transformed numbers of voters ({\tt nvoters}), options ({\tt noptions}), and polling rounds ({\tt npolls});
shares of LCP ({\verb!rshare_LCP!}), HCP ({\verb!rshare_HCP!}), 
strategic ({\verb!sshare_S!}), trial-and-error ({\verb!sshare_T!}), heuristic ({\verb!sshare_H!}), and factional ({\verb!sshare_F!}) voters;
a dummy indicating whether the first option was a constructed compromise option ({\verb!with_compromise!});
and the parameters of the preference model ({\tt log(Bmr), Bmh, Bmiota} or {\tt dim, log(omega), rho}).
The regression analysis shows that the case number was large enough to distinguish
the influences of all explanatory variables since almost all estimated coefficients were significantly different from zero.

\subsubsection{Social welfare}

\paragraph{Absolute welfare metrics}

\begin{figure}\centering
\includegraphics[width=0.3\textwidth]{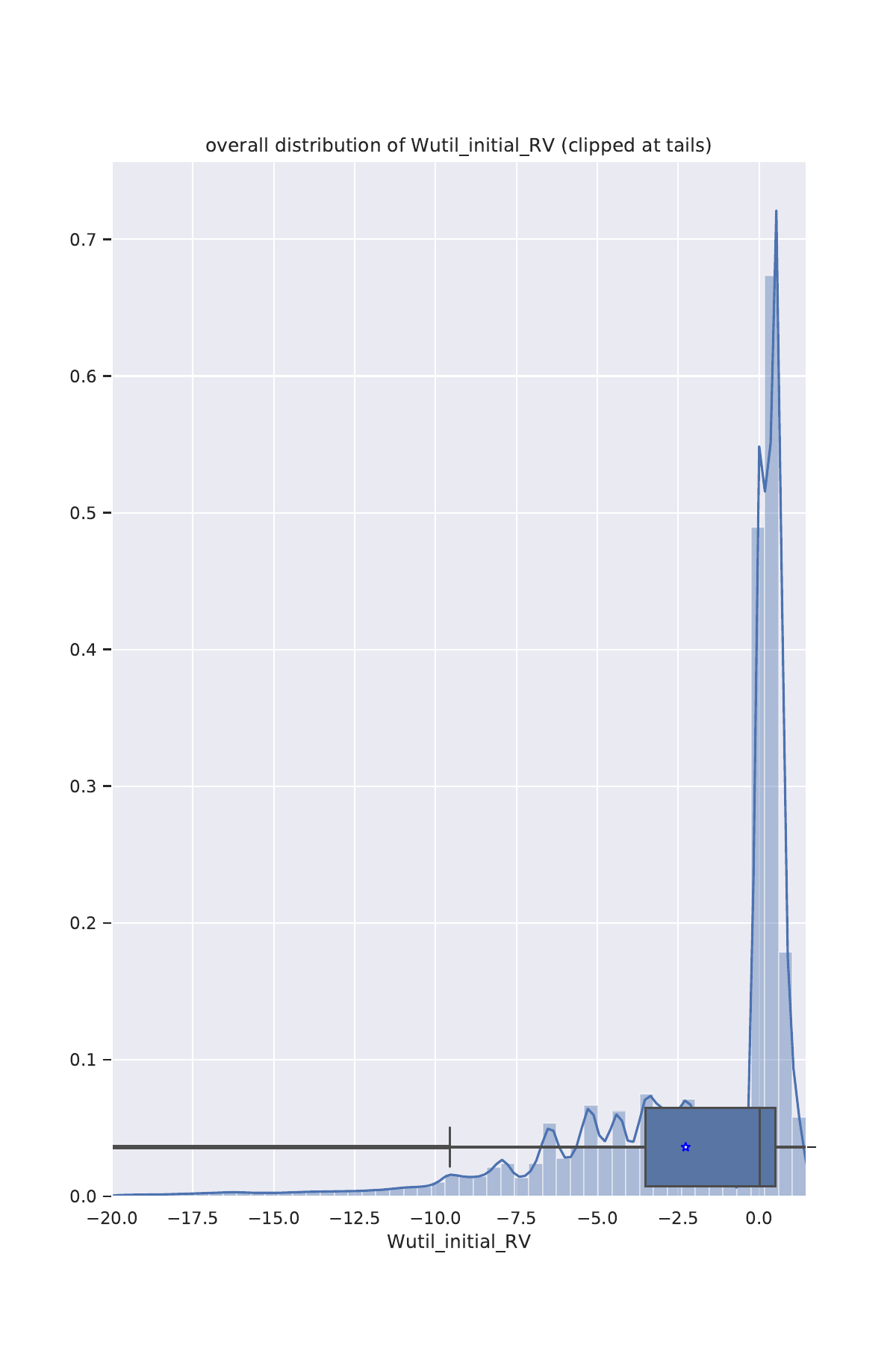}
\includegraphics[width=0.3\textwidth]{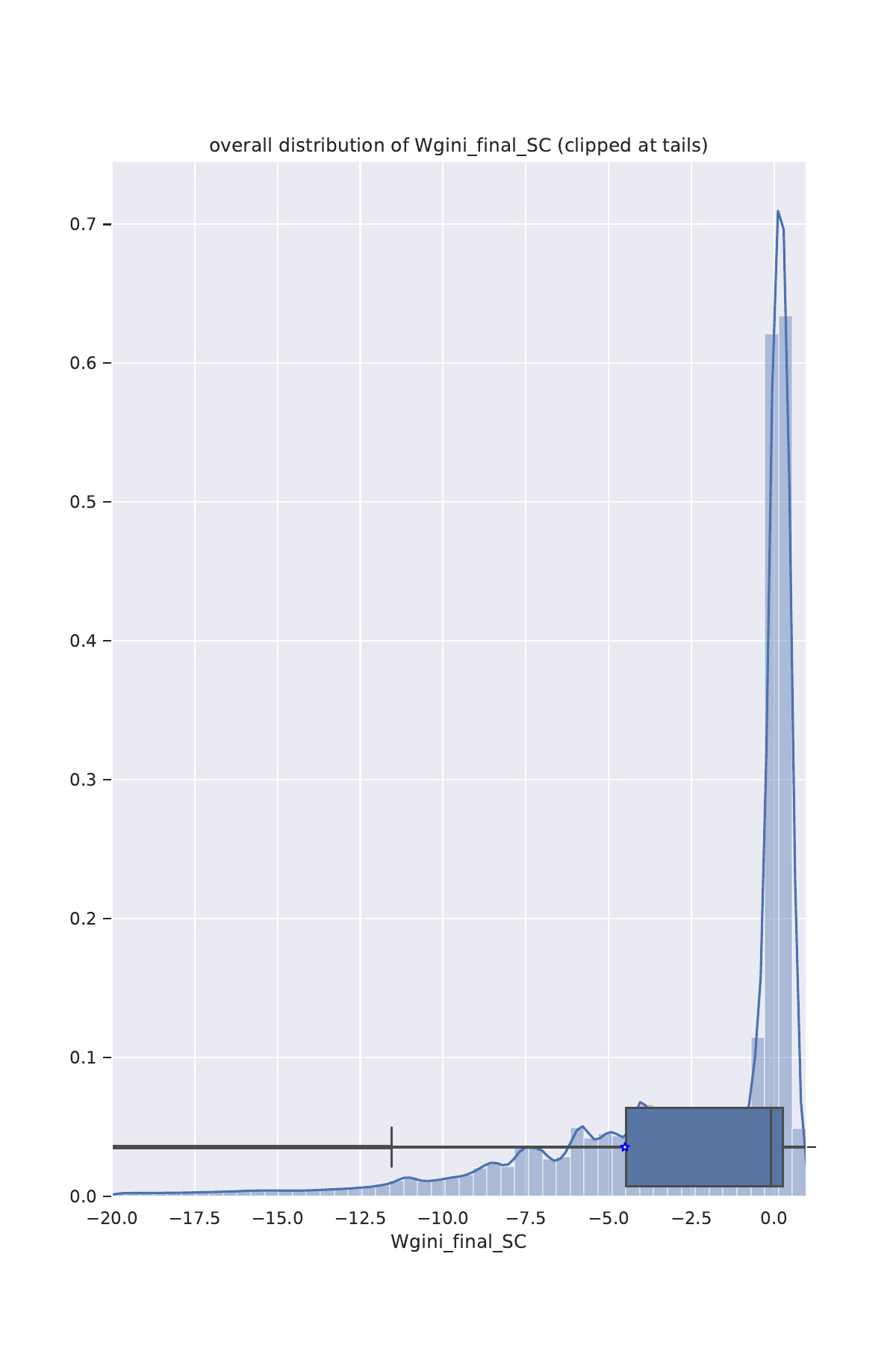}
\includegraphics[width=0.3\textwidth]{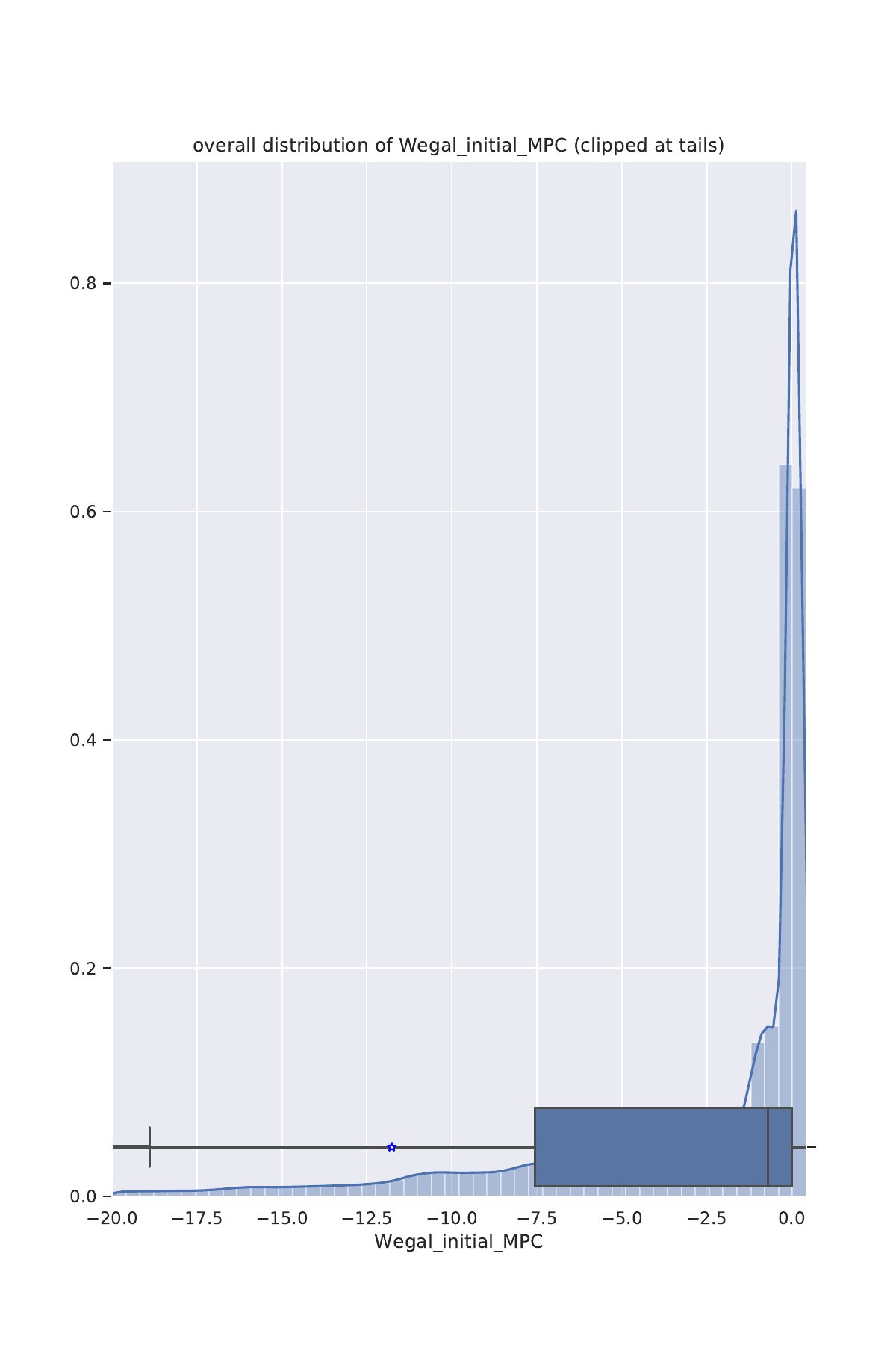}\\
\includegraphics[width=0.3\textwidth]{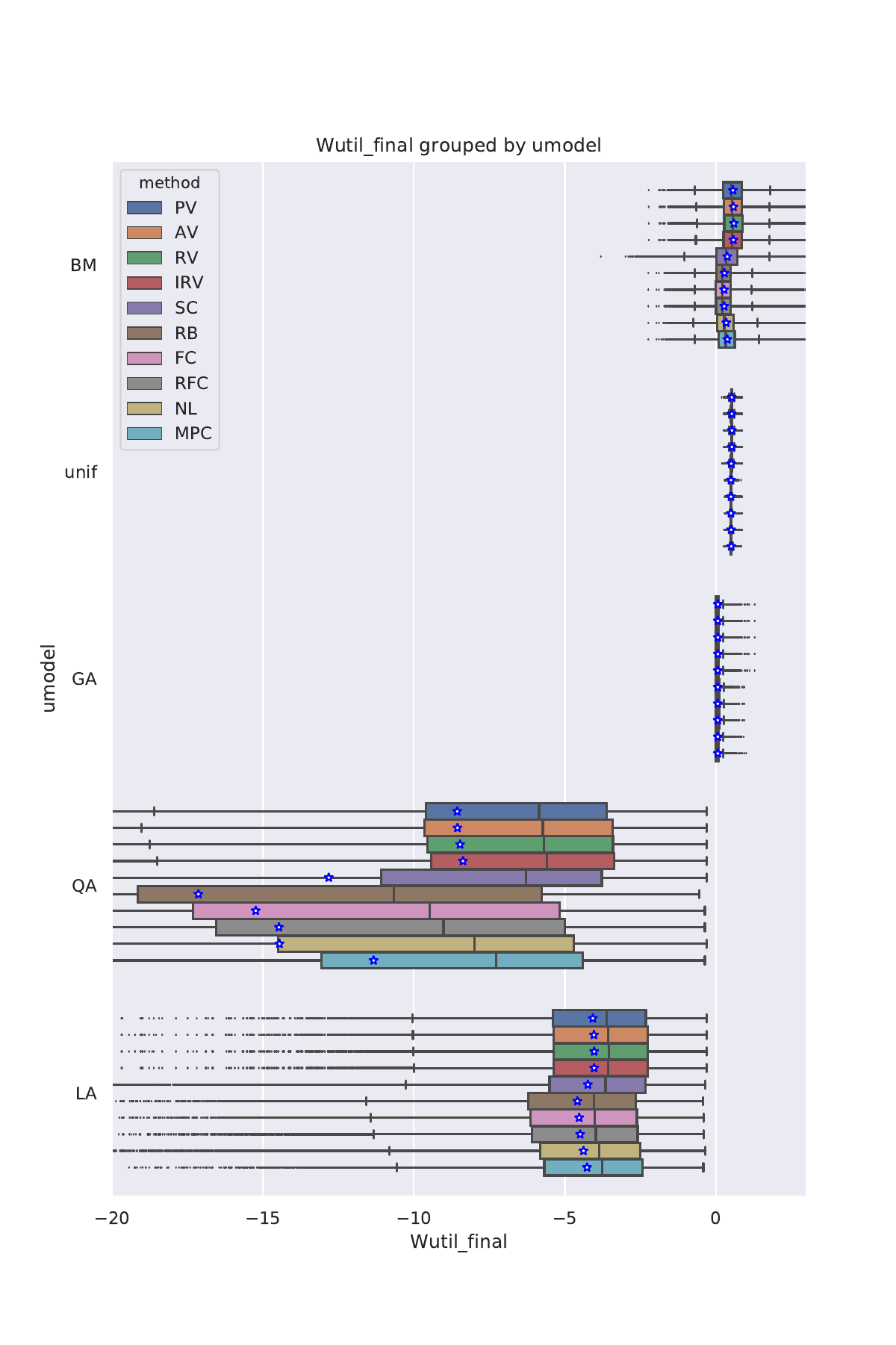}
\includegraphics[width=0.3\textwidth]{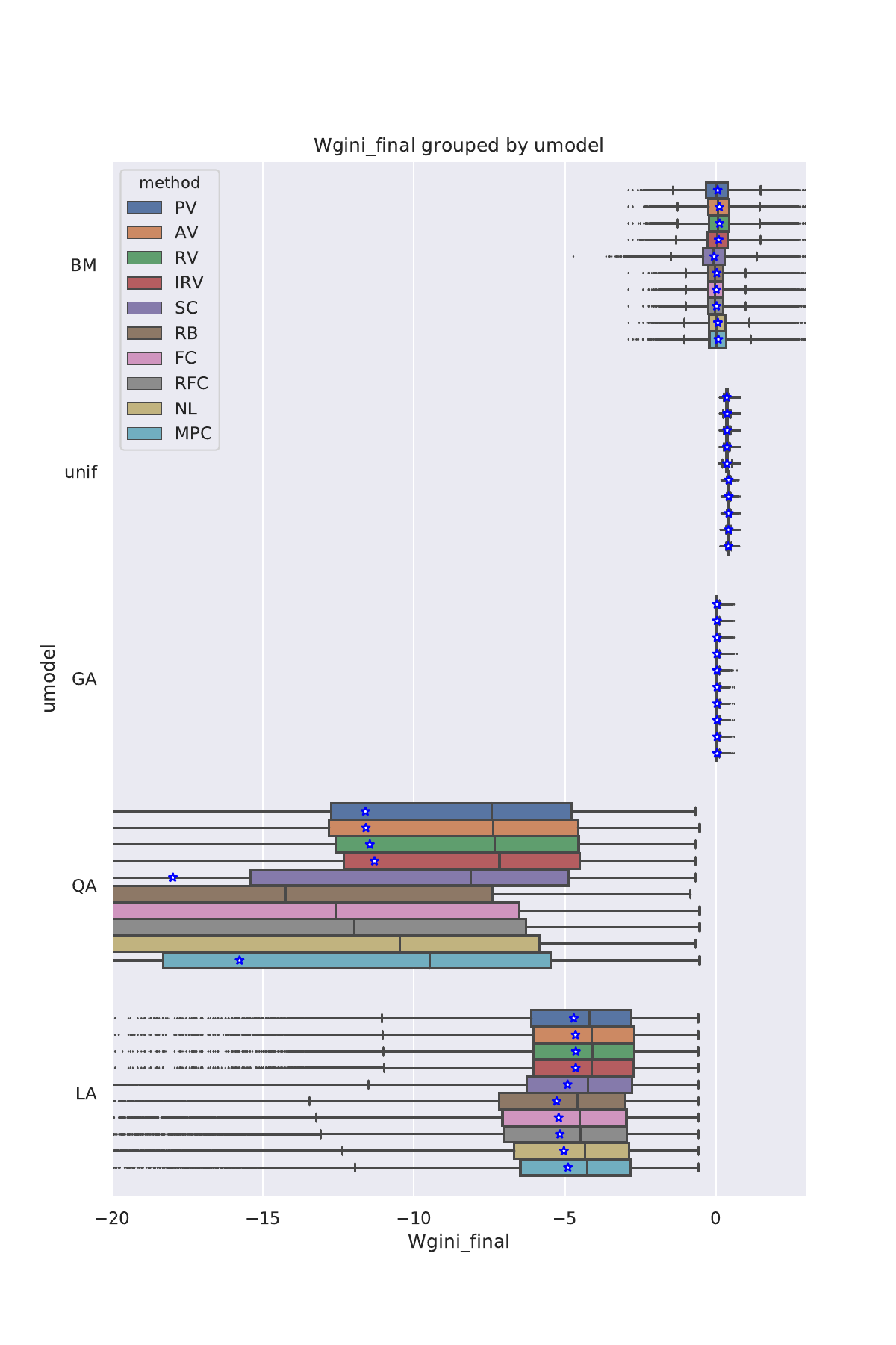}
\includegraphics[width=0.3\textwidth]{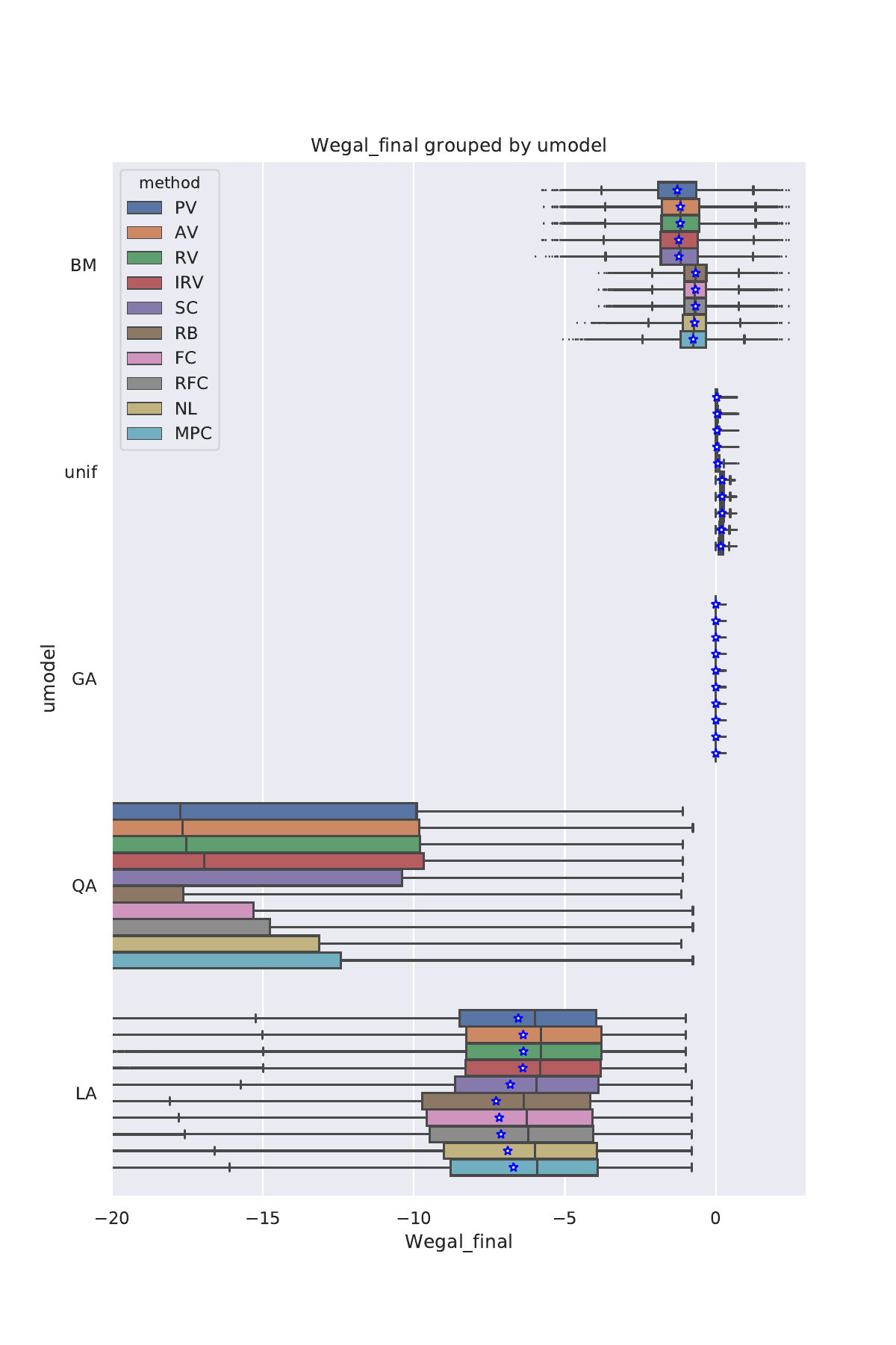}
\caption{\label{fig:abs_welf}%
Distribution of absolute social welfare across decision problems. 
Top: histograms, kernel density estimators, and boxplots with means for three example metrics/methods.
Bottom: distribution of final absolute welfare by preference model and method.
}\end{figure}

All six absolute welfare metrics 
({\verb!Wutil_initial!, \verb!Wutil_final!, \verb!Wgini_initial!, \verb!Wgini_final!, \verb!Wegal_initial!, \verb!Wegal_final!})
had considerably left-skewed distributions across problems.
When distinguishing by preference model ({\tt umodel}), one can see 
that this is due to the spatial preference models, 
and that their location depends strongly on the preference model (Fig.\ \ref{fig:abs_welf}).

In the block ({\tt BM}) and uniform ({\tt unif}) preference models, 
the majoritarian methods ({\tt PV, AV, RV, IRV, SC}) generated slightly larger utilitarian 
and slightly smaller egalitarian absolute welfare than the proportional methods ({\tt RB, FC, RFC, NL, MPC}),
being roughly equivalent on the intermediate Gini-Sen welfare metric.
In the QA and LA models, the majoritarian methods also outperformed the proportional ones 
in the Gini-Sen and egalitarian absolute welfare metrics,
most significantly in the QA model, less so in the LA model.
In the GA model, the differences between methods were still statistically significant (e.g., Tbl.\ \ref{tbl:regr:Wgini_final_GA}) 
but negligible in comparison to the overall dispersion of welfare across problems.

Throughout the regression models, 
more voter blocks (larger {\tt BMr}), 
larger policy-space dimension ({\tt dim}),
and larger spatial voter heterogeneity ({\tt omega}) 
decreased welfare,
and so did a larger number of voters except in the {\tt Wgini/GA} case.
More options 
and larger block size heterogeneity ({\tt BMh}) 
increased welfare. 
Larger individuality ({\tt Bmiota}) increased utilitarian but decreased Gini-Sen and egalitarian welfare.
For the spatial option broadness parameter ({\tt rho}) and shares of non-EUT voters, there was no clear pattern.
More pre-voting polling rounds had a statistically significant but very small negative influence.

Larger shares of sincere, trial-and-error, and factional voters
and lower shares of lazy voters tended to increase welfare,
the share of heuristic voters had no clear influence.
Surprisingly, 
adding a constructed compromise option only increased welfare in the spatial models
though not in the block and uniform models.

\paragraph{Relative welfare metrics}

\begin{figure}\centering
\includegraphics[width=0.3\textwidth]{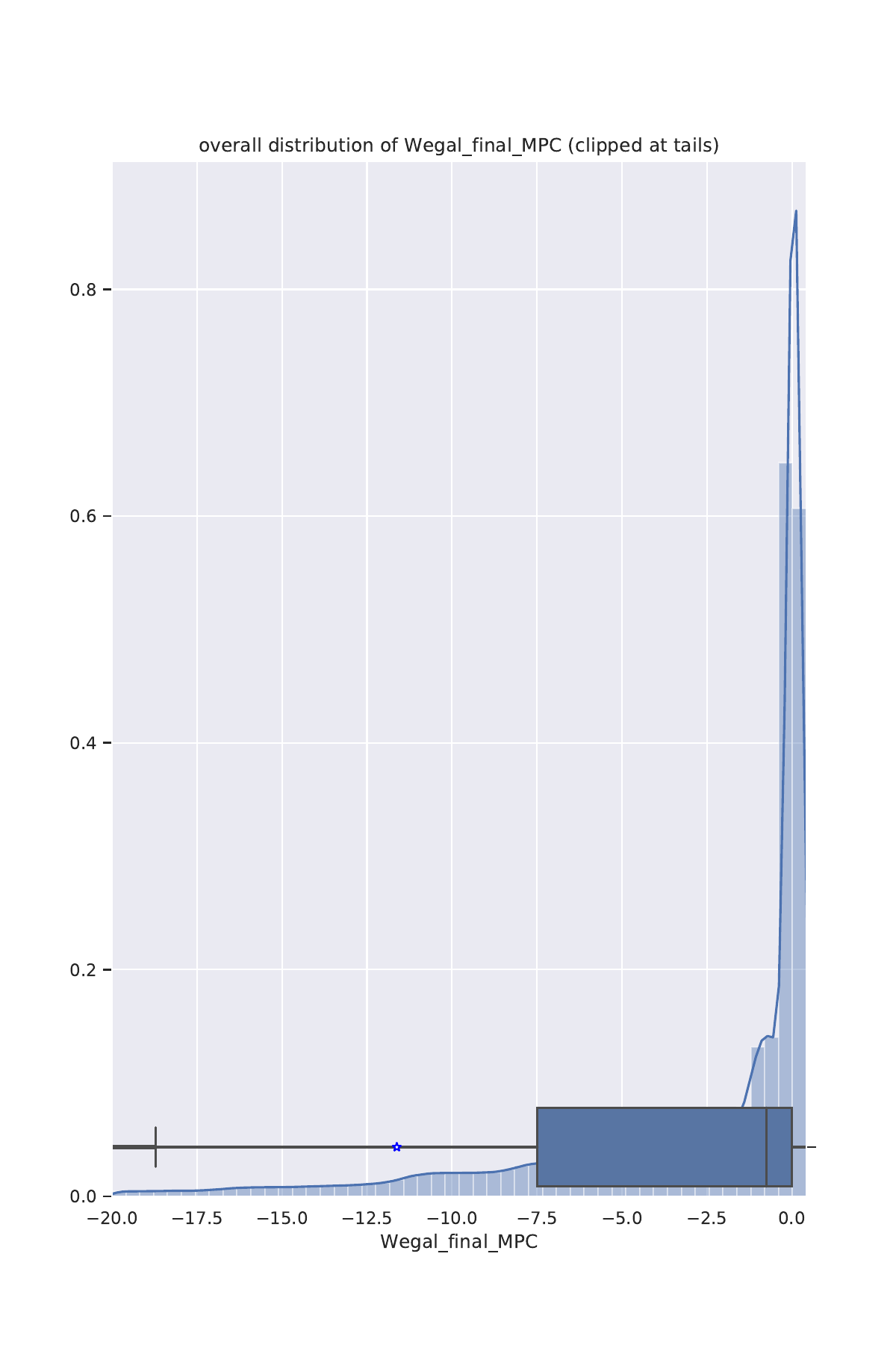}
\includegraphics[width=0.3\textwidth]{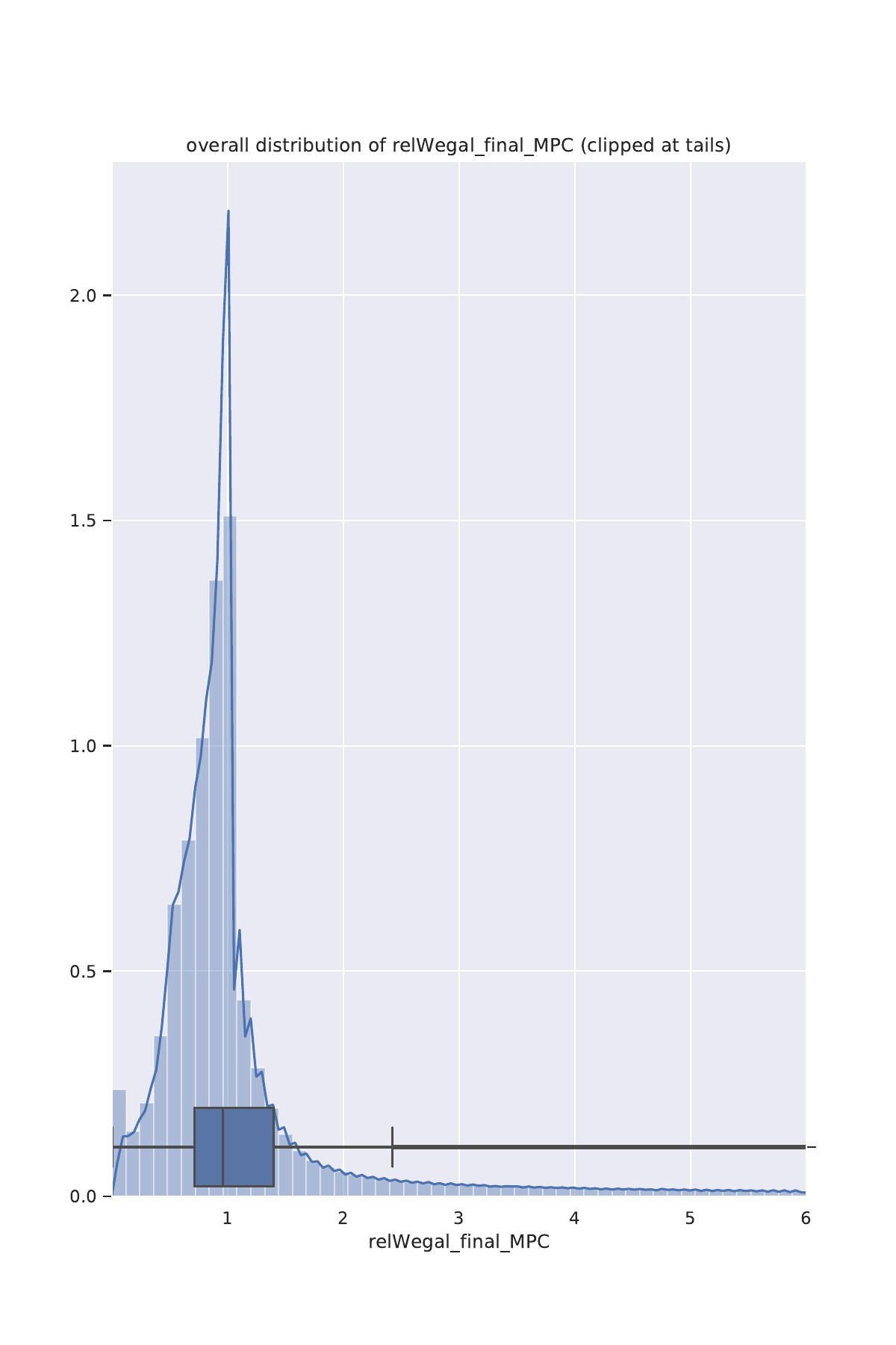}
\includegraphics[width=0.3\textwidth]{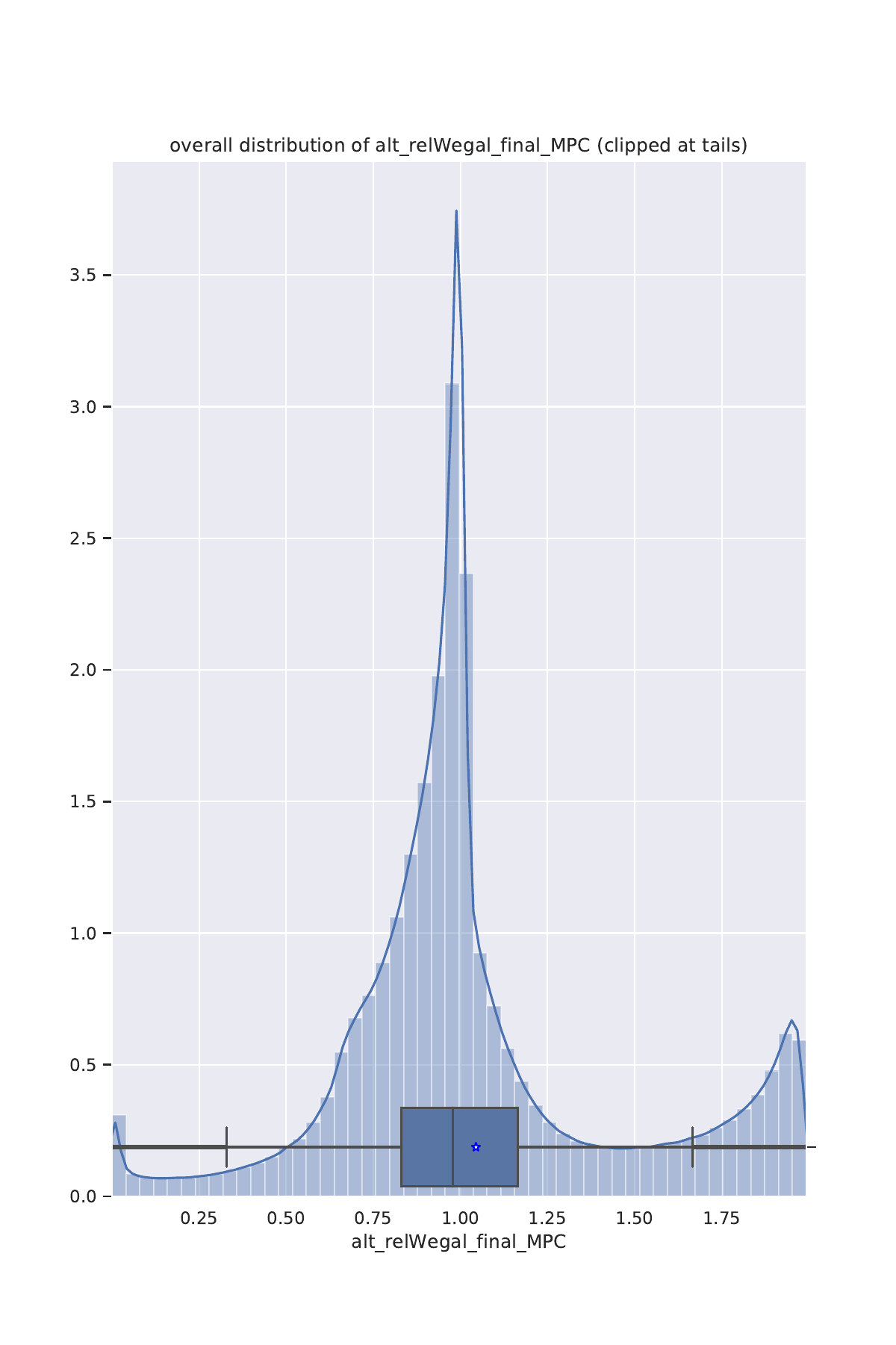}\\
\includegraphics[width=0.3\textwidth]{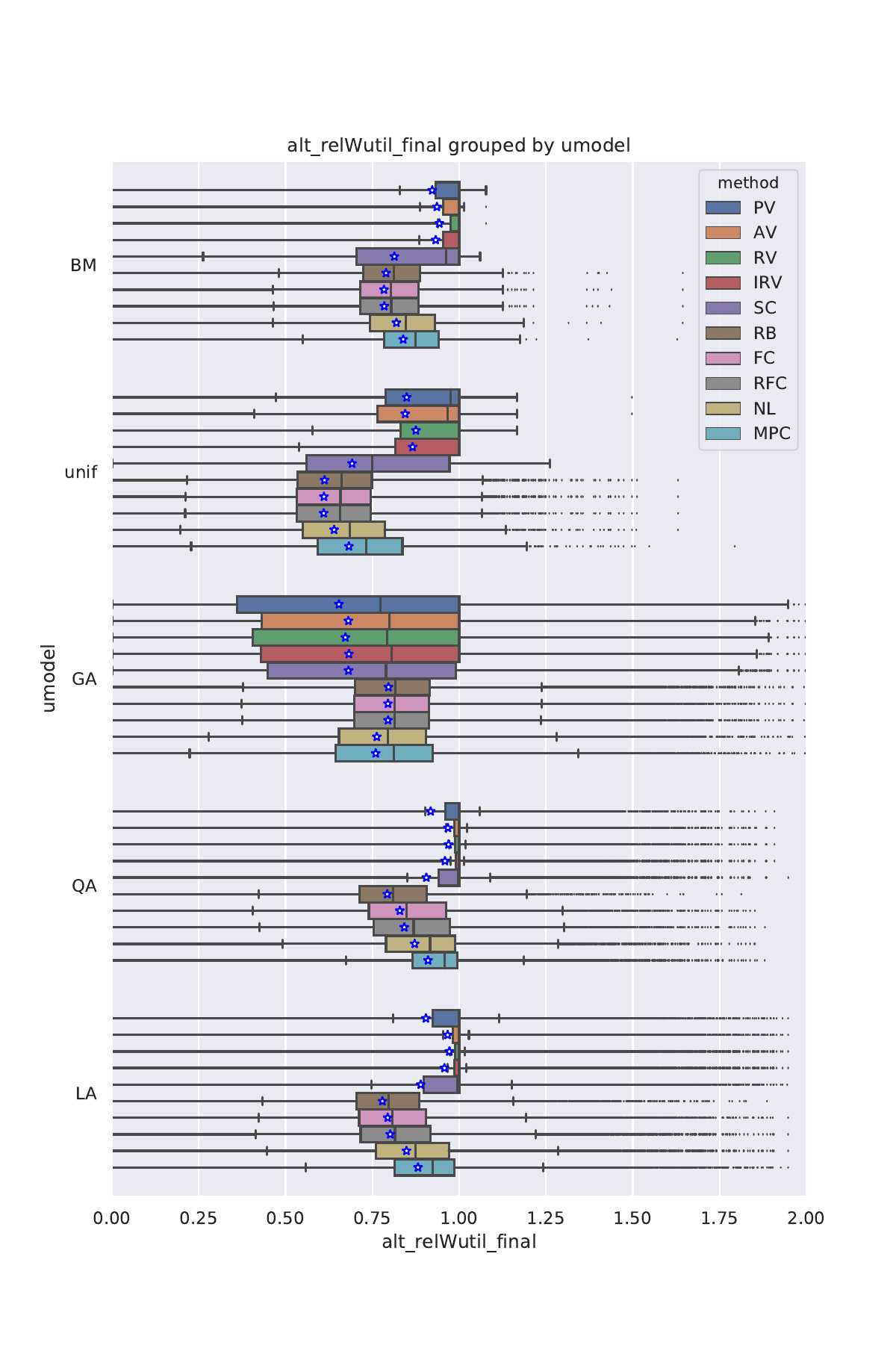}
\includegraphics[width=0.3\textwidth]{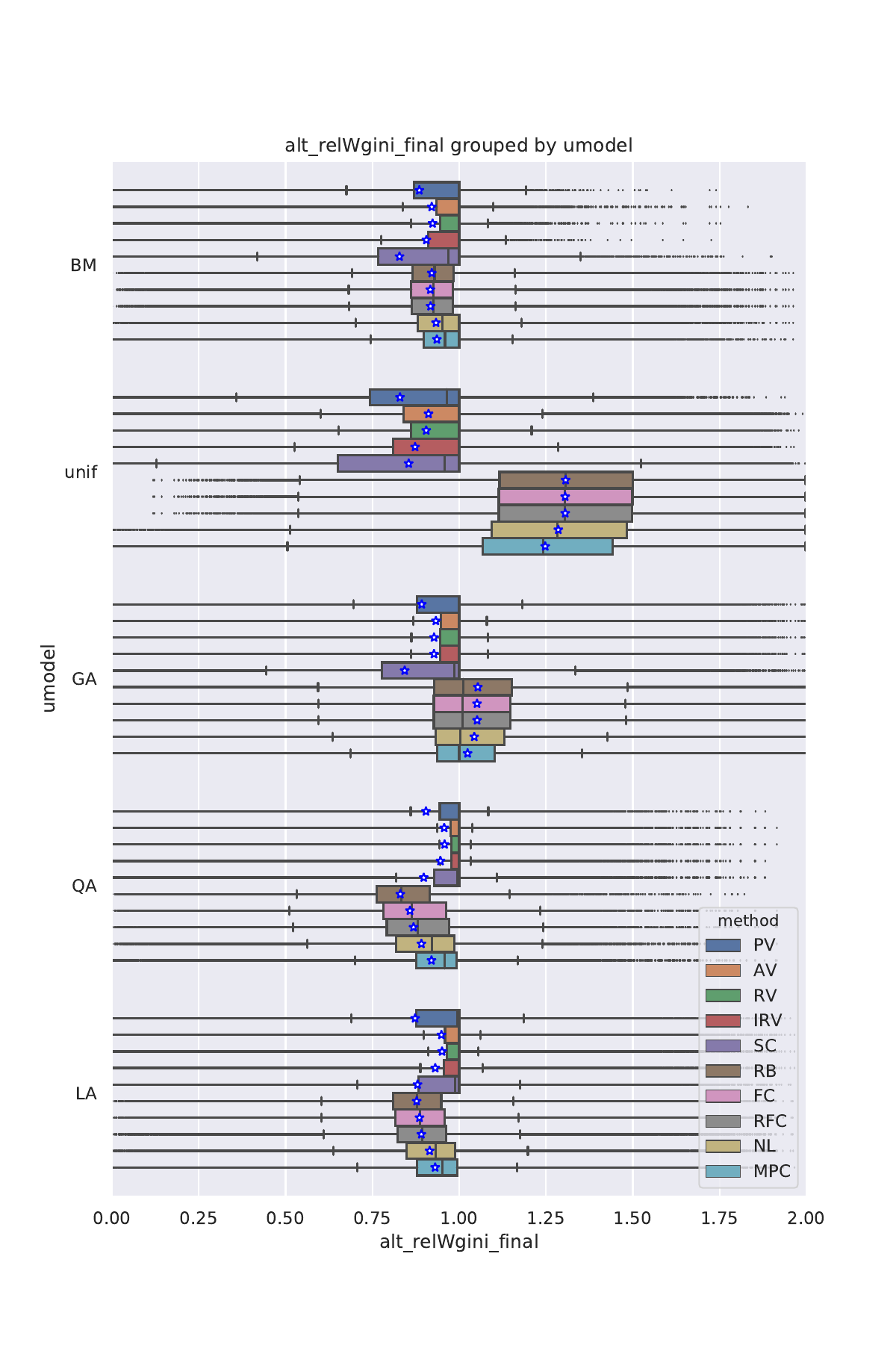}
\includegraphics[width=0.3\textwidth]{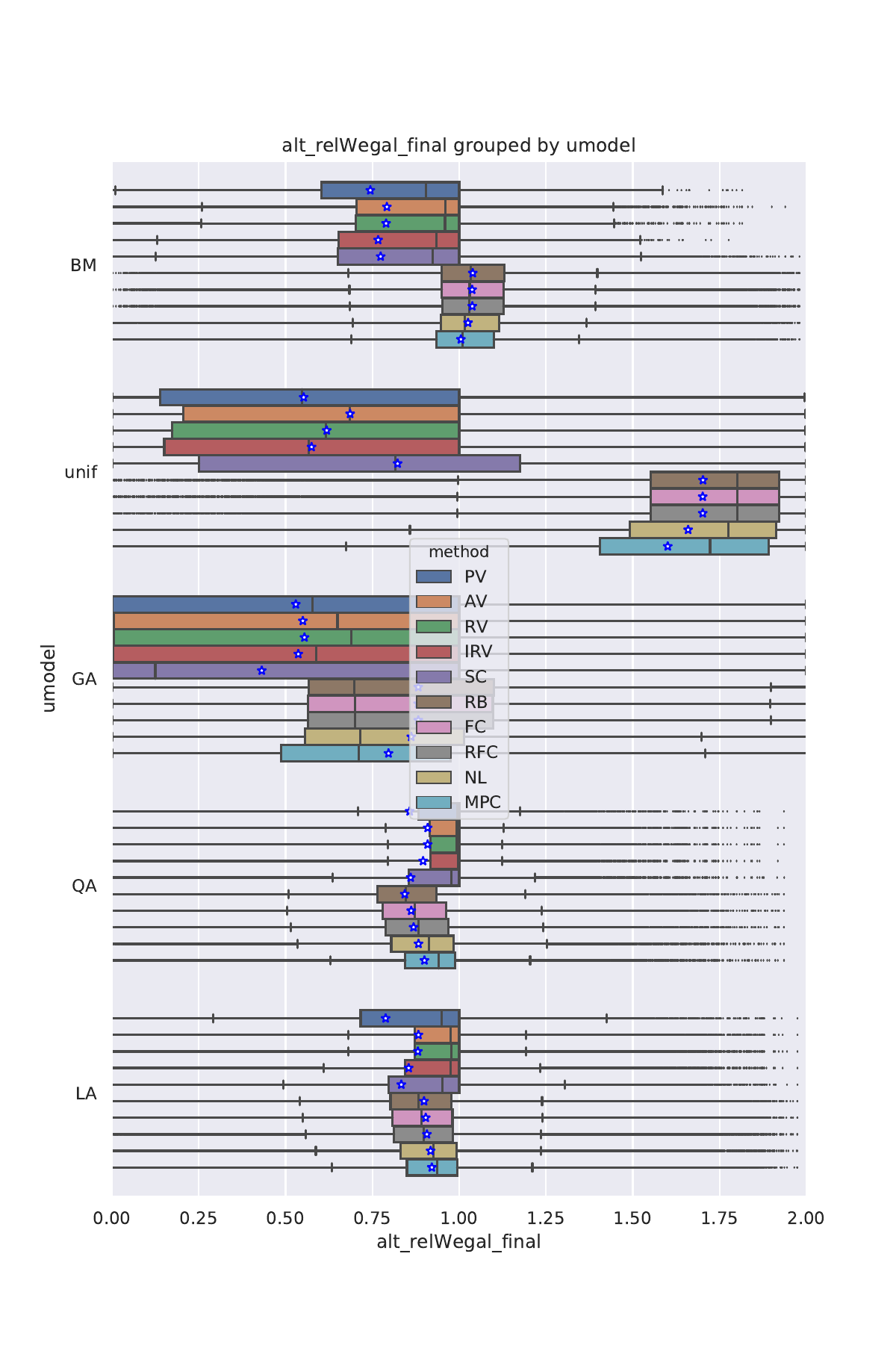}
\caption{\label{fig:rel_welf}%
Distribution of relative social welfare across decision problems. 
Top: comparison of the two versions of relative welfare with absolute welfare for one combination of metric and method.
Bottom: distribution of final alternative relative welfare metrics by preference model and method.
}\end{figure}

Because of the skewed distributions of absolute welfare,
we also analysed the two versions of relative welfare metrics,
Since the first of these was skewed in the other direction in case of the nondeterministic methods and inequality-averse metrics,
we focus on the second, alternative version of relative welfare metrics here, which were much more balanced (Fig.\,\ref{fig:rel_welf}).

Interestingly, in the uniform preference model, 
{\verb!alt_relWgini!} and {\verb!alt_relWegal!} had a particular trimodal distribution
for the proportional methods,
which performed similarly to the deterministic methods in most cases,
but much better in a somewhat smaller cluster of cases
and much worse in a still smaller cluster of cases.

Looking at the fairly inequality-averse Gini-Sen welfare metric in its ``middle'' version {\verb!alt_relWgini!} 
more closely in a regression analysis, we see that this metric typically
increased with
the no.\ of options (except for the uniform preference model);
it decreased with
the no.\ of polling rounds,
the share of lazy voters;
the addition of a constructed compromise option,
the risk attitude scenario,
and the spatial broadness parameter
had no consistent influence across preference models.
In contrast to their influence on the absolute welfare metric {\verb!Wgini!},
a larger no.\ of voter blocks {\tt BMr}, 
individuality {\tt BMiota},
policy space dimension {\tt dim}
increased {\verb!alt_relWgini!}.

\begin{figure}\centering
\includegraphics[width=0.4\textwidth]{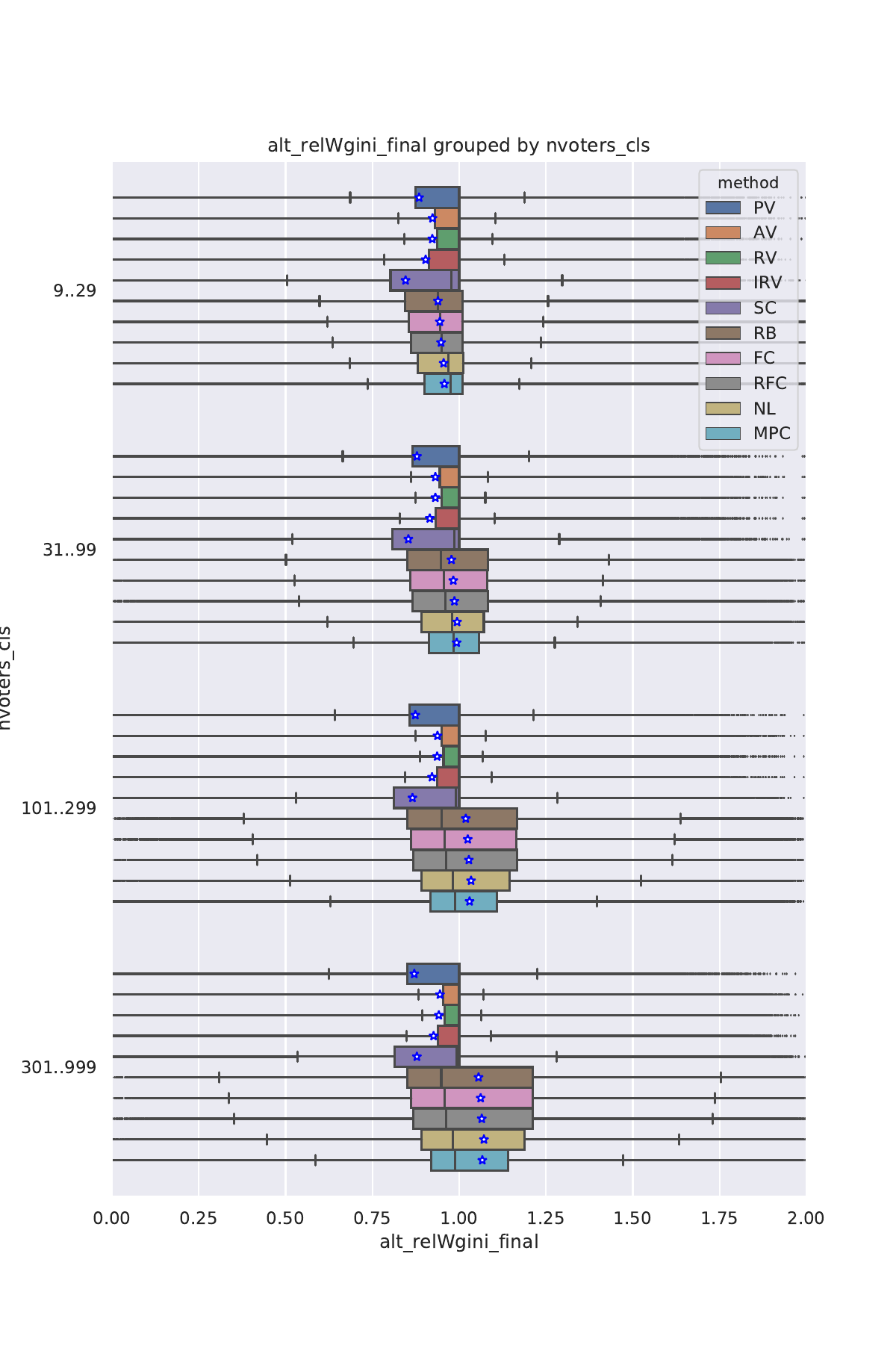}
\includegraphics[width=0.4\textwidth]{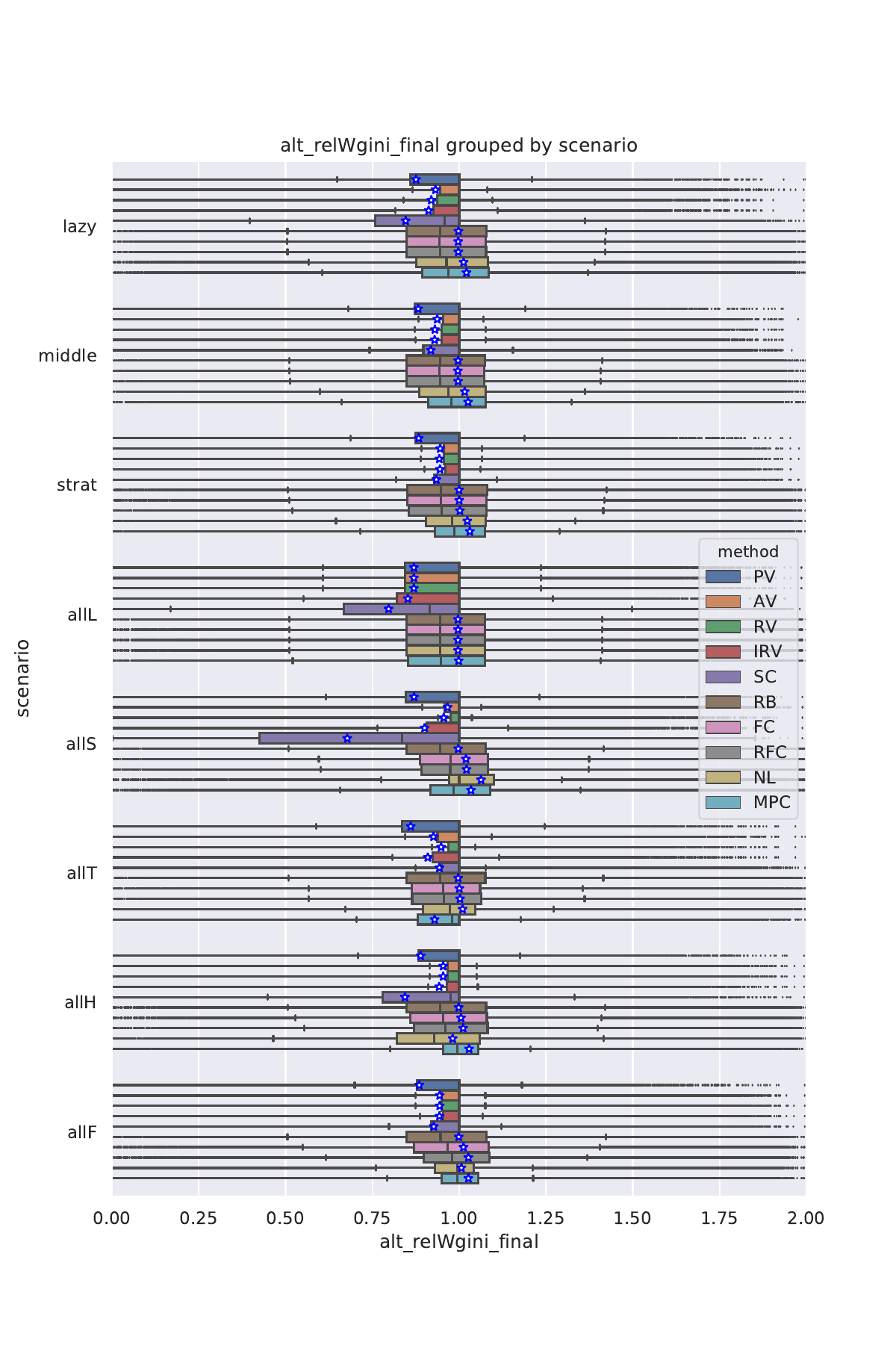}\\
\includegraphics[width=0.4\textwidth]{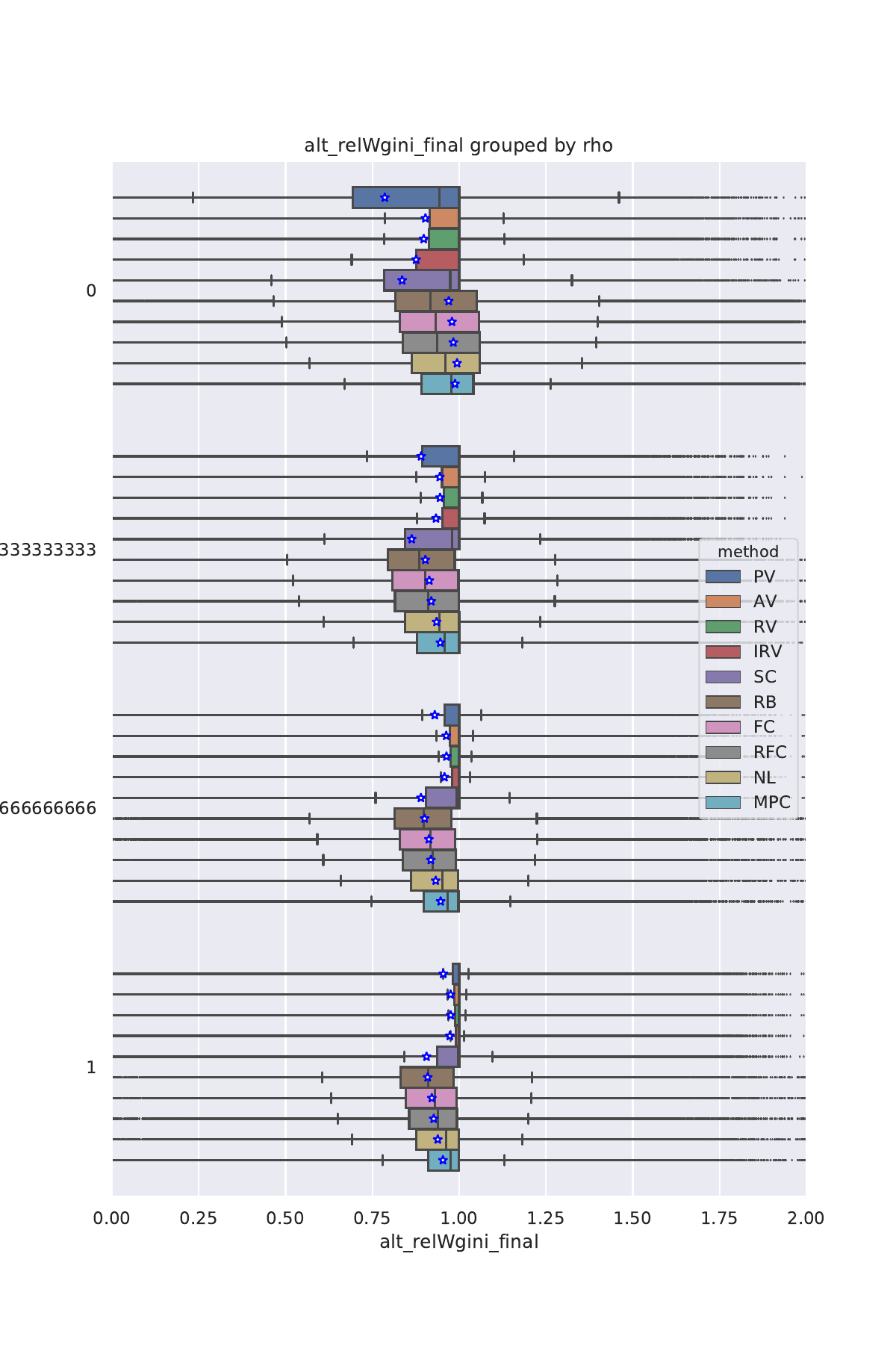}
\includegraphics[width=0.4\textwidth]{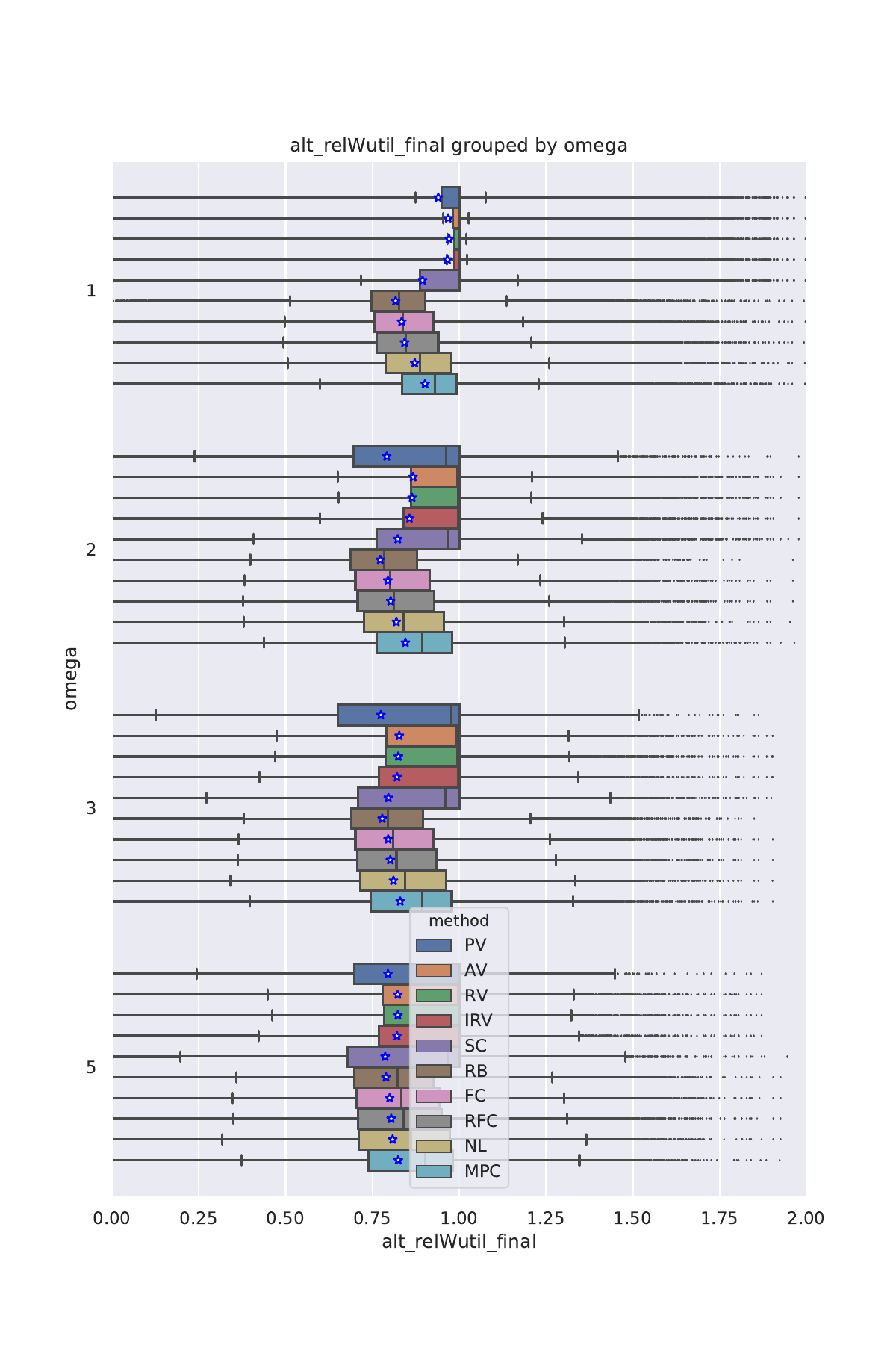}
\caption{\label{fig:rel_welf2}%
Distribution of relative social welfare across decision problems. 
Specific influences of no.\ of voters, behavioural type scenario, 
spatial option broadness, and spatial voter heterogeneity.
}\end{figure}

In addition, the grouped boxplots in Fig.\,\ref{fig:rel_welf2} show some specific influences.
A larger no.\ of voters increased {\verb!alt_relWgini!} considerably for the proportional, but not for the majoritarian methods.
The behavioural type scenario influenced {\verb!alt_relWgini!} under the Simple Condorcet method much more than the other methods:
interestingly, the all-sincere scenario produced much less welfare than the all-lazy one.
Larger spatial option broadness {\tt rho} had opposite effects 
for the majoritarian methods (increasing {\verb!alt_relWgini!})
and the proportional methods (decreasing {\verb!alt_relWgini!}):
when options had narrow appeal (or candidates a narrow platform), 
proportional methods performed considerably better,
with broadly appealing options (or platforms) it was the other way around.
Finally, {\verb!alt_relWutil!} decreased with increasing spatial voter heterogeneity {\tt omega}
more strongly for majoritarian methods, their clearest advantage mostly restricted to narrow spatial voter distributions.

\paragraph{Frequency of best-performing methods}

\begin{figure}\centering
\def\doplot#1{\includegraphics[width=0.25\textwidth,trim=0 40 80 40,clip]{plots/2020_03_28_1mio_all_#1.png}}
\begin{tabular}{rccc}
{\tt umodel}
    &\doplot{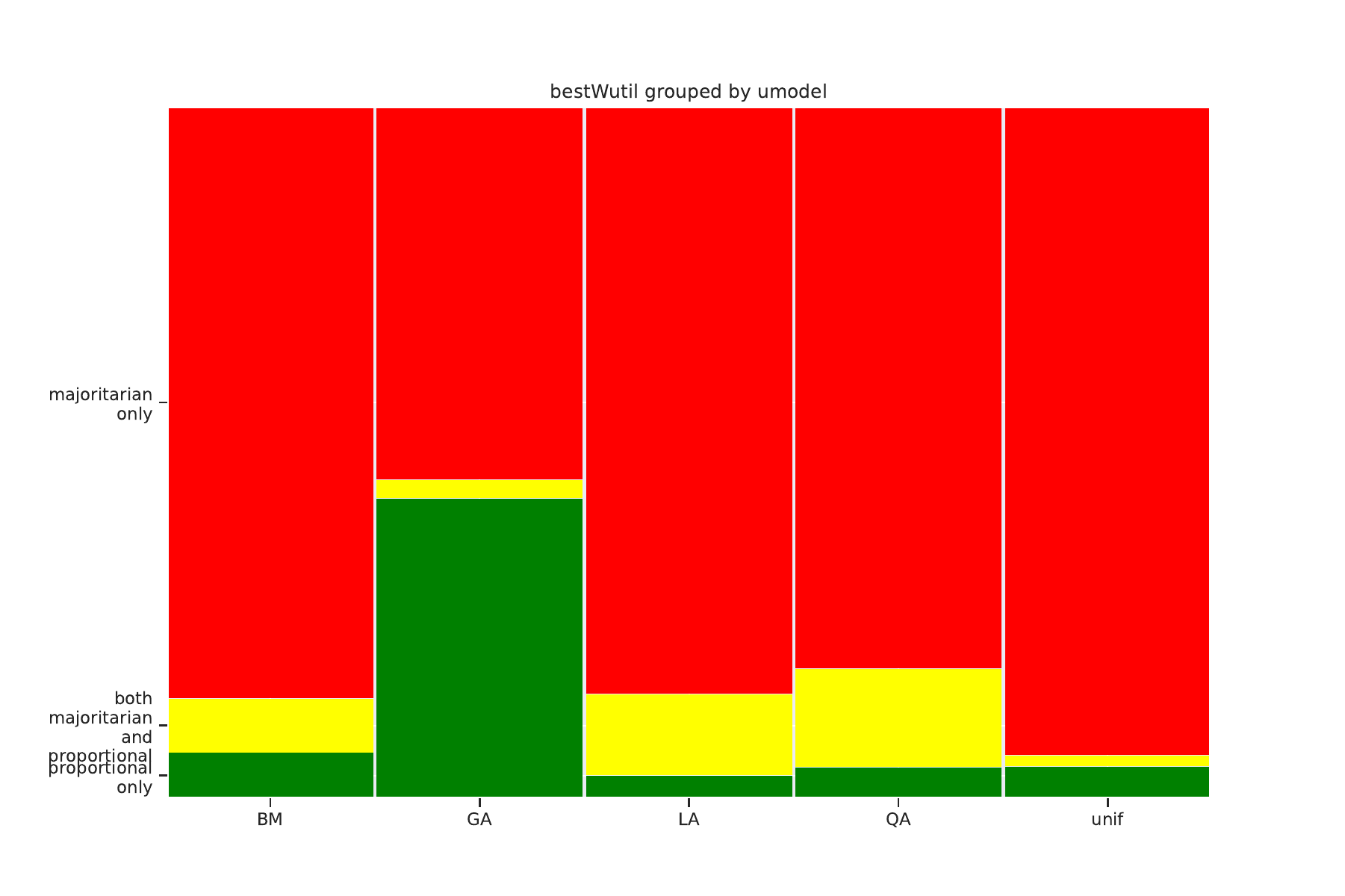}
    &\doplot{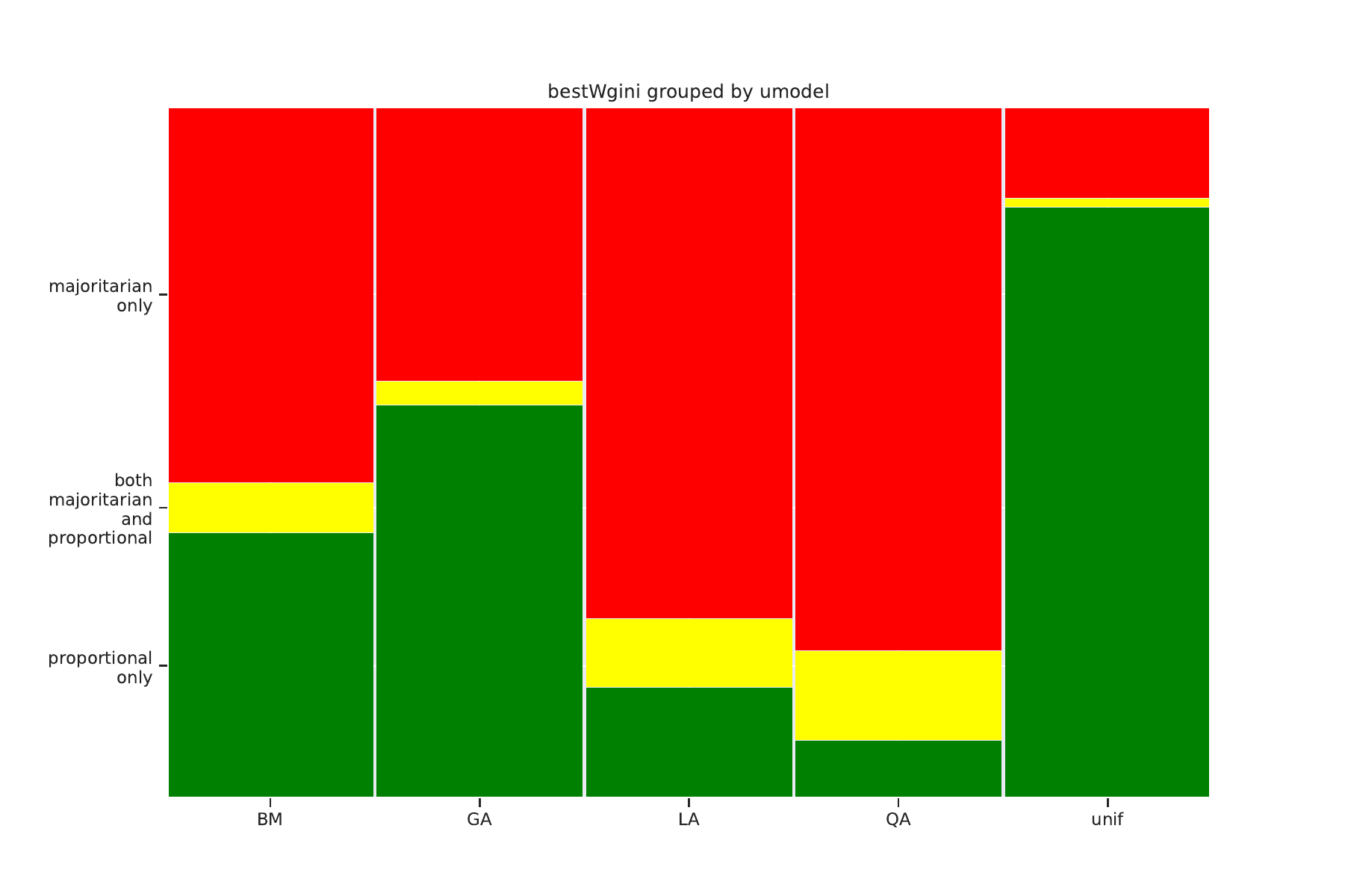}
    &\doplot{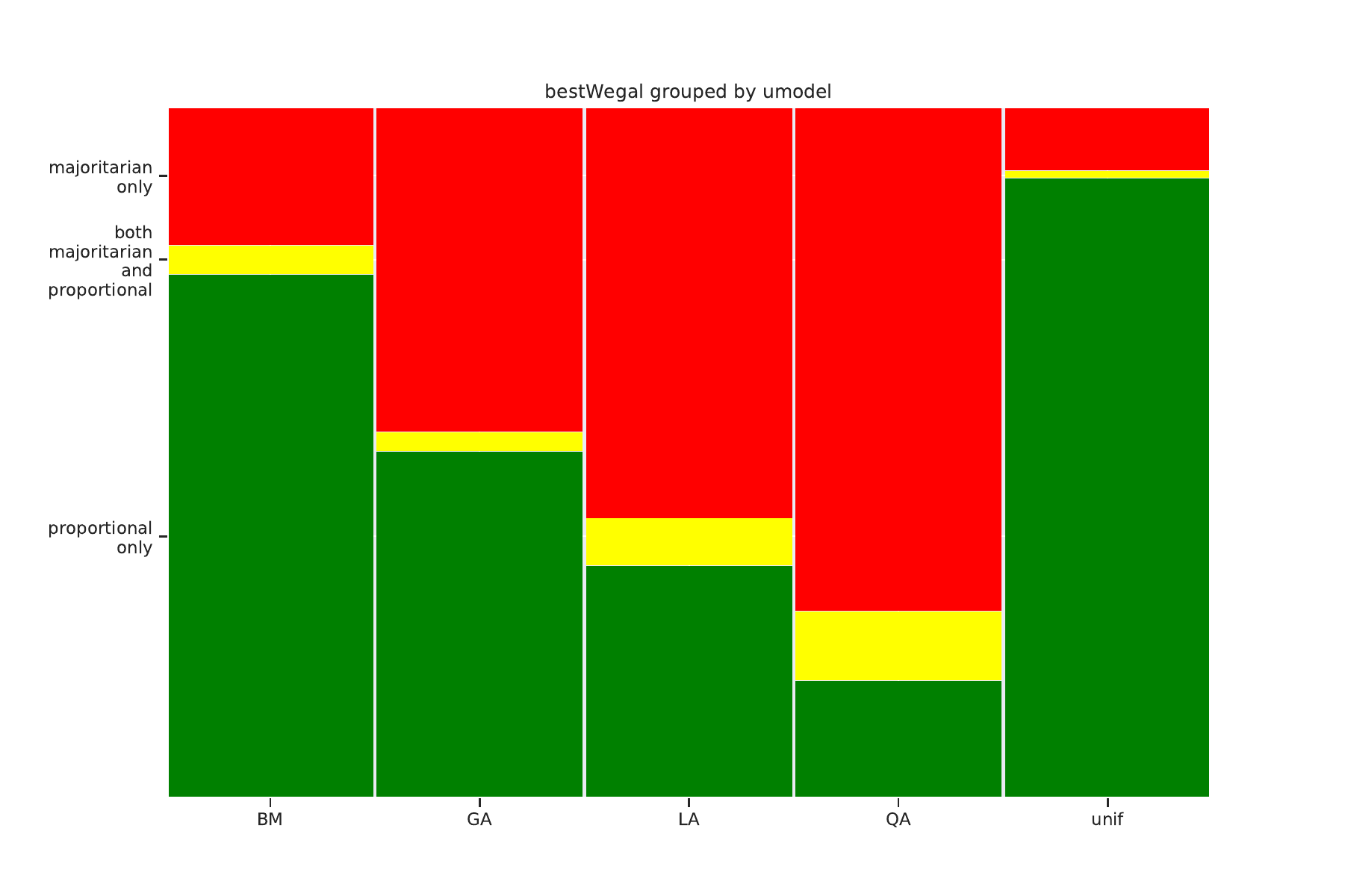}\\
{\tt BMr}
    &\doplot{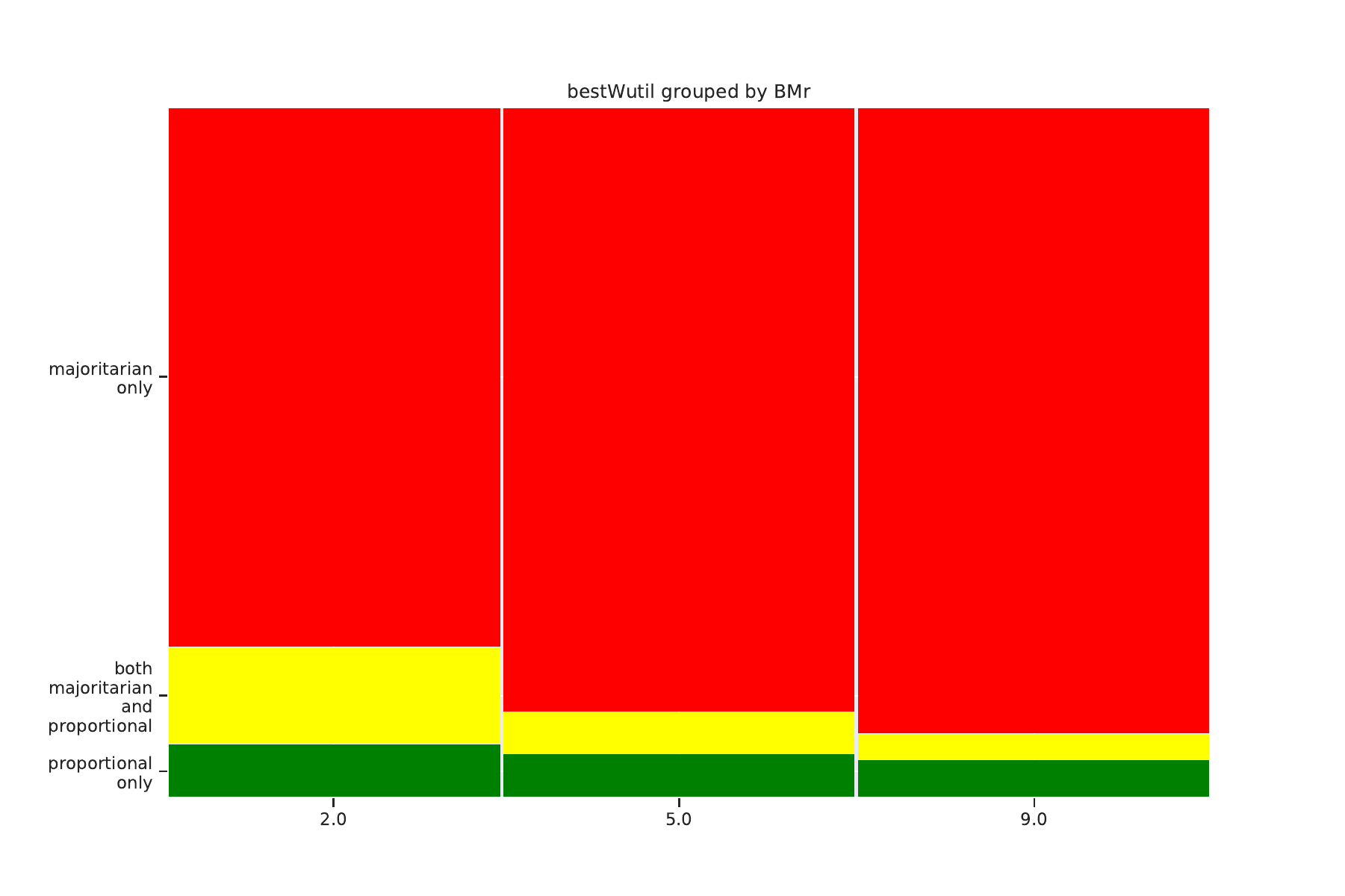}
    &\doplot{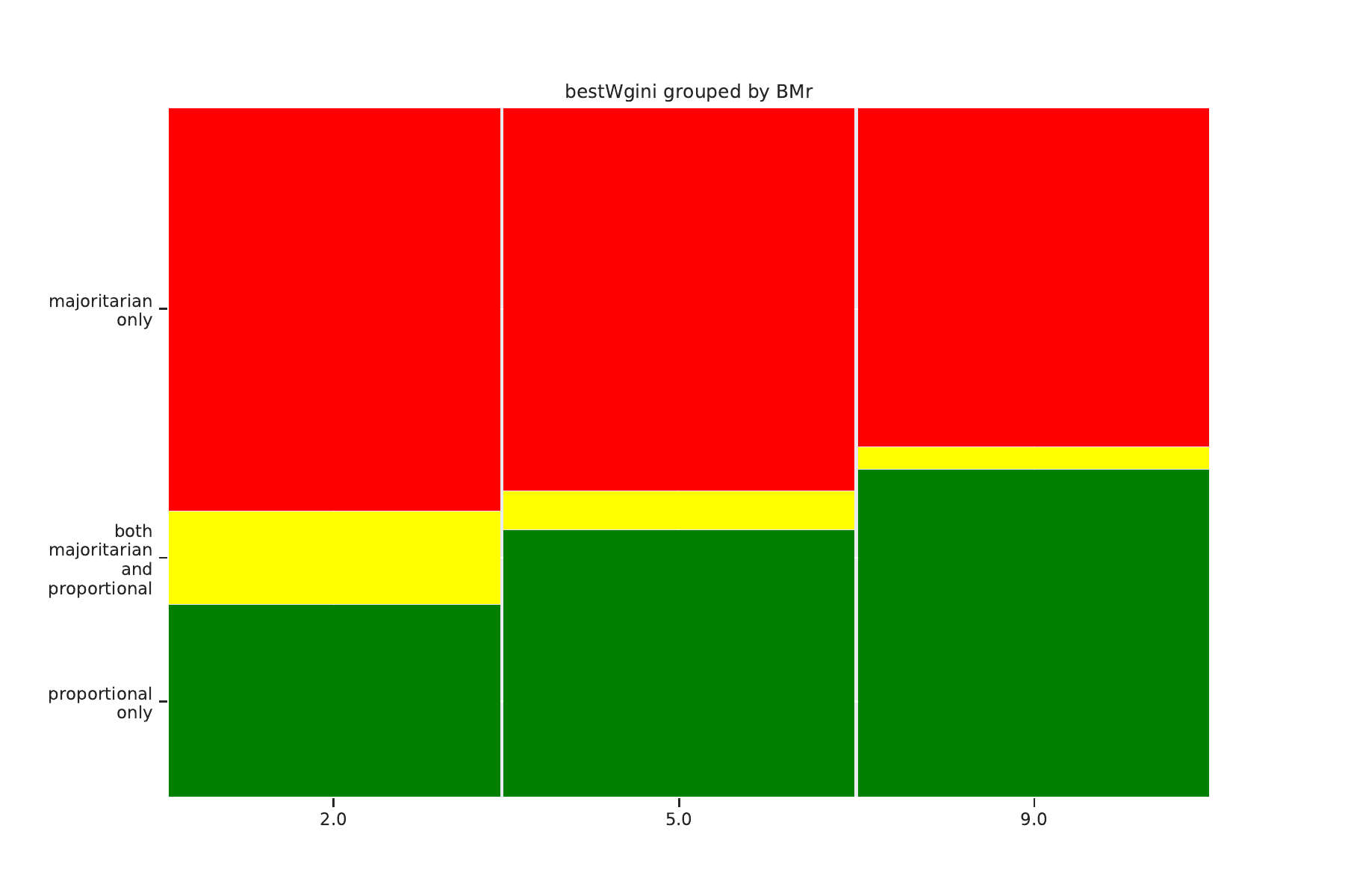}
    &\doplot{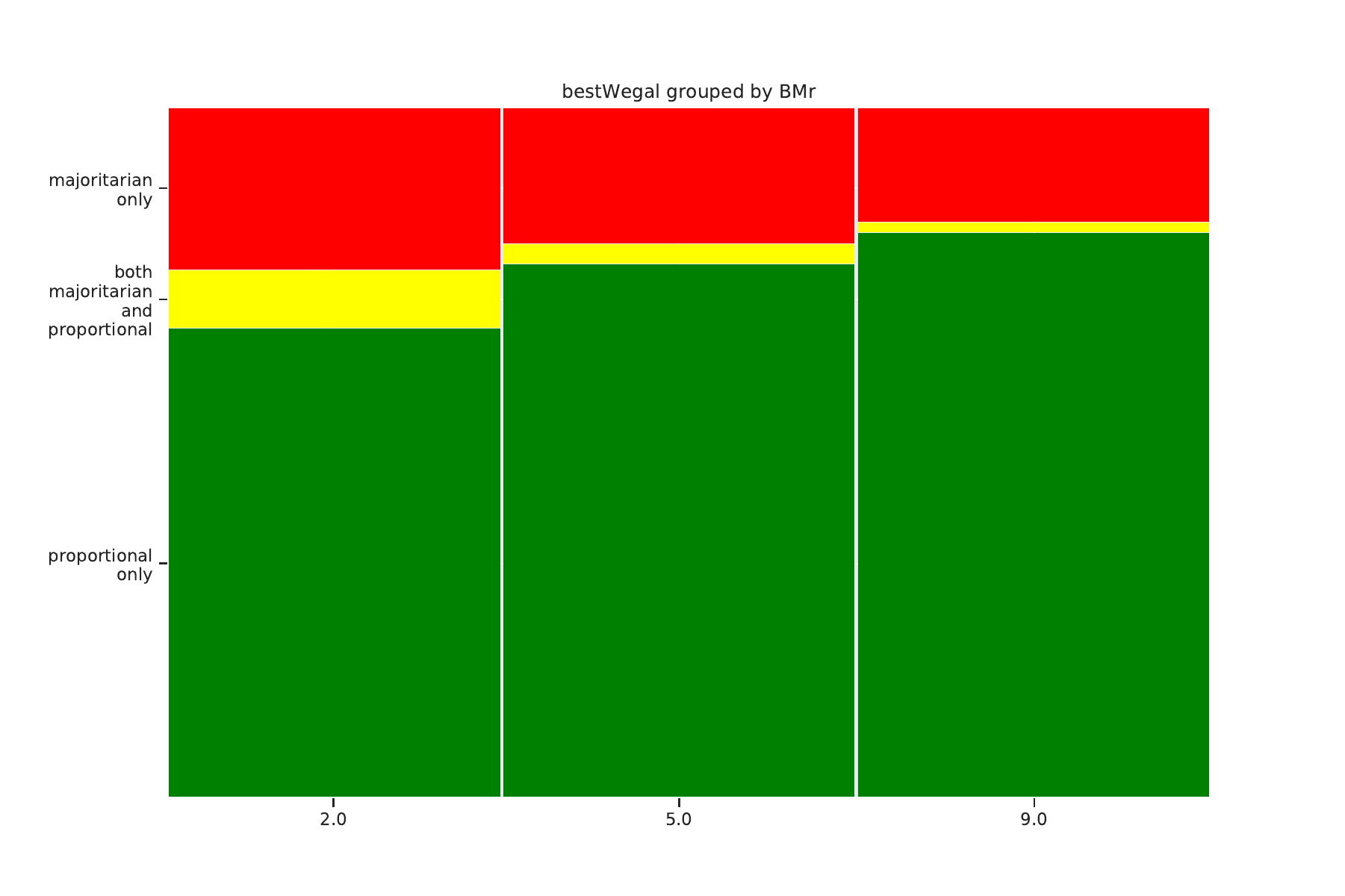}\\
{\tt BMiota}
    &\doplot{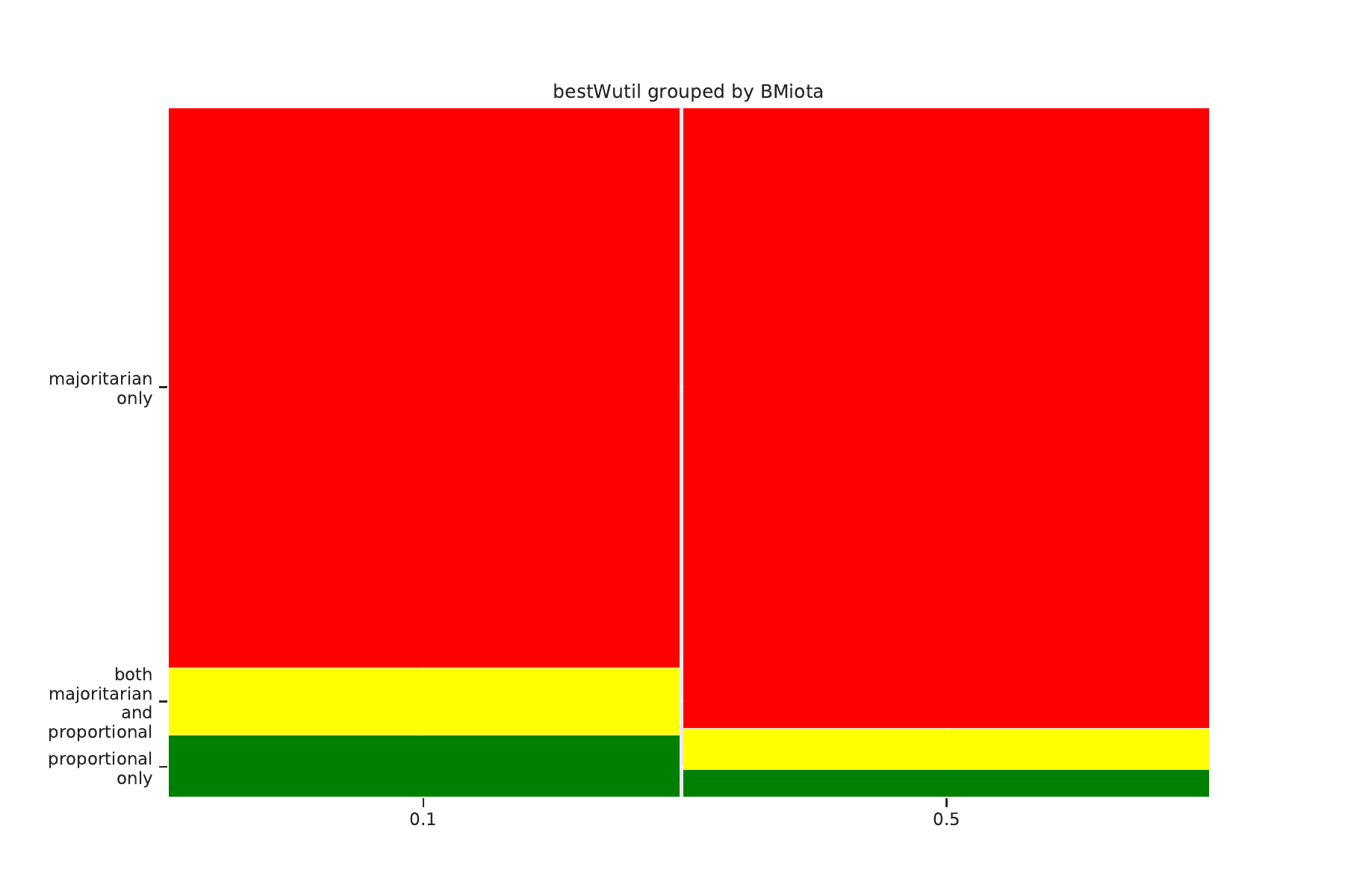}
    &\doplot{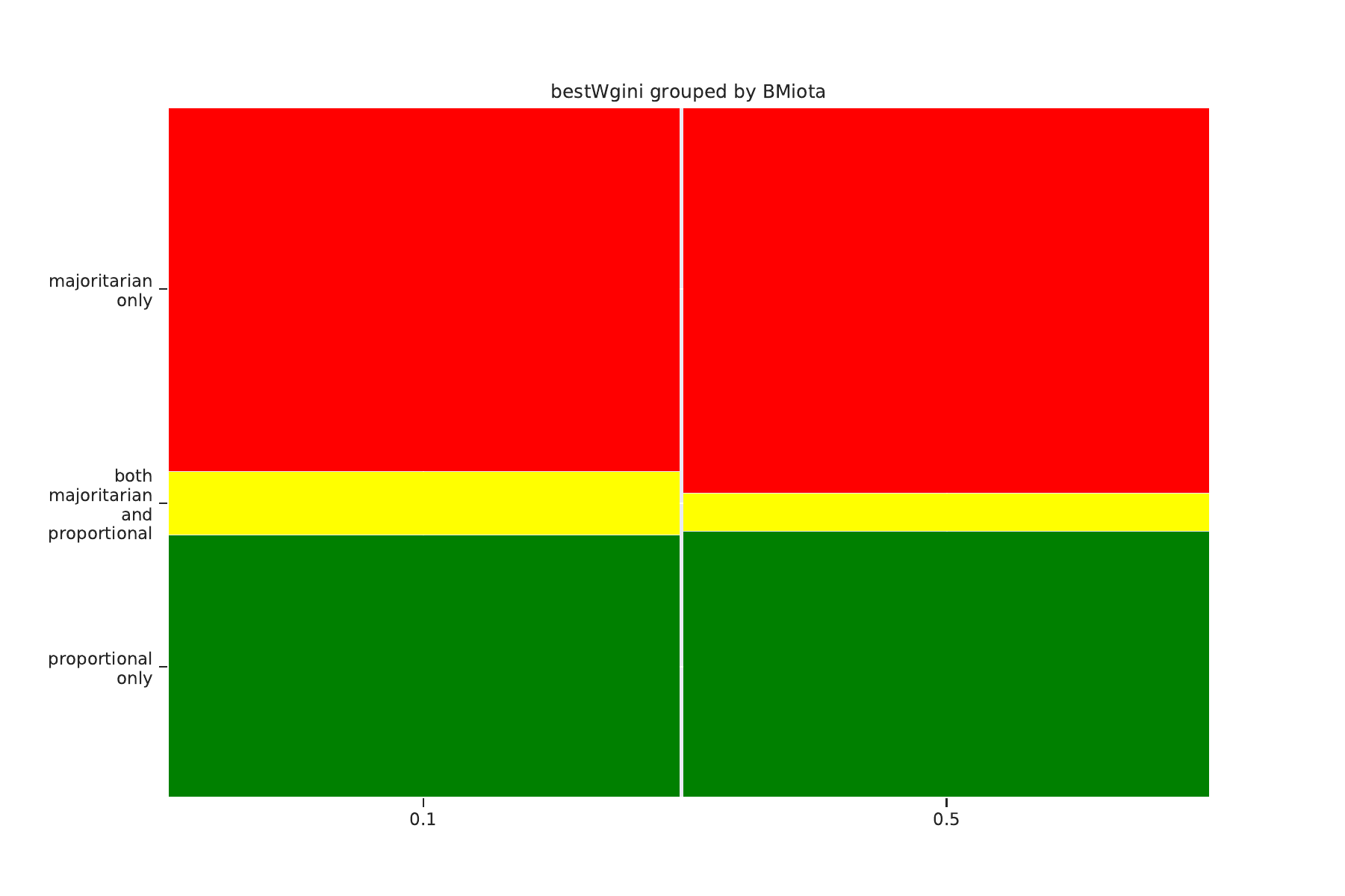}
    &\doplot{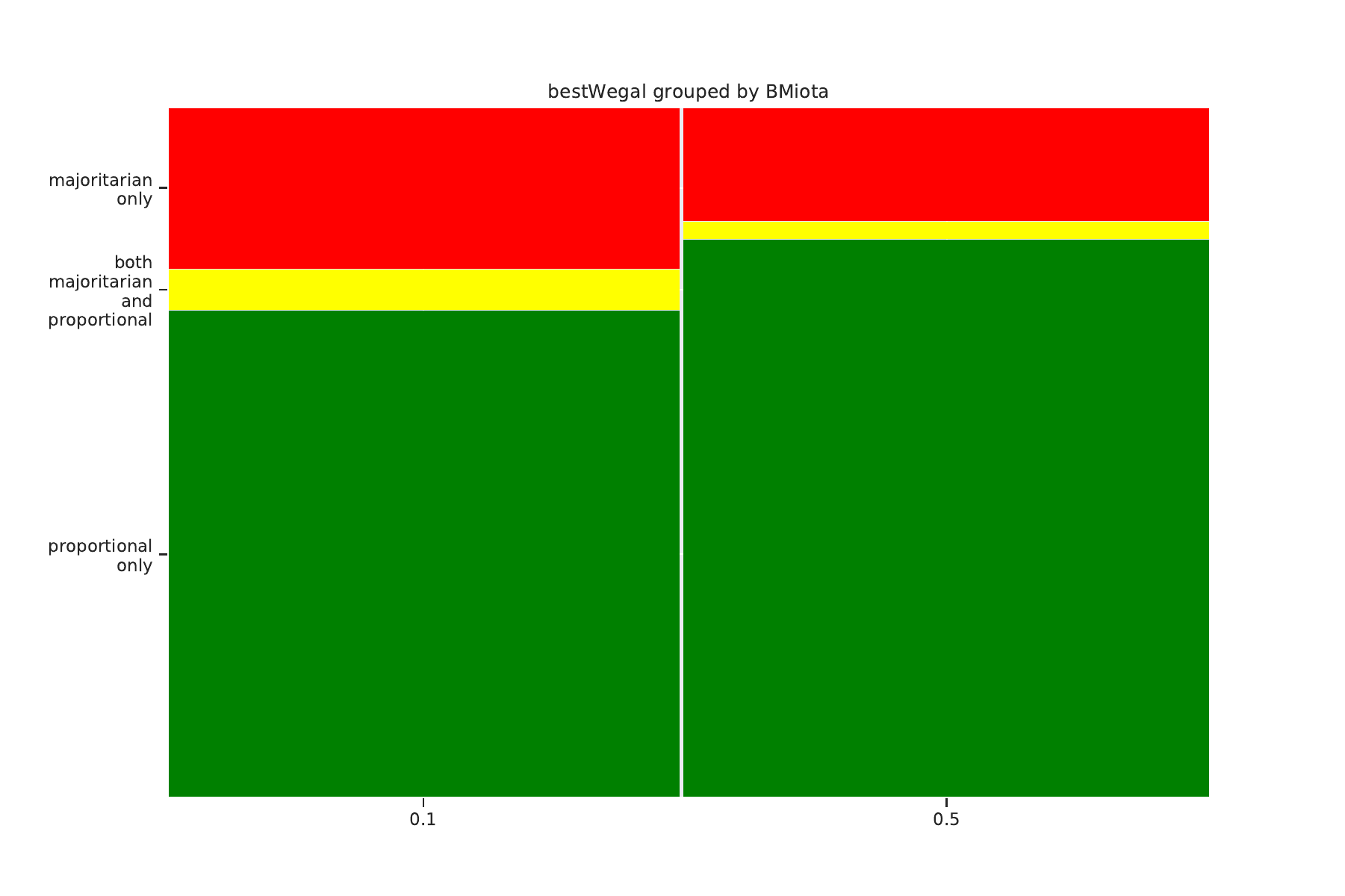}\\
{\tt omega}
    &\doplot{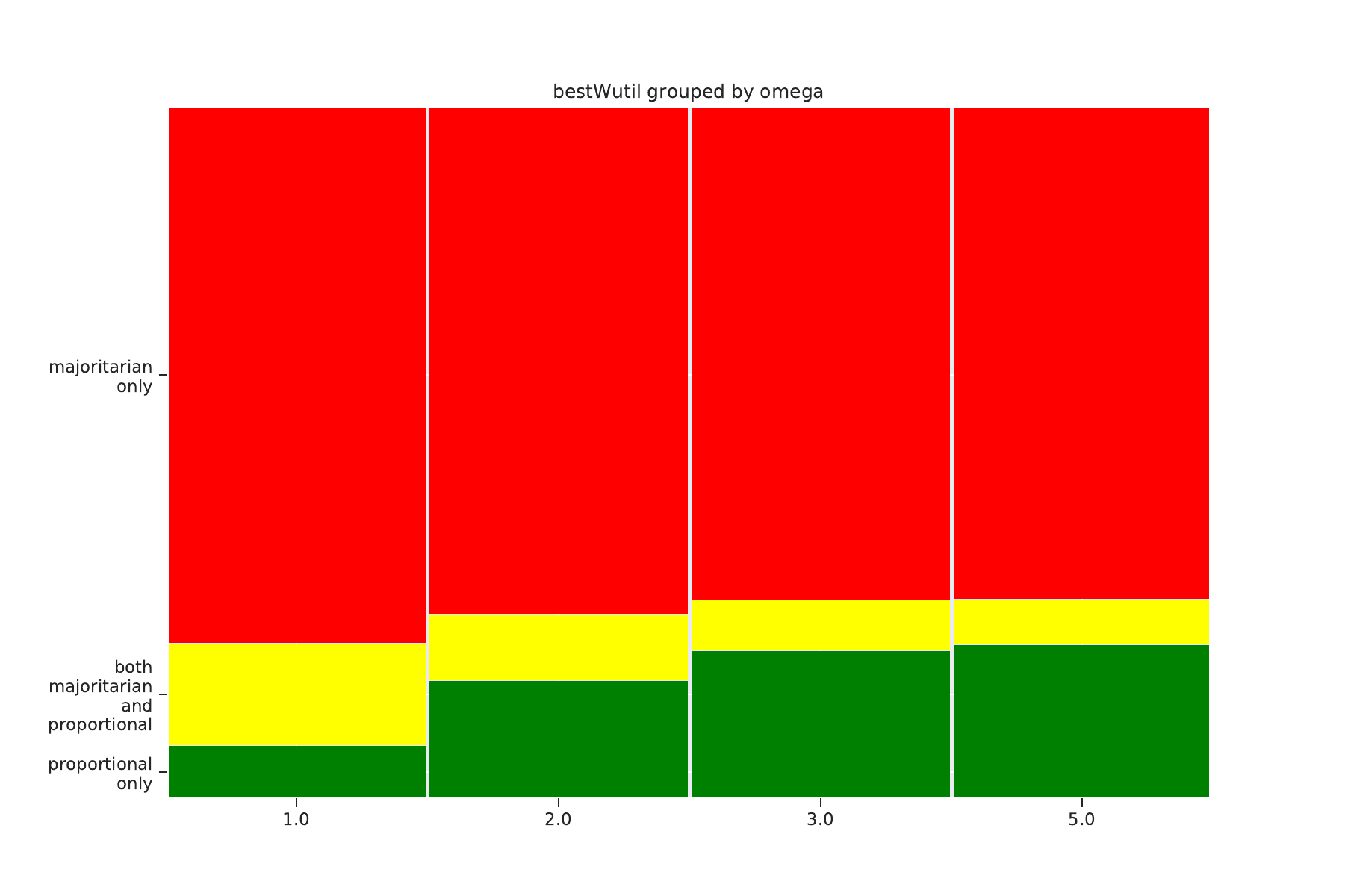}
    &\doplot{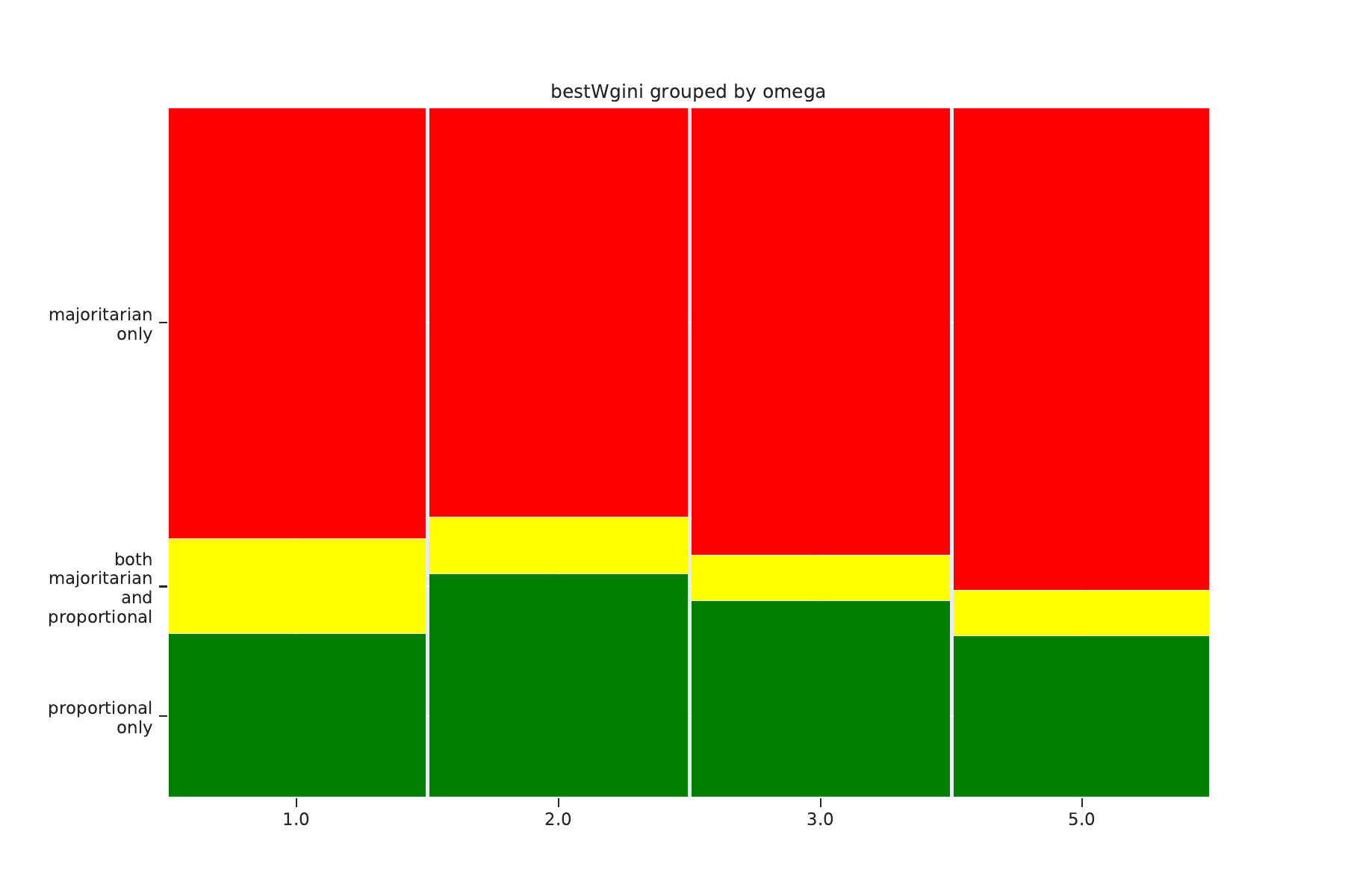}
    &\doplot{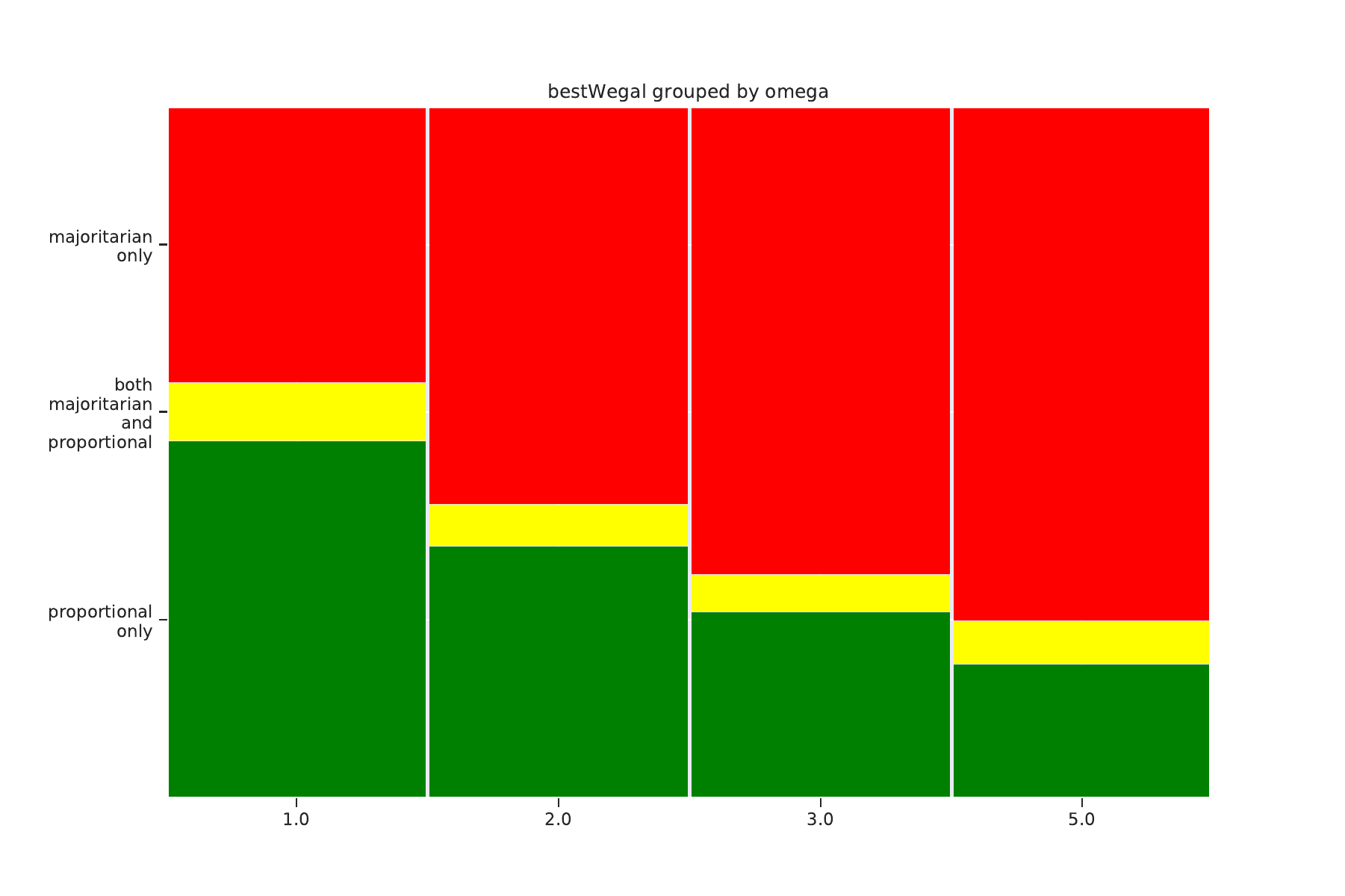}\\
{\tt rho}
    &\doplot{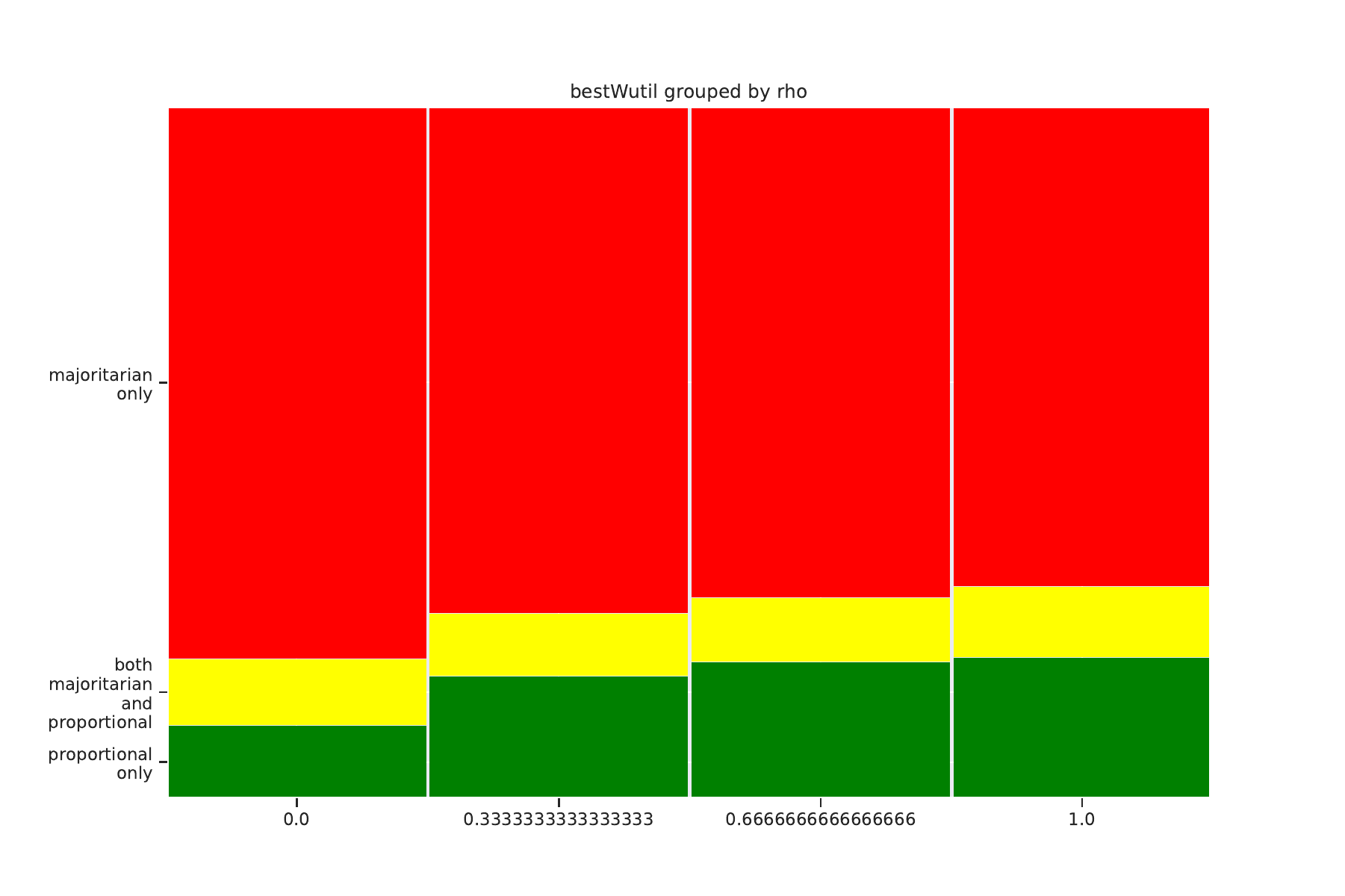}
    &\doplot{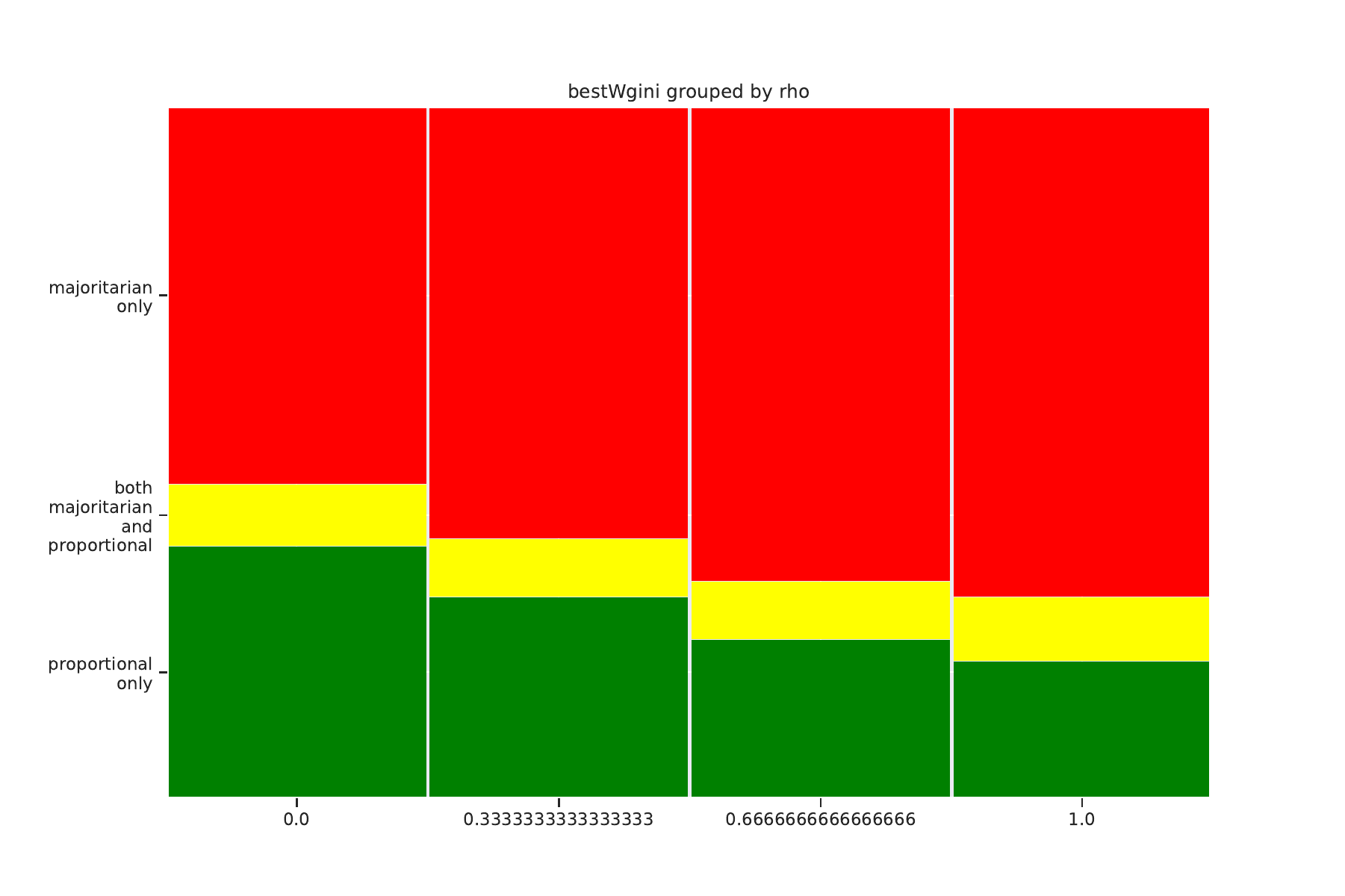}
    &\doplot{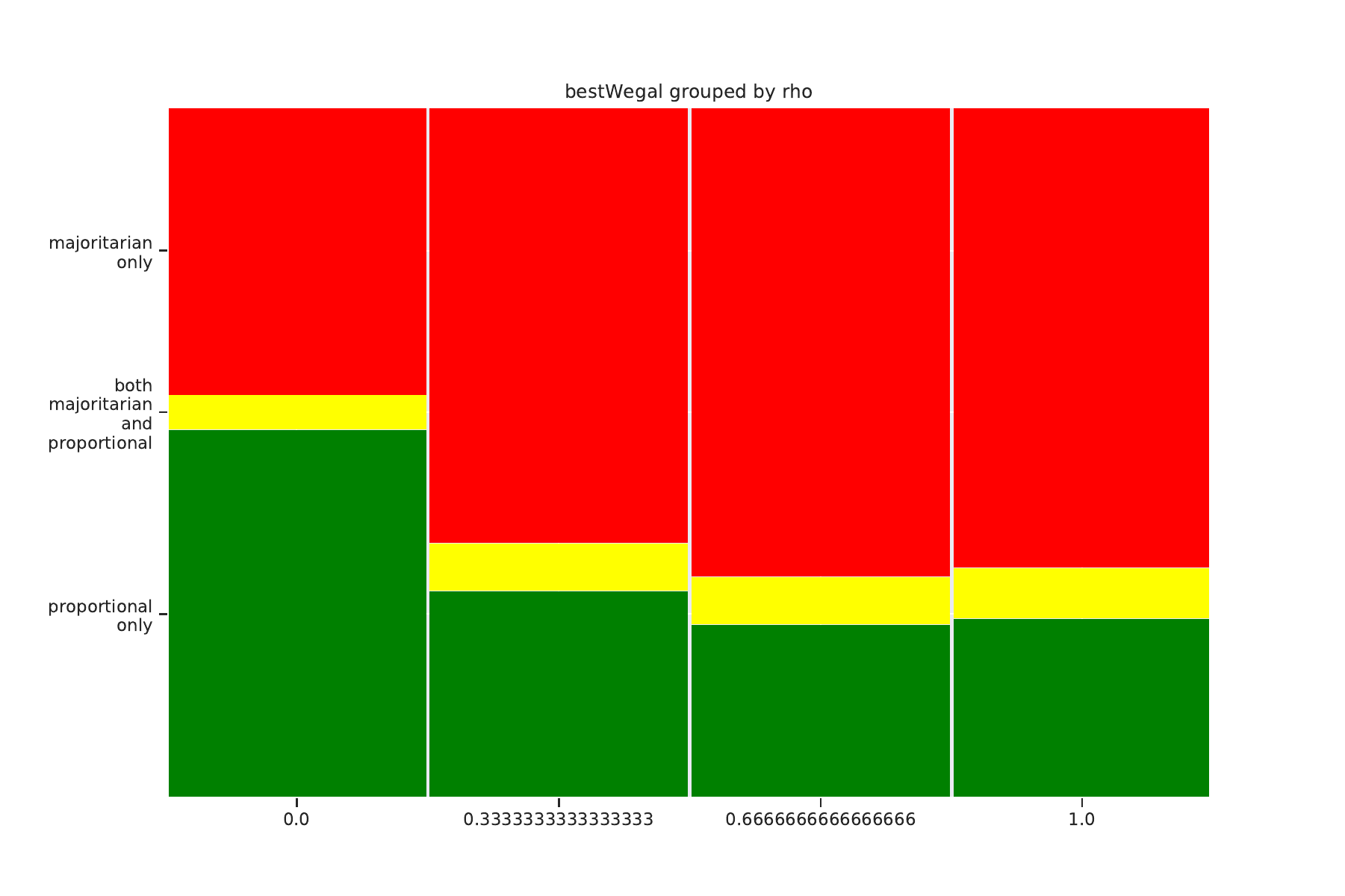}\\
{\tt riskmodel}
    &\doplot{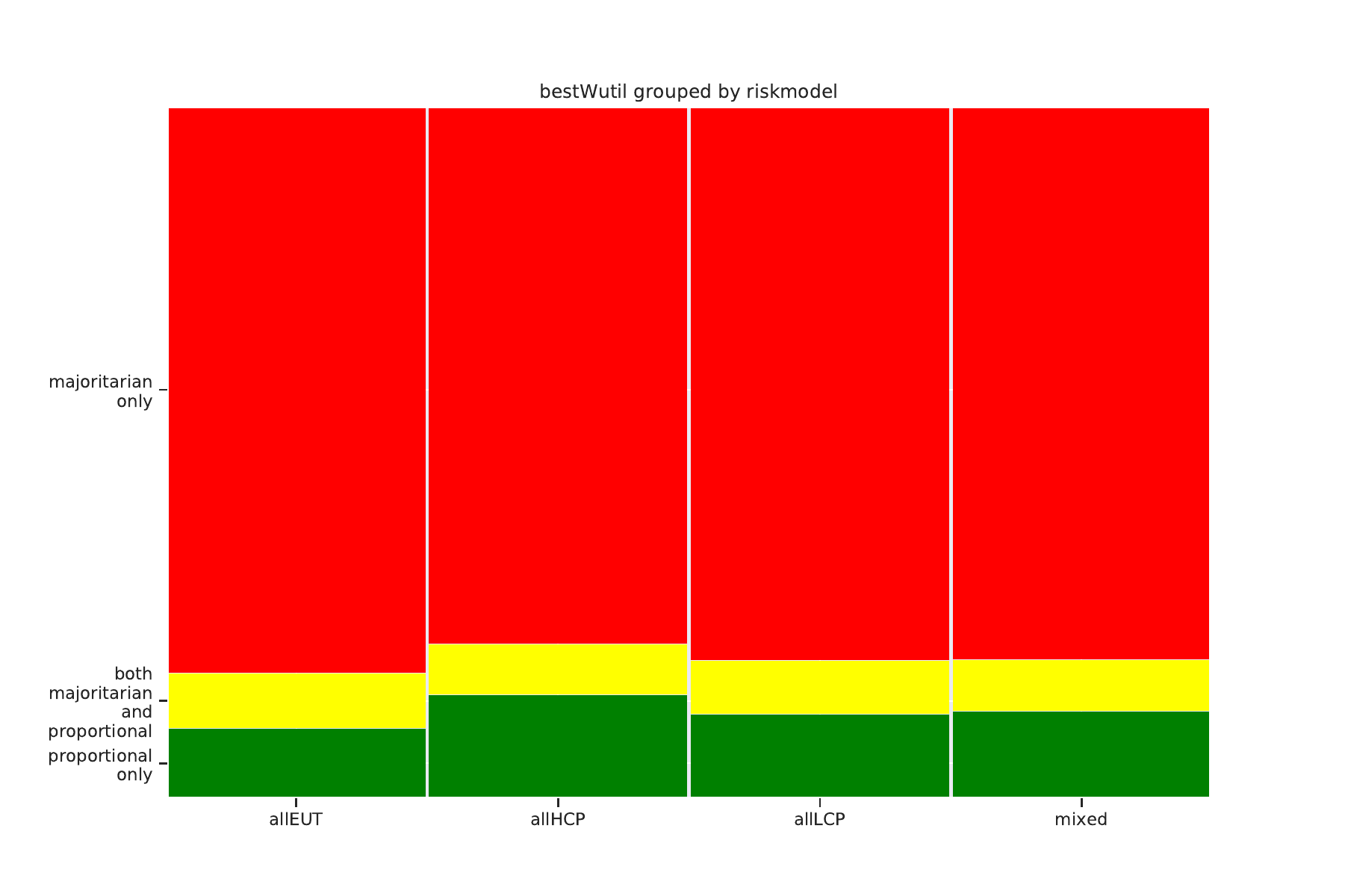}
    &\doplot{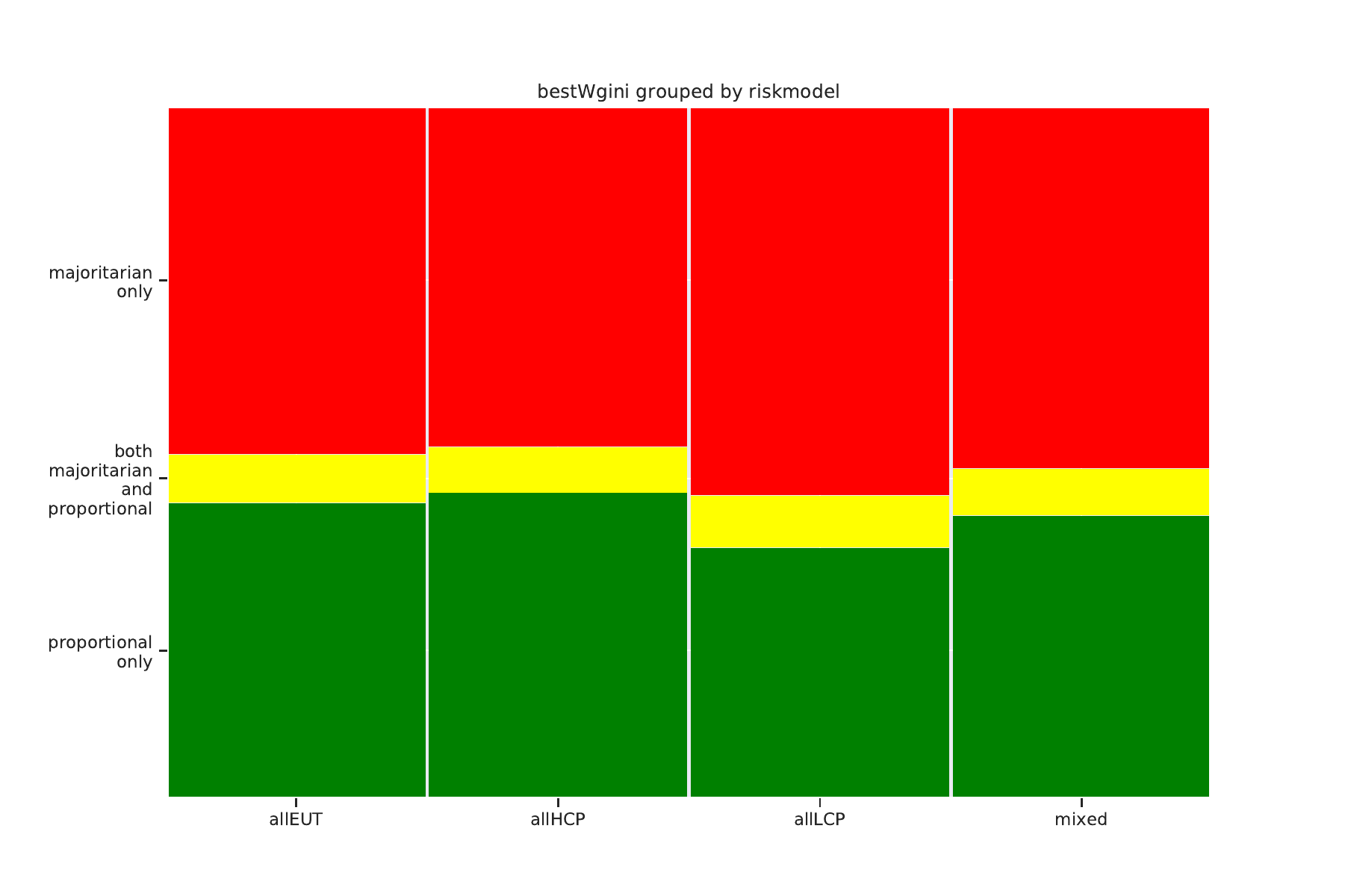}
    &\doplot{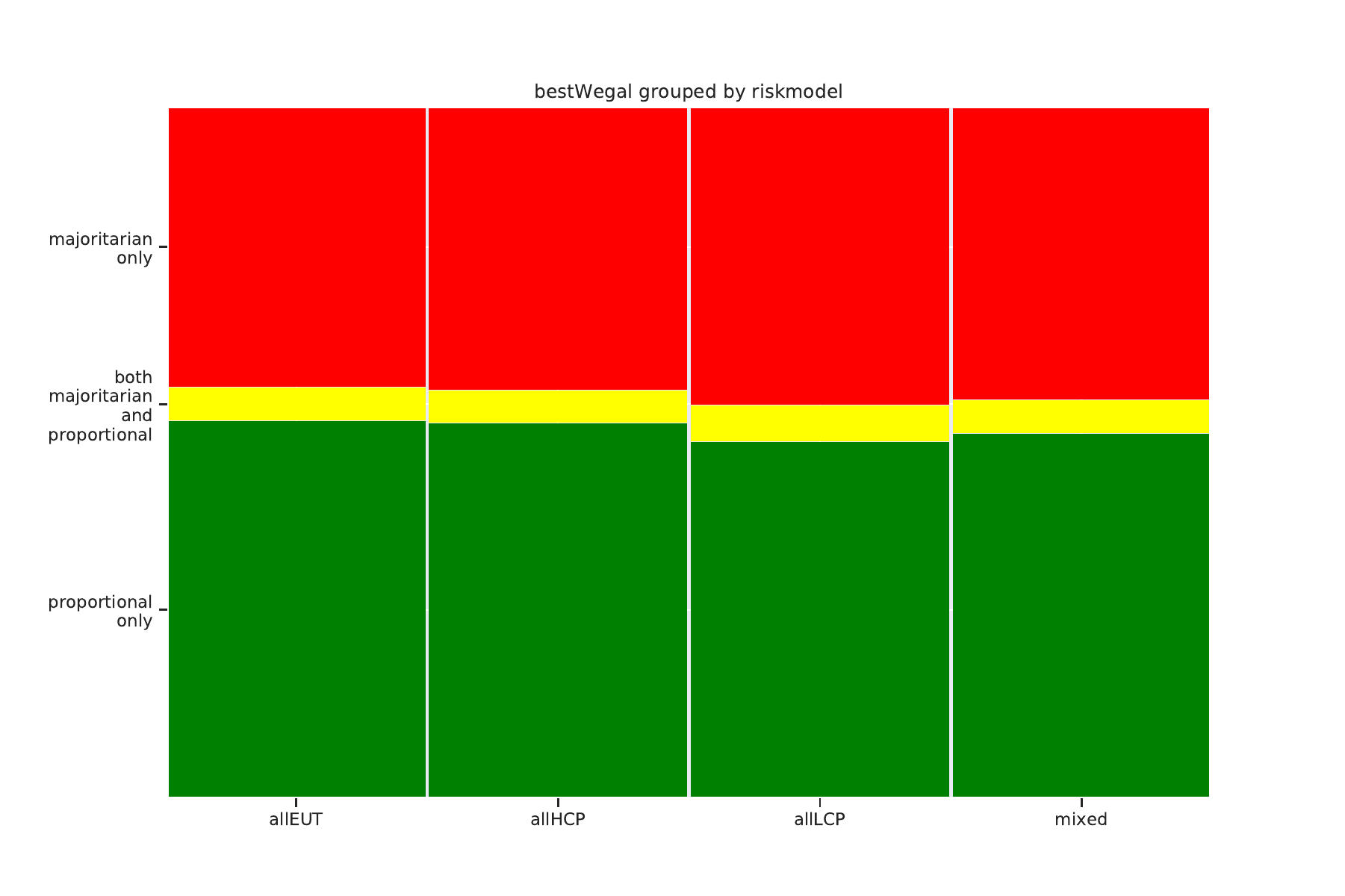}\\
{\tt scenario}
    &\doplot{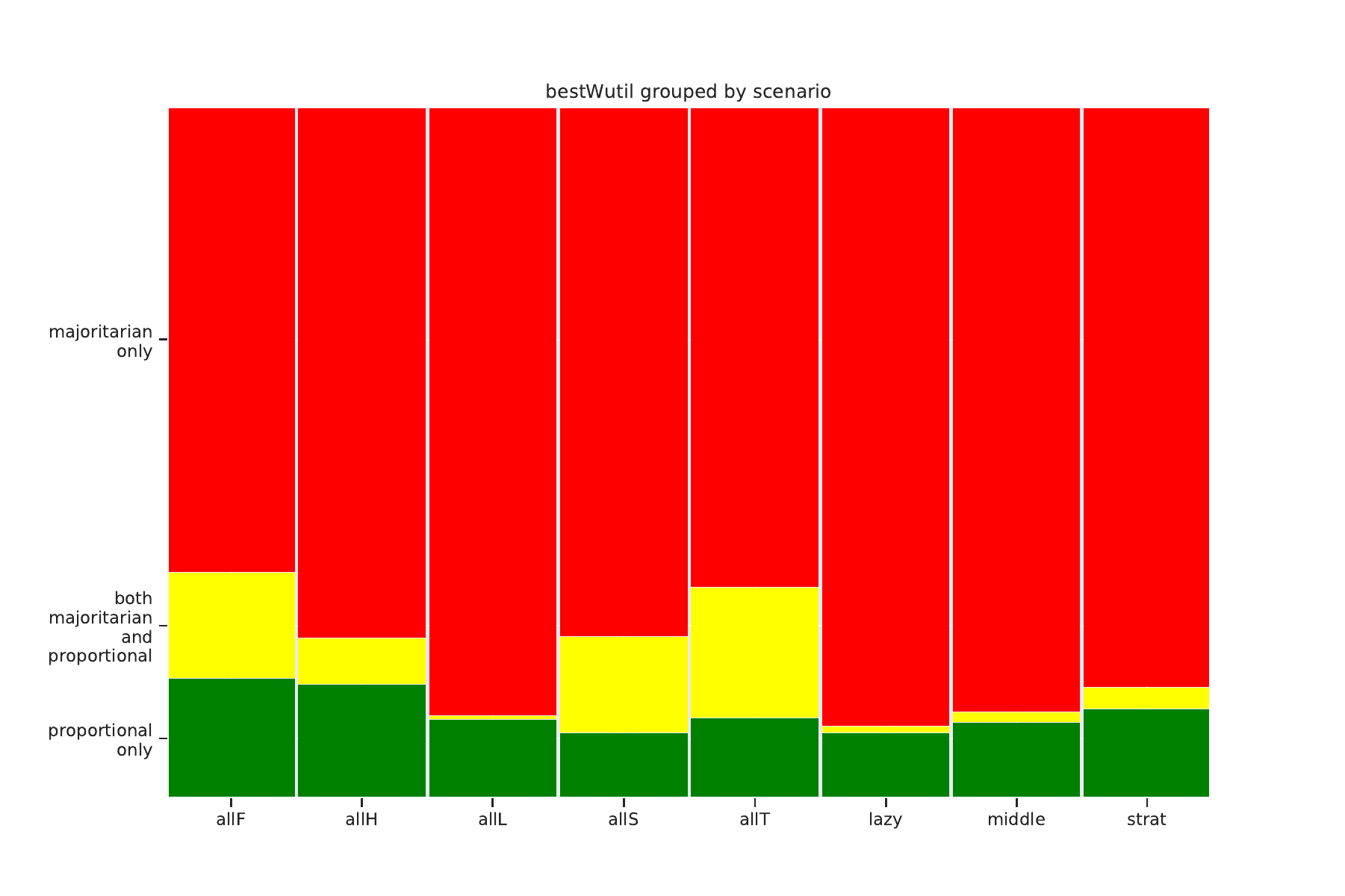}
    &\doplot{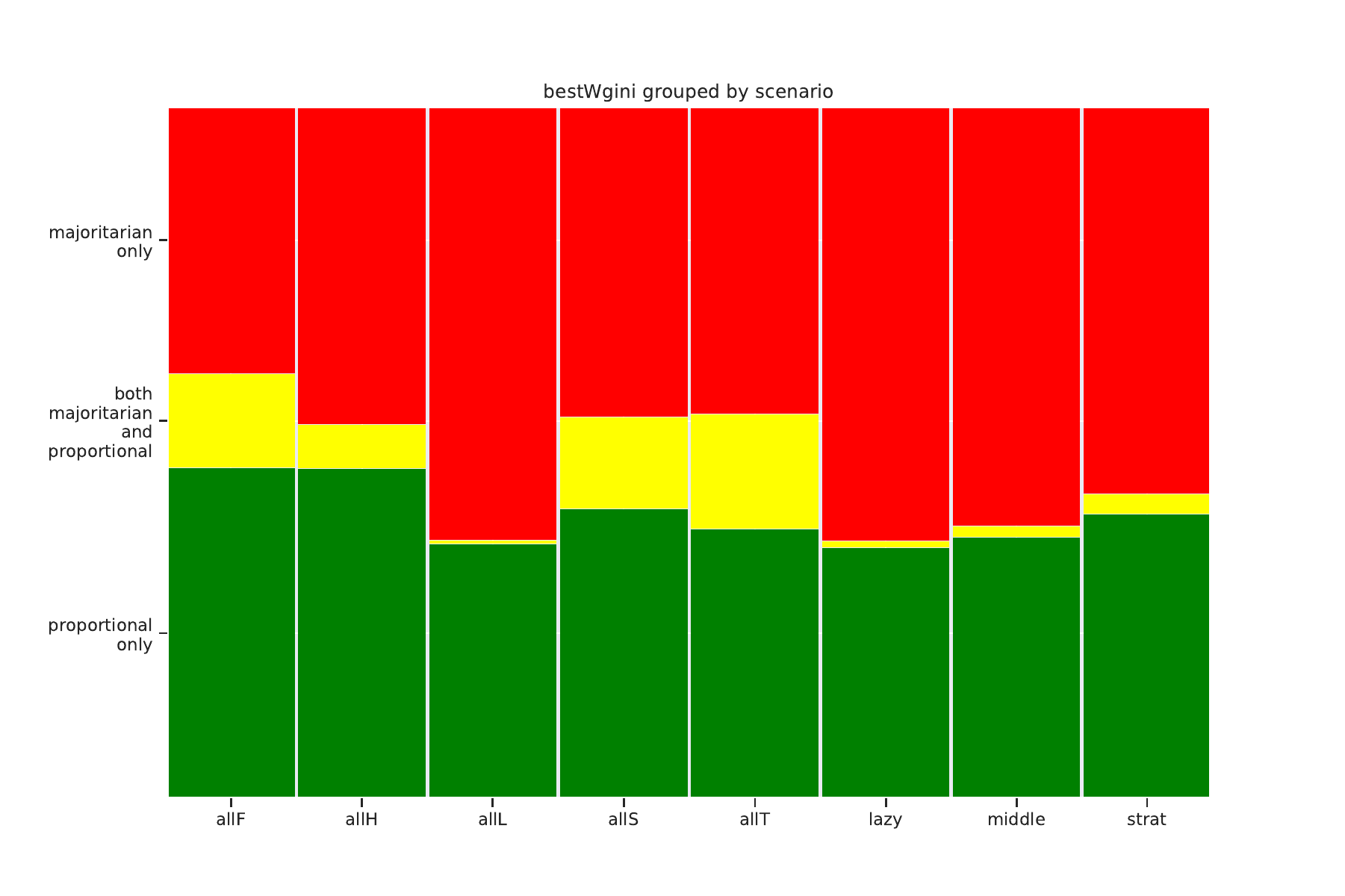}
    &\doplot{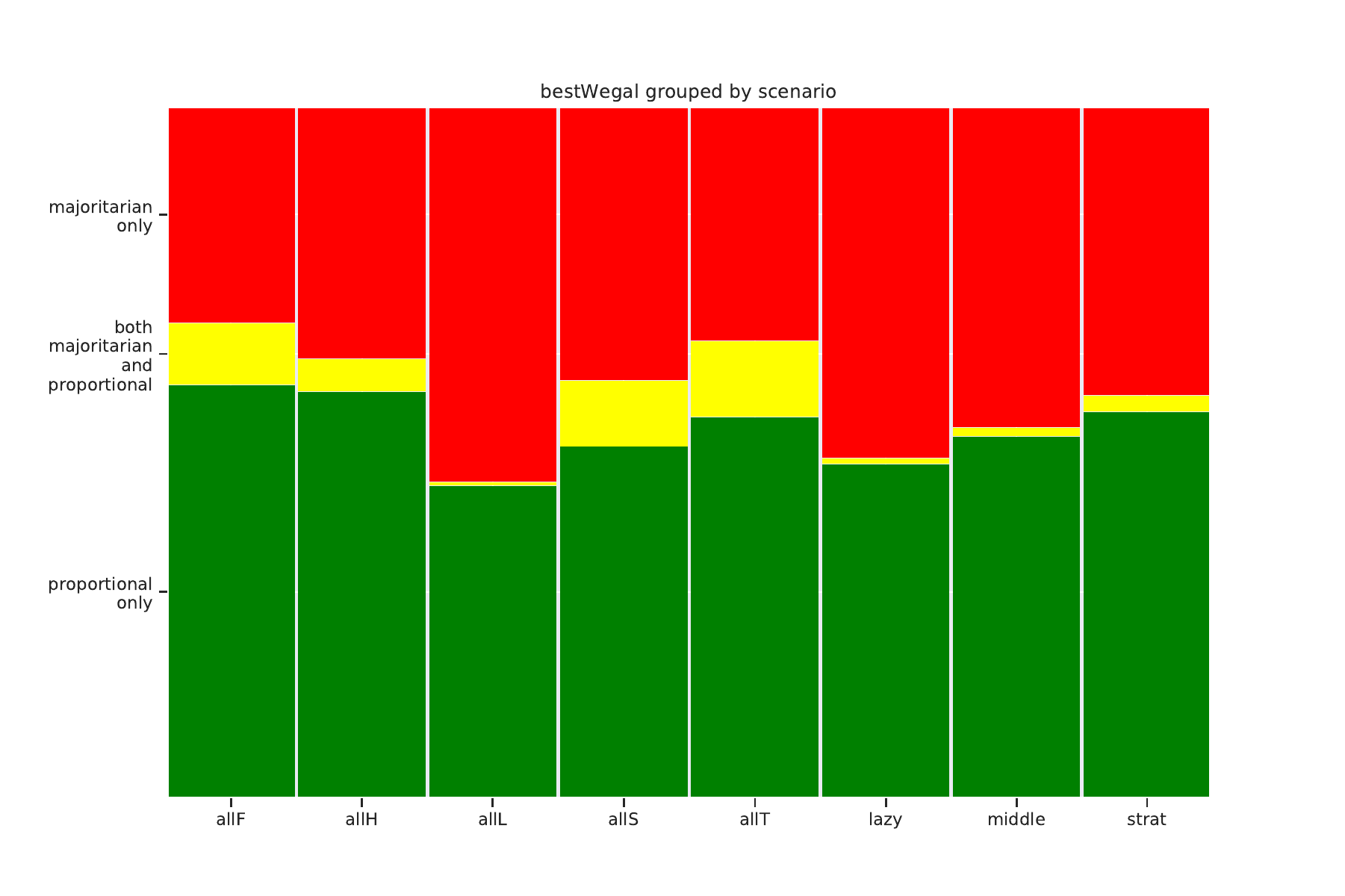}\\
&{\verb!Wutil_final!}
&{\verb!Wgini_final!}
&{\verb!Wegal_final!}
\end{tabular}
\caption{\label{fig:best}%
Frequency of majoritarian and proportional methods being best according to final 
utilitarian (left column), Gini-Sen (middle), and egalitarian (right) social welfare,
by various parameters (rows).
}\end{figure}

If we focus on the more qualitative question of which methods perform ``best'' how often,
we can study the share of decision problems in which the largest welfare 
was (i) only provided by one or more majoritarian methods (red in Fig.\,\ref{fig:best}),
(ii) only provided by one or more proportional methods (green),
or (iii) provided by at least one method from both groups (yellow).
In Fig.\,\ref{fig:best}, this is shown for all three final welfare metrics ({\tt Wutil, Wgini, Wegal}),
grouped by parameters that made a significant difference.

One can see that across all three welfare metrics,
proportional methods were performing best according to this statistic 
in the Gaussian allotment and uniform preference models,
and with fewer lazy voters.

W.r.t.\ Gini-Sen and egalitarian welfare, they performed best with 
larger no.\ of voter blocks,
more individuality,
larger spatial voter heterogeneity,
and lower spatial option broadness;
w.r.t.\ utilitarian welfare, this was the other way around.

\paragraph{``Cost of fairness''}

\begin{figure}\centering
\includegraphics[width=0.4\textwidth]{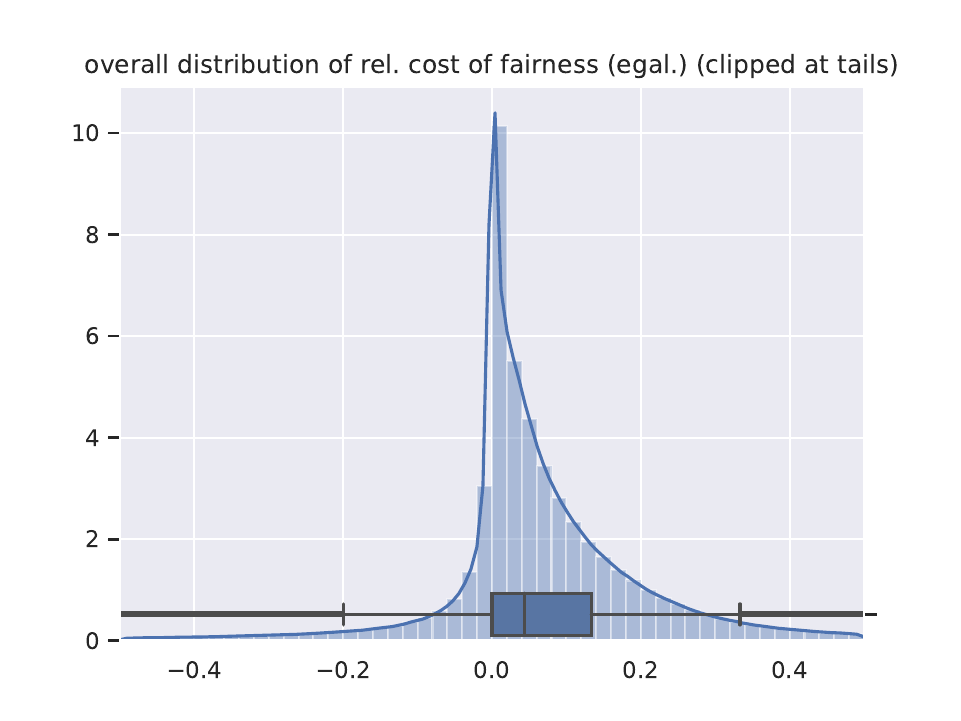}
\caption{\label{fig:cost}%
Distribution of the ``relative cost of fairness'' across decision problems.
}\end{figure}

As a final indicator of the social welfare effects of using proportional instead of majoritarian methods,
we show the distribution of the above-defined cost of fairness measure across all simulated decision problems
in Fig.\,\ref{fig:cost}.
It shows that, typically,
the decrease (if any) in average voter utility one gets by switching from the best majoritarian method (Range Voting)
to the best proportional (and thus ``fair'') method in our study (MaxParC),
was about an order of magnitude smaller than 
the difference between the utilitarian and egalitarian absolute social welfare resulting from using Range Voting
(which can be seen as a natural measure of absolute inequality in voters' utility).
The 2\%-trimmed mean of this ``relative cost of fairness'' was 0.08.

\paragraph{Summary}

Overall, one can conclude that proportional methods can well compete with majoritarian ones regarding social welfare effects
and that the welfare assessment depends strongly on the choice of welfare metric used:
the more inequality-averse the welfare metric, the more it favours proportional methods.
In addition, welfare effects of method choice depend strongly on the distribution of voters' preferences,
which suggests that proportional methods may particularly well perform in situations 
with heterogeneous voters and when many options have a comparatively narrow appeal.

Since proportional methods achieve this by randomization, 
and hence only over sequences of several decisions 
(which is why we used `ex-ante'/long-run welfare metrics here),
we need to look next at the amount of randomization actually used.

\subsubsection{Randomization}

In ``deterministic'' methods, randomization is only used to resolve the odd tie,
hence Shannon entropy (and similarly R\'enyi entropy) is mostly zero, sometimes $\log 2$, and rarely larger,
and maximal option probability is mostly one, sometimes $1/2$, and rarely smaller.
Fig.\,\ref{fig:probabilities} compares this to the level of randomization in the nondeterministic methods used in this study.
Under RB, entropy is distributed in a left-skewed distribution with a peak at $\log$ {\tt noptions},
leading to a mixture distribution with mean around 1.3.
Under FC, this distribution is further mixed with a peak at 0 representing the cases where the full compromise was found.
RFC was similar to FC regarding randomization.

Under NL, this probability of finding a full compromise was almost twice as large;
while it still shows a mixture of left-skewed distributions with peaks at $\log k$ for some integer $k$,
these $k$ are now generally smaller than for RB since the optimization of the Nash sum typically leaves some options with zero probability.
The mean entropy for NL is thus smaller, 1.05.
For MPC, this is even more pronounced, with two clear peaks at 0 and $\log 2$
and mean entropy initially 0.9.
For MPC, the interactive phase changed entropy the most, bringing it down to 0.8 on average.

For RB and FC, the 25\% quantile [and mean] of the largest option probability were only at around 0.3 [0.43],
for NL is was at around 0.35 [0.53], and for MPC initially around 0.45 [0.6] and finally (after the interactive phase) around 0.5 [0.65].

Regression analysis reveals that Shannon entropy increases (and max.\ probability decreases) with
the no.\ of options, 
no.\ of voters (slightly), 
no.\ of voter blocks,
and share of lazy voters (considerably).
Adding a constructed compromise option decreased entropy (and increased max.\ probability) even if it had no clear welfare effects.
Surprisingly, the share of non-expected-utility, risk- (and hence randomization-) averse voters 
had no clear effect on the level of randomization,
and neither had the no.\ of pre-voting polls.

Fig.\,\ref{fig:randomization} shows some further parameter influences on individual methods' level of randomization.
With more options and more trial-and-error voters, MPC's advantage becomes more pronounced,
while NL randomizes less if all voters are sincere or are factionally strategic, 
the latter reflecting the fact that the factional strategy in NL was implemented as an optimization problem in our simulations
which however might be hard to solve in reality.

In the next subsection we will see whether strategizing pays off and whether it gives a relative advantage.

\subsubsection{Satisfaction}

In order to see whether and when different behavioural types of voters have advantages,
we study the distribution of the average satisfaction of all voters of a certain type across our simulations.
Fig.\,\ref{fig:satisfaction} shows that typical shape of these distributions depends very much on the method,
but is almost identical for the two polar behavioural types of lazy voters and factionally strategic voters
(and also for the other three types). 
In other words, voting strategically does not so much give a comparative advantage to the strategic voters over the lazy ones,
but rather increases overall welfare (as seen above). 

\begin{figure}\centering
\includegraphics[width=0.49\textwidth]{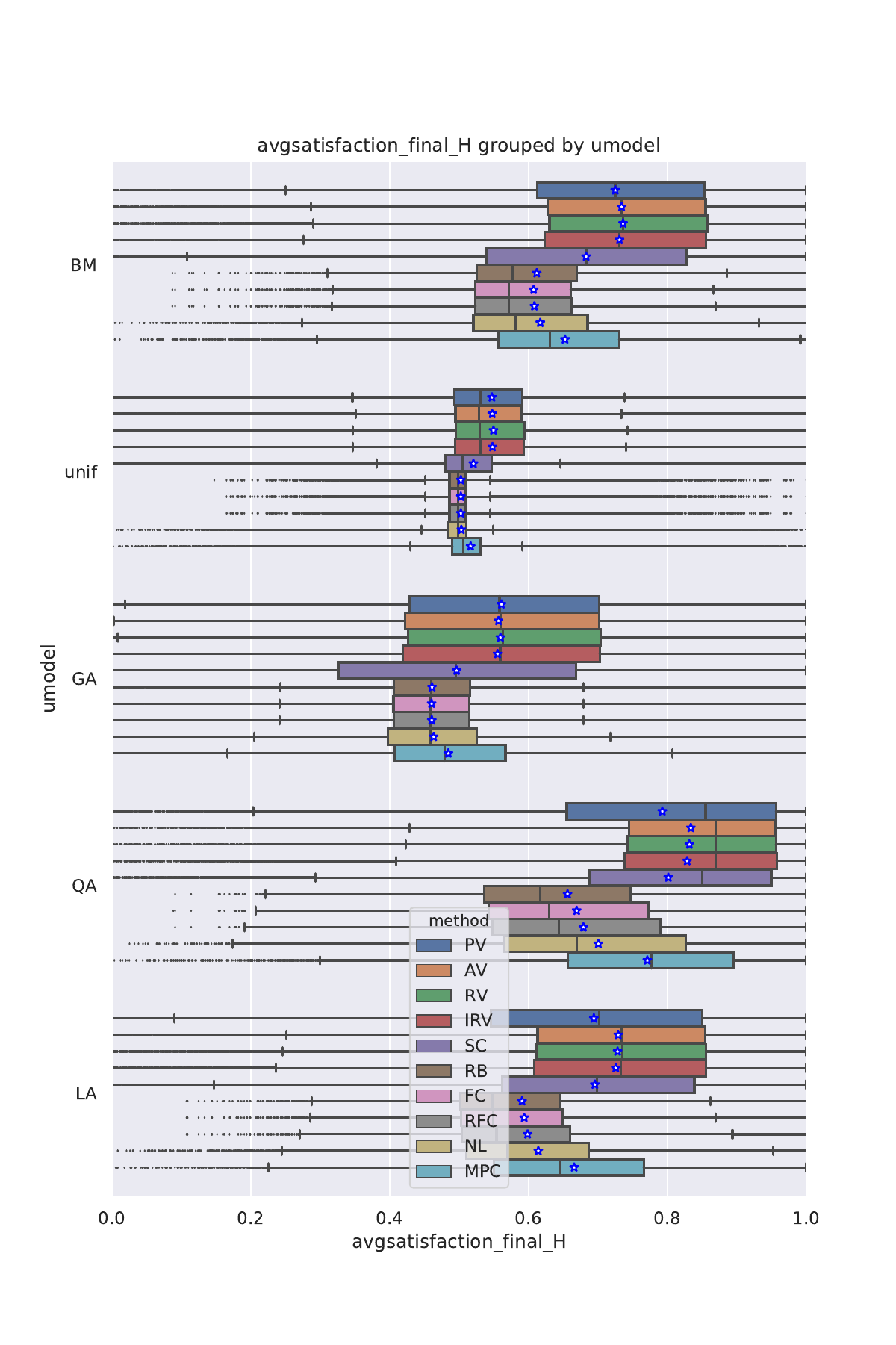}
\includegraphics[width=0.49\textwidth]{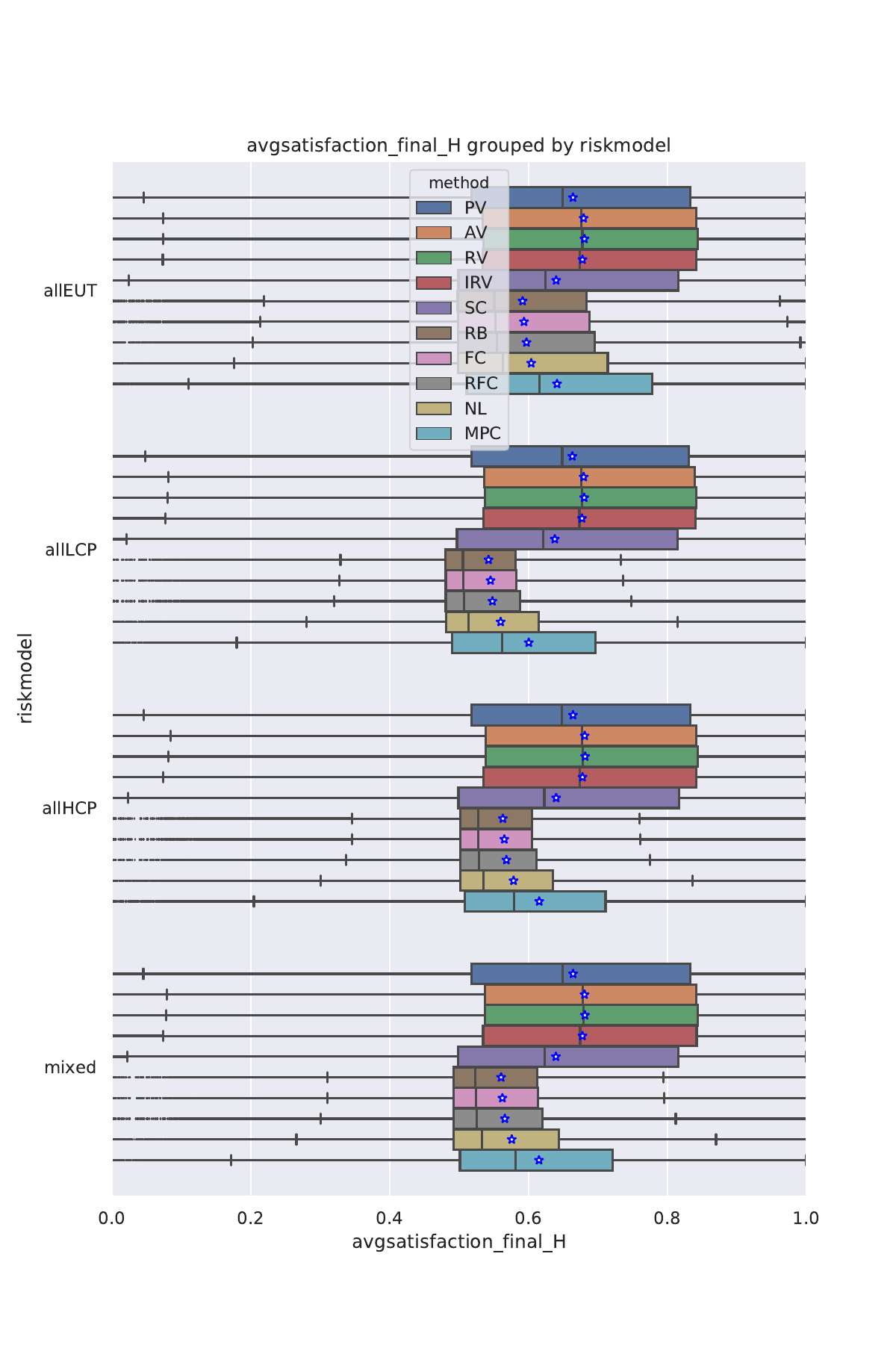}
\caption{\label{fig:satisfaction2}%
Distribution of the final average satisfaction of heuristic voters across decision problems, by method.
}\end{figure}

\begin{sidewaysfigure}\centering
\def\doplot#1{\includegraphics[width=0.13\textwidth,trim=40 40 40 40,clip]{plots/2020_03_28_1mio_all_#1.png}}
\begin{tabular}{ccccccc}
\doplot{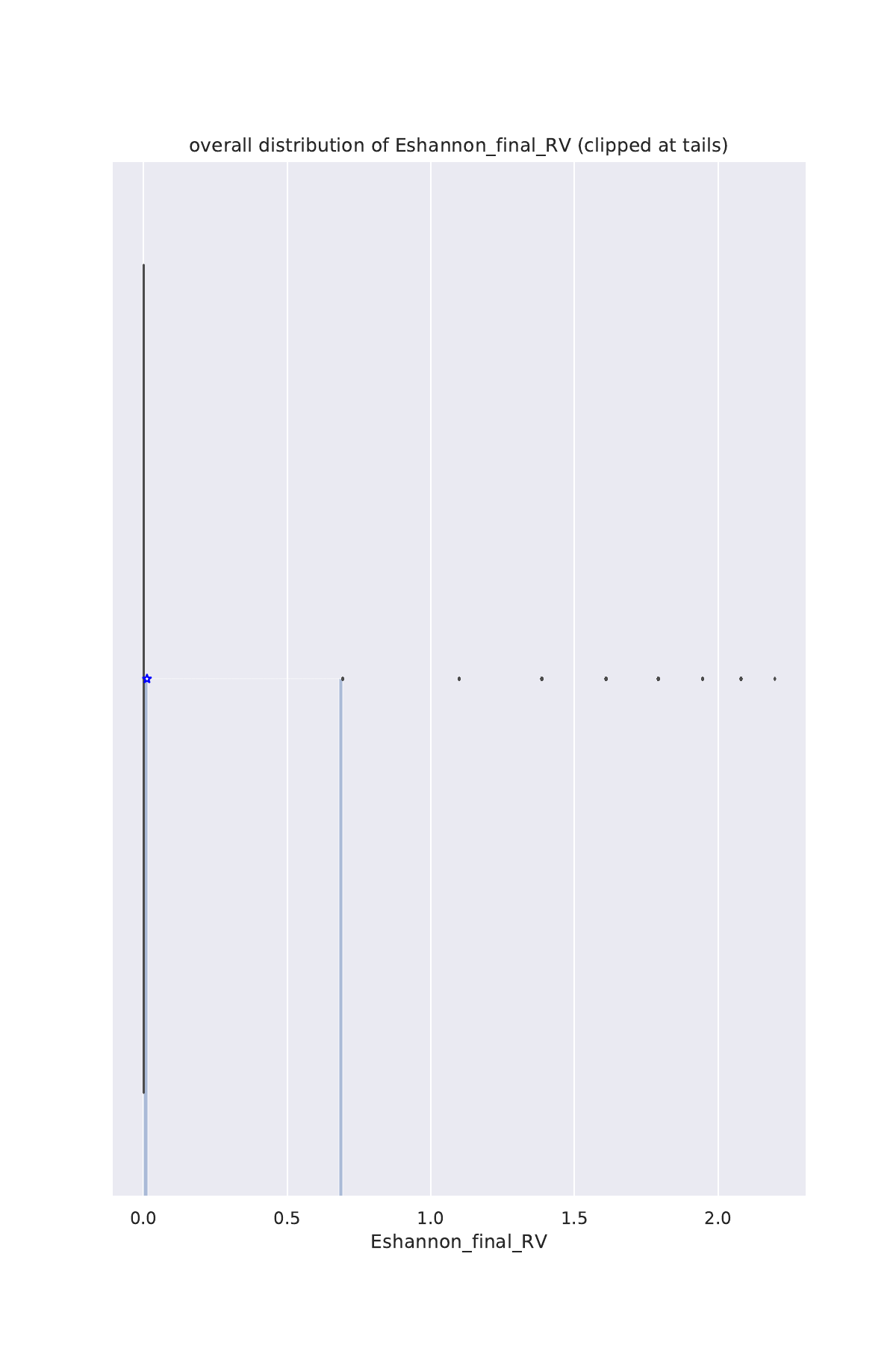}
&\doplot{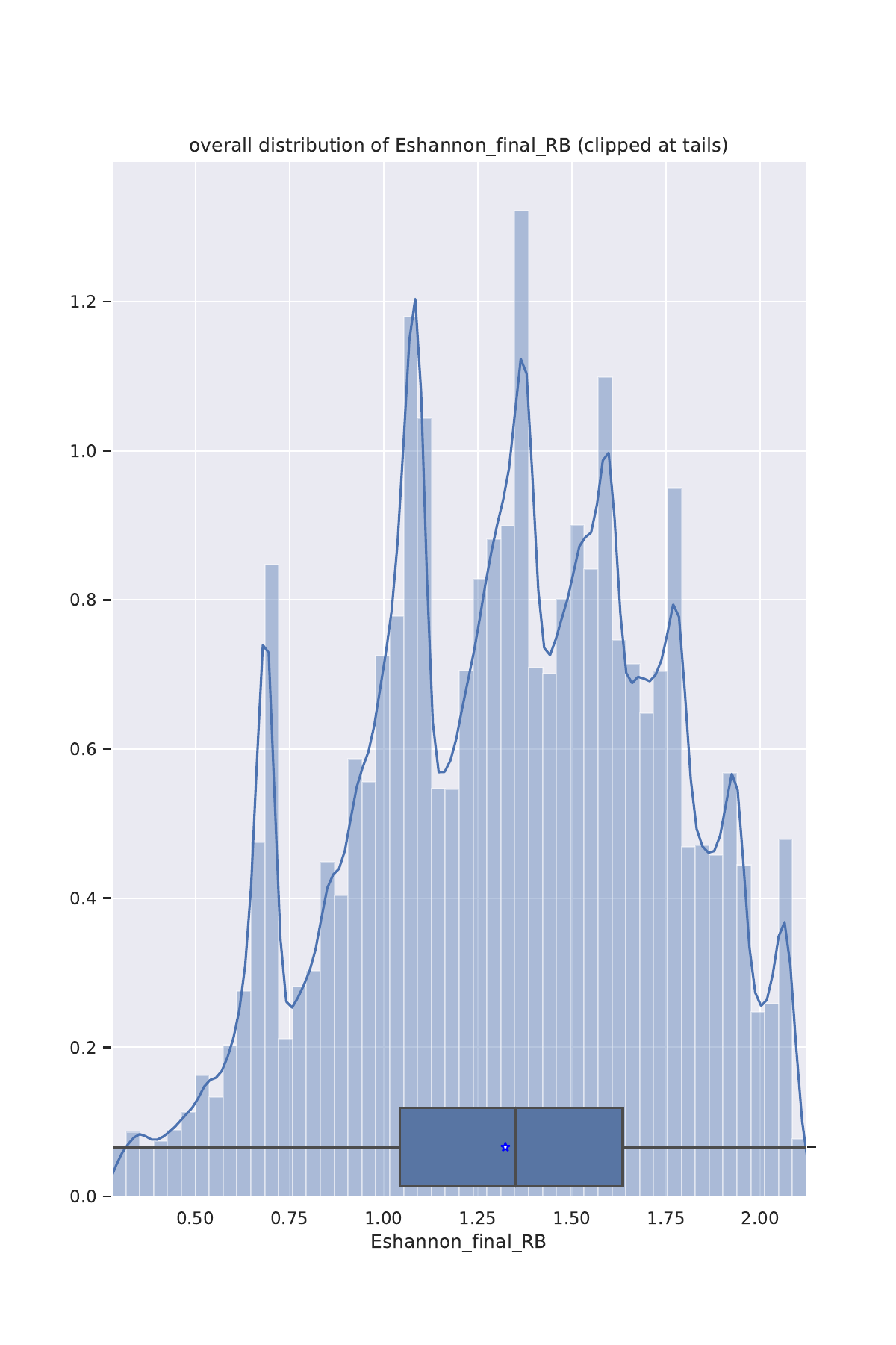}
&\doplot{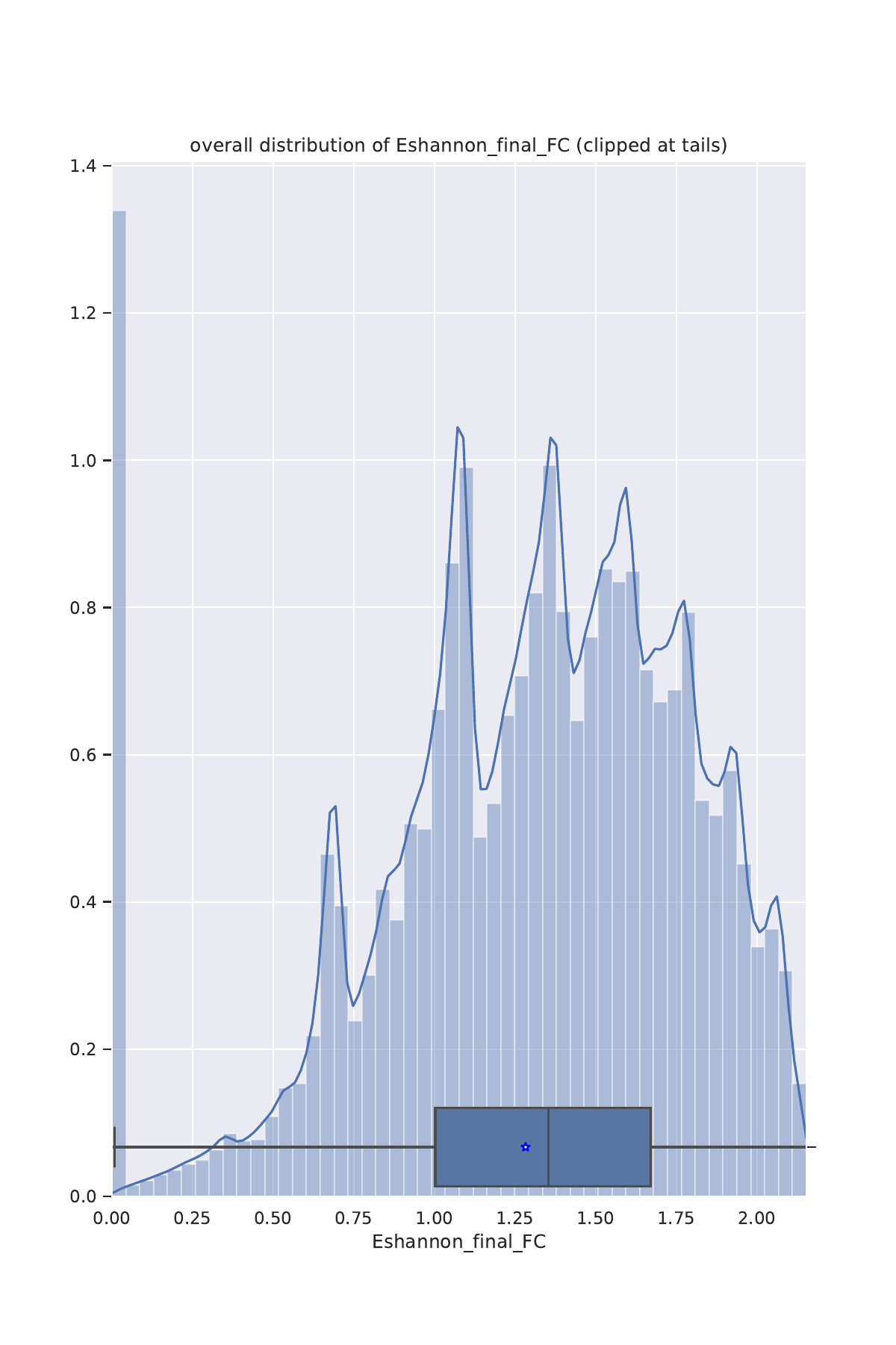}
&\doplot{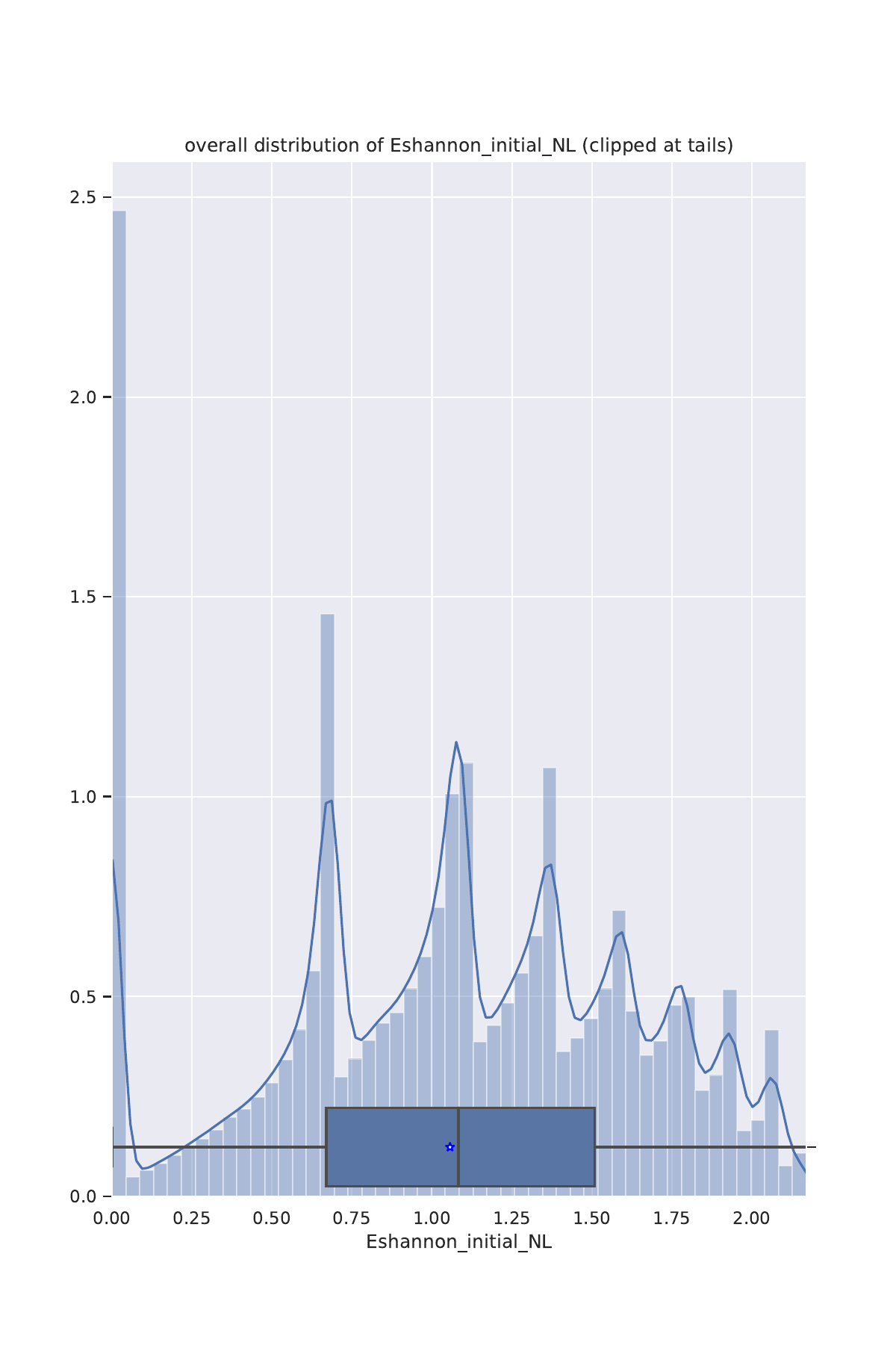}
&\doplot{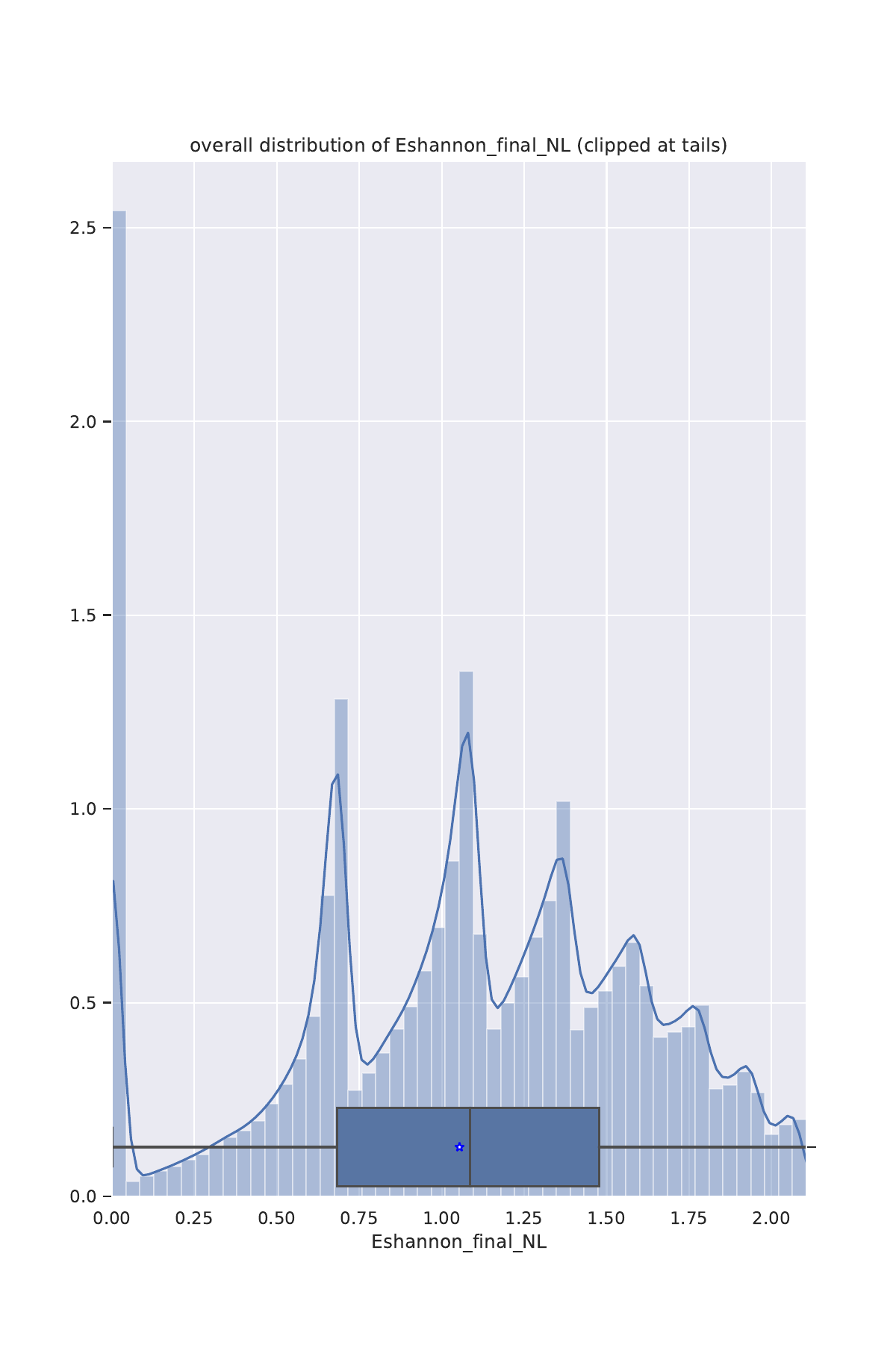}
&\doplot{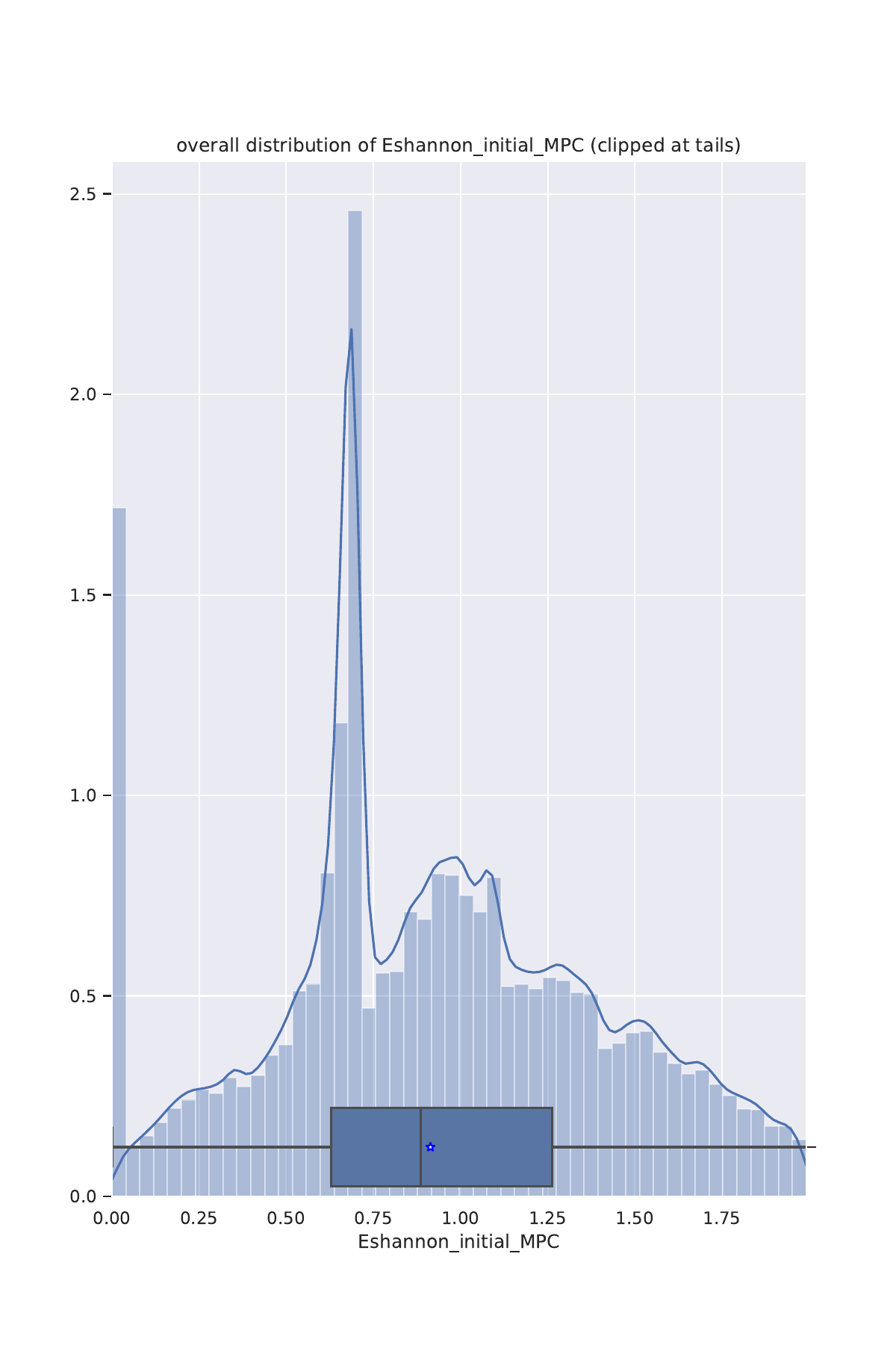}
&\doplot{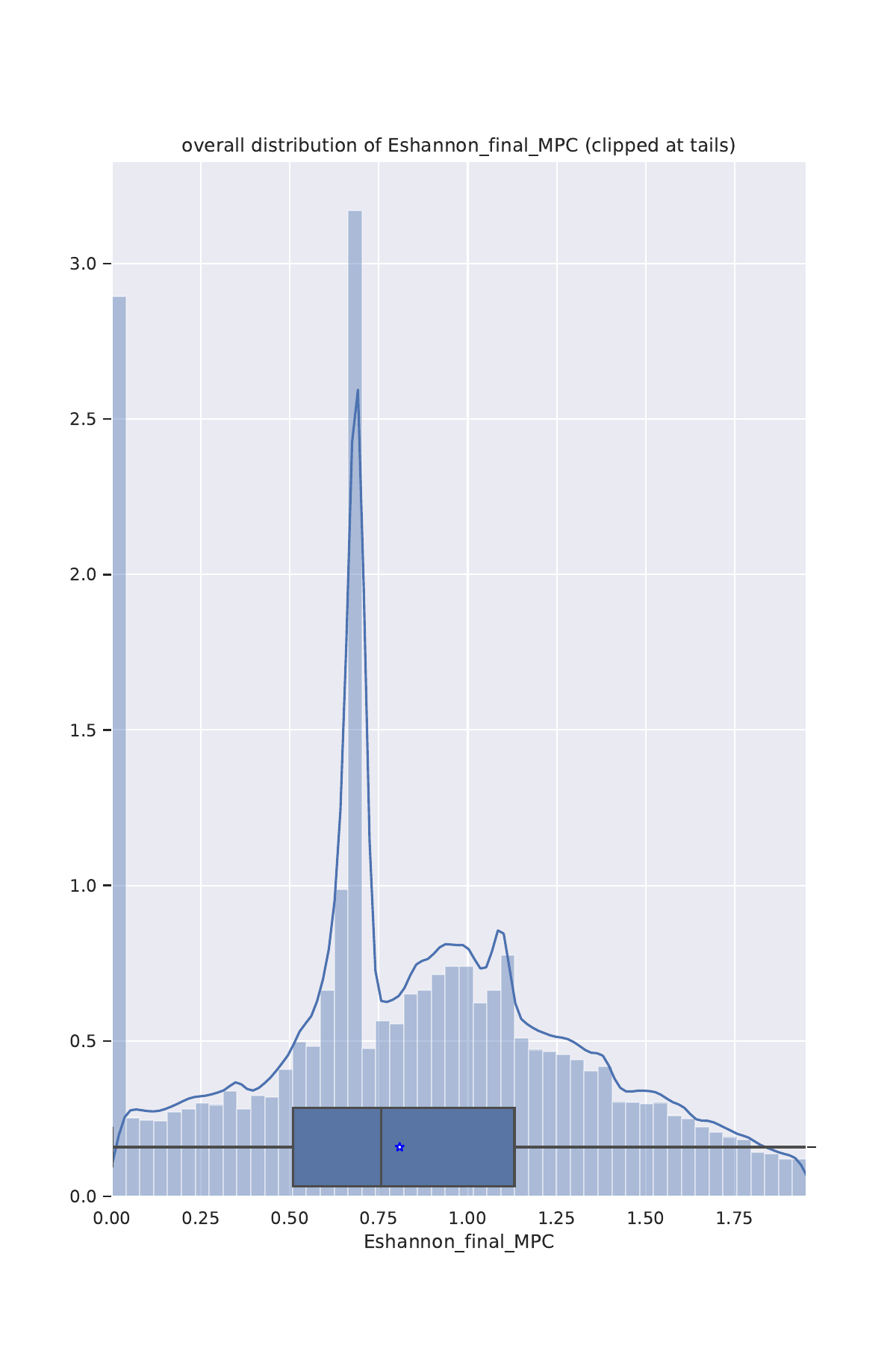}\\
\doplot{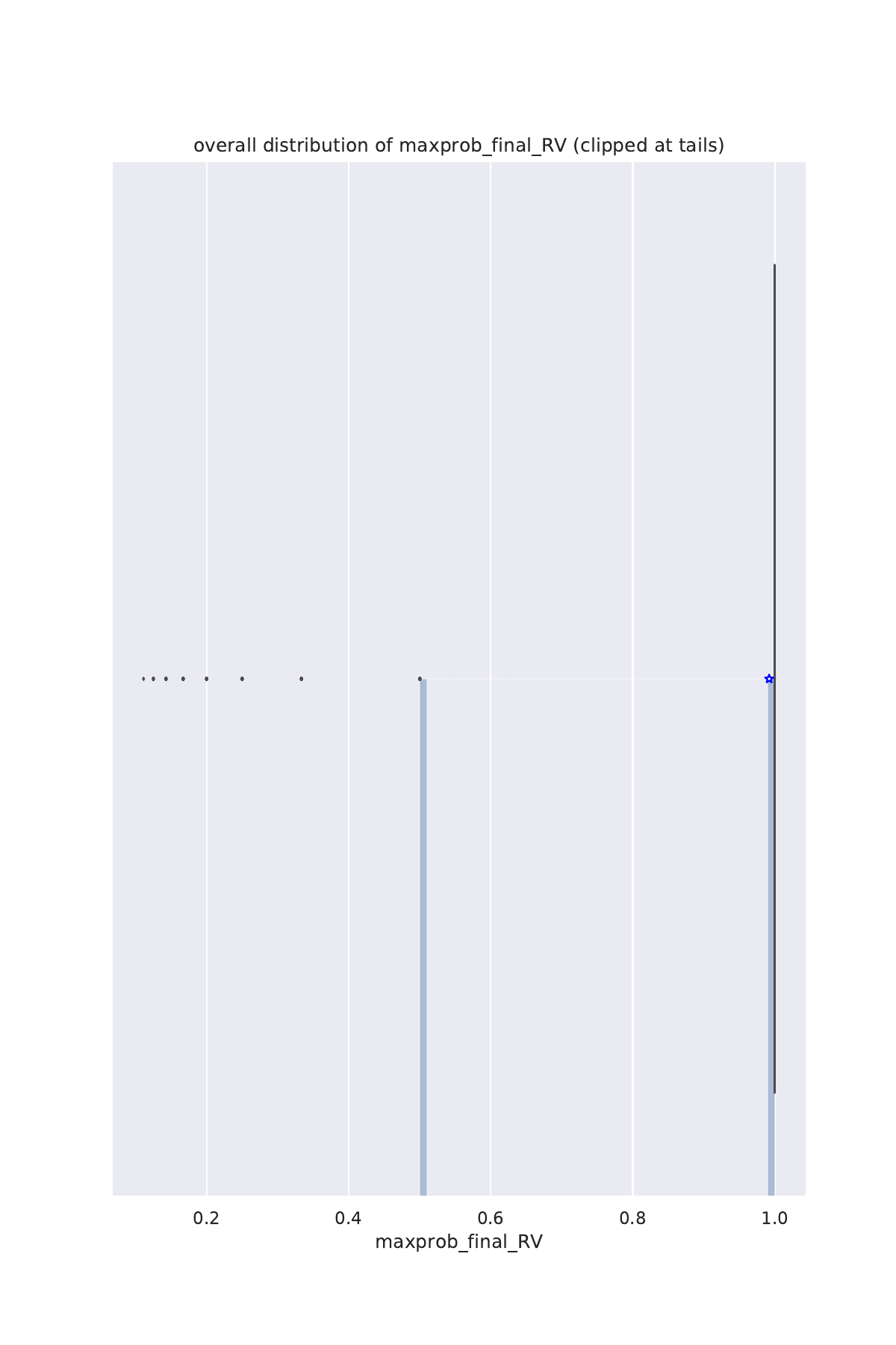}
&\doplot{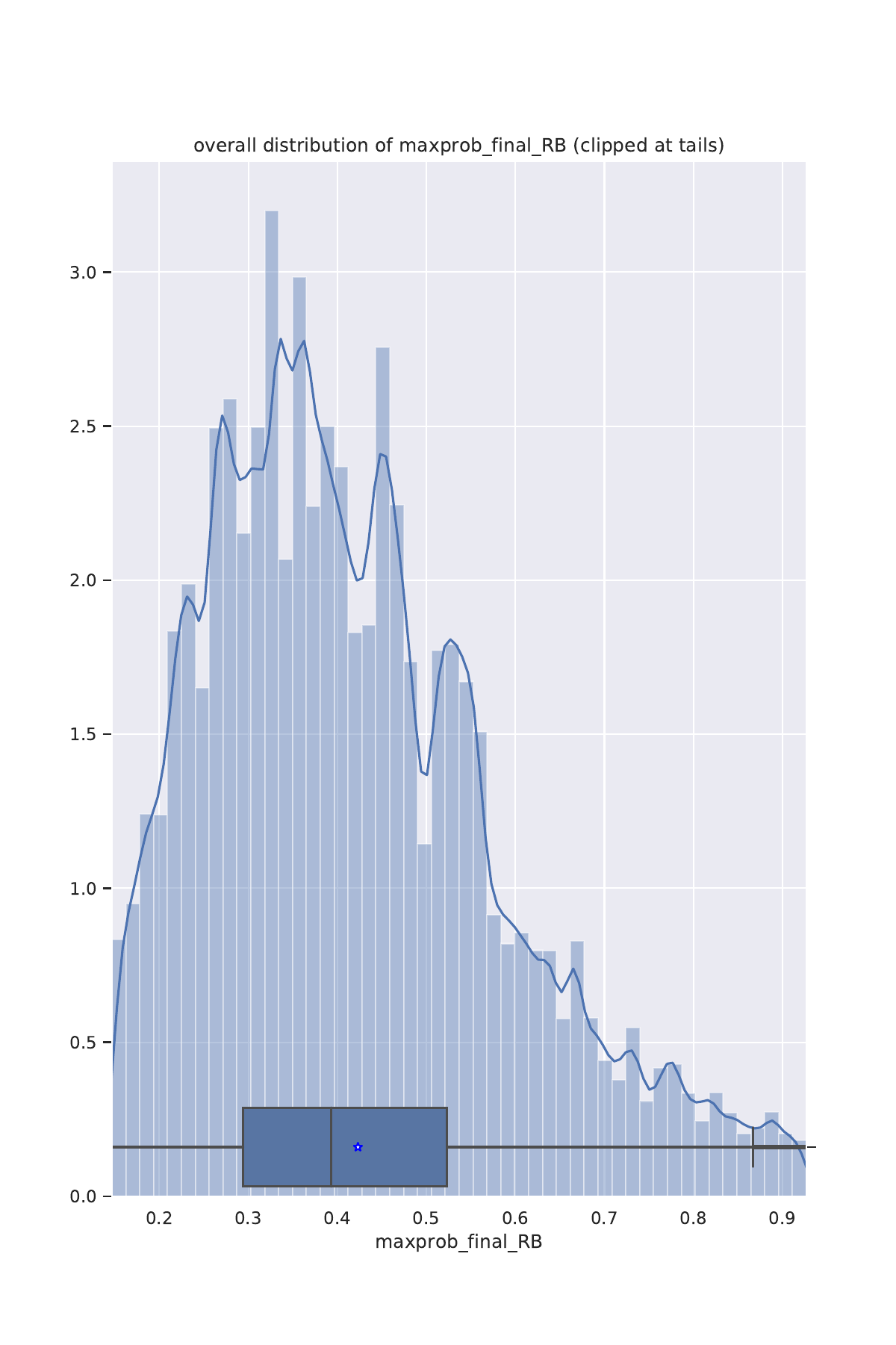}
&\doplot{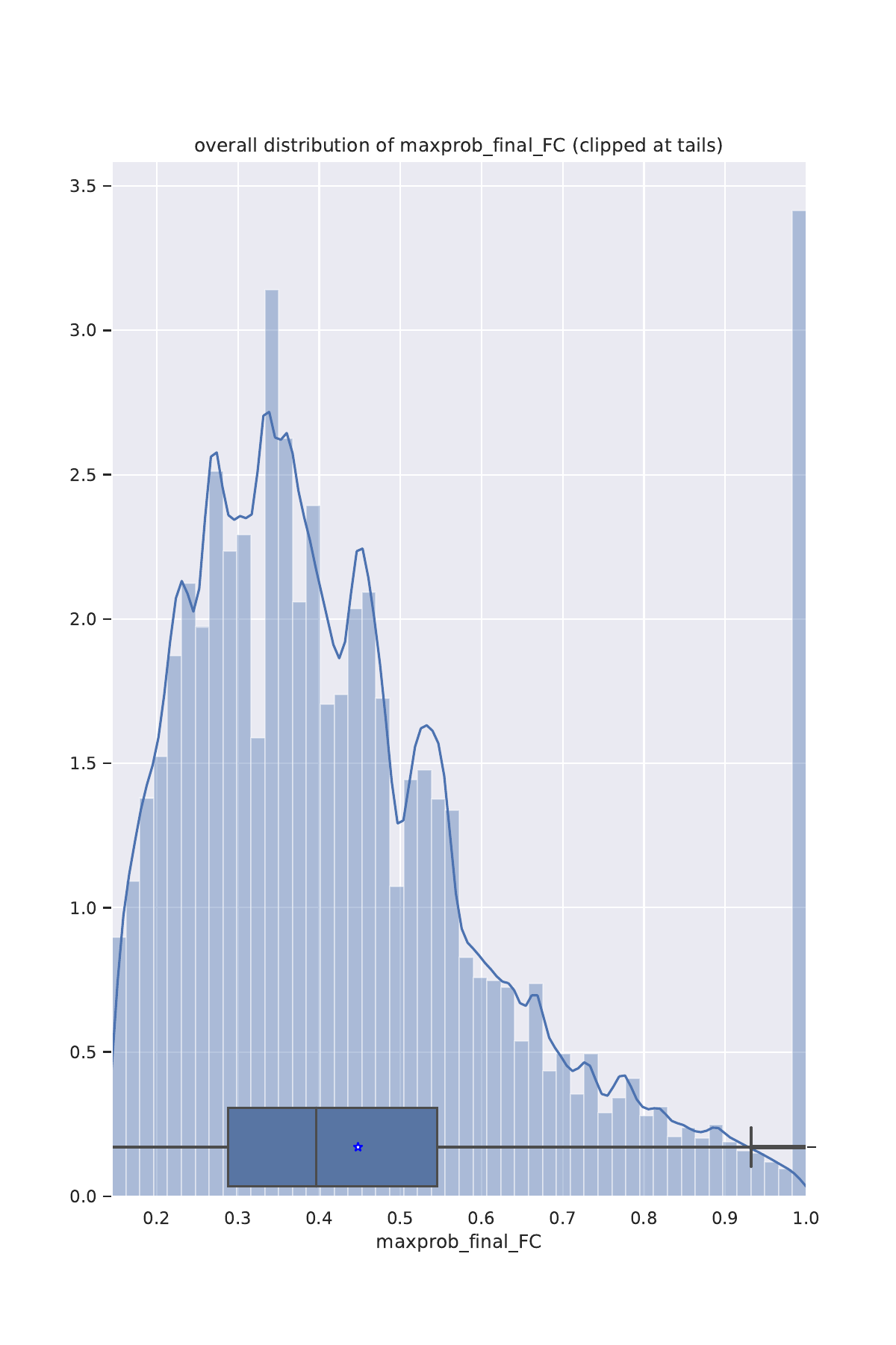}
&\doplot{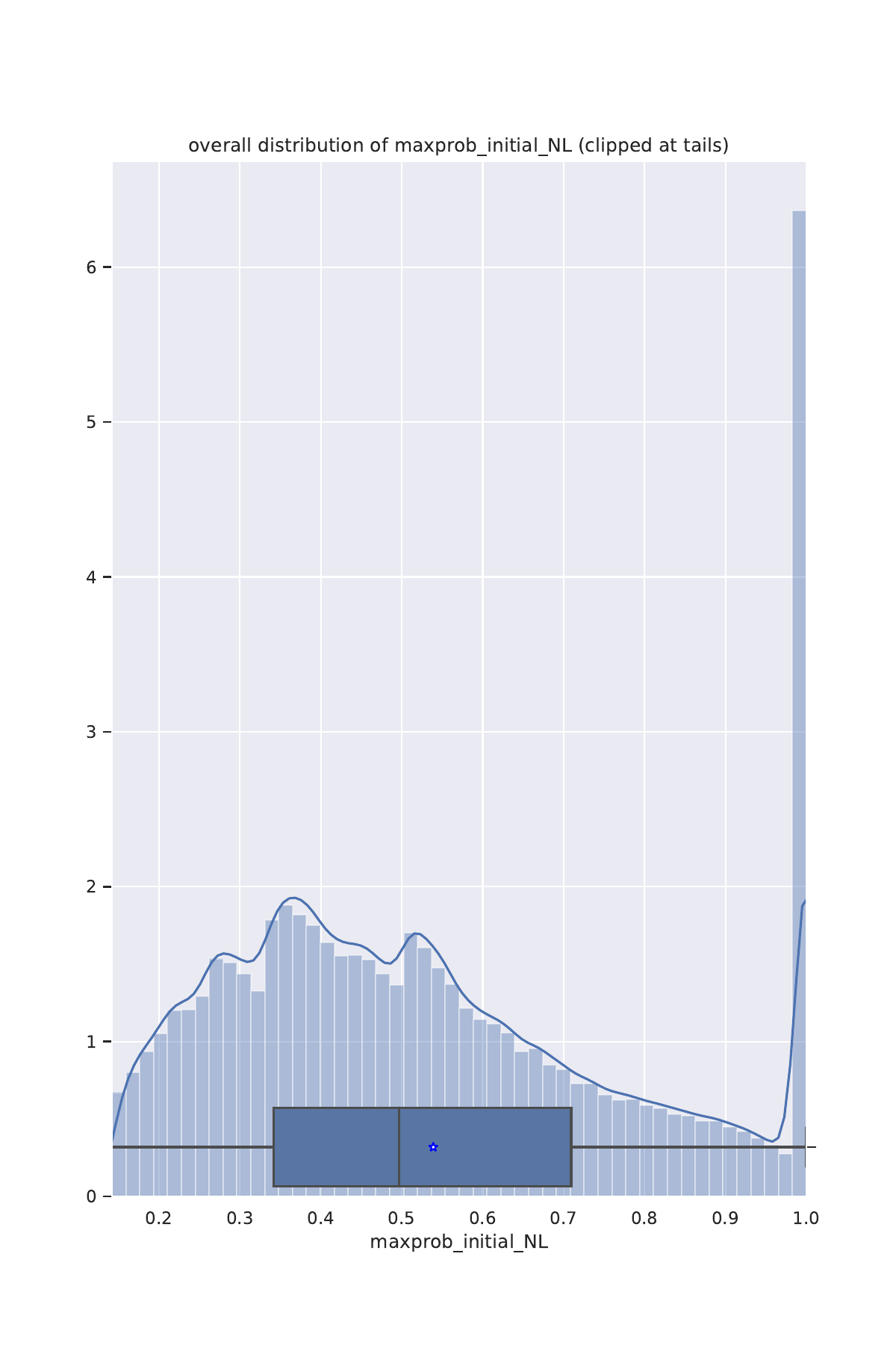}
&\doplot{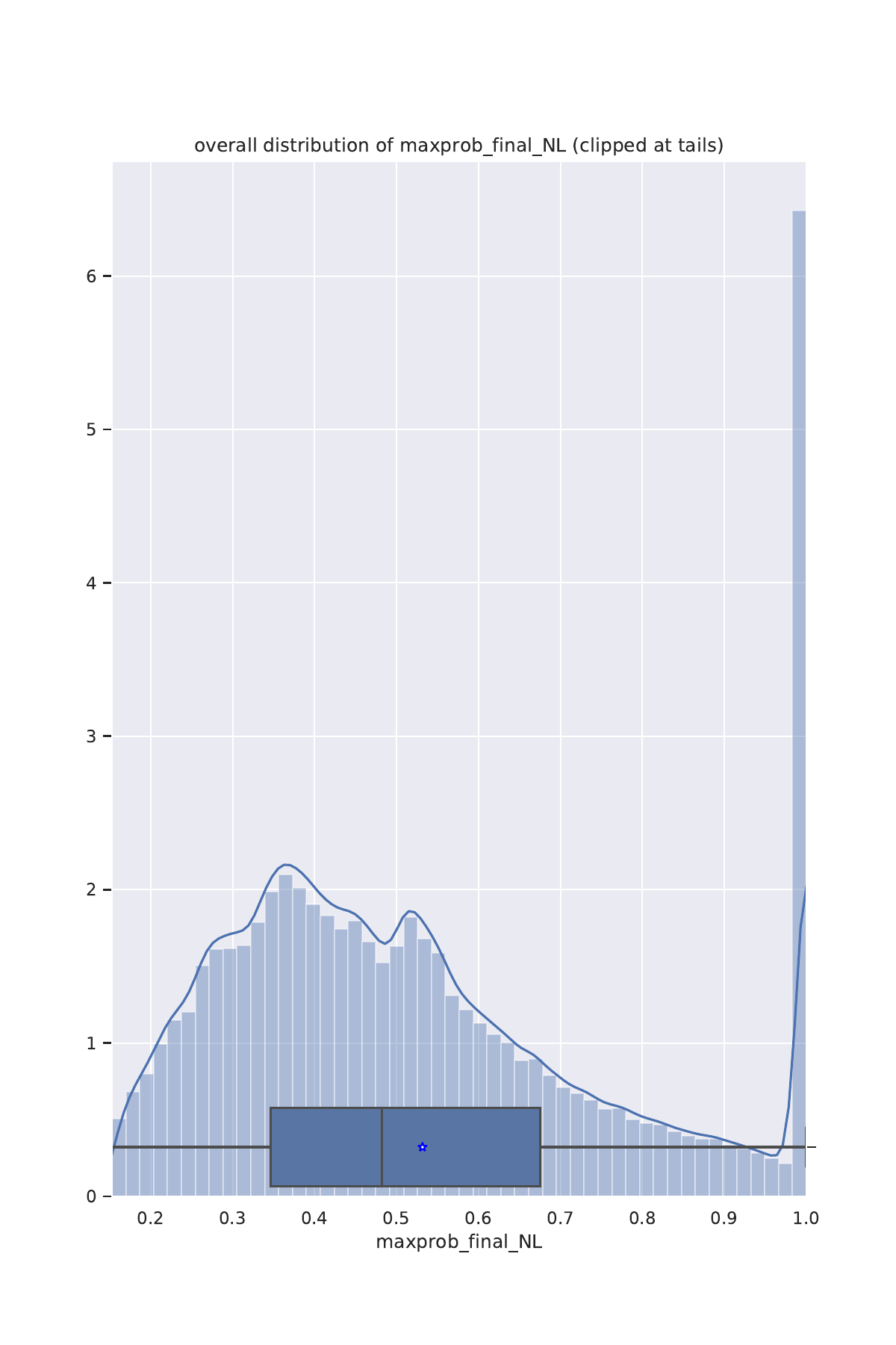}
&\doplot{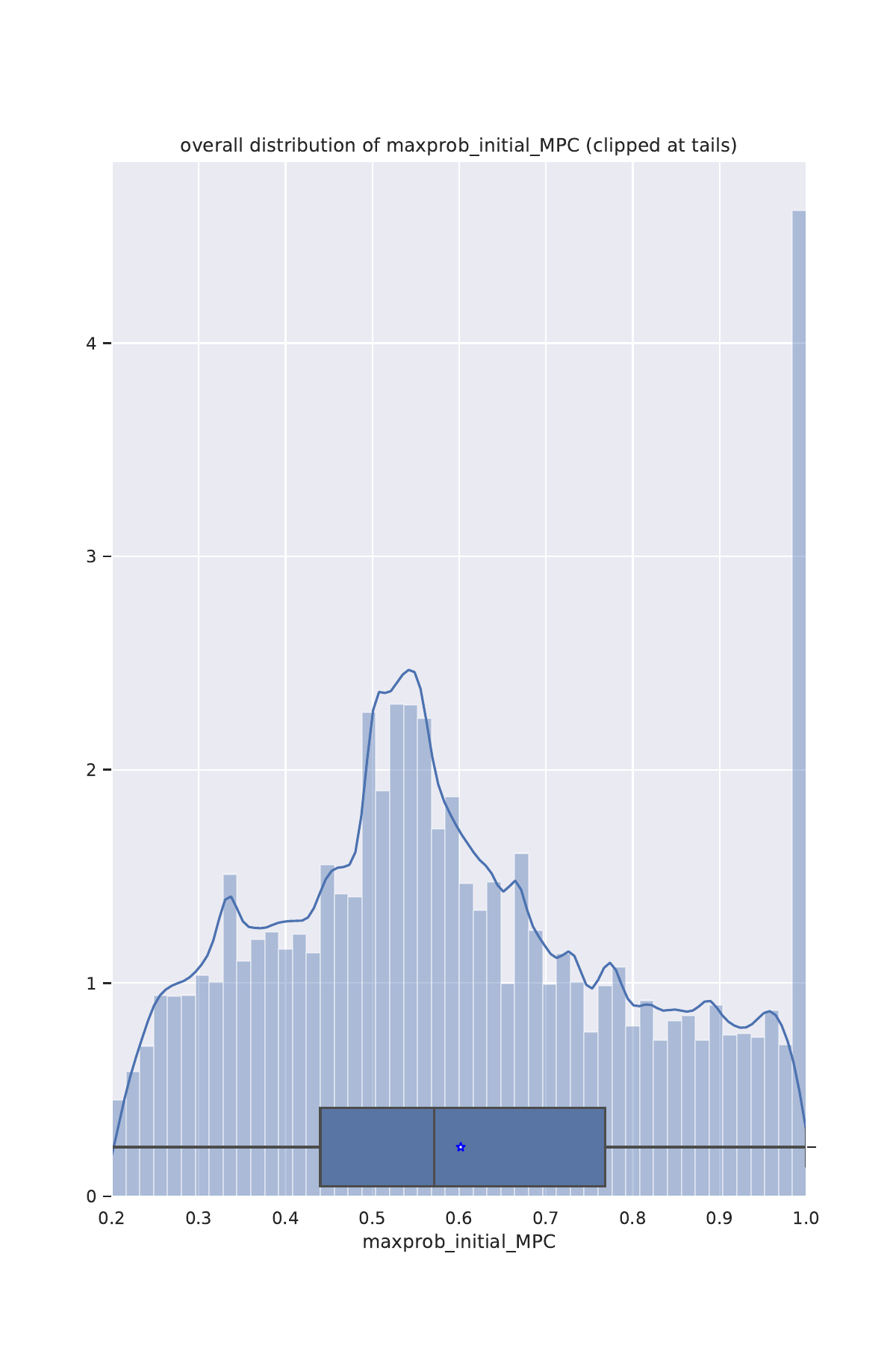}
&\doplot{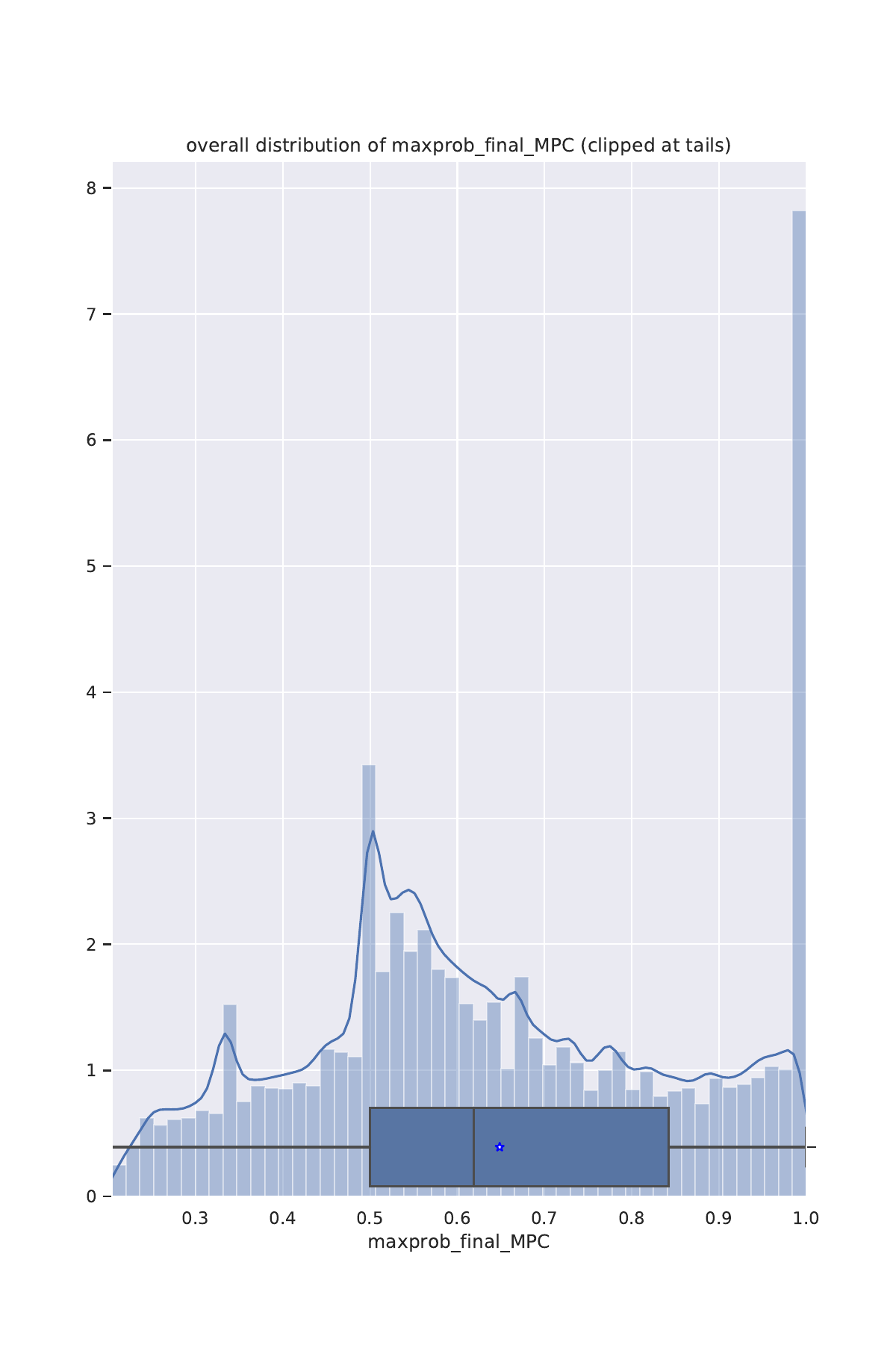}\\
RV (final) & RB (final) & FC (final) & NL (initial) & NL (final) & MPC (initial) & MPC (final)
\end{tabular}
\caption{\label{fig:probabilities}%
Initial and final distribution of Shannon entropy (top) and maximal option probability (bottom) across decision problems for selected methods. 
}\end{sidewaysfigure}

\begin{sidewaysfigure}\centering
\def\doplot#1{\includegraphics[width=0.19\textwidth,trim=40 40 40 40,clip]{plots/2020_03_28_1mio_all_#1.png}}
\begin{tabular}{ccccc}
\doplot{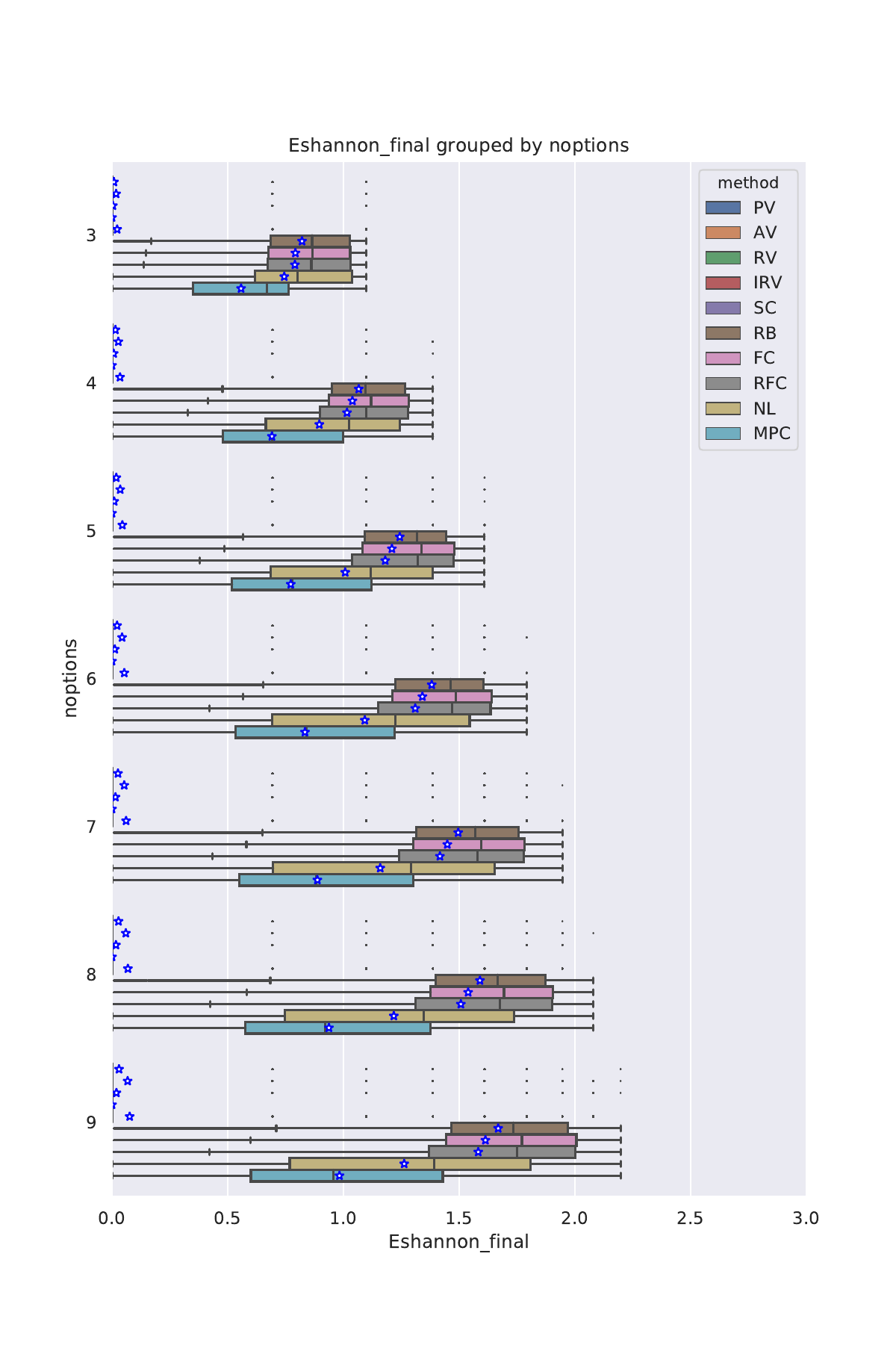}
&\doplot{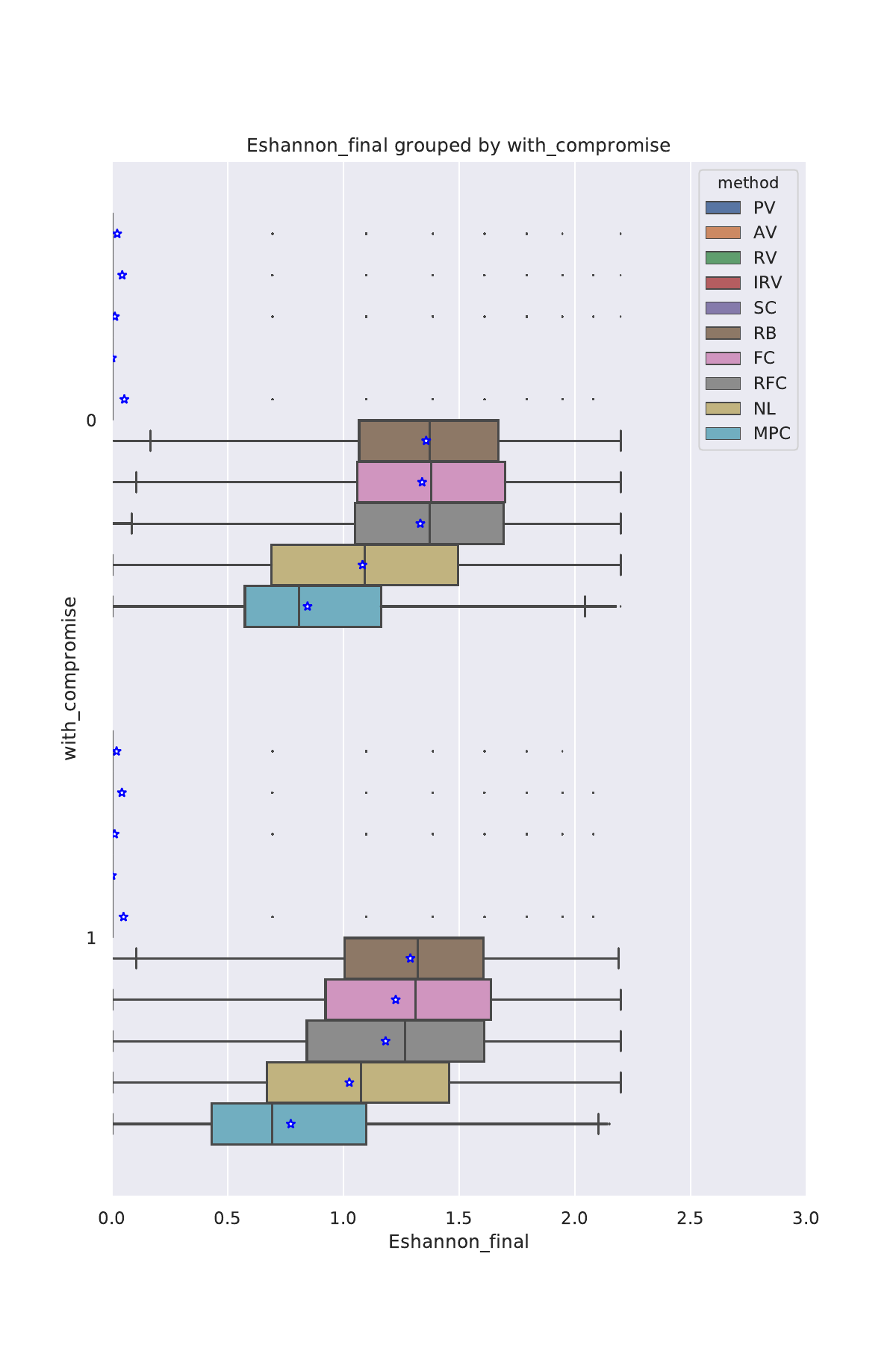}
&\doplot{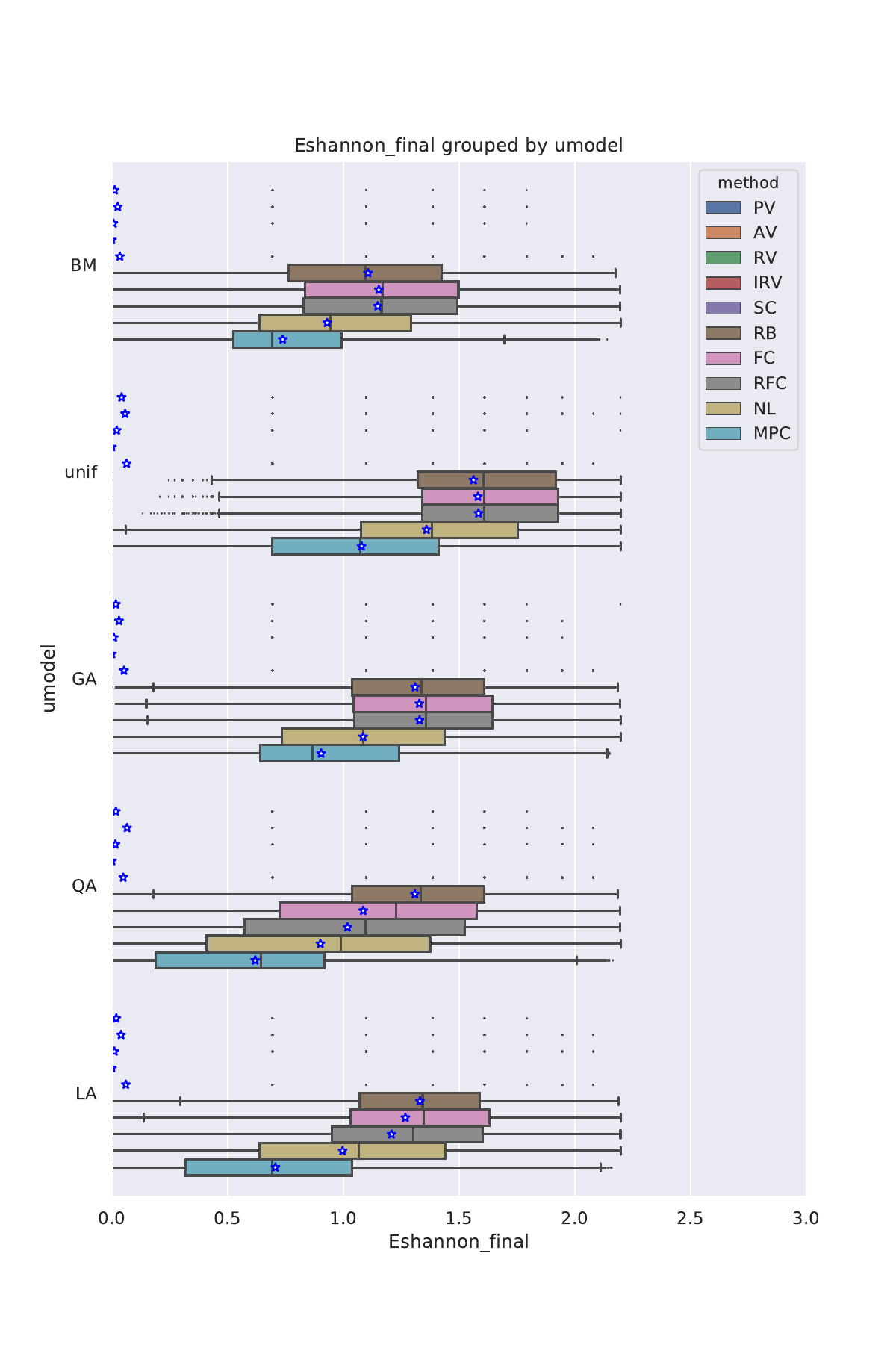}
&\doplot{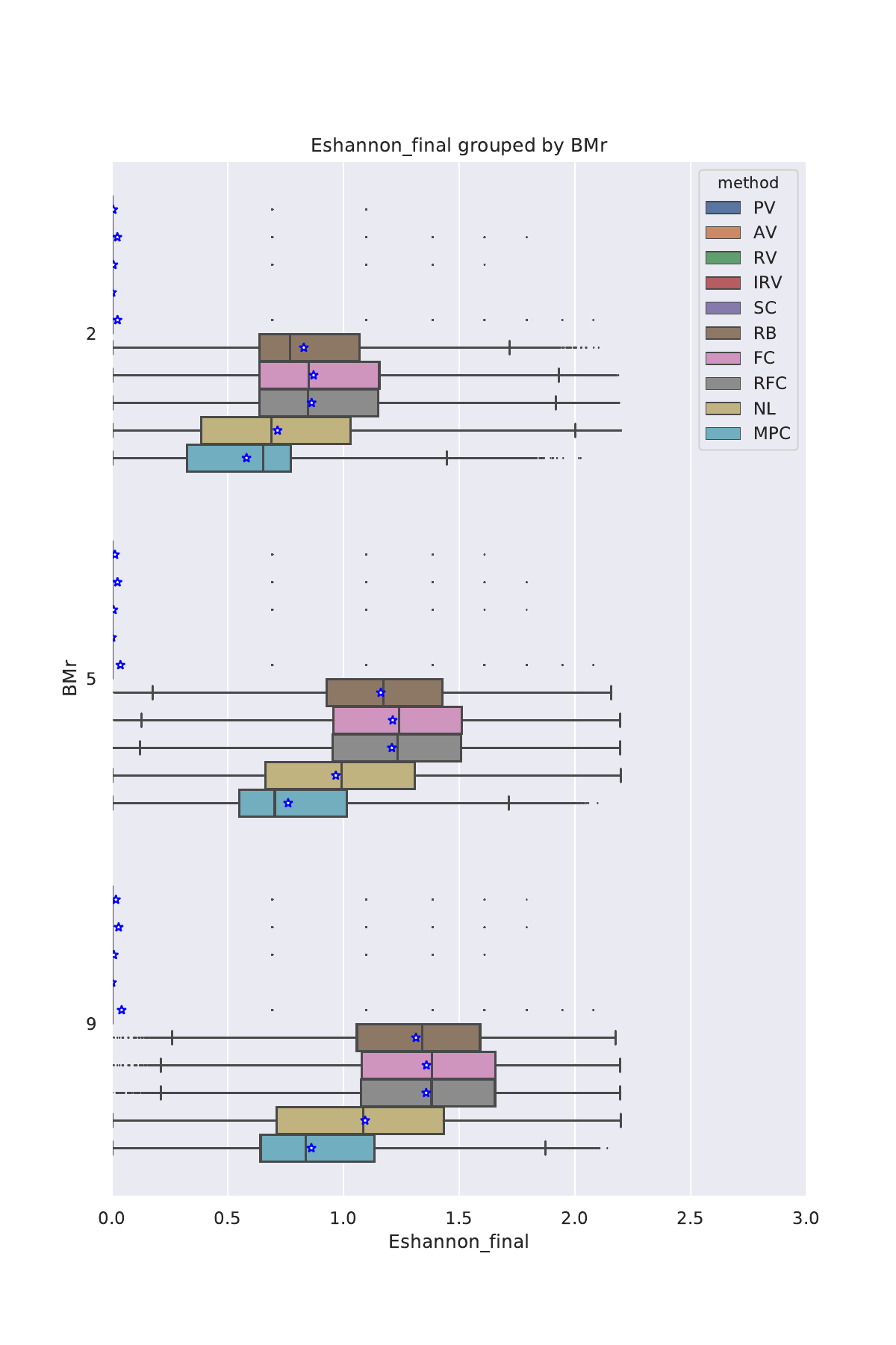}
&\doplot{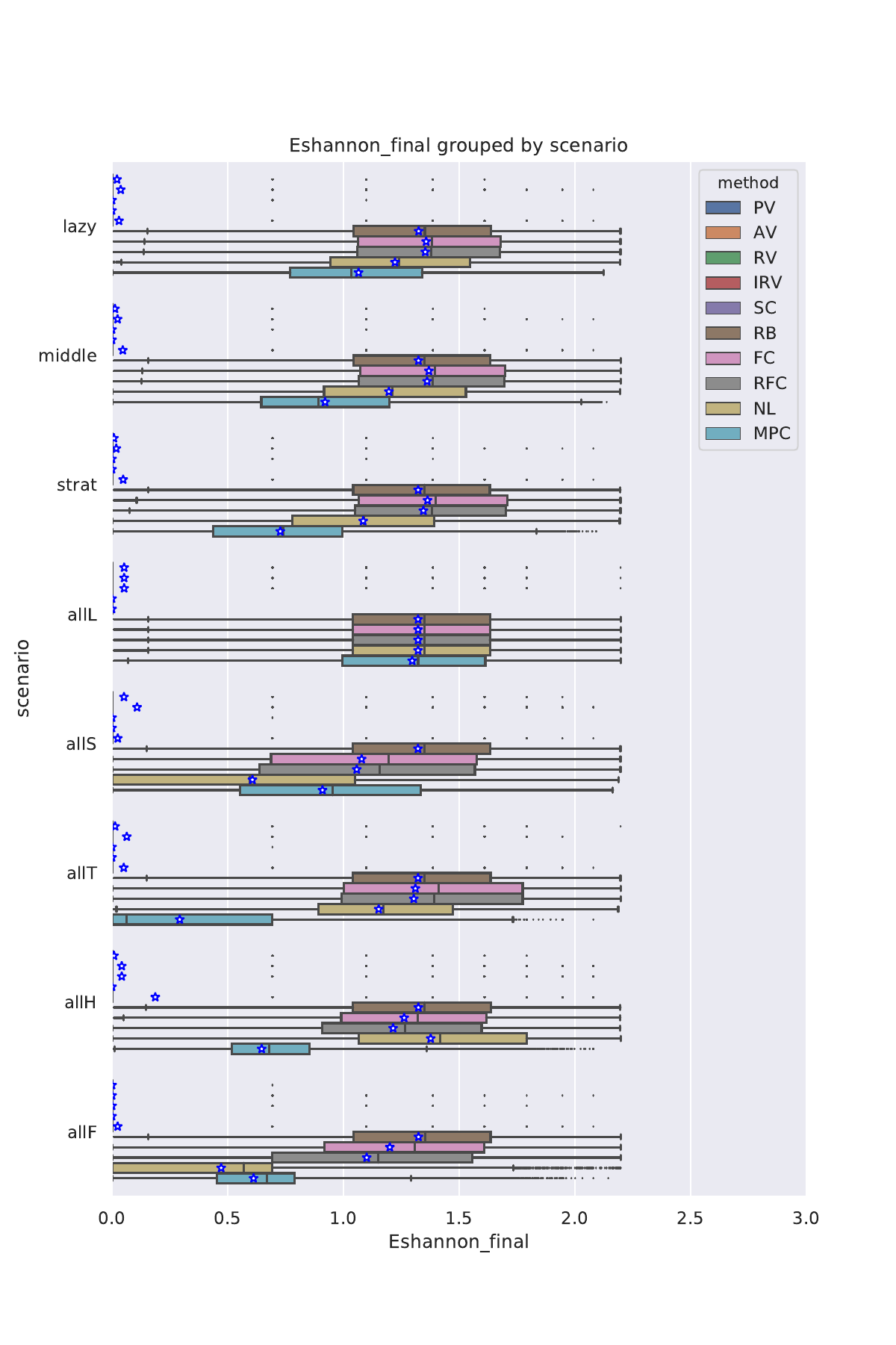}\\
\doplot{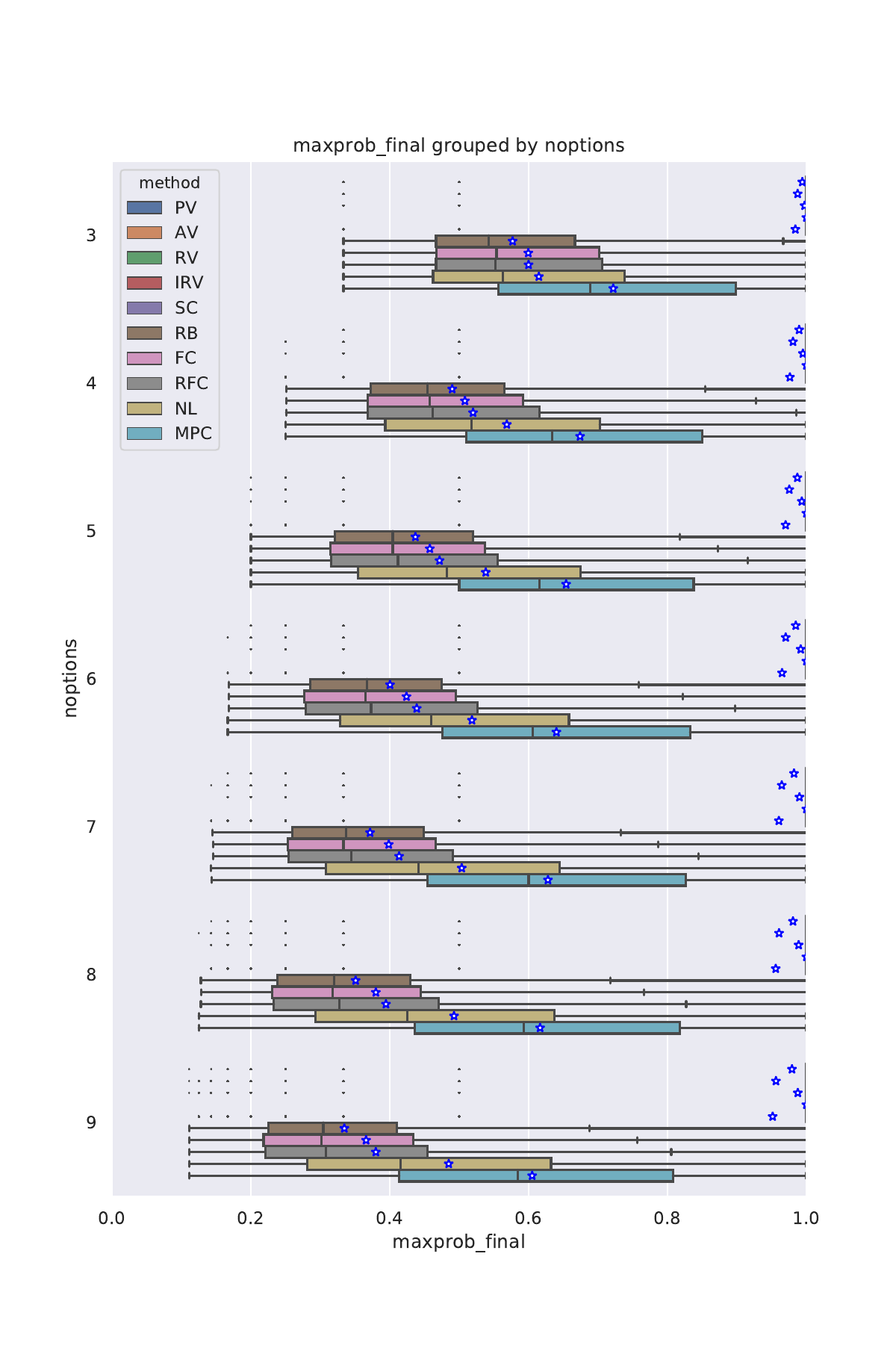}
&\doplot{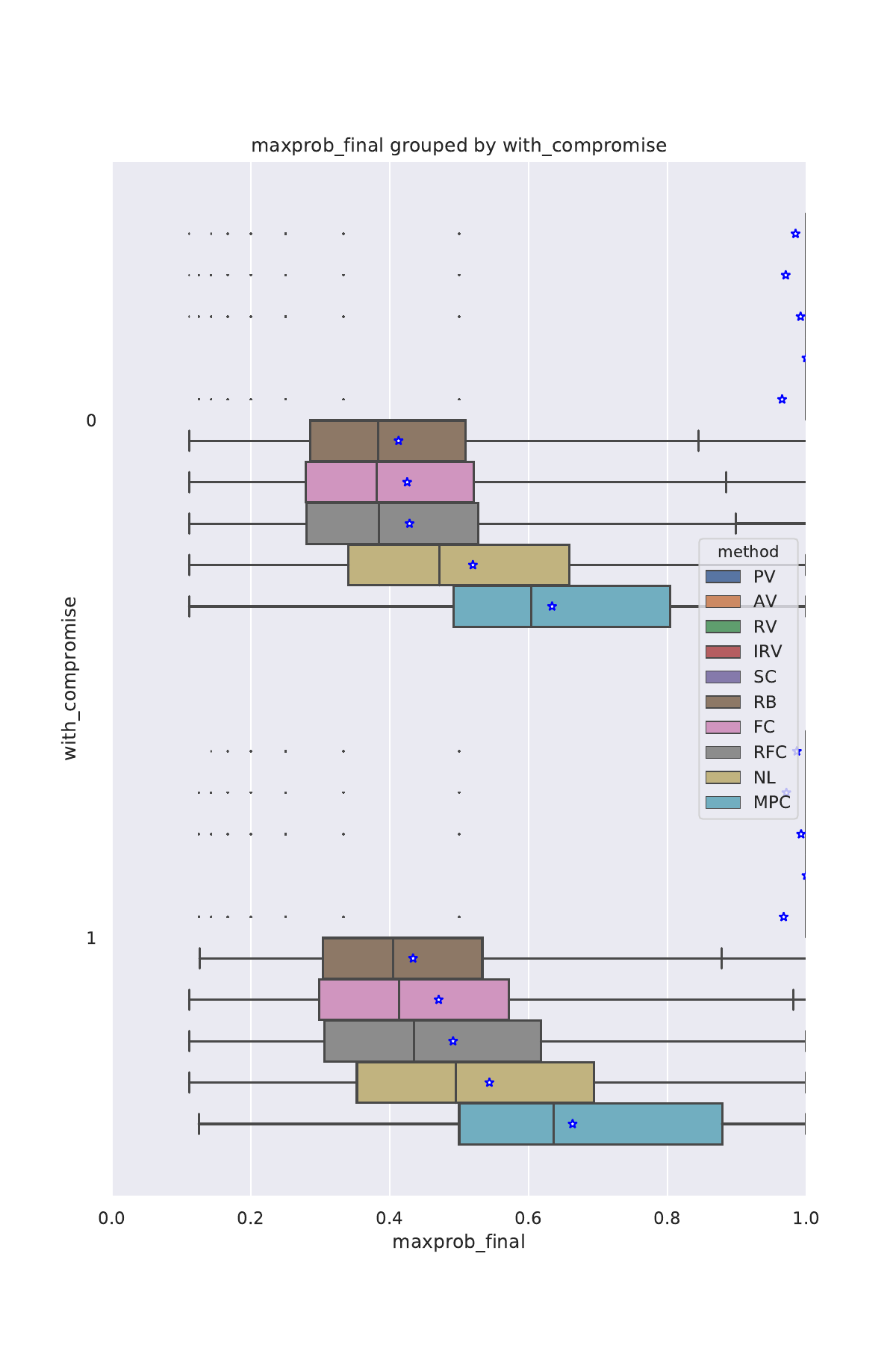}
&\doplot{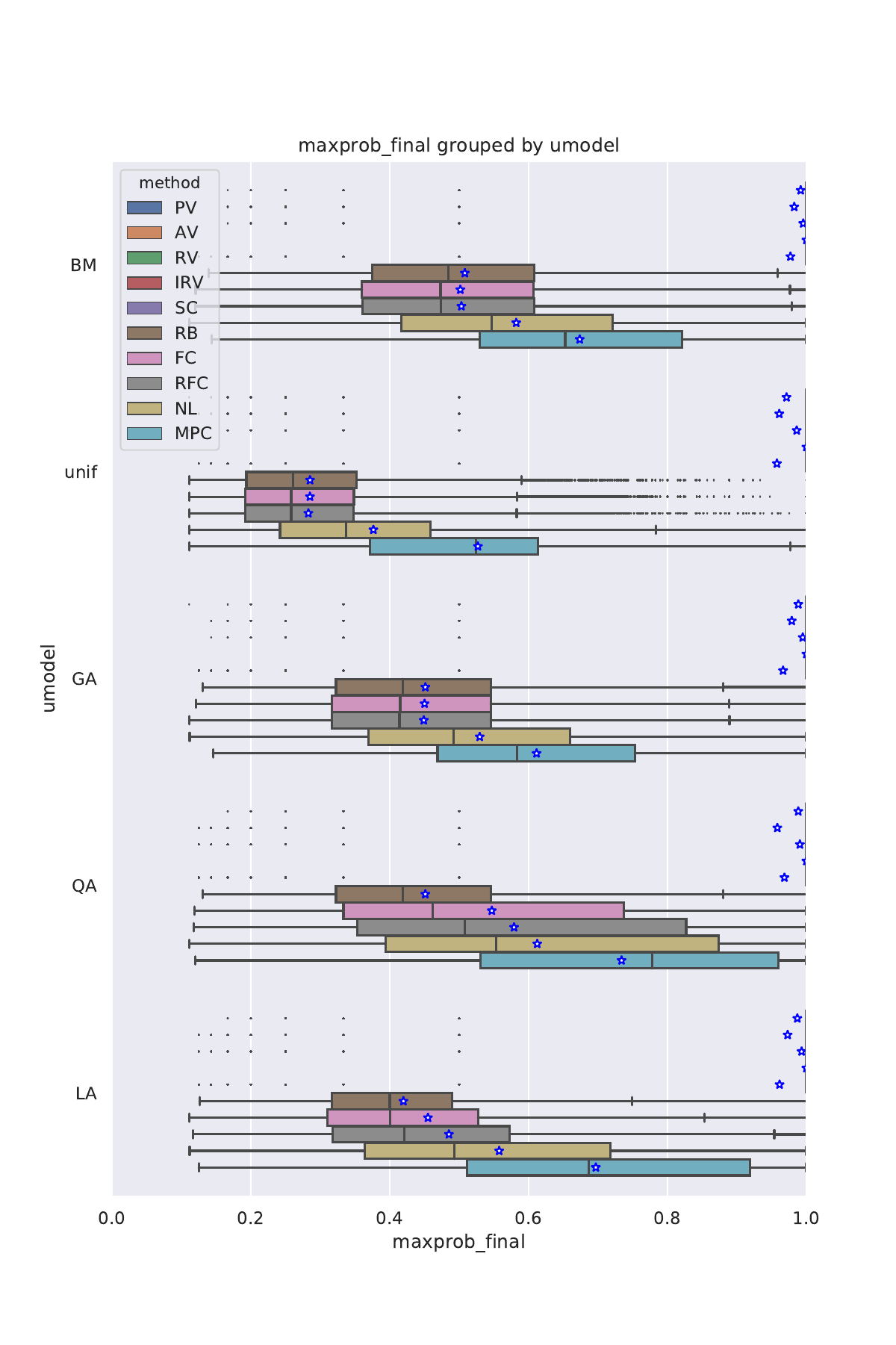}
&\doplot{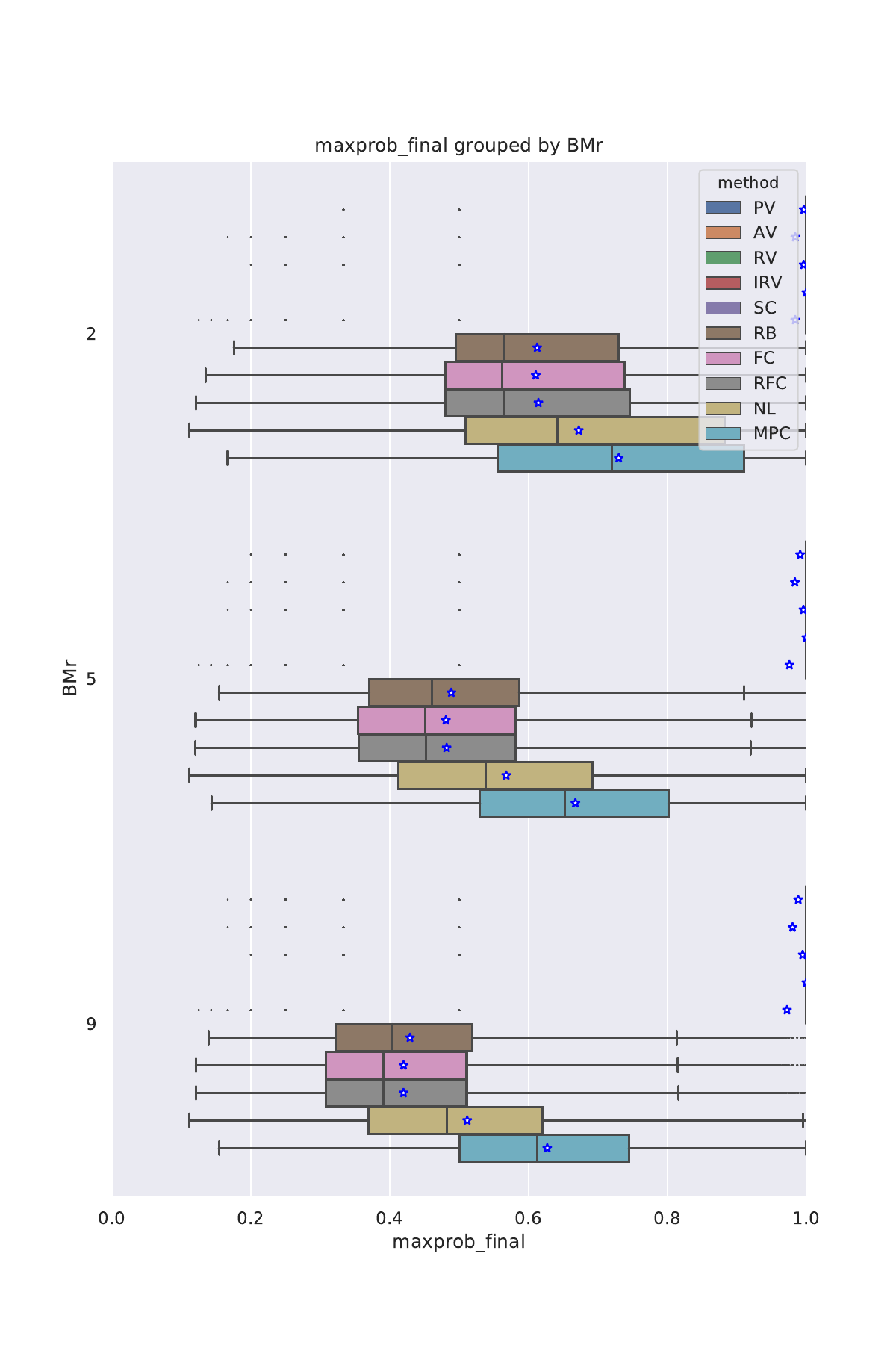}
&\doplot{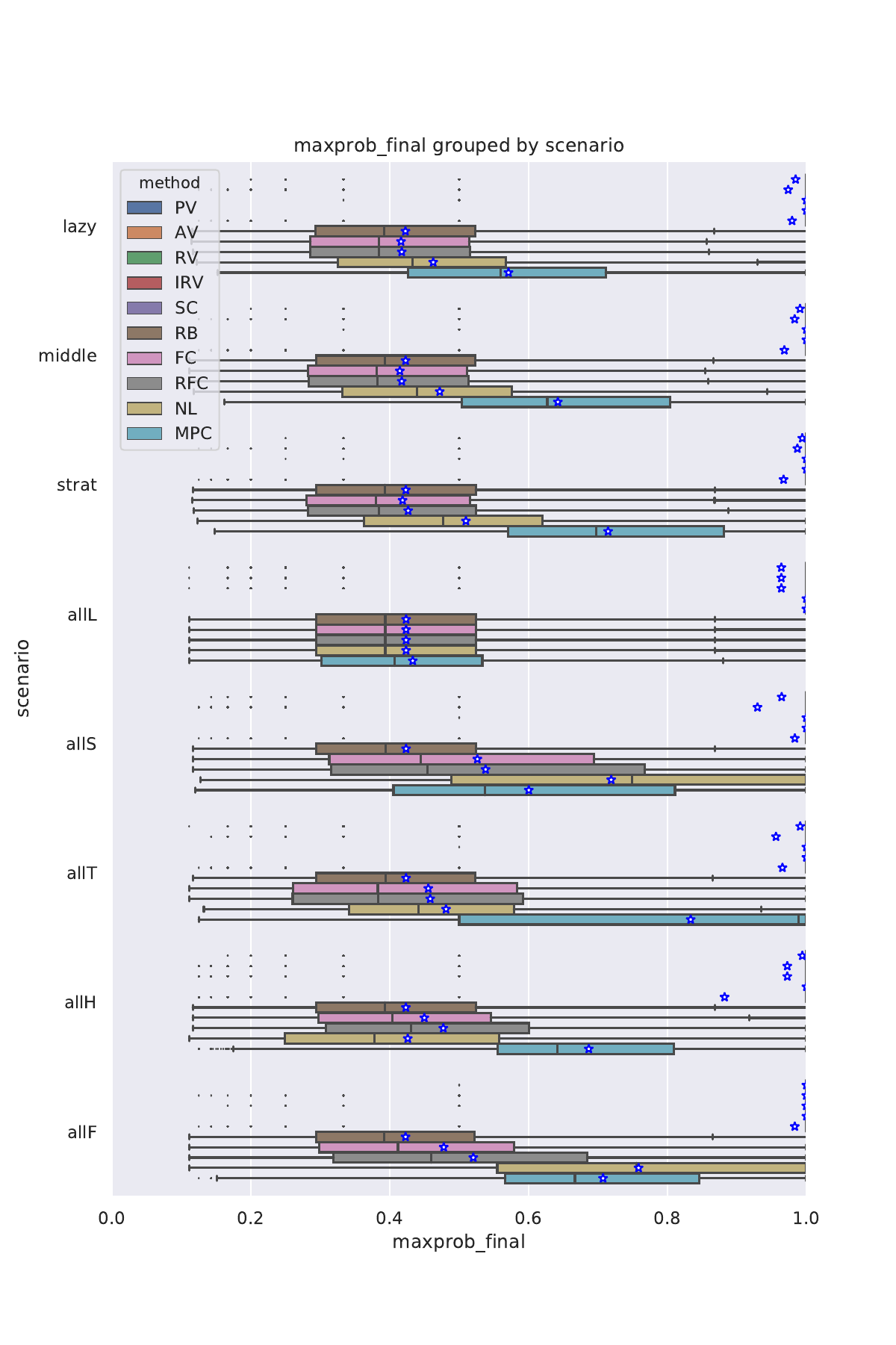}\\
{\tt noptions} & {\verb!with_compromise!} & {\tt umodel} & {\tt BMr} & {\tt scenario}
\end{tabular}
\caption{\label{fig:randomization}%
Statistics for final Shannon entropy (top) and maximal option probability (bottom) across decision problems for all methods,
grouped by parameters with considerable influence. 
}\end{sidewaysfigure}

\begin{sidewaysfigure}\centering
\def\doplot#1{\includegraphics[width=0.19\textwidth,trim=40 40 40 40,clip]{plots/2020_03_28_1mio_all_#1.png}}
\begin{tabular}{ccccc}
\doplot{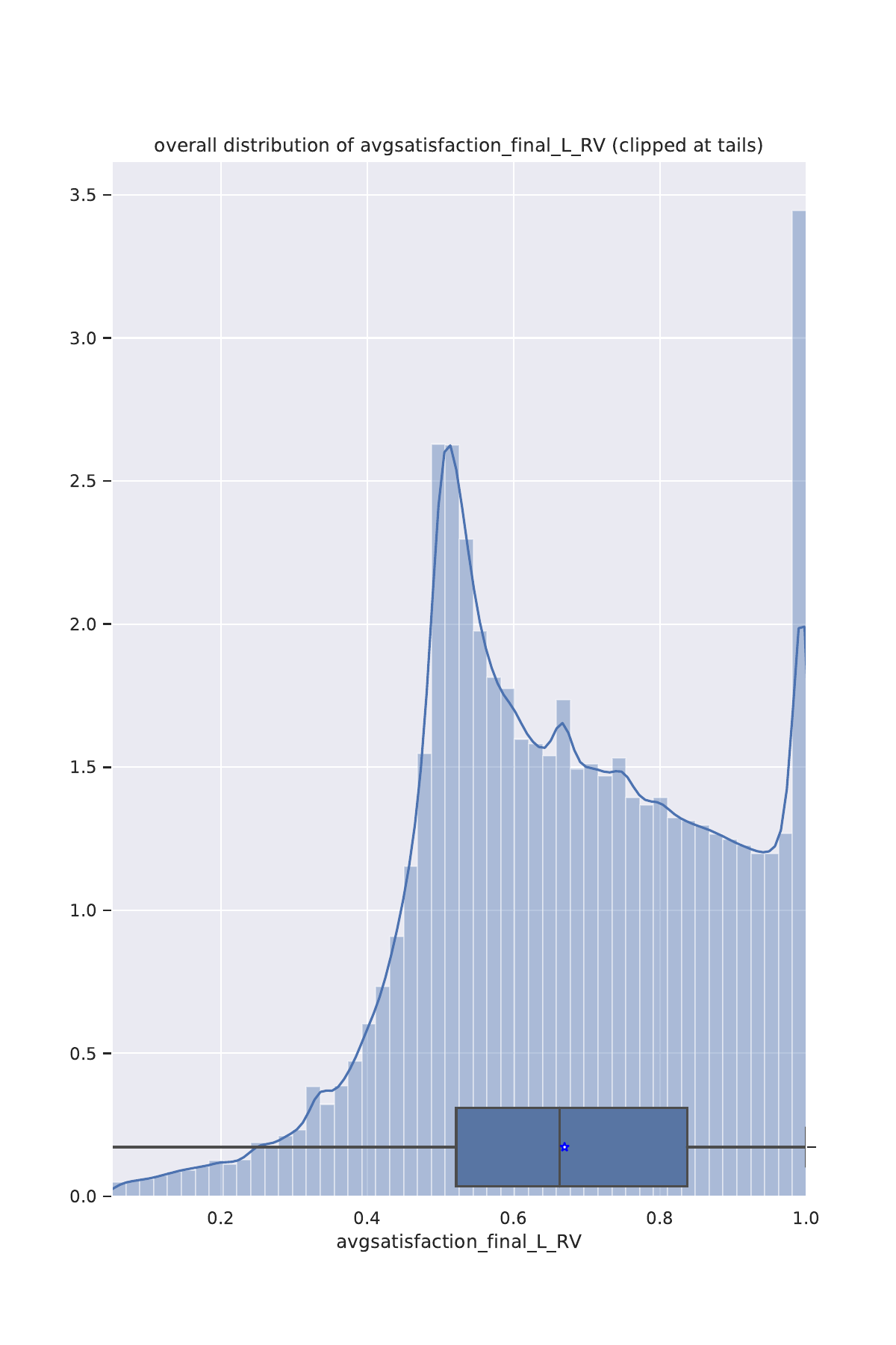}
&\doplot{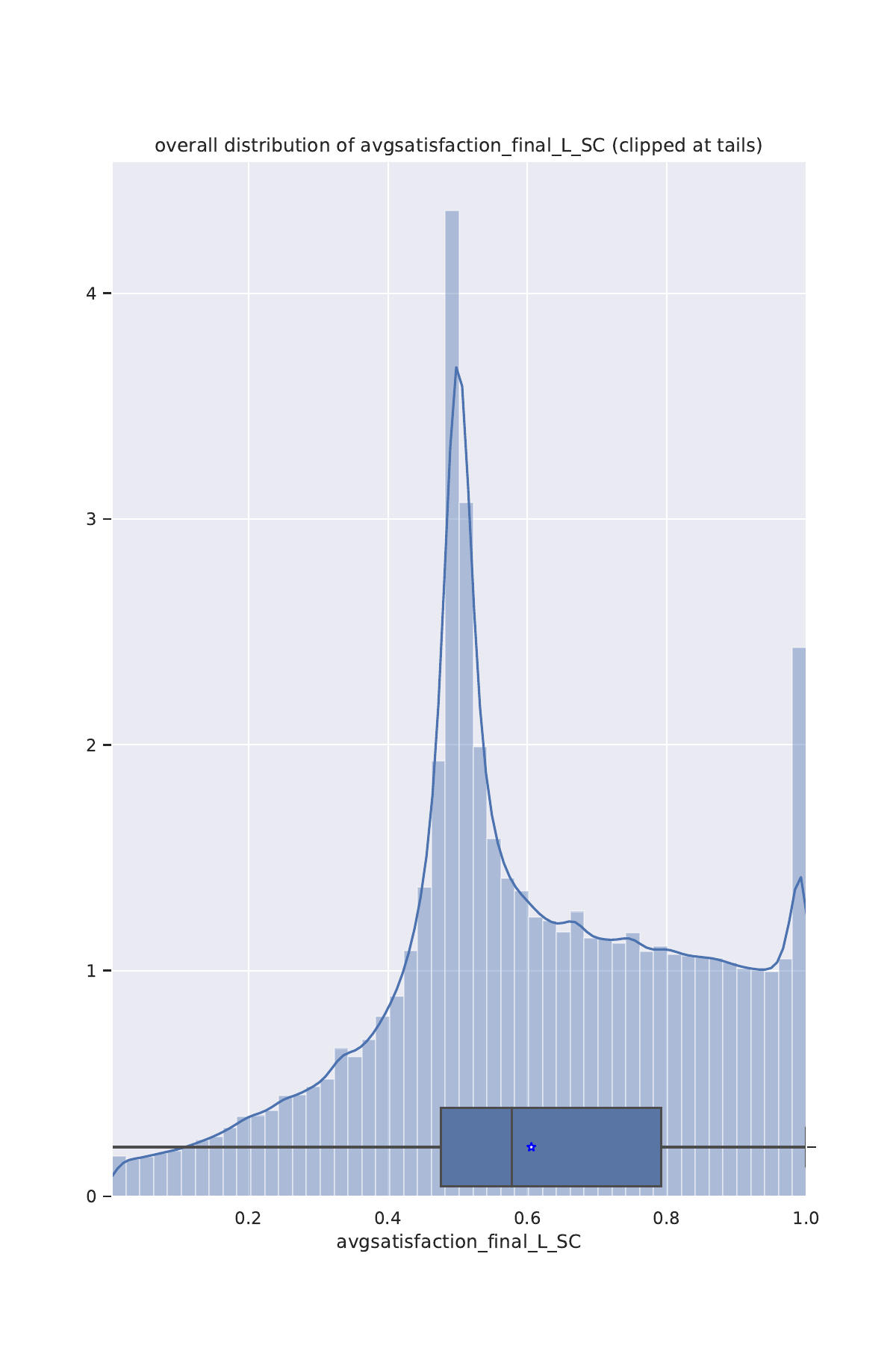}
&\doplot{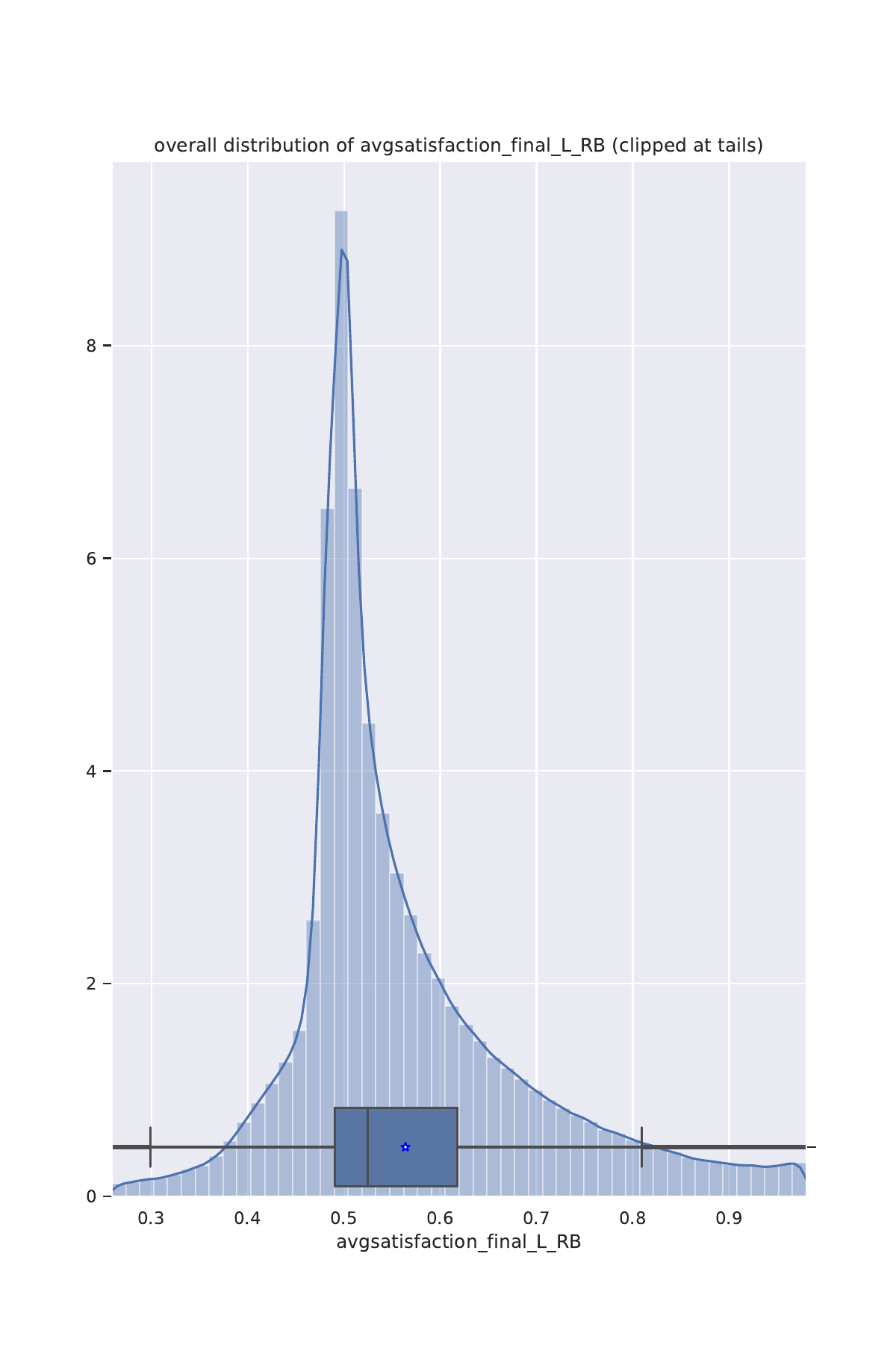}
&\doplot{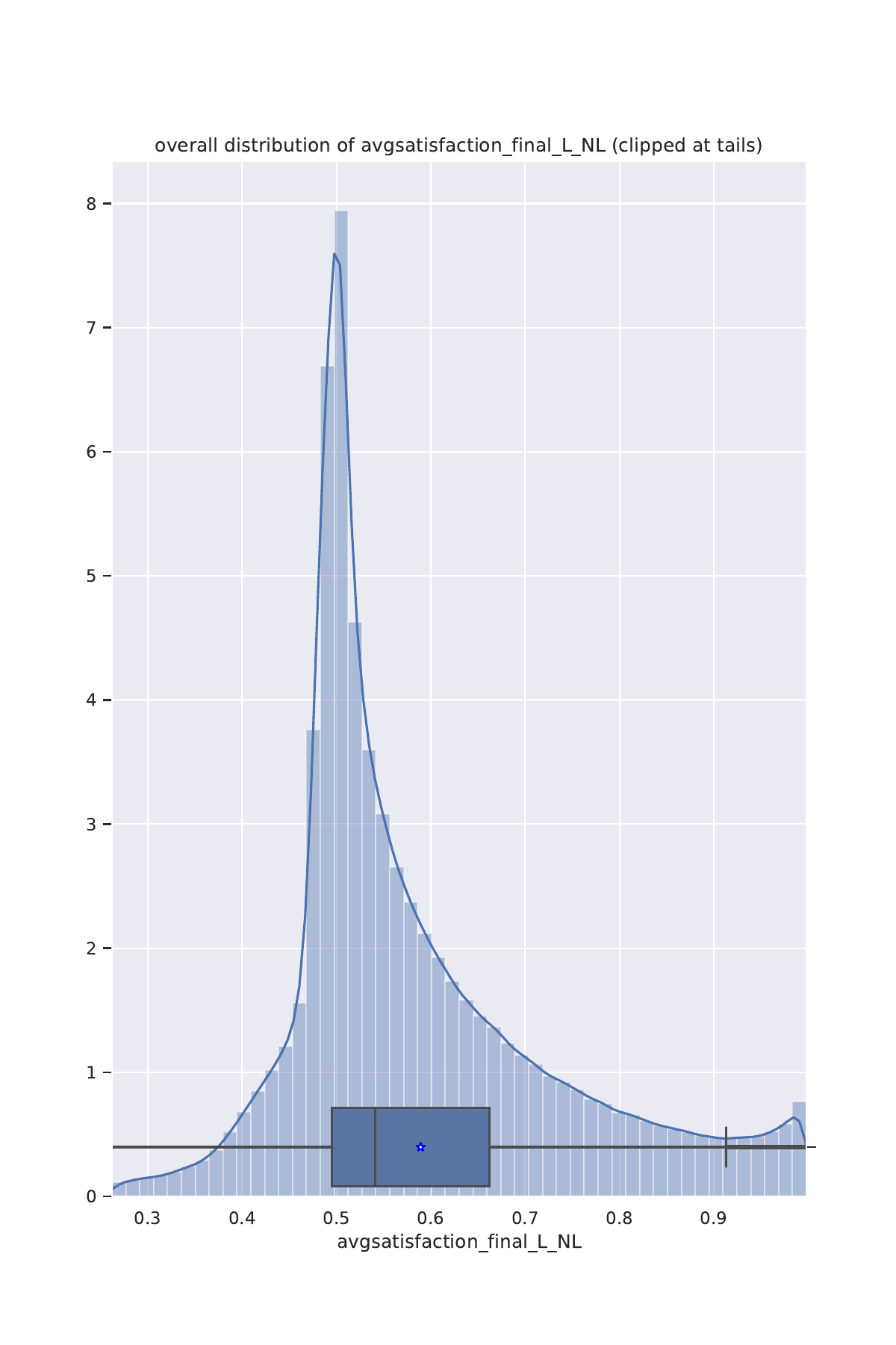}
&\doplot{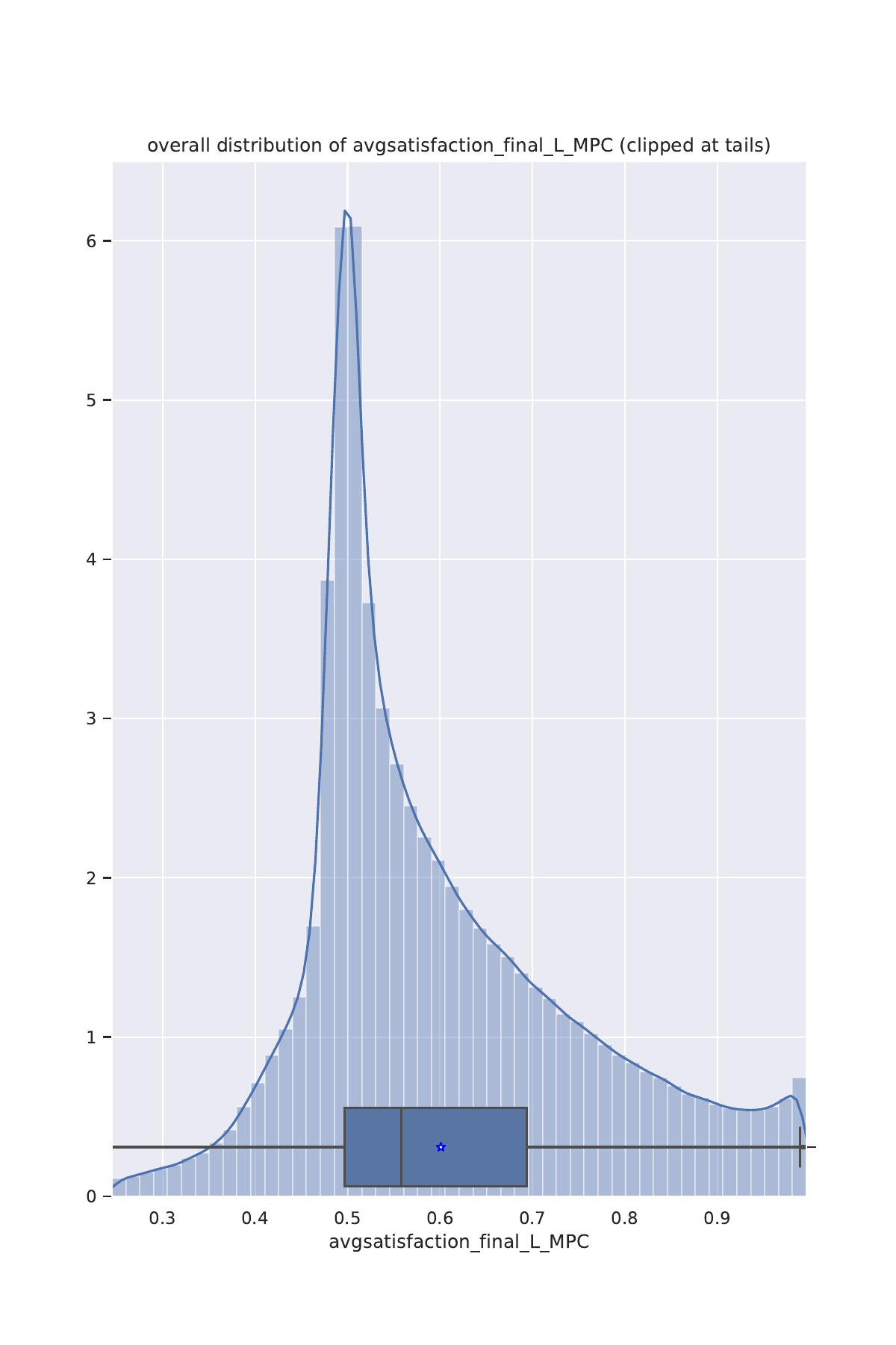}\\
\doplot{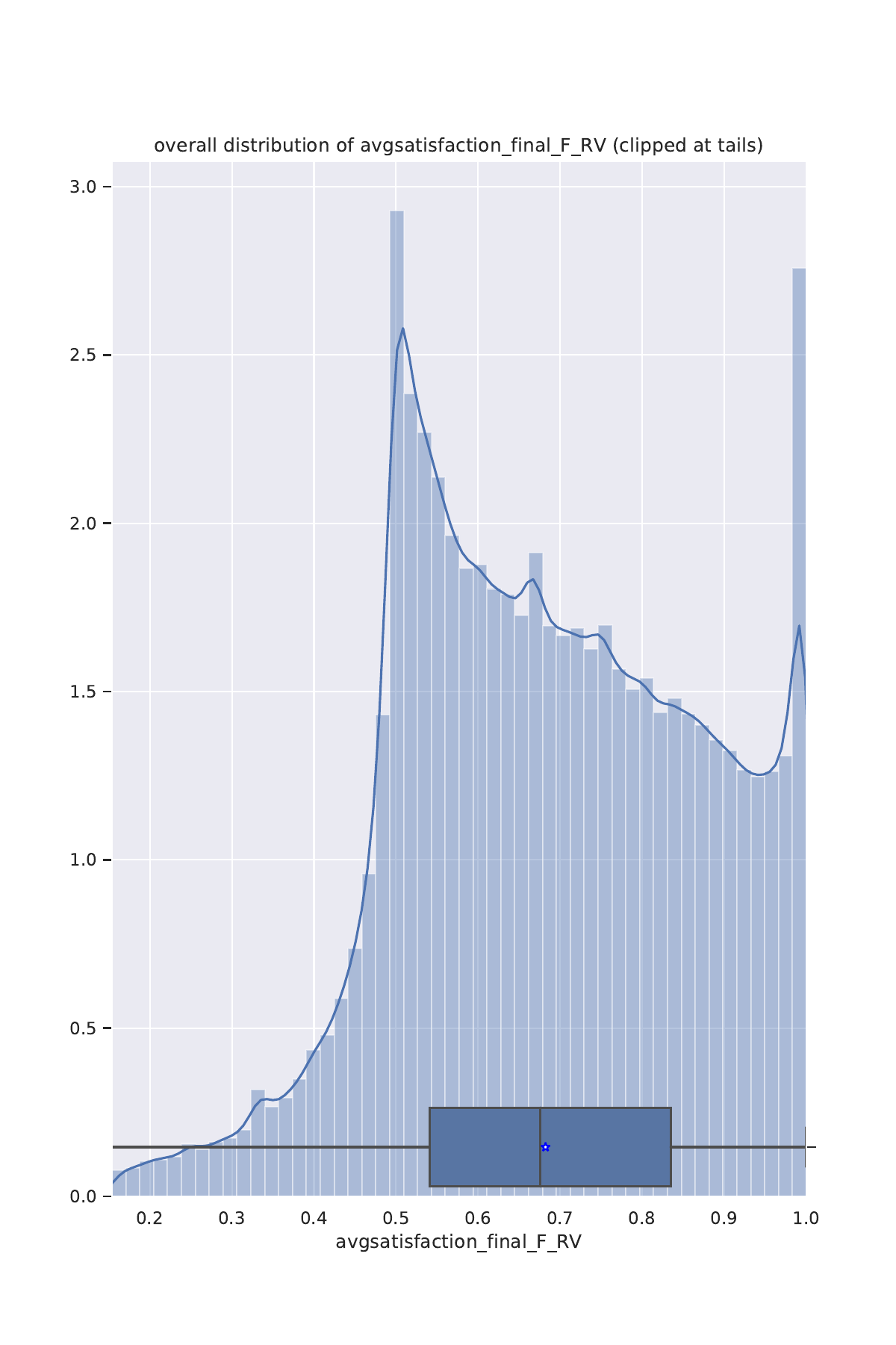}
&\doplot{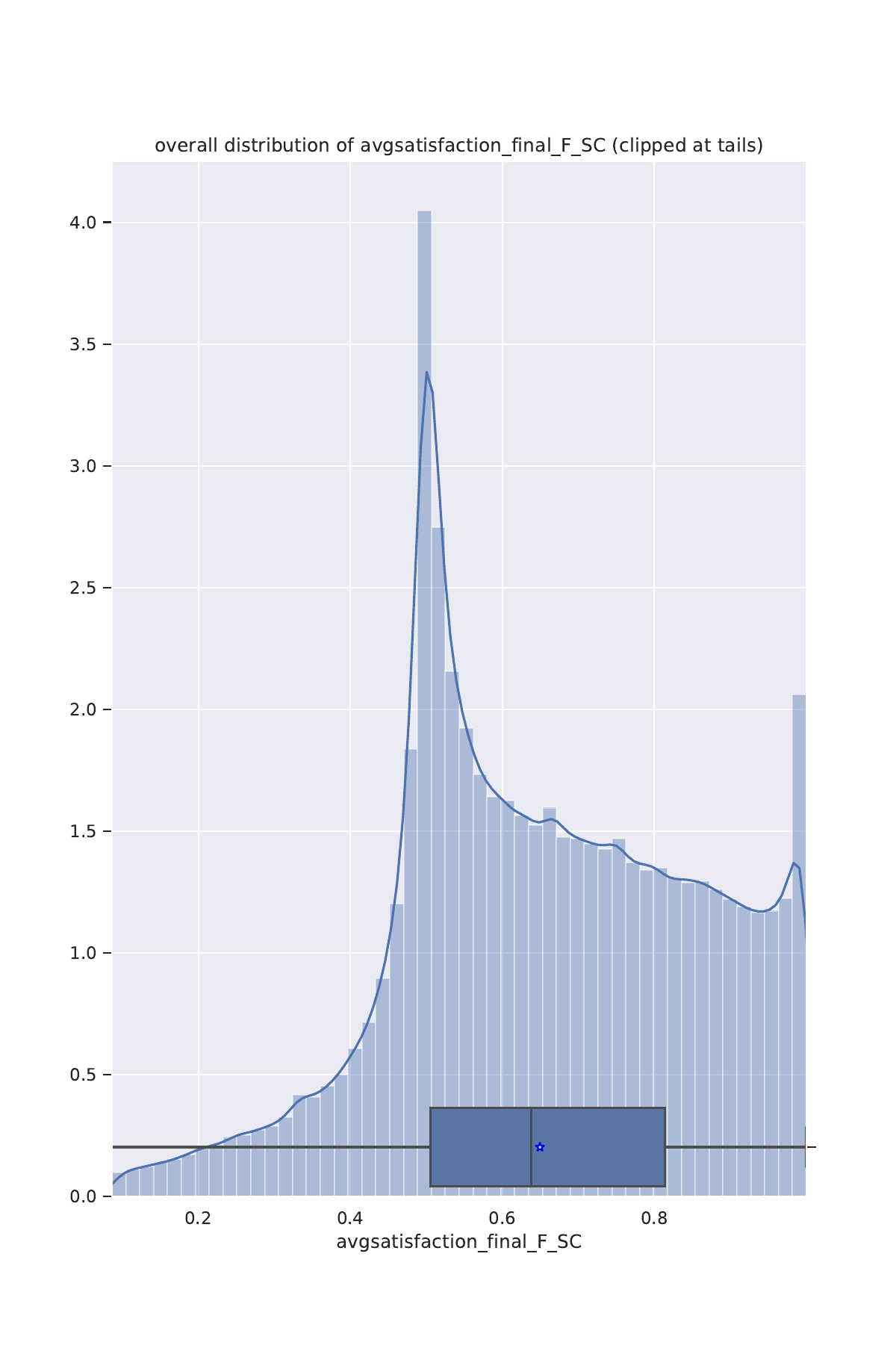}
&\doplot{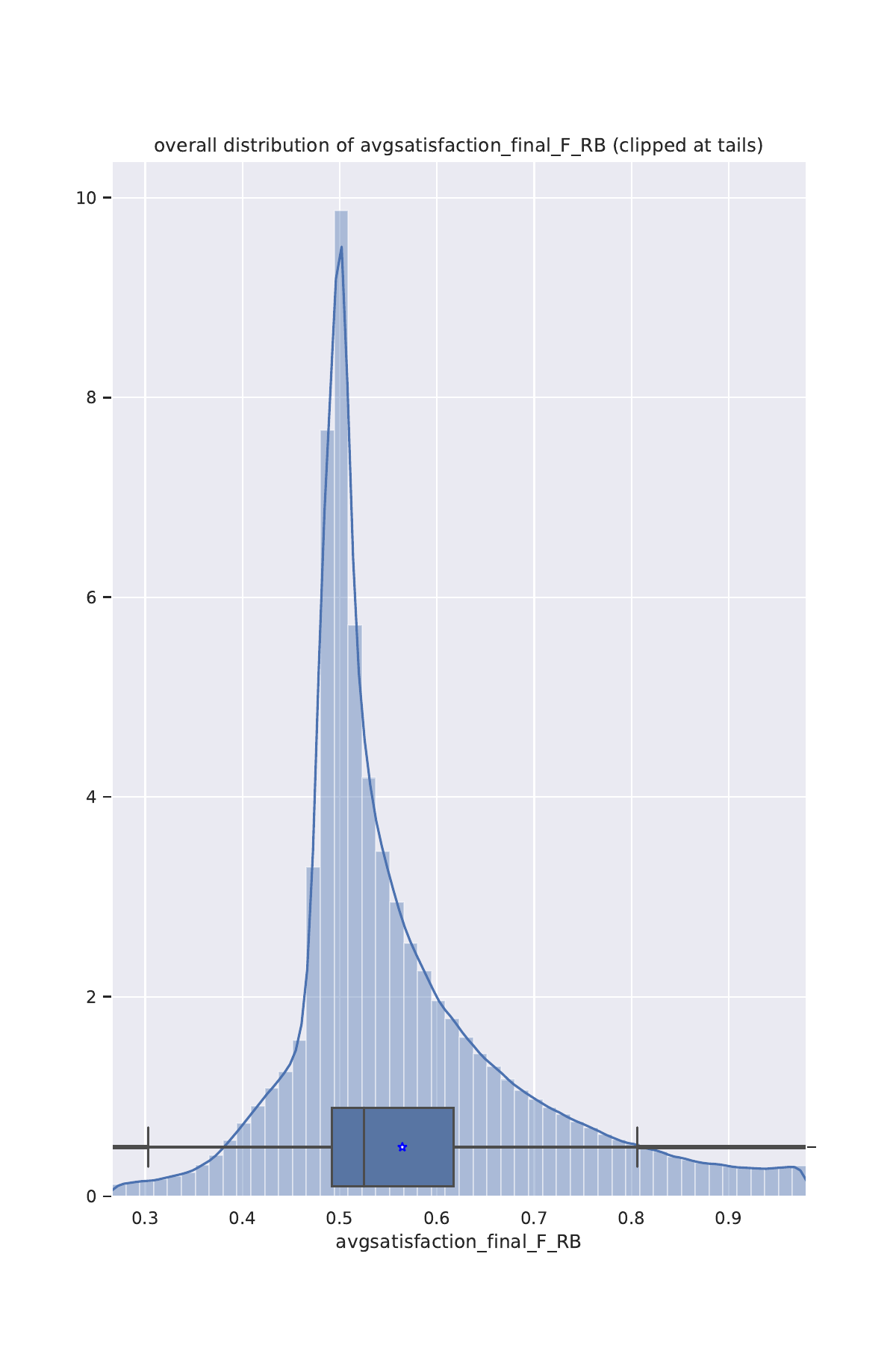}
&\doplot{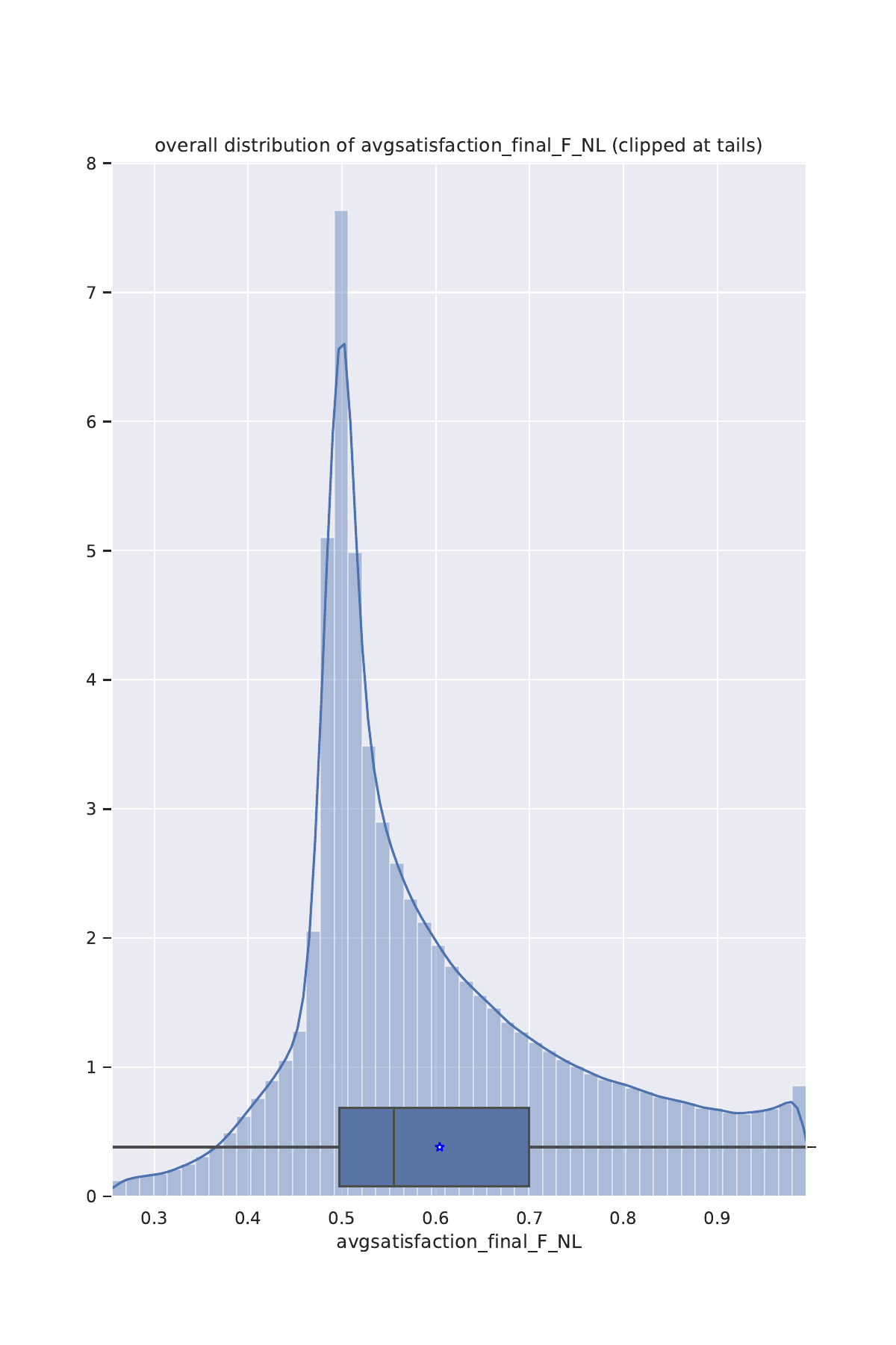}
&\doplot{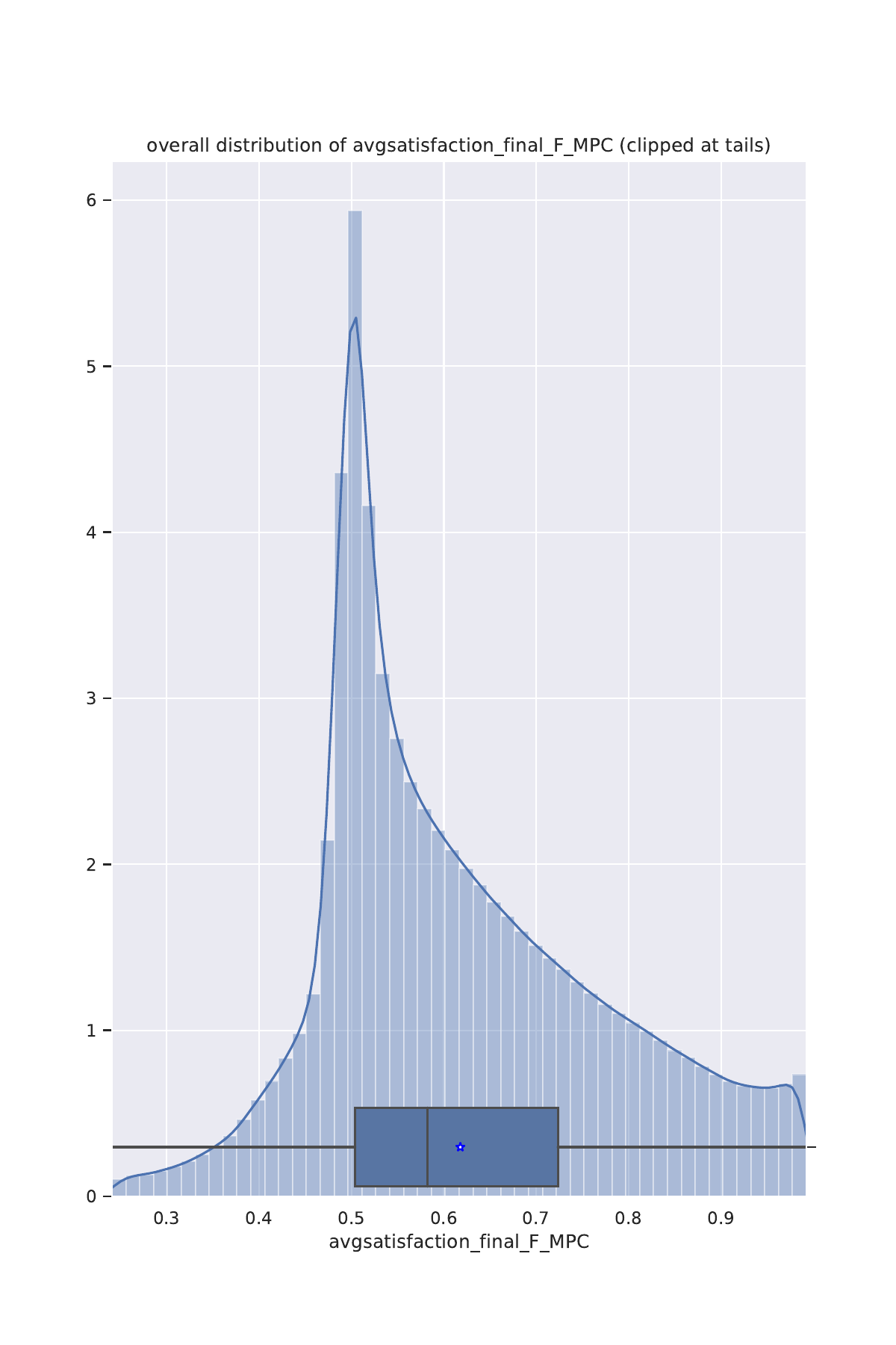}\\
RV & SC & RB & NL & MPC
\end{tabular}
\caption{\label{fig:satisfaction}%
Distribution of final average satisfaction of lazy (top) and factionally strategic (bottom) voters across decision problems for selected methods.
}\end{sidewaysfigure}

As can be seen in Fig.\,\ref{fig:satisfaction2}, the majoritarian methods fare somewhat better regarding average voter satisfaction 
with mean values around 0.66 compared to MPC's mean of around 0.61,
again quite much depending on the preference model.
As can be expected, for the proportional methods the risk-averse non-EUT voters were less satisfied.

\subsubsection{Preferences over methods}

If voters were to decide between the ten voting methods
and would use for this ``meta-decision'' a pairwise comparison method such as Simple Condorcet or any other majoritarian method,
which voting method would win?

If one assumes that voters are purely consequentialist and judge a method only by its generated utility,
the surprising answer seems to be that they would then end up with Instant-Runoff Voting.
As table \ref{tbl:compliance} shows, IRV is the Condorcet Winner of this meta-decision
since it would win a pairwise decision against all other nine methods.\footnote{%
    Which method would win if the meta-decision was made using a proportional method,
    we can only speculate here since our current results do not provide us a way to predict this.}

However, in an actual decision about a future voting method, 
people would however probably also have non-consequentialist preference components
related to fairness, consistency, and other criteria such as those discussed in section \ref{sec:properties}.

\subsubsection{Further conclusions}

Our experiments indicate that assessments via agent-based simulations involving individual preferences
will usually depend very much on the particular assumptions about voters' preference distributions,
whether from a spatial or other model (or ``culture''),
and on the particular functional forms (e.g., linear, quadratic, or Gaussian) 
and parameter values used in these models.

They also seem to suggest that using an interactive phase only rarely has any considerable effect on the most important metrics, 
with the decrease in randomization under MaxParC being an exception.
This might however be due to our very restricted assumptions on what agents can do during the interactive phase.
Future work, whether empirical or numerical, should therefore consider the possibilities
of information gathering, communication and other forms of social dynamics during the interactive phase.

To this end, we are currently developing a social app 
that offers an interactive version of MaxParC for making everyday group decisions,
which we plan to use in empirical studies to assess the real-world performance 
and social dynamics of nondeterministic proportional consensus decision making methods.

\subsection{Derivation of heuristics and strategies}

\subsubsection{Sincere voting under MaxParC}
\label{sec:mpsincere}

Since the MaxParC ballot explicitly asks for a quantity (the level ``willingness'' to approve of an option)
whose meaning ultimately depends on ones beliefs about the other voters' preferences,
voting under MaxParC always incorporates some form of ``strategic'' thinking in some sense,
so it is in a way pointless to ask what ``the'' sincere way of filling in a MaxParC ballot is.
Rather, one may apply any of a number of different heuristics that all lead to a sincere ballot in the sense that
more-preferred options are assigned higher willingness values.

\paragraph{Conservative satisficing heuristic.}

This sincere voting heuristic is based on the idea to assign to any option $y$ a willingness $b(y)$
just small enough to guarantee that
if I end up approving of $y$ and all who don't approve of $y$ approve of their favourite only,
the resulting lottery will not be worse
than the benchmark lottery $\ell$ that would result if all approve of their favourite only.

To find this willingness value $b(y)$, a EU-type voter would proceed as follows.
Assume $\ell(x)>0$ for all $x\in C$ (otherwise ignore those $x$ for which $\ell(x)=0$ in the following).
Sort the options into an ordering $x_1,x_2,\ldots$ by descending utility, so that $u_i(x_1) > u_i(x_2) > \ldots > u_i(x_k)$
and $x_1=f_i$ is the favourite of $i$.
For all $a=0\ldots k$, let $F_a = \sum_{c=a+1}^k \ell(x_c)$ and $U_a = \sum_{c=a+1}^k \ell(x_c) u_i(x_c)$,
noting that $1 - \ell(f_i) = F_1 > \ldots > F_k = 0$, $U_0 = u_i(\ell)$ and $U_k = 0$.
For each $\beta\in[0,1-\ell(f_i)]$, let $a(\beta)$ the smallest $a$ with $F_a\le\beta$,
noting that $a(0) = k$ and $a(1-\ell(f_i)) = 1$.
For each $y\in C\setminus\{f_i\}$ with $u_i(y)\ge u_i(\ell)$, let
\begin{equation}
    V_y(\beta) = (1 - \beta) u_i(y) + U_{a(\beta)} + (\beta - F_{a(\beta)})u_i(x_{a(\beta)}),
\end{equation}
noting that $V_y(0) = u_i(y) \ge u_i(\ell)$ and $V_y(1-\ell(f_i)) = u_i(\ell) - \ell(f_i)(u_i(f_i)-u_i(y)) < u_i(\ell)$.
$V_y(\beta)$ is the evaluation of the lottery that results
if the $\beta N$ voters whose favourites I like least (this is the ``conservative'' aspect of the heuristic)
approve of their favourite only while the rest (including me) approve also of $y$
so that $y$ will win with probability $1-\beta$
while the rest of the winning probability goes to options I rather don't like.
The ``satisficing'' aspect of the heuristic is to be satisfied if this evaluation is not worse than that of the benchmark lottery.
Hence one lets $b(y) = 100\beta$ for the largest $\beta$ with $V_y(\beta) \ge u_i(\ell)$,
so that $0 \le b(y) < 100(1-\ell(f_i))$,
and completes the ballot by putting $b(f_i) = 100$
and $b(y) = 0$ for all $y\in C$ with $u_i(y) < u_i(\ell)$.

LCP-type voters may use the same formula based on the probability weights $w_j$ instead of
the actual probabilities $\ell(x_j)$.

Also for HCP-type voters, one can easily derive a similar formula.

Note that this heuristic indeed produces a sincere ballot since $u_i(y')>u_i(y)$ will imply $b(y)>b(y')$.
If one has to expect that a fraction $\alpha\in[0,1]$ of all voters is lazy,
one would adjust $b(y)$ to $b(y) = 100(\alpha + (1-\alpha)\beta)$ if $\beta > 0$.

\paragraph{Informed satisficing heuristic.}

If more information about the other voters' preferences is available or can be estimated,
one may rather want to apply this heuristic in which $i$ assumes that
if a fraction of voters $j$ will eventually approve of some option $y$,
it will be those $j$ with the largest relative utility $\rho_j(y) := \frac{u_j(y) - u_j(\ell)}{u_j(f_j) - u_j(\ell)}$.
Hence let us assume $i$ has beliefs regarding the distribution of $\rho_j(y)$ inside each faction $F_x$
and hence can sort the voters into an ordering $j_1,\dots,j_N$ by descending $\rho_j(y)$,
so that $\rho_{j_1}(y) > \ldots > \rho_{j_N}(y)$ and so that she knows $f_{j_a}$ for all $a$.
If the first $n\le N$ voters in this ordering rather assign their winning probability to $y$ than to $f_{j_a}$,
$i$'s utility becomes
\begin{equation}
    V_y(n) = (n u_i(y) + \sum_{a=n+1}^N u_i(f_{j_a})) / N.
\end{equation}
Now if $i$ is satisfied if this is no smaller than $u_i(\ell)$, she would seek
the smallest $n$ with
$\rho_{j_n}(y)\le \rho_i(y)$ and
$V_y(n') \ge u_i(\ell)$ for all $n' > n$ and put $b(y) = 100(1 - n/N)$,
or, if there is no such $n$, put $b(y) = 0$.
Note that also this heuristic has $b(y)>0$ iff $u_i(y)\ge u_i(\ell)$ (since then $n<N$).

However, this heuristic may produce insincere ballots in which $b(y') < b(y)$ despite $u_i(y') > u_i(y)$
since the voter ordering used for $b(y')$ may be completely different than the one used for $b(y)$.
Still, one can argue that in many situations,
the ballot will be approximately sincere.
This is because often
(i) $\rho_{j_n}(y)\le \rho_i(y)$ will imply $V_y(n') \ge u_i(\ell)$ for all $n' > n$,
and hence $b(y) \approx 100(1 - |\{j:\rho_j(y)\ge \rho_i(y)\}|)$,
and (ii) the distribution of $\rho_j(y)$ in the electorate will be similar for all relevant options $y$,
so that $b(y)$ is approximately monotonic in $\rho_i(y)$.
In a spatial model with concave utilities $u_i(y) = f(||\eta_i-\xi_y||)$ (such as the LH and QH models)
and smoothly and widely distributed voter and option positions,
$i$ will indeed prefer $y$ to a lottery of favourites of those voters $j$ with $\rho_j(y)\ge \rho$ for any $\rho$
since those voters are distributed approximately uniformly and symmetrically around $\eta_y$,
so the average distance from $\eta_i$ to their favourites is at least $||\eta_i-\xi_y||$,
translating into an expected utility from the lottery that is below $u_i(y)$ since $f$ is concave.
More particularly, both in the $1$-dimensional $LH$ and the 2-dimensional $QH$ model,
the number of voters $j$ with $\rho_j(y)\ge \rho_i(y)$ scales roughly linearly
with $1 - \rho_i(y)$, hence $b(y)$ will scale roughly linearly with $\rho_i(y)$,
whereas in a higher-dimensional model, $b(y)$ will become a concave function of $\rho_i(y)$.

Our next two heuristics mimic this linear or concave behaviour to some extent with much simpler formulae.


\paragraph{Linear heuristic.}

A much simpler heuristic is the one we assume in our simulations,
where $b(y) = 100\left(\alpha + (1-\alpha) \frac{u_i(y) - u_i(\ell)}{\max_{x\in C} u_i(x) - u_i(\ell)}\right)$
for all $y\in C$ with $u_i(y)\ge u_i(\ell)$,
i.e., one assigns a willingness of $0$ to options worse than the benchmark, $100$ to one's favourite,
and interpolates linearly between $100\alpha$ and $100$ based on the options' utilities,
where $\alpha\in[0,1]$ is the expected share of lazy voters in the electorate.

One motivation for this heuristic is that under certain assumptions,
it can be interpreted as an approximation of the conservative satisficing heuristic.
Assume the number of options is large,
their utilities $u_i(y)$ for $i$ are distributed uniformly,
say (without loss of generality) between $0$ and $u_i(f_i) = 1$,
and their benchmark winning probabilities $\ell(y)$ are not correlated to $i$'s evaluations $u_i(y)$.
Then $u_i(\ell)\approx 1/2$,
$F_a$ and $a(\beta)$ decrease approximately linearly in $a$ or $\beta$, respectively,
$U_{a(\beta)}\approx \beta^2/2$,
and $\beta - F_{a(\beta)}$ is small.
Hence $V_y(\beta) \approx (1-\beta)u_i(y) + \beta^2/2$,
which equals $u_i(\ell)$ for $\beta \approx 2 u_i(y) - 1$,
hence $b(y)\approx 100\left(\alpha + (1-\alpha) \frac{u_i(y) - u_i(\ell)}{\max_{x\in C} u_i(x) - u_i(\ell)}\right)$
for all $y$ with $u_i(y)\ge u_i(\ell)$.
The same derivation can be made under the weaker assumption that only
those options $y$ with $u_i(y)\ge u_i(\ell)$ are numerous,
have uniformly distributed $u_i(y)$
and have $\ell(y)$ uncorrelated to $u_i(y)$.
These assumptions are, e.g., approximately fulfilled if $k$ is large and utility follows the $LH$ model.
If, instead, utility depends more concavely on distance, as in the $QH$ model,
the linear heuristic will tend to produce larger willingness values than the conservative satisficing heuristic,
hence will produce more compromise outcomes which however
may sometimes be worse than the benchmark lottery for some voters.
On the contrary, if utility depends more convexly on distance, as in the tails of the $GH$ model,
the linear heuristic will tend to produce smaller willingness values than the conservative satisficing heuristic,
hence may sometimes not produce a partial consensus when there is a potential one.

\paragraph{Hyperbolic heuristic.}

A little less simple is the heuristic that puts
$b(f_i) = 100$,
$b(y) = 0$ for all $y$ with $u_i(y) < u_i(\ell)$, and
$b(y) = 100 (1 - \frac{u_i(\ell) - \min_j u_i(f_j)}{u_i(y) - \min_j u_i(f_j)})$
for all other $y$, which has a hyperbolical rather than a linear dependency on $u_i(y)$,
growing fast for $u_i(y)$ slightly above $u_i(\ell)$ and much slower for $u_i(y)$ approaching $u_i(f_i)$.

Also this formula can be derived as an approximation of the conservative satisficing heuristic,
under different assumptions on the distribution of utility.
Assume that $i$ considers all other options than $f_i$ that occur as favourites of any voter as approximately equally bad,
so that we can assume $u_i(f_i) = 1$ and $u_i(f_j) \approx 0$ for all $j$ with $f_j\neq f_i$.
Then $u_i(\ell)\approx \ell(f_i)$,
$V_y(\beta)\approx (1-\beta)u_i(y)$,
hence $b(y)\approx 100 (1 - \frac{u_i(\ell)}{u_i(y)})
\approx 100 (1 - \frac{u_i(\ell) - \min_j u_i(f_j)}{u_i(y) - \min_j u_i(f_j)})$.
Since these assumptions on utility are even more extremely ``convex'' than in the GA model,
the hyperbolic heuristic may be a better choice in Gaussian utility situations than the linear heuristic.

\subsubsection{Heuristic Nash Lottery strategy}
\label{heurnashlott}

Assume $N \gg 1$
and $C = \{1,\ldots,k\}$,
put $m = k - 1$,
$e = (1,\dots,1) \in \reals^m$,
$p = (\ell_1, \ldots, \ell_m)$,
$v = u_{1k}$,
$w = (u_{11} - v, \ldots, u_{1m} - v)$.
We focus on voter 1's choice of ratings $r_{1*} = \beta_i$
and consider
$s = r_{1k} \ge 0$,
$t = (r_{11} - s, \ldots, r_{1m} - s)$
with $t_x \ge - s$
the control variables,
all vectors being column vectors.
Then the Nash sum (= log of Nash lottery target function) is
\begin{equation}
    f(p|s,t) = g(p|s,t) + h(p)
\end{equation}
with
\begin{equation}
    g(p|s,t) = \log (s + p^\top t),\\
    h(p) = \sum_i \log (r_{ik} + \sum_x p_x r_{ix}),
\end{equation}
where
summation over $i$ means $i = 2\ldots N$ (likewise for $j$)
and
summation over $x$ means $x = 1\ldots m$ (likewise for $y,z$).

Since $N\gg 1$, we can approximate
\begin{equation}
    f(p|s,t) = g(q|s,t) + d^\top G(s,t) + h(q) + d^\top H d / 2,
\end{equation}
where
\begin{eqnarray}
    q &=& \arg\max_p h(p),\\
    d &=& p - q,\\
    G(s,t)_x &=& \partial_{p_x} g(p|s,t) |_{p=q} = \frac{t}{s + q^\top t} = \gamma(s,t) t,\\
    H_{xy} &=& \partial_{p_x} \partial_{p_y} h(p) |_{p=q}
        = \sum_i \partial_{p_x} \frac{r_{iy}}{r_{ik} + \sum_z p_z r_{iz}}
        = - \sum_i \frac{r_{ix} r_{iy}}{(r_{ik} + \sum_z p_z r_{iz})^2}
\end{eqnarray}
with
\begin{eqnarray}
    \gamma(s,t) &=& \frac{1}{s + q^\top t} > 0,\\
    \partial_s \gamma(s,t) &=& - \frac{1}{(s + q^\top t)^2} = - \gamma(s,t)^2 < 0,\\
    \nabla_t \gamma(s,t) &=& - \frac{q}{(s + q^\top t)^2} = - \gamma(s,t)^2 q.
\end{eqnarray}
Assume that $H$ is nonsingular with
\begin{equation}
    I = H^{-1},
\end{equation}
and note that $H,I$ are symmetric and negative semidefinite.
Assume that $q_x > 0$ for all $x$ and $t\neq 0$, so that $s + q^\top t > 0$ and $\gamma(s,t) < \infty$.

The Nash lottery $p^\ast (s,t)$ is that $p\in[0,1]^m$ with $e^\top p\le 1$ which maximizes $f(p|s,t)$.
Assume this is an interior solution (e.g., since there is at least one bullet voter for each option),
then the first-order condition is
\begin{equation}
    0 = \nabla_p f(p|s,t) = \nabla_d (d^\top G(s,t) + d^\top H d / 2)  = G(s,t) + H d,
\end{equation}
hence
\begin{eqnarray}
    d^\ast (s,t) &=& - I G(s,t) = - \gamma(s,t) I t,\\
    \partial_s d^\ast (s,t) &=& \gamma(s,t)^2 I t,\\
    \nabla_t d^\ast (s,t)^\top &=& \gamma(s,t)^2 q t^\top I - \gamma(s,t) I.
\end{eqnarray}
Voter 1's expected utility is then
\begin{equation}
    U(s,t) = v + p^\ast (s,t)^\top w = v + q^\top w + d^\ast (s,t)^\top w = v + q^\top w - \gamma(s,t) t^\top I w.
\end{equation}
If she considers abstaining (which is equivalent to putting $t\equiv 0$ and an arbitrary $s > 0$, w.l.o.g.\ $s = 1$)
and wonders what small change in ratings $\Delta r$ would improve her utility most,
she would calculate
\begin{eqnarray}
    \partial_{r_{1x}} U(s,t)|_{t\equiv 0} &=& - \partial_{t_x} (\gamma(s,t) t^\top I w)|_{t\equiv 0}
        = - (\partial_{t_x} t^\top|_{t\equiv 0}) I w = - (I w)_x,\\
    \partial_{r_{1k}} U(s,t)|_{t\equiv 0} &=& - (\partial_s - \sum_x\partial_{t_x})(\gamma(s,t) t^\top I w)|_{t\equiv 0}
        = \sum_x (I w)_x.
\end{eqnarray}

Not knowing $H$ and hence $I$,
voter 1 might use the following heuristic to estimate an approximate $H$ from the latest favourite polling data,
simply assuming every other voter $j>1$ is lazy and submits a bullet vote $r_{jx} = 1$ for their favourite option $x$,
putting all others to $r_{jy} = 0$.
In that case, assuming $f^p(x) > 0$ for all $x$,
\begin{eqnarray}
    H_{xy} &\approx& N^2\left(1/f^p(k) + \delta_{xy}/f^p(x)\right)
\end{eqnarray}
where $\delta_{xy}=1$ iff $x=y$, else $\delta_{xy}=0$.
Hence $H$ is a matrix filled with equal positive entries plus some positive diagonal.
Its inverse then has
\begin{eqnarray}
    I_{xx} &\approx& \zeta f^p(x) \left(N - f^p(x)\right) \\
    I_{xy} &\approx& - \zeta f^p(x) f^p(y)
\end{eqnarray}
for some $\zeta > 0$ and all $x\neq y$.
This would imply that voter 1's utility grows fastest in the direction $\Delta r_1$ with
\begin{eqnarray}
    \Delta r_{1x} &=& - (I w)_x \approx \zeta f^p(x) \left( N w_x - \sum_y f^p(y) w_y \right)
        = \zeta N f^p(x) ( u_{1x} - \upsilon ),\\
    \Delta r_{1k} &=& \sum_x (I w)_x \approx \zeta N \sum_x f^p(x) \left( \upsilon - u_{1x} \right)
        = \zeta N f^p(k) ( u_{1k} - \upsilon ),
\end{eqnarray}
where $\upsilon = \sum_{x=1}^k u_{1x} f^p(x) / N$ is voter 1's expected utility of the benchmark lottery
based on the latest favourite polling data.
A natural heuristic is then that voter 1 moves her ratings from $r_{1x}\equiv r_{1k}=1$
as much in the above direction as is possible without any rating getting negative, i.e.,
putting
\begin{equation}
    r_{1x} = 1 + \rho f^p(x) ( u_{1x} - \upsilon ) \ge 0
\end{equation}
for all $x = 1\ldots k$, where
\begin{eqnarray}
    \rho &=& 1 / f^p(y) ( \upsilon - u_{1y} ) > 0,\\
    y &=& \arg\min_{x=1}^k f^p(x) ( u_{1x} - \upsilon ),
\end{eqnarray}
so that $r_{1y} = 0$.
In the special case where $f^p(x)\equiv N / k$ (e.g.\ before the first poll),
we get $y = \arg\min_{x=1}^k u_{1x}$, $\rho = k / N ( \upsilon - u_{1y} )$, and
$r_{1x} = \frac{u_{1x} - \min_{z=1}^k u_{1z}}{\upsilon - \min_{z=1}^k u_{1z}}
\propto u_{1x} - \min_{z=1}^k u_{1z}$,
i.e., voter 1 would then vote sincerely.
If, however, some options appear to have much higher chances than others,
she would exaggerate her stated preferences regarding those options that appear to have higher chances (high $f^p(x)$)
while playing down her stated preferences regarding those options that appear to have lower chances,
which can result in rating some promising well-liked compromise option higher than her favourite if the latter has low chances,
or rating some lurking less-liked compromise option lower than a very improbable least-liked option.

\subsubsection{Factional unanimous best response in IRV}
\label{factionalirv}

We show that w.l.o.g., one can restrict the analysis on the described set $A$ of ballots.
First, assume some ballot ranks some option $y$ which however gets eliminated before all higher-ranked options are eliminated.
Then submitting a shorter ballot with $y$ left out instead leads to the exact same tally process.
Second, assume $y$ is ranked but gets eliminated at the same point as when submitting the shorter ballot with $y$ left out.
Then submitting the shorter ballot also leads to the exact same tally process.
Hence we can restrict our focus on ballots ranking only options that survive the elimination process strictly longer
than when not ranked, and don't get eliminated before any higher-ranked option.
For any ballot $b=(x_1,x_2,\dots,x_\ell)\in A$,
let $a(b)\in\{0,1,\dots,k-1\}$ be the number of options eliminated strictly before $x_\ell$ when submitting $b$,
and assume that also $b'=(x_1,x_2,\dots,x_\ell,y)\in A$.
Note that if submitting $b$, $y$ is eliminated after at least $a(b) + 1$ many options, but is not the winner (otherwise $b'\notin A$),
hence there are at most $k - a(b) - 2$ many different $y\in C$ such that $(x_1,x_2,\dots,x_\ell,y)\in A$.
Thus the number $a'(b,y)$ of options eliminated strictly before $y$ when submitting $b$ (not $b'$!)
is one of the numbers in $\{a(b) + 1,\ldots,k - 2\}$ and is different for all $y$ for which $(x_1,x_2,\dots,x_\ell,y)\in A$.
If submitting $b'$, $y$ must survive longer than when submitting $b$,
hence
\begin{equation}
    a(b') > a'(b,y) \ge a(b) + 1 > a'((x_1,x_2,\dots,x_{\ell-1}),x_\ell) + 1,
\end{equation}
i.e., $a(b') \ge a(b) + 2$.
This implies that any ballot $b=(x_1,x_2,\dots,x_\ell)\in A$ can be uniquely encoded via a
sequence of integers $(a'(\emptyset,x_1)$, $a'((x_1),x_2)$, $a'((x_1,x_2),x_3)$, \dots, $a'((x_1,x_2,\dots,x_{\ell-1}),x_\ell))$
that fulfils
\begin{equation}
    a'((x_1,\dots),x_i) + 1 < a'((x_1,\dots),x_{i+1}
\end{equation}
for all $i$.
There are less than $2^{k}$ such sequences in $0,\ldots,k-1$, hence $|A| \le 2^{k}$.

\bibliographystyle{Science} 
\bibliography{library}

\begin{table}\begin{center}\begin{tabular}{lllllllll}
     & \multicolumn{3}{l}{\bf behavioural types scenario}\\
     \bf type & \bf lazy & \bf middle & \bf strategic & \bf all-L & \bf all-S & \bf all-T & \bf all-H & \bf all-F \\\hline
     L (lazy)            & 1/3 & 1/6 & 1/20 & 1 & 0 & 0 & 0 & 0 \\
     S (sincere)         & 1/3 & 1/6 & 1/20 & 0 & 1 & 0 & 0 & 0 \\
     T (trial-and-error) & 1/9 & 1/6 & 1/5  & 0 & 0 & 1 & 0 & 0 \\
     H (heuristic)       & 1/9 & 1/6 & 1/5  & 0 & 0 & 0 & 1 & 0 \\
     F (factional)       & 1/9 & 1/3 & 1/2  & 0 & 0 & 0 & 0 & 1 \\\hline
\end{tabular}\caption{\label{tbl:bt}Distribution of behavioural types in different behavioural type scenarios}\end{center}\end{table}

\begin{table}\tiny\begin{center}\begin{tabular}{lcccccccccc}\toprule
                                & \bf PV    & \bf AV    & \bf RV    & \bf IRV   & \bf SC    & \bf RB    & \bf FC    & \bf RFC   & \bf NL    & \bf MPC \\\hline
anonymous                       & yes       & yes       & yes       & yes       & yes       & yes       & yes       & yes       & yes       & yes  \\  
neutral                         & yes       & yes       & yes       & yes       & yes       & yes       & yes       & yes       & yes       & yes  \\  
Pareto-efficient 
  w.r.t.\ stated preferences    & yes       & yes       & yes       & yes       & yes       & yes       & no        & no        & yes       & yes  \\
strongly mono-raise monotonic   & yes       & yes       & yes       & no        & yes       & yes       & no        & no        & no        & yes  \\
weakly mono-raise monotonic     & yes       & yes       & yes       & no        & yes       & yes       & yes       & no        & yes ?     & yes  \\
weakly mono-raise-abstention 
  mon.                          & yes       & yes       & yes       & yes       & yes       & yes       & yes       & yes       & yes       & yes  \\
independent from Pareto-dominated
  alternatives                  & partial   & full      & full      & full      & full      & partial   & no        & no        & full      & full \\   
independent from losing options & no        & full      & full      & no        & no        & partial   & no        & no        & full      & full \\   
independent from exact clones   & no        & yes       & yes       & yes       & yes       & yes       & no        & yes       & yes       & yes  \\  
stronger forms of 
  clone-proofness               & no        & yes       & yes       & no        & yes       & yes       & no        & ?         & ?         & ? \\  
strategy-freeness               & no        & no        & no        & no        & no        & yes       & no        & no        & no        & no \\  
reveals preferences             & no        & some      & some      & no        & no        & fav.      & fav.      & utility   & fav.      & fav. \\
allocates power proportionally  & no        & no        & no        & no        & no        & yes       & yes       & yes       & yes       & yes  \\  
supports full consensus
  with sincere voters           & no        & yes       & yes       & no        & no        & no        & yes       & yes       & yes       & yes  \\
supports full consensus
  with strategic voters         & no        & no        & no        & no        & no        & no        & yes       & yes       & yes       & yes  \\
supports partial consensus
  with strategic voters         & no        & no        & no        & no        & no        & no        & no        & no        & yes       & yes  \\
\midrule
                                    &\bf PV &\bf AV &\bf RV &\bf IRV&\bf SC &\bf RB &\bf FC &\bf RFC&\bf NL &\bf MPC \\\hline
\verb!moverate!                     & 0.15  & 0.178 & 0.193 & 0.0257& 0.0768& 0     & 0.00145&0.013 & 0.488 & 0.327\\
\verb!keeprate!                     & 0.289 & 0.315 & 0.248 & 0.293 & 0.356 & 0     & 0.208 & 0.211 & 0.198 & 0.282\\
\verb!interactivechanged!           & 0.0697& 0.192 & 0.113 & 0.0934& 0.307 & 0     & 0.403 & 0.423 & 0.62  & 0.482\\
\hline
\verb!Eshannon_initial!             & 0.0344& 0.0703& 0.0173& 0     & 0.0795& 1.32  & 1.25  & 1.23  & 1.06  & 0.914\\
\verb!Eshannon_final!               & 0.0211& 0.0434& 0.0119& 0     & 0.0515& 1.32  & 1.28  & 1.26  & 1.05  & 0.809\\
\verb!Erenyi2_initial!              & 0.0344& 0.0703& 0.0173& 0     & 0.0795& 1.21  & 1.14  & 1.12  & 0.958 & 0.792\\
\verb!Erenyi2_final!                & 0.0211& 0.0434& 0.0119& 0     & 0.0515& 1.21  & 1.18  & 1.15  & 0.961 & 0.689\\
\verb!maxprob_initial!              & 0.976 & 0.954 & 0.988 & 1     & 0.949 & 0.423 & 0.456 & 0.466 & 0.539 & 0.602\\
\verb!maxprob_final!                & 0.985 & 0.971 & 0.992 & 1     & 0.966 & 0.423 & 0.448 & 0.46  & 0.532 & 0.649\\
\hline
\verb!pcompromise_initial!          & 0.0883& 0.308 & 0.236 & 0.171 & 0.309 & 0.0833& 0.147 & 0.186 & 0.214 & 0.257\\
\verb!pcompromise_final!            & 0.0921& 0.27  & 0.234 & 0.192 & 0.431 & 0.0833& 0.156 & 0.193 & 0.221 & 0.278\\
\hline
\verb!Wutil_initial!                & -2.3  & -2.36 & -2.27 & -2.25 & -4.06 & -4.17 & -3.8  & -3.7  & -4.04 & -2.97\\
\verb!Wutil_final!                  & -2.29 & -2.28 & -2.26 & -2.24 & -3.22 & -4.17 & -3.78 & -3.62 & -3.58 & -2.93\\
\verb!Wgini_initial!                & -3.18 & -3.26 & -3.13 & -3.09 & -5.76 & -5.86 & -5.3  & -5.15 & -5.66 & -4.07\\
\verb!Wgini_final!                  & -3.17 & -3.14 & -3.11 & -3.09 & -4.51 & -5.86 & -5.27 & -5.03 & -4.98 & -4.02\\
\verb!Wegal_initial!                & -8.43 & -8.77 & -8.32 & -8.15 & -16.8 & -18.6 & -16.5 & -15.9 & -17.1 & -11.8\\
\verb!Wegal_final!                  & -8.4  & -8.33 & -8.24 & -8.13 & -12.7 & -18.6 & -16.3 & -15.4 & -14.8 & -11.6\\
\hline
\verb!relWutil_initial!             & 0.813 & 0.857 & 0.87  & 0.857 & 0.681 & 0.637 & 0.653 & 0.657 & 0.694 & 0.723\\
\verb!relWutil_final!               & 0.813 & 0.849 & 0.861 & 0.851 & 0.747 & 0.637 & 0.651 & 0.657 & 0.69  & 0.729\\
\verb!relWgini_initial!             & 0.857 & 0.968 & 0.931 & 0.896 & 0.765 & 2.1e+20 & 2.1e+20 & 2.1e+20 & 2.1e+20 & 2.1e+20\\
\verb!relWgini_final!               & 0.851 & 0.947 & 0.934 & 0.908 & 0.885 & 2.1e+20 & 2.1e+20 & 2.1e+20 & 2.1e+20 & 2.1e+20\\
\verb!relWegal_initial!             & 1.93e+20 & 3.6e+26 & 1.91e+20 & 1.91e+20 & 1.91e+20 & 2.2e+67 & 2.2e+67 & 2.2e+67 & 2.2e+67 & 2.21e+67\\
\verb!relWegal_final!               & 1.92e+20 & 1.35e+59 & 1.91e+20 & 1.91e+20 & 1.31e+21 & 2.2e+67 & 2.2e+67 & 2.2e+67 & 2.2e+67 & 2.21e+67\\
\hline
\verb!alt_relWutil_initial!         & 0.85  & 0.886 & 0.893 & 0.884 & 0.739 & 0.754 & 0.764 & 0.767 & 0.792 & 0.813\\
\verb!alt_relWutil_final!           & 0.849 & 0.879 & 0.886 & 0.879 & 0.796 & 0.754 & 0.762 & 0.767 & 0.788 & 0.814\\
\verb!alt_relWgini_initial!         & 0.881 & 0.948 & 0.937 & 0.915 & 0.782 & 0.998 & 1.01  & 1.01  & 1.02  & 1.02\\
\verb!alt_relWgini_final!           & 0.877 & 0.934 & 0.933 & 0.917 & 0.861 & 0.998 & 1     & 1.01  & 1.01  & 1.01\\
\verb!alt_relWegal_initial!         & 0.701 & 0.781 & 0.749 & 0.717 & 0.68  & 1.07  & 1.08  & 1.08  & 1.08  & 1.07\\
\verb!alt_relWegal_final!           & 0.694 & 0.764 & 0.75  & 0.726 & 0.744 & 1.07  & 1.08  & 1.08  & 1.07  & 1.04\\
\hline
\verb!avgsatisfaction_initial_F!    & 0.664 & 0.684 & 0.684 & 0.68  & 0.616 & 0.564 & 0.568 & 0.57  & 0.588 & 0.614\\
\verb!avgsatisfaction_final_F!      & 0.665 & 0.68  & 0.682 & 0.678 & 0.649 & 0.564 & 0.569 & 0.575 & 0.604 & 0.618\\
\verb!avgsatisfaction_initial_H!    & 0.664 & 0.684 & 0.684 & 0.679 & 0.615 & 0.564 & 0.568 & 0.57  & 0.588 & 0.614\\
\verb!avgsatisfaction_final_H!      & 0.664 & 0.68  & 0.681 & 0.677 & 0.639 & 0.564 & 0.566 & 0.57  & 0.579 & 0.618\\
\verb!avgsatisfaction_initial_L!    & 0.657 & 0.672 & 0.672 & 0.667 & 0.581 & 0.564 & 0.564 & 0.565 & 0.6   & 0.597\\
\verb!avgsatisfaction_final_L!      & 0.657 & 0.669 & 0.67  & 0.666 & 0.606 & 0.564 & 0.563 & 0.564 & 0.589 & 0.601\\
\verb!avgsatisfaction_initial_S!    & 0.657 & 0.683 & 0.684 & 0.677 & 0.577 & 0.564 & 0.574 & 0.574 & 0.617 & 0.608\\
\verb!avgsatisfaction_final_S!      & 0.658 & 0.68  & 0.681 & 0.676 & 0.602 & 0.564 & 0.572 & 0.574 & 0.608 & 0.613\\
\verb!avgsatisfaction_initial_T!    & 0.658 & 0.68  & 0.684 & 0.678 & 0.578 & 0.564 & 0.573 & 0.574 & 0.619 & 0.609\\
\verb!avgsatisfaction_final_T!      & 0.661 & 0.676 & 0.682 & 0.677 & 0.65  & 0.564 & 0.567 & 0.569 & 0.594 & 0.616\\
\hline
\verb!pctprefer_PV_over!            & ---   & 18    & 15.4  & 13.3  & 34.3  & 65.4  & 63.4  & 63.1  & 61.2  & 58.7\\
\verb!pctprefer_AV_over!            & 19.2  & ---   & 9.47  & 12.5  & 33.5  & 69.2  & 65.2  & 65.2  & 62.8  & 59.1\\
\verb!pctprefer_RV_over!            & 17.3  & 10.4  & ---   & 10.8  & 33.3  & 69.1  & 65.3  & 65.2  & 63.2  & 59.4\\
\verb!pctprefer_IRV_over!           & 15.8  & 14.1  & 11.6  & ---   & 32.8  & 68    & 65.1  & 64.9  & 63.1  & 60.1\\
\verb!pctprefer_SC_over!            & 26.8  & 24.2  & 23.5  & 22.8  & ---   & 61.7  & 59.3  & 58.9  & 57    & 53.5\\
\verb!pctprefer_RB_over!            & 34.4  & 30.6  & 30.7  & 31.9  & 38.1  & ---   & 20.6  & 20.4  & 37.4  & 26.9\\
\verb!pctprefer_FC_over!            & 35.1  & 31.1  & 31.1  & 32.4  & 38.8  & 24.3  & ---   & 21.4  & 39.2  & 27.8\\
\verb!pctprefer_RFC_over!           & 35.5  & 31.4  & 31.4  & 32.7  & 39.2  & 28.5  & 25.8  & ---   & 40    & 28.9\\
\verb!pctprefer_NL_over!            & 36.9  & 34    & 33.4  & 34.2  & 41    & 62.4  & 59.5  & 58.7  & ---   & 42.3\\
\verb!pctprefer_MPC_over!           & 38.4  & 34.1  & 34.4  & 35.5  & 42.4  & 63.6  & 59.5  & 58.7  & 56.4  & ---\\
\hline
& \bf PV    & \bf AV    & \bf RV    & \bf IRV   & \bf SC    & \bf RB    & \bf FC    & \bf RFC   & \bf NL    & \bf MPC \\
\bottomrule\end{tabular}\caption{\label{tbl:compliance}
Level of compliance with voting method consistency criteria, 
and average performance metrics from agent-based simulations. 
}\end{center}\end{table}

\begin{table}\footnotesize\centering
\begin{verbatim}
                            OLS Regression Results                            
==============================================================================
Dep. Variable:            Wgini_final   R-squared:                       0.661
Model:                            OLS   Adj. R-squared:                  0.661
Method:                 Least Squares   F-statistic:                 2.663e+05
Date:                Sat, 04 Apr 2020   Prob (F-statistic):               0.00
Time:                        15:34:58   Log-Likelihood:             1.0244e+07
No. Observations:             5124153   AIC:                        -2.049e+07
Df Residuals:                 5124130   BIC:                        -2.049e+07
Df Model:                          22                                         
Covariance Type:                  HC1                                         
===================================================================================
                      coef    std err          z      P>|z|      [0.025      0.975]
-----------------------------------------------------------------------------------
Intercept           0.1581      0.000   1223.248      0.000       0.158       0.158
PV                 -0.0018   6.48e-05    -27.023      0.000      -0.002      -0.002
AV               5.993e-05   6.55e-05      0.916      0.360   -6.84e-05       0.000
IRV                -0.0002   6.53e-05     -3.193      0.001      -0.000   -8.05e-05
SC                 -0.0030   6.58e-05    -44.931      0.000      -0.003      -0.003
RB                  0.0019   6.47e-05     29.673      0.000       0.002       0.002
FC                  0.0018   6.47e-05     28.243      0.000       0.002       0.002
RFC                 0.0018   6.47e-05     28.405      0.000       0.002       0.002
NL                  0.0017    6.5e-05     26.489      0.000       0.002       0.002
MPC                 0.0014   6.52e-05     22.143      0.000       0.001       0.002
log(nvoters)        0.0003   1.06e-05     25.800      0.000       0.000       0.000
log(noptions)       0.0046   4.01e-05    115.265      0.000       0.005       0.005
with_compromise     0.0004    2.9e-05     12.443      0.000       0.000       0.000
rshare_LCP          0.0004   4.05e-05     10.453      0.000       0.000       0.001
rshare_HCP          0.0009   4.08e-05     22.671      0.000       0.001       0.001
log(npolls)      6.738e-05   1.87e-05      3.607      0.000    3.08e-05       0.000
sshare_S            0.0005   5.79e-05      9.036      0.000       0.000       0.001
sshare_T            0.0007   5.69e-05     13.150      0.000       0.001       0.001
sshare_H            0.0012   5.68e-05     20.744      0.000       0.001       0.001
sshare_F            0.0013   5.42e-05     23.297      0.000       0.001       0.001
dim                -0.0465   1.95e-05  -2384.367      0.000      -0.047      -0.046
log(omega)         -0.0447   3.11e-05  -1437.194      0.000      -0.045      -0.045
rho                -0.0018   3.89e-05    -46.644      0.000      -0.002      -0.002
==============================================================================
Omnibus:                   909575.095   Durbin-Watson:                   1.060
Prob(Omnibus):                  0.000   Jarque-Bera (JB):          1549242.812
Skew:                           1.174   Prob(JB):                         0.00
Kurtosis:                       4.319   Cond. No.                         63.3
==============================================================================
\end{verbatim}
\caption{\label{tbl:regr:Wgini_final_GA}%
Generalized linear model for Gini-Sen absolute social welfare in the GA preference model.
}
\end{table}

\clearpage
\pagebreak[4]
\global\pdfpageattr\expandafter{\the\pdfpageattr/Rotate 90}

\end{document}